\documentclass[twocolumn,fleqn]{svjour2}    
\smartqed  
\usepackage{graphicx}
\usepackage{mathptmx}      
\usepackage{latexsym}
\usepackage{amsfonts}
\usepackage{bm}

\newcommand{\la}{\left<}
\newcommand{\ra}{\right>}
\newcommand{\Acal}{\ensuremath{{\cal A}}}
\newcommand{\Bcal}{\ensuremath{{\cal B}}}
\newcommand{\Ccal}{\ensuremath{{\cal C}}}
\newcommand{\Fcal}{\ensuremath{{\cal F}}}
\newcommand{\Lcal}{\ensuremath{{\cal L}}}
\newcommand{\Ocal}{\ensuremath{{\cal O}}}
\newcommand{\ehat}{\ensuremath{\hat{e}}}
\newcommand{\fhat}{\ensuremath{\hat{f}}}
\newcommand{\Fhat}{\ensuremath{\hat{F}}}
\newcommand{\Ghat}{\ensuremath{\hat{G}}}
\newcommand{\Shat}{\ensuremath{\hat{S}}}
\newcommand{\Gtilde}{\ensuremath{\tilde{G}}}

\newcommand{\kvec}{\ensuremath{\underline{k}}}
\newcommand{\lvec}{\ensuremath{\underline{l}}}
\newcommand{\qvec}{\ensuremath{\underline{q}}}
\newcommand{\rvec}{\ensuremath{\underline{r}}}
\newcommand{\uvec}{\ensuremath{\underline{u}}}
\newcommand{\vvec}{\ensuremath{\underline{v}}}
\newcommand{\deltavec}{\ensuremath{\underline{\delta}}}
\newcommand{\Fscal}{\ensuremath{\tilde{f}}}
\newcommand{\Fscalhat}{\ensuremath{\hat{f}}}
\newcommand{\Sscal}{\ensuremath{\tilde{s}}}
\newcommand{\Sscalhat}{\ensuremath{\hat{s}}}
\newcommand{\lonevec}{\ensuremath{\underline{l}_1}}
\newcommand{\ltwovec}{\ensuremath{\underline{l}_2}}
\newcommand{\ronevec}{\ensuremath{\underline{r}_1}}
\newcommand{\rtwovec}{\ensuremath{\underline{r}_2}}

\newcommand{\qonevec}{\ensuremath{\underline{q}_1}}
\newcommand{\qtwovec}{\ensuremath{\underline{q}_2}}
\newcommand{\kcut}{\ensuremath{k_c}}

\newcommand{\nmon}{\ensuremath{n_\mathrm{mon}}}
\newcommand{\Poper}{{\bf P}}
\newcommand{\fDB}{f_{\mathrm{DB}}}
\newcommand{\overlap}{\ensuremath{\varepsilon}}
\newcommand{\overlapghost}{\ensuremath{\varepsilon_\mathrm{sg}}}
\newcommand{\Nov}{\ensuremath{N_\mathrm{ov}}}
\newcommand{\Novghost}{\ensuremath{N_\mathrm{sg}}}

\newcommand{\kB}{\ensuremath{k_\mathrm{B}}}
\newcommand{\kBT}{\ensuremath{k_\mathrm{B}T}}
\newcommand{\kappaT}{\ensuremath{\kappa_\mathrm{T}}}
\newcommand{\gT}{\ensuremath{g}}
\newcommand{\gN}{\ensuremath{g_\mathrm{N}}}
\newcommand{\gex}{\ensuremath{g_\mathrm{ex}}}

\newcommand{\eself}{\ensuremath{e_\mathrm{self}}}

\newcommand{\cV}{\ensuremath{c_\mathrm{V}}}

\newcommand{\ds}{\ensuremath{d_\mathrm{s}}}
\newcommand{\thetatwo}{\ensuremath{\theta_2}}
\newcommand{\cinf}{\ensuremath{c_{\infty}}}
\newcommand{\lpersist}{\ensuremath{l_\mathrm{p}}}
\newcommand{\spersist}{\ensuremath{s_\mathrm{p}}}
\newcommand{\rhotagged}{\ensuremath{\rho_\mathrm{l}}}
\newcommand{\rhoN}{\ensuremath{\rho_\mathrm{N}}}
\newcommand{\rhostarN}{\ensuremath{\rho^*_\mathrm{N}}}
\newcommand{\rhostars}{\ensuremath{\rho^*_\mathrm{s}}}
\newcommand{\Ustar}{\ensuremath{u^*}}
\newcommand{\UstarN}{\ensuremath{u^*_\mathrm{N}}}
\newcommand{\Ustars}{\ensuremath{u^*_\mathrm{s}}}
\newcommand{\veff}{{\rm v}}
\newcommand{\vpot}{\tilde{\rm v}}
\newcommand{\vvir}{{\rm v}_\mathrm{2}}
\newcommand{\Gi}{\ensuremath{G_\mathrm{z}}}
\newcommand{\Giref}{\ensuremath{G_\mathrm{z0}}}
\newcommand{\Giite}{\ensuremath{G_\mathrm{i}}}
\newcommand{\Ntest}{\ensuremath{N_\mathrm{t}}}
\newcommand{\rN}{\ensuremath{\rvec_\mathrm{N}}}
\newcommand{\rs}{\ensuremath{\rvec_\mathrm{s}}}
\newcommand{\RN}{\ensuremath{R_\mathrm{N}}}
\newcommand{\Rs}{\ensuremath{R_\mathrm{s}}}

\newcommand{\Pone}{\ensuremath{P_\mathrm{1}}}
\newcommand{\Ponehat}{\ensuremath{\hat{P}_\mathrm{1}}}
\newcommand{\Poneasym}{\ensuremath{P_\mathrm{\infty}}}
\newcommand{\Ponecyc}{\ensuremath{P_\mathrm{r}}}

\newcommand{\Ptwo}{\ensuremath{P_\mathrm{2}}}

\newcommand{\Rendz}{\ensuremath{R_\mathrm{e,z}}}
\newcommand{\Rgyr}{\ensuremath{R_\mathrm{g}}}
\newcommand{\Rgyrz}{\ensuremath{R_\mathrm{g,z}}}
\newcommand{\aref}{\ensuremath{a_0}}
\newcommand{\bref}{\ensuremath{b_0}}
\newcommand{\dref}{\ensuremath{e}}

\newcommand{\bratio}{\ensuremath{b_\mathrm{N}}}
\newcommand{\ce}{\ensuremath{c_\mathrm{s}}}
\newcommand{\cP}{\ensuremath{c_P}}

\newcommand{\fDebye}{\ensuremath{f_{\rm D}}}
\newcommand{\Kp}{\ensuremath{K_p}}
\newcommand{\alphap}{\ensuremath{\alpha_p}}

\newcommand{\Iinner}{\ensuremath{I_\mathrm{i}}}
\newcommand{\Iouter}{\ensuremath{I_\mathrm{o}}}
\newcommand{\Iplus}{\ensuremath{I_{+}}}
\newcommand{\Iminus}{\ensuremath{I_{-}}}

\newcommand{\rhotwo}{\ensuremath{\rho_2}}

\newcommand{\gstarup}{\mbox{$g^{\star}$}}
\newcommand{\gstardo}{\mbox{$g_{\star}$}}

\newcommand{\freqB}{\ensuremath{\omega_B}}
\newcommand{\ks}{\ensuremath{k_s}}
\newcommand{\kr}{\ensuremath{k_r}}
\newcommand{\muchainN}{\ensuremath{\mu_\mathrm{N}}}
\newcommand{\dmuchainN}{\ensuremath{\delta\mu_\mathrm{N}}}
\newcommand{\muchainn}{\ensuremath{\mu_\mathrm{n}}}
\newcommand{\dmuchainn}{\ensuremath{\delta\mu_\mathrm{n}}}
\newcommand{\cep}{\ensuremath{c_\mathrm{\mu}}}
\newcommand{\wep}{\ensuremath{w_p}}
\newcommand{\Nav}{\ensuremath{\la N \ra}}
\newcommand{\Npav}{\ensuremath{\la N^p \ra}}
\newcommand{\nonexp}{\ensuremath{L_p}}
\newcommand{\pchain}{\ensuremath{p_\mathrm{N}}}
\newcommand{\pchaintwo}{\ensuremath{p_\mathrm{2N}}}
\newcommand{\rhochain}{\ensuremath{\rho_N}}
\newcommand{\upot}{\ensuremath{u_n}}
\newcommand{\utau}{\ensuremath{u_t}}
\newcommand{\rstar}{\ensuremath{r_*}}
\newcommand{\ctwo}{\ensuremath{c_{\star}}}
\newcommand{\Icyc}{\ensuremath{I_\mathrm{r}}}
\newcommand{\Ione}{\ensuremath{I_\mathrm{l}}}
\newcommand{\Itwo}{\ensuremath{I_{\star}}}

\newcommand{\MSDall}{\ensuremath{h}}
\newcommand{\MSDmon}{\ensuremath{h}}
\newcommand{\MSDcmN}{\ensuremath{h_\mathrm{N}}}
\newcommand{\MSDcms}{\ensuremath{h_\mathrm{s}}}
\newcommand{\Dcurv}{\ensuremath{D_\mathrm{c}}}

\newcommand{\DN}{\ensuremath{D_\mathrm{N}}}
\newcommand{\Ds}{\ensuremath{D_\mathrm{s}}}
\newcommand{\Ts}{\ensuremath{T_\mathrm{s}}}
\newcommand{\TN}{\ensuremath{T_\mathrm{N}}}
\newcommand{\Te}{\ensuremath{T_\mathrm{e}}}
\newcommand{\Ne}{\ensuremath{N_\mathrm{e}}}

\newcommand{\de}{\ensuremath{d_\mathrm{e}}}
\newcommand{\CN}{\ensuremath{C_\mathrm{N}}}
\newcommand{\CNstar}{\ensuremath{C_\mathrm{T_N}}}
\newcommand{\Cs}{\ensuremath{C_\mathrm{s}}}
\newcommand{\Csstar}{\ensuremath{C_\mathrm{T_s}}}
\newcommand{\tstar}{\ensuremath{t^*}}
\newcommand{\xidyn}{\ensuremath{\xi_\mathrm{d}}}
\newcommand{\ftot}{\ensuremath{f_\mathrm{t}}}
\newcommand{\ftotvec}{\ensuremath{\underline{f}_\mathrm{t}}}
\newcommand{\fext}{\ensuremath{f_\mathrm{e}}}

\newcommand{\fran}{\ensuremath{f_\mathrm{r}}}
\newcommand{\franvec}{\ensuremath{\underline{f}_\mathrm{r}}}
\newcommand{\fcol}{\ensuremath{f_\mathrm{c}}}
\newcommand{\fcolvec}{\ensuremath{\underline{f}_\mathrm{c}}}
\newcommand{\pjump}{\ensuremath{\underline{u}(0)}}
\newcommand{\deltat}{\ensuremath{\delta t}}
\journalname{J Stat Phys}
\begin{document}

\title{Scale-free static and dynamical correlations in melts of monodisperse and Flory-distributed homopolymers}
\subtitle{A review of recent bond-fluctuation model studies}

\titlerunning{Scale-free correlations in dense polymer solutions}   

\author{J.~P. Wittmer  \and
        A.~Cavallo \and H.~Xu \and J.~E.~Zabel \and P.~Poli\'nska \and N.~Schulmann
        \and H.~Meyer \and J.~Farago \and A.~Johner \and S.P.~Obukhov \and J.~Baschnagel 
}

\authorrunning{J.~P. Wittmer {\em et al.}} 

\institute{J.~P. Wittmer \and J.~E.~Zabel \and P.~Poli\'nska \and N.~Schulmann
           \and H.~Meyer \and J.~Farago \and A.~Johner \and J.~Baschnagel \at
              Institut Charles Sadron, Universit\'e de Strasbourg, CNRS,
              23 rue du Loess, 67037 Strasbourg Cedex, France \\
           \and
           A. Cavallo \at
           Dipartimento di Fisica, Universit\`a degli Studi di Salerno, 
           via Ponte don Melillo, I-84084 Fisciano, Italy 
           \and
           H. Xu \at
              LPMD, ICPM, Universit\'e Paul Verlaine Metz, 1 bd Arago, 57078 Metz Cedex 03, France
           \and
           S. Obukhov \at
              Department of Physics, University of Florida, Gainesville FL 32611, USA
}

\date{Received: date / Accepted: date}

\maketitle

\begin{abstract}
It has been assumed untill very recently that all long-range correlations are screened in 
three-di\-men\-sion\-al melts of linear homopolymers on distances beyond the correlation 
length $\xi$ characterizing the decay of the density fluctuations.
Summarizing simulation results obtained by means of a variant of the bond-fluctuation model 
with finite mono\-mer excluded volume interactions and topology violating local and global Monte Carlo moves,
we show that due to an interplay of the chain connectivity and the 
incompressibility constraint, both static and dynamical correlations arise on distances $r \gg \xi$.
These correlations are scale-free and, surprisingly, do not depend explicitly on the compressibility 
of the solution.
Both monodisperse and (essentially) Flory-dis\-tri\-buted equilibrium polymers are considered. 
\end{abstract}

\section{Introduction}
\label{intro}

\subsection{General context}
\label{intro_context}
Solutions and melts of ma\-cro\-mo\-le\-cular polymer chains are disordered con\-dens\-ed-matter 
systems \cite{ChaikinBook} of great complexity and richness of both their physical and chemical properties
\cite{FloryBook,DegennesBook,DoiEdwardsBook,BenoitBook,DescloizBook,SchaferBook,RubinsteinBook}. 
Being of great industrial importance and playing a central role in biology and biophysics 
\cite{RubinsteinBook,OozawaBook}, they represent one relatively well-understood fundamental 
example of the vast realm of so-called ``soft matter" systems \cite{WittenPincusBook} 
comprising also, e.g., colloids \cite{DhontBook}, liquid crystals \cite{ChaikinBook} and 
self-assembled surfactant systems \cite{WittenPincusBook}.\footnote{A soft matter system may 
be defined as a fluid in which large groups of the elementary molecules have been 
permanently or transiently connected together, 
e.g. by reversibly bridging oil droplets in water by telechelic polymers 
or similar systems of autoassociating polymers \cite{ANS95,ANS00,ANS01}, 
so that the {\em permutation freedom} of the liquid state is lost for the time window
probed experimentally \cite{WittenPincusBook}. 
The thermal fluctuations which dominate the liquid state must thus coexist with constraints 
reminiscent of the solid state. Reflecting this solid state characteristics the dynamic 
shear-modulus $\mu(t)$ thus often remains finite up to macroscopic times $t$
and the associated shear viscosity $\eta \sim \int_0^{\infty} {\rm d}t \mu(t)$ may become huge.
Most soft matter systems behave as liquids for very long times,
i.e. $\mu(t) \to 0$ for $t \to \infty$ and $\eta$ remains finite.}
The notion ``polymer" is not limited to the hydrocarbon macromolecules 
of organic polymer chemistry {\em \`a la} H.~Staudinger or W.H.~Carothers
\cite{PolymerPioneersBook,InventingPolymerBook} but refers also to biopolymers such as DNA or 
corn starch \cite{WittenPincusBook,RubinsteinBook} and various self-assembled essentially 
linear chain-like supramolecular structures such as, e.g., actin filaments \cite{OozawaBook} or 
wormlike giant micelles formed by some surfactants \cite{WittenPincusBook,CC90}.  

Obviously, dense polymeric systems are quite complicated, 
but since their large scale properties are dominated by the interactions of many polymers, 
each of these interactions should only have a small (both static and dynamical) effect. 
A sound theoretical starting point is thus to add up these small effects independently and 
to correct then {\em self-consistently} for the deviations due to correlations between the 
interactions \cite{DoiEdwardsBook}.\footnote{As seen in Sec.~\ref{theo}, 
it is, e.g., necessary to ``renormalize" \cite{DegennesBook} the local bond length $l$ to the 
``effective bond length" $b$ \cite{DoiEdwardsBook} or the second virial coefficient $\vvir$ 
of the monomers to the ``effective bulk modulus" $\veff$ \cite{ANS05a} to take into account the 
coupling of a reference chain to the bath.} 
Obviously, the number of chains a reference chain interacts with, and thus the success of 
such a mean-field approach, depends an the spatial dimension $d$ of the problem considered.
Let $N$ be the number of monomeric repeat units per polymer chain (with $N \gg 100$), 
$\RN \sim N^{\nu}$ their typical size, 
$\rhostarN \approx N/\RN^d$ their self-density (also called ``overlap density") and  
$\nu$ their inverse fractal dimension \cite{MandelbrotBook}. 
Assuming the monomer number density $\rho$ to be $N$-in\-de\-pen\-dent, 
a mean-field theoretical approach may be hoped to be successful if
\cite{DegennesBook}
\begin{equation}
\frac{\rhostarN}{\rho} \approx \frac{N}{\rho \RN^d} \sim N^{1-d\nu} \ll 1.
\label{eq_intro_rhostarN}
\end{equation}
For to leading order Gaussian chains ($\nu = 1/2$) this implies that typically 
$\rho/\rhostarN \sim N^{d/2-1} = \sqrt{N} \gg 1$ chains interact in $d=3$. 
Hence, dense three-di\-men\-sion\-al (3D) polymeric liquids should
be much easier to understand than normal liquids, say benzene and let alone water, 
where each molecule has only a few, say ten, directly interacting neighbors \cite{Israela}.
 
\subsection{Coarse-grained lattice models for polymer melts}
\label{intro_lattice}
\begin{figure}[t]
\centerline{\resizebox{0.8\columnwidth}{!}{\includegraphics*{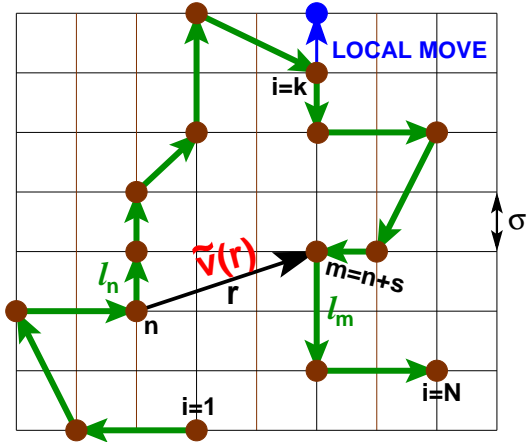}}}
\caption{
Two-dimensional sketch of one chain of an idealized polymer melt 
represented by self-avoiding walks on a simple cubic lattice ($d=3$). 
As notations we use $\rvec_i$ for the position of a monomer $i$,
$\lvec_i = \rvec_{i+1}-\rvec_i$ for its bond vector,
$l$ for the root-mean-squared bond vector,
$\rvec_{nm}=\rvec_m-\rvec_n$ for the end-to-end vector of the subchain between
the monomers $n$ and $m=n+s$ and $r=|\rvec_{nm}|$ for its length.
Taking advantage of the translational invariance along long chain contours, 
subchain properties, such as the $2p$-th moment of $\rvec_{nm}$, 
are generally sampled here by averaging over all possible pairs of monomers $(n,m)$.
The bond vectors between monomers are taken from a set of allowed lattice vectors.
The compressibility of the melt may be imposed by means of a Lagrange multiplier $\overlap$ 
conjugated to the local density fluctuations (Sec.~\ref{theo_incompress}). 
Two monomers of a chain at a distance $r$ interact through an effective potential $\vpot(r)$ 
as predicted by Edwards \cite{DoiEdwardsBook} and further discussed in Sec.~\ref{theo_perturb}.
\label{fig_intro_lattice}
}
\end{figure}

Focusing on the generic statistical properties of flexible and neutral homopolymer melts 
in $d=3$,\footnote{Although we focus on 3D bulks the general $d$-dependence
is often indicated since this may help to clarify the intrinsic structure of the relations.
The reader is invited to replace $d$ by $d=3$.
For similar reasons we often make explicit the inverse fractal dimension $\nu$.
It should be replaced by its value $\nu=1/2$.} 
one of the simplest idealizations consists, 
as illustrated in Fig.~\ref{fig_intro_lattice}, in replacing the
intricate chemical chain structure by {\em self-avoiding walks} on a 
periodic lattice \cite{FloryBook,DegennesBook,VanderzandeBook}. 
Let us assume canonical ensemble statistics 
\cite{AllenTildesleyBook,ThijssenBook,FrenkelSmitBook,BinderHerrmannBook,LandauBinderBook,KrauthBook}
where the total mo\-no\-mer number $\nmon = V\rho$, the volume $V=L^d$ of the $d$-dimensional simulation box 
and the temperature $T$ are imposed. (Boltzmann's constant $\kB$ is set to unity throughout the paper.)
A bond vector $\lvec_i = \rvec_{i+1}-\rvec_i$ connecting the mo\-no\-mers $i$ and $i+1$ is not necessarily 
restricted to a step of length $l=\sigma$ (with $\sigma$ being the lattice constant) to the next-nearest
lattice site. As in the well-known bond-fluctuation model (BFM) \cite{BFM,BWM04,BFMreview05} 
presented in detail in Sec.~\ref{bfm}, 
one may in general take bonds from a suitably chosen set of allowed lattice vectors to achieve a better 
representation of the continuum space symmetries of real polymers \cite{ChaikinBook,Wang09b}.
We focus on systems of high monomer number density $\rho \approx \sigma^{-d}$. 
In principle, although not in practice as seen below, one could sample the configuration space of such a 
lattice polymer melt using a Monte Carlo (MC) algorithm \cite{LandauBinderBook,KrauthBook} with local 
hopping move attempts to the $2d$ next-nearest neighbor sites on the lattice, as shown for the monomer $k$ 
in panel (a). The specific local MC move illustrated is an example for the more general local and global 
rearrangements of the monomers \cite{KrauthBook,BWM04,Kron65,Wall75,dePablo94,pakula02,Theodorou02,dePablo03} 
(respecting the set of allowed bond vectors) one may implement as discussed in Sec.~\ref{bfm_equil}.
Due to their simplicity and computational efficiency, such coarse-grain\-ed lattice models have become
valuable tools for the numerical verification of theoretical concepts and predictions 
\cite{FloryBook,VanderzandeBook,BFMreview05}, 
especially focusing on scaling properties in terms of the chain length $N$
or the arc-length $s = m - n \le N-1$ of subchains \cite{DegennesBook}.

\subsection{Flory's ideality hypothesis for polymer melts}
\label{intro_Flory}
One of the most central concepts of polymer physics is that in a polymer melt {\em all} long-range 
static and dynamical correlations are generally assumed to be negligible beyond the screening length $\xi$ 
(defined properly in Sec.~\ref{theo_incompress} and tested numerically in Sec.~\ref{TD_compress})
characterizing the decay of the density fluctuations \cite{DoiEdwardsBook}. 
The polymer chains are thus expected to adopt random-walk
sta\-tistics as already stated by Flory in the 1940th \cite{FloryBook,Flory45,Flory49} and
worked out more systematically in terms of a perturbation theory
by Edwards two decades later \cite{DoiEdwardsBook,Edwards65,Edwards66,Edwards75,Edwards82}.
The (normalized) probability distribution $G(r,s)$ of the end-to-end distance $r$ of a subchain 
of a sufficiently large arc-length $s$ should thus be a Gaussian,
\begin{equation}
G(r,s) = \left( \frac{d}{2\pi b^2s} \right)^{d/2} \exp\left(- \frac{d}{2} \frac{r^2}{b^2 s}\right)
\label{eq_intro_Grs_gauss}
\end{equation}
for $l s \gg r \gg \xi$ and $N \ge s \gg \gT$ with $b$ being the ``effective bond length" and 
$\gT$ the arc-length spanning the correlation length $\xi$.
Note that in general the effective bond length $b$ differs for various reasons from the 
root-mean-squared bond length $l$ between mo\-no\-mers and ultimately must be fitted 
as will be shown in Sec.~\ref{conf_RNRs}. 
One immediate consequence of Eq.~(\ref{eq_intro_Grs_gauss}) is of course that the second moment of $G(r,s)$ 
must increase linearly, 
\begin{equation}
\Rs^2 \equiv \la (\rvec_{m=n+s} - \rvec_n)^2 \ra  = b^2 s \mbox{ for } \  s \gg  \gT,
\label{eq_intro_Rs_gauss}
\end{equation}
and the same holds for the mean-squared total chain end-to-end distance 
$\RN^2 \equiv \la (\rvec_N-\rvec_1)^2 \ra$ with $s=N-1 \approx N$,
i.e. $\nu=1/2$ for the inverse fractal dimension. Since the bond-bond correlation function $\Pone(s)$
measures the curvature of the mean-squared subchain size $\Rs^2$,
\begin{equation}
\Pone(s) \equiv \frac{\la \lvec_{n} \cdot \lvec_{m=n+s} \ra}{\la \lvec_n^2 \ra} 
\approx \frac{1}{2l^2} \frac{d^2}{ds^2} \Rs^2,
\label{eq_intro_PonesRsrelated}
\end{equation}
it follows from Eq.~(\ref{eq_intro_Rs_gauss}) that
$\Pone(s) = 0$ for $s \gg \gT$.\footnote{This lack of memory of the bond orientations 
may be taken as the defining statement of Flory's ideality hypothesis. Using the central limit 
theorem \cite{DhontBook} one immediately gets back to Eq.~(\ref{eq_intro_Grs_gauss}).}
This suggests to test for possible deviations from $\Rs^2 \sim s$ by computing directly the 
bond-bond correlation function $\Pone(s)$ as we shall do in Sec.~\ref{conf_bondbond}.
That Eq.~(\ref{eq_intro_PonesRsrelated}) holds follows from the general relation for displacement 
correlations Eq.~(\ref{eq_app_ht2Ct}) demonstrated in Appendix~\ref{app_ht2Ct}.\footnote{It
is also common to define the bond-bond (or angular) correlation function as the first 
Legendre polynomial $\Ponehat(s) \equiv \la \ehat_n \cdot \ehat_m \ra$ 
using the normalized tangent vector $\ehat_n = \lvec_n/|\lvec_n|$.
The difference between $\Pone(s)$ and $\Ponehat(s)$ is irrelevant from the scaling point of view
and negligible quantitatively for all computational models we have studied.
Higher order angular correlations may be described in terms of general Legendre
polynomials $\hat{P}_i$ \cite{abramowitz} 
sampled as a function of the arc-length $s$ between the bond pair or their spatial distance $r$.
The polynomials $\hat{P}_{2i}(r)$ may be used to characterize the interchain
angular correlations, a question out of the scope of the present study.}

\subsection{Rouse model hypothesis for polymer melts}
\label{intro_Rouse}
Based on Flory's ideality hypothesis for the chain conformations, the celebrated reptation model 
suggested by Edwards and de Gennes \cite{DegennesBook,DoiEdwardsBook} provides a 
widely accepted theoretical description for the dynamical properties of 
entangled polymer melts explaining qualitatively the observed increase of the melt viscosity $\eta$
with respect to the chain length and other related observables \cite{Lodge90}. 
The central idea is that chains cannot cross each other and must due to this topological constraint
``reptate" along an effective ``primative path" set by their own contour.\footnote{In its original
formulation the reptation model is a single chain mean-field theory, the constraints due to other chains
being accounted for by the effective topological tube \cite{DoiEdwardsBook}. Due to density
fluctuations generated by the motion of the chains and the finite compressibility of the elastic
mesh, different chains must be coupled, however. As suggested by A.N. Semenov {\em et al.}
\cite{ANS97,ANS98}, the relaxational dynamics of long reptating chains ($N\gg \Ne^3$ in $d=3$) 
is ultimately ``activated", i.e. the longest relaxation time and the viscosity increase 
exponentially with chain length $N$ and not as simple power laws \cite{Deutsch85,MWBBL03}.}
Using local topology conserving MC moves various authors have thus 
attempted in the past to verify the reptation predictions by lattice MC models as the one sketched 
in Fig.~\ref{fig_intro_lattice} \cite{Paul91a,Paul91b,WPB92,WPB94,Shaffer94,Shaffer95,KBMB01,HT03}.
Similar tests have also been performed by means of molecular dynamics (MD) \cite{FrenkelSmitBook},
Brownian dynamics (BD) \cite{AllenTildesleyBook} and 
MC simulations of various off-lattice bead-spring models 
\cite{KG86,KG90,LAMMPS,Briels01,Briels02,EveraersScience,MKCWB07,Wang09a}.\footnote{The computational activities have
focused recently on the latter off-lattice models, one essential advantage being that the 
experimentally relevant shear-stress correlation function $\mu(t)$ 
can be computed directly without using additional input from theory.} 

We remind that reptational motion along the primative path is assumed to be the dominant relaxational process 
only for chains much longer than the (postulated) entanglement length $\Ne$ and for times $t$ larger than 
the entanglement time $\Te \sim \Ne^2$ \cite{DegennesBook}. In the opposite limit the dynamics is thought 
to be described by the Rouse model proposed in the 1950th, i.e. by a simple (position) Langevin equation 
for a single chain with all interactions from the bath being dumped into a friction term and random 
uncorrelated forces \cite{DegennesBook}. Thus the reptation model ``sits on top" of the Rouse model
and should the latter model prove not to be sufficient this must be relevant 
for long chains in the entangled regime.\footnote{As a consequence of the ``hydrodynamic screening
hypothesis" for dense polymer solutions \cite{DoiEdwardsBook}, all Zimm-like long-range correlations 
are generally regarded to be negligible beyond the static screening length $\xi$ 
\cite{Duenweg93,Duenweg01,Spenley00}. The monomer momentum is not lost, of course, but 
transmitted to the bath and for large times and distances the melt is still described by the 
Navier-Stokes equation \cite{ChaikinBook}. Although these hydrodynamic effects are beyond the scope 
of this paper focusing on lattice MC simulations, a short hint on scale-free corrections to the 
hydrodynamic screening hypothesis is given at the end of Sec.~\ref{conc_outlook}.}

Interestingly, it is relatively easy to construct for the lattice models sketched in 
Fig.~\ref{fig_intro_lattice} local monomer moves switching off the topological constraints 
without changing the excluded volume and other interactions and conserving thus all static properties. 
This has allowed to demonstrate that topological constraints are paramount for the relaxation process 
\cite{Shaffer94,Shaffer95}.
Since the local MC hopping attempts correspond to the random white forces
of the Rouse model and since all other long-range correlations of the bath are supposed to be 
negligible beyond $\xi$, one generally assumes that such a topology non-conserving lattice model 
should be described by the Rouse model for chain lengths $N \gg \gT$.
Since the forces acting on the chain center-of-mass (CM) $\rN(t)$ are uncorrelated,
the probability distribution of the CM displacements must be a Gaussian just as 
Eq.~(\ref{eq_intro_Grs_gauss}) with the CM displacement $\rN(t) - \rN(0)$ 
replacing $\rvec$, the time $t$ replacing the arc-length $s$ and $2 d \DN$ replacing $b^2$. 
The self-diffu\-sion coefficient $\DN$ and the longest 
Rouse relaxation time $\TN$ of chains of length $N$ should scale as \cite{DoiEdwardsBook}
\begin{equation}
\DN \approx b^2 W/N, \ \TN \approx \RN^2/\DN \approx N^{2\nu+1}/W
\label{eq_intro_DNTN_Rouse}
\end{equation}
with $W$ being a convenient local monomer mobility defined properly in Sec.~\ref{dyna_jean} \cite{Paul91b}.
In analogy to Eq.~(\ref{eq_intro_Rs_gauss}), $\DN$ may be obtained by
fitting the mean-square displacement (MSD) of the chain CM 
\begin{equation}
\MSDcmN(t) \equiv \la (\rN(t)-\rN(0) )^2 \ra \approx 2 d \DN t. 
\label{eq_intro_hN_Rouse}
\end{equation}
The chain relaxation time $\TN$ corresponds to the motion of the CM over the typical chain size
and may be defined by setting $\MSDcmN(t = \TN) = \RN^2$ \cite{Paul91b}.
Note that Eq.~(\ref{eq_intro_hN_Rouse}) should hold for {\em all} times $t$ 
(taken apart very short times corresponding to displacements of order of $\xi$ or the lattice constant $\sigma$)
and not only for $t \gg \TN$ as for reptating chains.
In analogy to Eq.~(\ref{eq_intro_PonesRsrelated}) one may characterize the CM displacements
by a four-point correlation function in time, the ``velocity correlation function" (VCF)
\begin{equation}
\CN(t) \equiv 
\la \frac{\uvec(t)}{\deltat} \cdot \frac{\uvec(0)}{\deltat} \ra
\approx \frac{1}{2} \frac{\partial^2 \MSDcmN(t)}{\partial t^2} \sim \deltat^0 \mbox{ for } t \gg \deltat,
\label{eq_intro_CNdef}
\end{equation}
which measures directly the curvature of $\MSDcmN(t)$ with respect to time with 
$\uvec(t)=\rN(t+\deltat)-\rN(t)$ being the CM displacement for a time window $\deltat$.
Since in a MC algorithm there is no velocity in the sense of an Euler-Lagrange equation, ``velocity" 
refers to the displacement per time increment $\deltat$. As shown in Appendix~\ref{app_ht2Ct},
Eq.~(\ref{eq_intro_CNdef}) states that $\CN(t)$ becomes independent of $\deltat$ for $t \gg \deltat$.
Since Eq.~(\ref{eq_intro_hN_Rouse}) implies that $\CN(t)$ must vanish for all times, 
this begs for a direct numerical test as will be discussed in Sec.~\ref{dyna}.
Interestingly, the above statements do not only hold for the displacement field of the total chain CM 
$\rN(t)$ but also for the displacements of the CM $\rs(t)$ of arbitrary subchains of arc-length $s$ 
for times $t \ll \Ts \approx s^{2\nu+1}/W$ with $\Ts$ being the Rouse relaxation time of the 
subchain.\footnote{For times $t > \Ts$ a subchain feels the connectivity with
the rest of the chain and must follow the typical monomers motion for $t \ll \TN$.} We can thus state 
more generally that according to the Rouse model one expects 
\begin{equation}
\Cs(t) = 0 \mbox{ for } t \ll \Ts \mbox{ and } \gT \ll s \le N-1
\label{eq_intro_Cs_Rouse}
\end{equation}
with $\Cs(t)$ being the VCF associated to the subchain CM displacements $\uvec(t) = \rs(t+\deltat)-\rs(t)$.

\subsection{Aim of this study and key results}
\label{intro_key}
Summarizing recent theoretical and numerical work made by the Strasbourg polymer theory group
but also by other authors 
\cite{SchaferBook,Obukhov81,Schweizer91,MM00,Wang02,Auhl03,Rubinstein08,Morse09,Morse09b,Paul98,Smith00,PaulGlenn}, 
we address and question in this review the validity of the general screening assumption for dense polymer 
solutions and melts both concerning Flory's ideality hypothesis for the chain conformations 
\cite{WMBJOMMS04,WBCHCKM07,WBJSOMB07,BJSOBW07,WBM07,MWK08,WCK09,WJO10,WJC10,ANS10a}
and the Rouse model assumption for the dynamics of systems with irrelevant or switched off topology 
\cite{WPC11,papFarago,ANS11a,ANS11b}. 
We argue that due to the overall incompressibility of the polymer melt 
both the static and the dynamical fluctuations of chains and subchains are 
coupled on scales beyond the screening length $\xi$.
In $d=3$ dimensions this leads to weak but measurable scale-free deviations which do not depend 
explicitly on the (low but finite) isothermal compressibility $\kappaT$ of the solution.

The static correlations are made manifest by the finding \cite{WMBJOMMS04,WBM07,MWK08,WCK09}
that the intrachain bond-bond correlation does not vanish as implied by Flory's hypothesis,
but rather decays analytically in $d=3$ dimensions as
\begin{equation}
\Pone(s) = \frac{\cP}{s^{3/2}} 
\mbox{ with } \cP  = \cinf \ce /8
\mbox{ for } \gT \ll s \ll N
\label{eq_intro_keystat}
\end{equation}
the amplitude $\cP$ being given by the dimensionless stiffness coefficient $\cinf = (b/l)^2$ \cite{DoiEdwardsBook}
and the so-called ``swelling coefficient" $\ce = \sqrt{24/\pi^3}/\rho b^3$ \cite{WBM07}.
Due to Eq.~(\ref{eq_intro_PonesRsrelated}) this is consistent with a typical subchain size 
\begin{equation}
\Rs^2 = b^2 s \left(1- \ce/\sqrt{s} \right) \mbox{ for } \gT \ll s \ll N,
\label{eq_intro_Rs}
\end{equation}
i.e. $\Rs^2/s$ approaches the asymptotic limit $b^2$ monotonously from below
and the chain conformations are thus weakly swollen as shown in Sec.~\ref{conf_Rs}.
Note that generalizing Eq.~(\ref{eq_intro_keystat}) as demonstrated in Appendix~\ref{app_stat_Ps}
the bond-bond correlation function is predicted to decay in (effectively) $d$ dimensions as 
\cite{ANS03,CMWJB05,LFM11,thinfilmdyn} 
\begin{equation}
\Pone(s) = 
\frac{1}{2\omega} \left(\frac{\omega}{\pi}\right)^{\omega}
\frac{\cinf}{b^d \rho} \frac{1}{s^{\omega}} 
\mbox{ with } \omega = d \nu
\label{eq_intro_PsD}
\end{equation}
for $\gT \ll s \ll N$
with $\rho$ being the $d$-dimensional monomer density.\footnote{For $d \le 2$ the range of validity 
of Eq.~(\ref{eq_intro_PsD}) is further restricted as discussed in Sec.~\ref{conc_outlook_slit}.}

At variance to Eq.~(\ref{eq_intro_Cs_Rouse}) the displacement correlation functions $\CN(t)$ and $\Cs(t)$
of chains and subchains are observed not to vanish. Instead an algebraic long-time power-law tail
\begin{equation}
\Cs(t) \approx - \frac{b^2 W}{s} \frac{1}{b^d \rho} (W t)^{-\omega} \mbox{ with } 
\omega = \frac{2+d}{2+1/\nu} = 5/4
\label{eq_intro_keydyna}
\end{equation}
is found in $d=3$ for all $s \le N$ and $t \ll \Ts$ \cite{WPC11}.
Due to Eq.~(\ref{eq_intro_CNdef}) this finding implies for $s=N-1$
that the CM MSD of chains $\MSDcmN(t)$ takes an additional algebraic contribution which dominates
for short times such that
\begin{equation}
\MSDcmN(t) \sim t^{\beta} \mbox{ with } \beta = 2- \omega = 3/4
\mbox{ for } t \ll \tstar \sim N^0
\label{eq_intro_betadef}
\end{equation}
in qualitative agreement with the anomalous power-law exponent $\beta \approx 0.8$
obtained in various numerical 
\cite{Paul91b,Shaffer95,KBMB01,HT03,KG90,Briels01,Briels02,Skolnick87,Guenza02,Theodorou03,Theodorou07} 
and experimental studies \cite{Paul98,Smith00,PaulGlenn}. Our theoretical and numerical result, 
Eq.~(\ref{eq_intro_keydyna}), may thus help to clarify a long-lasting debate.\footnote{We
emphasize that Eq.~(\ref{eq_intro_PsD}) and Eq.~(\ref{eq_intro_keydyna}) are both scale-free and do not depend explicitly 
on the compressibility $\kappaT$ of the solution. Interestingly, both key findings depend on the subchain length $s$ 
considered and not on the total chain size $N$ as long as $s \ne N$.
}

\subsection{Outline}
\label{intro_outline}
We begin this review by summarizing in Sec.~\ref{theo} a few well-known theoretical concepts from 
polymer theory and outline the perturbation calculations which lead, e.g., to Eq.~(\ref{eq_intro_keystat}).

Since the predicted deviations Eq.~(\ref{eq_intro_keystat}) and Eq.~(\ref{eq_intro_keydyna}) 
are rather small in $d=3$ dimensions, the key challenge from the computational side is to obtain high
precision data for well-equilibrated polymer melts containing sufficiently long chains to avoid chain end effects. 
This review focuses on numerical results obtained by means of the 
BFM algorithm with finite excluded volume interactions \cite{WCK09} and 
topology violating local and global Monte Carlo moves \cite{WBM07} as described in 
Sec.~\ref{bfm}.\footnote{To check for caveats due to our lattice model we have verified our results 
by MD simulation of a standard bead-spring model with and without topological constraints 
\cite{WMBJOMMS04,WBJSOMB07,WBM07,MWK08,papFarago,ANS11a,ANS11b}. 
In order not to overburden the present summary, this story must be told elsewhere.}
Our numerical data for monodisperse melts are crosschecked \cite{WBCHCKM07,BJSOBW07,WJO10,WJC10}
using systems of annealed size-di\-stri\-bu\-tion, so-called ``equilibrium po\-ly\-mers" (EP)
\cite{CC90,Scott65,Wheeler80,Faivre86,Milchev95,WMC98a,WMC98b,Milchev00,HXCWR06,Kroeger96,Padding04},
where chains break and recombine constantly and for this reason the sampling becomes much 
faster than for monodisperse quenched chains (Sec.~\ref{bfm_EP}).

Static properties are discussed in Sec.~\ref{TD} and Sec.~\ref{conf}. 
In the first section we investigate some thermodynamic properties of dense BFM solutions such as 
the chemical potential $\muchainN$ in EP systems in Sec.~\ref{TD_muEP} \cite{WJC10}.
We verify in Sec.~\ref{TD_compress} the range of validity of the (static) ``Random Phase Approximation" (RPA) 
\cite{DoiEdwardsBook} describing the coupling of the total density fluctuations, encoded by the total structure 
factor $S(q)$ at wavevector $q$, with the degrees of freedom of tagged chains, 
encoded by the intrachain coherent single chain form factor $F(q)$ \cite{BenoitBook}.
Section~\ref{conf} contains the central part of our work related to conformational intrachain 
properties where we demonstrate Eq.~(\ref{eq_intro_keystat}).
The deviations from Flory's ideality hypothesis are also of experimental relevance,
as shown in Sec.~\ref{conf_Fq}, since the measured form factor $F(q)$ should differ from the 
corresponding Gaussian chain form factor $F_0(q)$ as \cite{WBJSOMB07,BJSOBW07}
\begin{equation}
\frac{1}{F(q)}-\frac{1}{F_0(q)} = \frac{|q|^3}{32\rho} \mbox{ for } 1/\RN \ll q \ll 1/\xi.
\label{eq_intro_Fq}
\end{equation}

Our work on dynamical correlations in Rouse-like systems without topological constraints  \cite{WPC11,papFarago}
is summarized in Sec.~\ref{dyna}. A simple perturbation theory argument leading to Eq.~(\ref{eq_intro_keydyna}) 
will be given in Sec.~\ref{dyna_perturb}. This argument uses the ``dynamical Random Phase Approximation" (dRPA)
\cite{ANS86,ANS97b,papFarago}
describing the coupling of the degrees of freedom of the bath (encoded by the dynamical structure factor $S(q,t)$)
to the degrees of freedom of a tagged test chain (encoded by the dynamical form factor $F(q,t)$).
We shall verify explicitly this important relation.

We conclude the paper in Sec.~\ref{conc} where we comment on related computational work 
focusing on dimensionality effects (Sec.~\ref{conc_outlook_slit}) \cite{ANS03,CMWJB05,Brochard79,LFM11},
collective interchain correlations (Sec.~\ref{conc_outlook_casimir}) \cite{ANS05a,ANS05b}
and long-range viscoelastic hydrodynamic interactions (Sec.~\ref{conc_outlook_offlattice}) 
\cite{ANS11a,ANS11b}.
Several theoretical issues are relegated to the Appendix.

\section{Some theoretical considerations}
\label{theo}

\subsection{Introduction}
\label{theo_intro}
Let us go back to the simple generic lattice model sketched in Fig.~\ref{fig_intro_lattice}.
Simplifying further, we begin in Sec.~\ref{theo_connectivity} by switching off {\em all} 
short- and long-range interactions between chains and monomers, the only remaining interaction 
being the connectivity of the mo\-no\-mers along the chain contours. 
Translational invariance along these contours is assumed.
No particular meaning is attached to the orientation of the monomer index $i$.\footnote{This
implies $i \leftrightarrow -i$ symmetry 
with respect to the monomer index $i$ 
which can be read as a time variable $t$ as seen in Appendix~\ref{app_ht2Ct}. Due to this 
reversibility the bond-bond correlation function can be expressed in terms of the 
non-Gaussian ``colored forces" \cite{vanKampenBook} acting on the monomers, Eq.~(\ref{eq_app_displ5}).}
To make this toy model more interesting let us on the other side introduce 
two additional features: local chain rigidity (Sec.~\ref{theo_connectivity_rigid}) 
and polydispersity (Sec.~\ref{theo_connectivity_poly}). 
Reminding some standard properties \cite{DegennesBook,DoiEdwardsBook,BenoitBook,SchaferBook,RubinsteinBook} 
we will then introduce for later reference the intrachain single chain form factor $F(q)$ (Sec.~\ref{theo_connectivity_Fq}) 
and the total monomer structure factor $S(q)$ (Sec.~\ref{theo_connectivity_Fq}).
Since the chains do not interact we can consider each chain independently.
The interaction between chains and monomers will be switched on again in Sec.~\ref{theo_incompress}
by means of a Lagrange multiplier $\overlap$ limiting the density fluctuations (Sec.~\ref{theo_incompress_overlap}). 
Note that assuming Flory's ideality hypothesis most intrachain properties discussed in Sec.~\ref{theo_connectivity}
should also hold rigorously in incompressible melts with the effective bond length $b$ being the only fit parameter.
The effective entropic correlation hole forces for chains and subchains arising due to the incompressibility 
constraint are introduced in Sec.~\ref{theo_corhole}. The generic scaling of the deviations from Flory's ideality hypothesis 
is motivated in Sec.~\ref{theo_corhole_perturbation} before we turn to the systematic perturbation calculation 
in Sec.~\ref{theo_perturb}.
Following Muthukumar and Edwards \cite{Edwards82} we argue that the reference length $\bref$ 
of the Gaussian reference chain of the calculation (Sec.~\ref{app_stat_Grs}) 
should be renormalized (Sec.~\ref{theo_perturb_brenorm}).
The bond-bond correlation function $\Pone(s)$ for asymptotically long chains is presented in 
Sec.~\ref{theo_perturb_bondbond} 
before we turn finally in Sec.~\ref{theo_perturb_Neffects} to finite chain size effects.

\subsection{Connectivity constraint}
\label{theo_connectivity}

\subsubsection{Local rigidity}
\label{theo_connectivity_rigid}
Keeping only the chain connectivity we switch off all other mono\-mer and chain interactions.
Let us apply a local stiffness energy proportional to the cosine of the bond angle $\theta$,
$\cos(\theta) = \ehat_n \cdot \ehat_{n+1}$, with $\ehat_n = \lvec_n/|\lvec_n|$ being the unit tangent vector. 
The stiffness energy is assumed to be not too large to avoid lattice artifacts \cite{WPB92}.
Due to the multiplicative loss of any information transferred recursively along the chain contour the 
bond-bond correlation function $\Pone(s)$ must decay exponentially with arc-length $s$ \cite{FloryBook},
\begin{equation}
\Pone(s) \approx \exp(-s/\spersist) 
\label{eq_theo_Pone_spersist}
\end{equation}
with $\spersist = \lpersist/l$ being the curvilinear persistence length, 
i.e. $\Pone(s) \approx 0$ for $s \gg \spersist$.
Using Eq.~(\ref{eq_intro_PonesRsrelated}) it follows of course that $\Rs^2 \approx b^2 s$ for subchains 
of arc-length $s \gg \spersist$. Assuming Eq.~(\ref{eq_theo_Pone_spersist}) this implies 
$b^2 = l^2 (2\spersist-1)$ for the effective bond length $b$ since more generally 
it is known that \cite{BWM04}
\begin{equation}
b^2 
= l^2 \left(2 \sum_{s=0}^{\infty} \Pone(s) -1 \right)
\equiv 2 l \lpersist - l^2.
\label{eq_theo_bdefintegral}
\end{equation}
The latter relation is consistent with the more common definition \cite{DoiEdwardsBook}\footnote{Definitions
based on such ``Einstein relations" \cite{ChandlerBook} are generally more robust numerically.}
\begin{equation}
b^2 \equiv 2 d a^2 \equiv \cinf l^2  \equiv \lim_{N\to \infty} \frac{\RN^2}{N}
\label{eq_theo_bdef}
\end{equation}
using the total chain mean-squared end-to-end distance \cite{DoiEdwardsBook}.
We have introduced here a convenient mono\-meric length $a$ allowing to simplify prefactors 
depending in a spurious manner on the spatial dimension $d$ and have reminded the dimensionless 
stiffness parameter $\cinf = (b/l)^2$. 
%

We note that a good approximation for a ``freely rotating chain" with local stiffness potential 
is given by \cite{DoiEdwardsBook}
\begin{equation}
\cinf = \frac{1+ \la \cos(\theta) \ra}{1- \la \cos(\theta) \ra } \ge 1,
\label{eq_theo_cinf_FR}
\end{equation}
which has been shown to be useful even for lattice models with discrete bond angles \cite{WPB92}.
In summary, a local stiffness energy does not change the scaling of the chain and subchain size
with, respectively, $N$ or $s$,  but merely increases its amplitude.
Note that a weak local chain stiffness with $\spersist \approx 1$ would be generated by disallowing chains 
to return after, say, two steps to the same lattice site 
(so-called ``non-reversal random walks" \cite{BinderHerrmannBook}) and similar local constraints. 
%
%
Since these are the only local stiffness contributions which may occur in the presented numerical work
these small rigidity effects can be safely disregarded below.\footnote{The only case where the small
local chain rigidity qualitatively matters is presented in Sec.~\ref{conf_Pr} for the
bond-bond correlation function $\Pone(r)$ for large distances $r \gg \rstar$ between bonds \cite{WJO10}.}

\subsubsection{Polydispersity}
\label{theo_connectivity_poly}
Let us relax the monodispersity constraint made above and assume a general (normalized) probability 
distribution $\pchain$ for chains of length $N$. We suppose that the mean chain length $\Nav$ is arbitrarily 
large and that all moments $\Npav = \int_0^{\infty} {\rm dN} \ N^p \ \pchain$ of the distribution exist.
For subchain properties everything remains as before. 
Let us call $\RN^2$ the typical mean-squared chain size of chains of length $N$.
Since one averages experimentally over chains of different length this total
average depends now on the moment $p$ of the distribution which is probed,
\begin{equation}
R_p^2 \equiv \frac{\la \RN^2 N^p \ra}{\la N^p \ra} = b^2 \frac{\la N^{p+1} \ra}{\la N^p\ra}
\equiv b^2 N_p,
\label{eq_theo_R2PN}
\end{equation}
with $N_p$ being the $p$-averaged chain length. 
Since for experimentally reasonable distributions $\Npav \sim \Nav^p \sim N_p^{\ p}$, we have normally 
$R_p^2 \sim b^2 \Nav$ with  numerical coefficients due to the moment taken. 
For the properties considered by us, it is the so-called ``$z$-average" for $p=2$ which matters most.
The $z$-average is for instance probed by sedimentation in an analytical ultracentrifuge
or in neutron scattering measurements of the intramolecular form factor $F(q)$ as further discussed 
in Sec.~\ref{theo_connectivity_Fq} \cite{BenoitBook}.
Note that all standard definitions and formulas \cite{DegennesBook,DoiEdwardsBook} are recovered 
for a monodisperse length distribution $\pchain = \delta(N^{\prime}-N)$.
%

For the important case of Flory-distributed polymers with 
\begin{equation}
\pchain = \mu \exp(- \mu N) \mbox{ with } \Nav = \mu^{-1}
\label{eq_theo_PN_Flory}
\end{equation}
we have $\Npav = p! \Nav^{p}$ and thus $R_p^2 = (p+1) b^2 \Nav$.
Such a Flory distribution is expected for systems of EP with an annealed length distribution
where a constant finite scission energy $E \ge 0$ has to be paid for the scission of each bond 
as described in Sec.~\ref{bfm_EP}. This can be seen by minimizing the Flory-Huggins
free energy functional \cite{OozawaBook,CC90,WMC98b}
\begin{equation}
F[\rhochain] = V \sum_N \rhochain \left(T \log(\rhochain) + \mu N + E + \dmuchainN \right)
\label{eq_theo_FloryHuggins}
\end{equation}
with respect to the density $\rhochain \equiv \rho \pchain/\Nav$ of chains of length $N$.
The first term on the right is the usual translational entropy.
The second term entails a Lagrange multiplier which fixes the total
monomer density 
\begin{equation}
\rho = \sum_{N=1}^{\infty} N \rhochain.
\label{eq_theo_FloryHuggins_constraint}
\end{equation}
All contributions to the chemical potential of the chain $\muchainN$ which are linear in $N$
can be absorbed within this Lagrange multiplier. The scission energy $E$ characterizes the imposed
enthalpic free energy cost for breaking a chain bond. The last term $\dmuchainN$ encodes the remaining 
non-linear contribution to the chemical potential $\muchainN$ which has to be paid for creating two 
new chain ends.\footnote{As shown in Sec.~\ref{TD_muEP} and Sec.~\ref{app_stat_muEP}, 
the free energy contribution $\dmuchainN$ may depend in general on the chain length $N$ 
\cite{DescloizBook,SchaferBook,WMC98a}.}
For non-interacting random walks on the 
lattice this contribution is just a (model depending) constant entropic factor which renormalizes the 
imposed scission energy to an $N$-independent effective scission free energy $E + \dmuchainN$ \cite{HXCWR06}. 
The minimization of Eq.~(\ref{eq_theo_FloryHuggins}) under the density constraint, 
Eq.~(\ref{eq_theo_FloryHuggins_constraint}), yields Eq.~(\ref{eq_theo_PN_Flory}) 
with a mean chain length
\begin{equation}
\mu^{-1}  = \Nav = \left(\Npav/p! \right)^{1/p} \approx \sqrt{\rho \exp(E/T)}
\label{eq_theo_Nav}
\end{equation}
as the reader will readily verify.

\subsubsection{Segmental size-distribution $G(r,s)$}
\label{theo_connectivity_Gqs}
Due to the translational invariance in space and along the chain contours most perturbation calculations outlined below 
\cite{BJSOBW07,WBM07,WJO10,WJC10} are more readily performed in reciprocal space. The Fourier transform of a function 
$f(\rvec)$ is denoted $f(\qvec) \equiv \Fcal[f(\rvec)] = \int  {\rm d}\rvec f(\rvec) e^{- {\rm i} \qvec\cdot \rvec}$
and we write $\fhat(t) \equiv \Lcal[f(s)] = \int_0^{\infty} f(s) e^{-s t}$ for the Laplace transform of a function $f(s)$ 
with $t$ being the Laplace variable conjugated to the arc-length $s$.
We have introduced in Sec.~\ref{intro_Flory} the probability distribution $G(r,s)$ of the end-to-end distance $r$ of a subchain 
of arc-length $s$ between the monomers $n$ and $m=n+s$ of a chain. The Fourier transform of this two-point intramolecular
correlation function is thus in general $G(q,s) \equiv \la \exp(- {\rm i} \qvec \cdot \rvec) \ra$ with the average being taken 
over all possible subchain vectors $\rvec = \rvec_{m}-\rvec_{n}$. 
For (infinite) Gaussian chains this becomes \cite{DoiEdwardsBook}
\begin{equation}
G(q,s) = G_0(q,s) \equiv \exp(-(a q)^2s).
\label{eq_theo_Gqs_gauss}
\end{equation}
The index $0$ has been introduced for the general case where $G(q,s)$ may differ from the Gaussian propagator $G_0(q,s)$ 
used in our perturbation calculations. 
Moments of the distribution $G(r,s)$ are readily obtained from derivatives of $G(q,s)$ 
taken at $q=0$ as recalled in Appendix~\ref{app_stat_moments} \cite{vanKampenBook}. 
It follows for instance for the $2p$-th moment of the distribution that
\begin{equation}
\la \rvec^{2p}\ra = \left. (-1)^p\Delta^p G_0(q,s)\right|_{q=0}
= \frac{(2p+1)! }{p!} \frac{(2d)^p}{6^p} s^p a^{2p}
\label{eq_theo_rsp_gauss}
\end{equation}
for Gaussian chains. We have thus
\begin{equation}
\Kp(s) \equiv 1 - \frac{p!6^p}{(2p+1)!} \frac{\la \rvec^{2p} \ra}{(b^2s)^p} = 0.
\label{eq_theo_Kps_gauss}
\end{equation}
A very closely related characterization of $G(r,s)$ is given by 
the standard non-Gaussianity parameter
\begin{equation}
\alphap(s) \equiv 
1 - \frac{p!6^p}{(2p+1)!} \frac{\la \rvec^{2p} \ra}{\la \rvec^2 \ra^p} 
\label{eq_theo_alphaps_gauss}
\end{equation}
which for Gaussian chains is identical to $\Kp(s)$. As  further discussed in Sec.~\ref{conf_moments},
$\alphap(s)$ has computationally the advantage that the effective bond length $b$ must not be 
known {\em a priori}. Obviously, the general distribution $G(r,s)$ is fully determined by either the 
dimensionless moments $\Kp(s)$ or $\alphap(s)$ \cite{vanKampenBook}.

In Fourier-Laplace space the Gaussian propagator reads
\begin{equation}
\Ghat_0(q,t) \equiv \frac{1}{(a q)^2 + t}.
\label{eq_theo_Gqt_gauss}
\end{equation} 
If one needs to average over all bond pairs at a given distance $r$ irrespective of their curvilinear distance $s$,
and one has thus to sum over all possible $s$ as in Sec.~\ref{app_stat_Pr}, 
this corresponds (for arbitrarily large chains) to setting $t=0$ for the corresponding Laplace variable.
The summed up Gaussian propagator for infinite chains is thus $\Ghat_0(q,t=0) = (a q)^{-2}$.
For Flory-distributed Gaussian chains of finite mean chain length $\Nav=1/\mu$ we have more
generally a summed up Gaussian propagator
\begin{equation}
\Gtilde(q) \equiv \Ghat_0(q,t=0) = \frac{1}{(a q)^2 + \mu}.
\label{eq_theo_Gqt0_Flory}
\end{equation}
Inverse Fourier transformation yields in $d=3$ the density 
\begin{equation}
\Gtilde(r) \equiv \Ghat_0(r,t=0) = \frac{1}{4\pi a^2 r} e^{-\sqrt{\mu}r/a}
\label{eq_theo_Grt0_gauss}
\end{equation} 
around a tagged reference monomer of all the monomers belonging to the same chain.
Obviously, for $\mu \to \infty$ one recovers the well-known density 
\begin{equation}
\Gtilde(r) \sim \frac{1}{r^{d-1/\nu}} \mbox{ for } d > 1/\nu
\label{eq_theo_Grt0_inf}
\end{equation}
for infinite objects of inverse fractal dimension $\nu$ \cite{DegennesBook,WittenPincusBook}.\footnote{The
reader may verify the indicated scaling by direct integration of Eq.~(\ref{eq_intro_Grs_gauss}) with respect to $s$
for monodisperse chains with $N \to \infty$.}
This power-law dependence of the local density means that our random-walk polymer is a fractal set in the sense 
of Mandelbrot \cite{MandelbrotBook}, i.e. the average mass (number of mono\-mers) within a distance $r$ of an 
arbitrary point of the set varies as a power of $r$. The $r$-dependence of the local density reflects 
a type of spatial order that is not connected to translation or rotation symmetries but to the ``dilation 
transformation" to a magnified system \cite{DegennesBook}. A structureless, uniform material
looks the same when magnified, provided the magnification is too weak to see the molecular constituents
(``lower cutoff"). The scaling exponent $d-1/\nu$ characterizes the dilation symmetry in the same way that
linear momentum characterizes translational symmetry and angular momentum characterizes rotational
symmetry \cite{WittenPincusBook}.

\begin{figure}[t]
\centerline{\resizebox{0.8\columnwidth}{!}{\includegraphics*{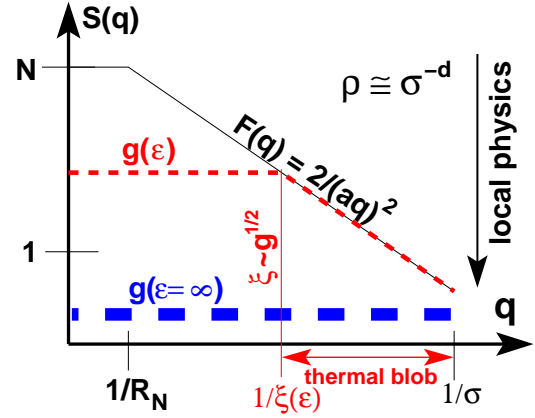}}}
\caption{Schematic representation of total structure factor $S(q)$ and
intrachain coherent form factor $F(q)$ for monodisperse chains (thin line) in double-logarithmic coordinates.
For a large Lagrange multiplier $\overlap \to \infty$ we have $S(q) \approx \gT(\overlap=\infty) \ll 1$
for $q \ll 1/\sigma$ (dashed bold line).
For finite $\overlap$ the incompressibility is only felt for
$q \ll 1/\xi(\overlap)$ with the screening length $\xi(\overlap)$
setting the size of the ``thermal blob" \cite{DegennesBook,DoiEdwardsBook}.
\label{fig_theo_SqFq}
}
\end{figure}
\subsubsection{Intramolecular coherent form factor $F(q)$}
\label{theo_connectivity_Fq}
The two-point correlation function $G(q,s)$ may be probed experimentally through the intramolecular form factor $F(q)$ 
which for monodisperse chains is defined by \cite{DoiEdwardsBook}
\begin{eqnarray}
F(q) & = &  \frac{1}{N} \sum_{n,m=1}^{N} \la \exp\left(- {\rm i}\qvec \cdot (\rvec_{n}-\rvec_{m}) \right) \ra
\label{eq_theo_Fq_mono}  \\
& = & \frac{1}{N} \la C^2 + S^2 \ra
\label{eq_theo_Fq_monocompu}
\end{eqnarray}
with $C = \sum_{n=1}^{N} \cos(\qvec\cdot\rvec_{n})$ and $S = \sum_{n=1}^{N} \sin(\qvec\cdot\rvec_{n})$.
The second representation being an operation linear in $N$ has obvious computational advantages 
for large chain lengths.
In principle, the characteristic polymer size may be obtained experimentally from $F(q)$ 
in the Guinier regime for small $q$ \cite{BenoitBook}. Expanding Eq.~(\ref{eq_theo_Fq_mono}) 
yields $F(q) = N (1 -(\Rgyr(N)q)^2/d)$ with $\Rgyr(N)$ being the gyration radius defined as
\begin{equation}
\Rgyr^2(N) \equiv \frac{1}{2N^2} \sum_{n,m=1}^{N} \la (\rvec_n-\rvec_m)^2 \ra.
\label{eq_theo_Rgyr}
\end{equation}
For chains following Gaussian statistics (as our lattice chains with switched off interactions)
where $\Rs^2 = \la  (\rvec_n-\rvec_m)^2 \ra = b^2 s$,
it follows by integration of Eq.~(\ref{eq_theo_Rgyr}) that $\Rgyr^2(N)= \RN^2/6 = d a^2 N/3$. 
For large wavevectors where the internal fractal chain structure is probed, the form factor decays as
\begin{equation}
F(q) = \frac{2}{(a q)^2} \sim N^0 \mbox{ for } q \Rgyr \gg 1 
\label{eq_theo_Fq_qlarge}
\end{equation}
as sketched by the thin line in Fig.~\ref{fig_theo_SqFq}.\footnote{The power-law scaling is obtained
directly by Fourier transformation of Eq.~(\ref{eq_theo_Grt0_inf}) which yields $F(q) \sim 1/q^{1/\nu}$.
This scaling only holds if the fractal object is ``open", i.e. $d > 1/\nu$,
and the scattering intensity is not dominated by the Porod scattering at the
(possibly fractal) ``surface" of a compact object \cite{BenoitBook,MSZ11}. 
The latter Porod scattering becomes relevant, e.g., for self-avoiding polymers in strictly $d=2$ dimensions \cite{MKA09,MWK10}
and for non-olympic and un-knotted rings in $d=3$ or thin films of finite width ($d=2^+$) \cite{MWC00,Vettorel09,MSZ11}. 
In both cases the polymers adopt compact configurations, $1/\nu=d$, with a fractal 
surface of well-defined fractal surface exponent $\ds \le d$. The form factor thus decays as
$F(q)/N \approx 1/(\RN q)^{2d-\ds}$ \cite{MSZ11}.}
More generally, the form factor of monodisperse Gaussian chains is given by $F(q) = N \fDebye(x)$ with
$x = (q \Rgyr(N))^2$ and the Debye function 
\begin{equation}
\fDebye(x) = \frac{2}{x^2} \left( \exp(-x)-1+x \right).
\label{eq_theo_Fq_Debye}
\end{equation}
For convenience of calculation, the Debye function is often replaced by the Pad\'e approximation 
\begin{equation}
F(q) \approx \frac{N}{1+ (q \Rgyr)^2/2}
\approx \frac{2}{(a q)^2 + 2/N}
\label{eq_theo_Fq_Pade}
\end{equation}
where the last step holds for $d=3$.

For a general mass distribution $\pchain$ the form factor is defined as \cite{BenoitBook,DescloizBook,BJSOBW07}
\begin{equation}
F(q) = \frac{1}{\Nav} \sum_{N=0}^{\infty} \pchain \sum_{n,m=1}^{N} 
\la \exp\left( - {\rm i} \qvec \cdot (\rvec_n-\rvec_m ) \right) \ra.
\label{eq_theo_Fq_poly}
\end{equation}
In practice, Eq.~(\ref{eq_theo_Fq_poly}) corresponds to the computation of the averaged sum 
over contributions $\la C^2 + S^2\ra$ for each chain which one divides finally
by the total number $\nmon$ of labeled mono\-mers \cite{BJSOBW07}.
In the small-$q$ regime one obtains again a Guinier relation
\begin{equation}
F(q) \approx \frac{\la N^2 \ra}{\Nav} \left(1 - \frac{(\Rgyrz \ q)^2}{d} \right)
\mbox{ for } \Rgyrz \ q \ll 1
\label{eq_theo_Fq_Guinier}
\end{equation}
where $\Rgyrz$ stands now for the $z$-averaged gyration radius \cite{DescloizBook} 
where the moment $p=2$ is taken over the standard radius of gyration $\Rgyr^2(N)$,
Eq.~(\ref{eq_theo_Rgyr}). Note that for Flory-distributed Gaussian chains we have 
$\Rgyrz^2 = d a^2/\mu$ and the
form factor becomes \cite{BJSOBW07}
\begin{equation}
F(q) = \frac{2}{(a q)^2 + \mu}.
\label{eq_theo_Fq_Flory}
\end{equation}
Eq.~(\ref{eq_theo_Fq_Flory}) reduces to Eq.~(\ref{eq_theo_Fq_qlarge})
for large wavevectors, as one expects, since in this limit $F(q)$ must become independent 
of the length distribution $\pchain$.
Note that Eq.~(\ref{eq_theo_Fq_Flory}) has the same form as the Pad\'e approximation 
for monodisperse chains, Eq.~(\ref{eq_theo_Fq_Pade}).

\subsubsection{Total monomer structure factor $S(q)$}
\label{theo_connectivity_Sq}
For non-interacting chains the local monomer density can freely fluctuate being re\-strict\-ed
only by the chain connectivity. The isothermal compressibility $\kappaT$ \cite{ChaikinBook}
is thus given by the osmotic contribution due to the density of chains,
i.e. $T \kappaT = 1/(\rho/\Nav)$ diverges linearly with the typical chain length $\Nav$. 
More generally, one may characterize the fluctuations of the total density 
$\rho(\qvec) = \sum_{n=1}^{\nmon} \exp(-i \qvec \cdot \rvec_n)$ 
in Fourier space by means of the total structure factor \cite{DoiEdwardsBook}
\begin{eqnarray}
S(q) & = & \frac{1}{\nmon} \sum_{n,m=1}^{\nmon} \la \exp\left(- {\rm i}\qvec \cdot (\rvec_n-\rvec_m) \right) \ra
\nonumber \\
& = & \frac{1}{\nmon} \la \left[\sum_n \cos(\qvec\cdot\rvec_n)\right]^2 + \left[\sum_n \sin(\qvec\cdot\rvec_n)\right]^2 \ra
\label{eq_theo_Sq}
\end{eqnarray}
with $\qvec$ being a  wavevector commensurate with the simulation box\footnote{If we use a cubic simulation 
box of linear dimension $L$ the smallest possible wavevector is $2\pi/L$.} 
and the thermal average $\la \ldots \ra$ being performed over all configurations of the 
ensemble and all possible wave\-vec\-tors of length $q = |\qvec|$.
Since we have switched off all mono\-mer interactions, mo\-no\-mers on different chains
are uncorrelated and, hence, 
\begin{equation}
\frac{1}{S(q)} = \frac{1}{F(q)} 
\label{eq_theo_sRPAnointer}
\end{equation}
as indicated for monodisperse chains by the thin line in Fig.~\ref{fig_theo_SqFq}.
Due to the chain connectivity the fluctuations of chains (measured by $\kappaT$) and the fluctuations of 
monomers (measured by $S(q)$) may differ for polydisperse non-inter\-act\-ing systems: 
While $S(q) \to \la N^2 \ra/\Nav$ in the small-$q$ limit as seen from Eq.~(\ref{eq_theo_Fq_Guinier}), 
we have $T \kappaT \rho = \Nav$ for the compressibility.\footnote{The dimensionless compressibility $\gT$ defined in 
Eq.~(\ref{eq_theo_gTdef}) for asymptotically long chains does not depend on this ideal chain contribution. 
Care is needed, however, if $\gT$ is determined numerically from $S(q)$ for polydisperse systems of finite 
$\la N^2 \ra/\Nav$ by extrapolation in analogy to the monodisperse case discussed in Sec.~\ref{TD_compress}.} 

\subsection{Incompressibility constraint}
\label{theo_incompress}

\subsubsection{Dimensionless compressibility $\gT$}
\label{theo_incompress_gT}

Obviously, dense polymer solutions and melts are essentially incompressible, i.e. $\kappaT \to 0$,
and the above assumption that the total density can freely fluctuate is not very realistic.
It is useful to introduce here a central dimensionless thermodynamic property characterizing the
degree of density fluctuations on large scales, the so-called ``dimensionless compressibility" 
\cite{BJSOBW07,WBM07,WCK09}
\begin{equation}
\gT \equiv \frac{1}{\veff \rho} \equiv \lim_{\Nav\to \infty} \left( T \kappaT \rho \right)
= \lim_{\Nav\to \infty} \left( \lim_{q\to 0} S(q) \right). 
\label{eq_theo_gTdef}
\end{equation}
At standard experimental polymer melt conditions $\gT$ remains of course finite, 
say $\gT \approx 0.1$, but typically well below unity as indicated by the bold dashed
line in Fig.~\ref{fig_theo_SqFq} \cite{RubinsteinBook}.
Note that we have defined $\gT$ in the limit of asymptotically long chains to take off the trivial 
compressibility contribution due to the translational invariance of the chains mentioned above. 
For later reference we have also introduced in Eq.~(\ref{eq_theo_gTdef}) the effective ``bulk modulus" 
$\veff$ \cite{ANS05b}. 
As indicated, the dimensionless compressibility can be determined directly in experiments or in a computer 
simulation from the low-$q$ limit of the total monomer structure factor $S(q)$.
This point is further elaborated in Sec.~\ref{TD_compress}.

\subsubsection{Lagrange multiplier $\overlap$}
\label{theo_incompress_overlap}

Physically, the incompressibility of dense polymer systems arises of course due to the short-range 
repulsion of the mono\-mers, i.e. it depends on non-uni\-ver\-sal physical and chemical properties. 
From the theoretical and computational point of view it is, however, inessential {\em how} the 
incompressibility at low wavevectors is imposed.\footnote{If the goal is to map a 
computational model onto a real polymer melt aiming to understand macroscopic properties, 
the starting point should be, in our opinion, to match the mechanical and thermodynamic properties
in the low-$q$ limit, e.g. the dimensionless compressibility $\gT$, 
rather than to fiddle with $S(q)$ on the monomer scale.} 
This constraint could be achieved, at least in principle, by ``simple sampling" \cite{LandauBinderBook} 
of only those configurations respecting the chosen $\gT$. In this sense it is thus the 
``throwing away" of configurations from the extended configuration ensemble containing all possible 
linear chain paths on the lattice which creates the repulsive forces between chains, subchains and 
monomers.\footnote{The scale-free correlations described in this paper are thus akin to the effective 
``anti-Casimir forces" which arise in dense polymer melts due to the throwing away of configurations 
containing closed loops from an extended configuration ensemble with both linear chains and rings 
\cite{ANS05a,ANS05b}.} Alternatively, one may design intricate local and global MC moves forcing 
the system through configuration space along a hyperplane of constant $\gT$ 
\cite{KrauthBook,BWM04}.

A more general and computationally more natural route is to use an extended ensemble
\cite{LandauBinderBook} and to impose the incompressibility constraint through an external field with 
a Lagrange multiplier $\overlap$ conjugated to the local monomer density fluctuations.
As may be seen in more detail in Sec.~\ref{bfm_soft} for a specific lattice model, 
this implies in practice that one has to pay an energy of order $\overlap$ for the overlap of two mono\-mers.
While in the low-$\overlap$ limit with $\gT(\overlap) \gg \Nav$ the chains do not interact, 
i.e. $S(q)=F(q)$ for all $q$, the incompressibility constraint $S(q) \approx \gT(\overlap=\infty) \ll 1$
holds for all wavevectors $q$ up to the monomeric scale ($q \approx 1/\sigma$)
in the opposite limit $\overlap \to \infty$. This is shown by the bold dashed line in Fig.~\ref{fig_theo_SqFq}.

\subsubsection{Thermal blobs of size $\xi$}
\label{theo_incompress_xi}

The situation is slightly more complicated for intermediate overlap penalties $\overlap$ 
with $1 \ll \gT(\overlap) \ll \Nav$ indicated by the thin dashed line. Since $\gT(\overlap)$ is 
now a well-defined characteristic chain length in curvilinear space along the chain contour, 
it corresponds to a characteristic scale in real space, the ``screening length" $\xi(\overlap)$
of the density fluctuations defined as \cite{DoiEdwardsBook}
\begin{equation}
\xi^2 \equiv \frac{1}{2} a^2 \gT = \frac{a^2}{2 \veff \rho}
\label{eq_theo_xidef}
\end{equation}
where we use the effective bulk modulus $\veff = 1/(\gT \rho)$ following 
Eq.~(\ref{eq_theo_gTdef}).\footnote{From the scaling point of view a curvilinear length $s \gg 1$ translates
quite generally to a spatial distance $r \gg \sigma$ and a wavevector $q \ll 1/\sigma$ according to 
$r \sim 1/q \sim s^{\nu}$ with $\nu$ being the inverse fractal dimension.} 
Generalizing Eq.~(\ref{eq_theo_sRPAnointer}) to systems with finite compressibility
the total structure factor is predicted to follow the so-called (static) 
``Random Phase Approximation" (RPA) \cite{DegennesBook,DoiEdwardsBook},
\begin{eqnarray}
\frac{1}{S(q)} & =  & \frac{1}{\gT} + \frac{1}{F(q)} \label{eq_theo_sRPA} \\
                 & \approx & \frac{1}{\gT} \left(1 + (\xi q)^2 \right) \label{eq_theo_sRPA2}
\end{eqnarray}
using Eq.~(\ref{eq_theo_Fq_qlarge}) in the second step.
The Ornstein-Zernike correlation equation \cite{DhontBook} Eq.~(\ref{eq_theo_sRPA2}) justifies 
the above definition of the ``screening length" $\xi$, i.e. density fluctuations
decay in $d=3$ exponentially as \cite{DoiEdwardsBook}
\begin{equation}
\la \rho(r) \rho(0) \ra - \rho^2 = \frac{3 \rho}{\pi b^2 r} \exp(-r/\xi).
\label{eq_theo_densitycorr}
\end{equation}
We remind that $\xi$ sets the size of the ``thermal blob" \cite{DegennesBook} corresponding to 
a free energy due to the effective monomer interaction of order $\kBT$.
If one considers short subchains of arc-length $s \ll \gT$ or small distances $r \ll \xi$, 
the (sub)chains behave as if they were barely interacting, i.e. $S(q) \approx F(q)$. 
If on the other side one focuses on the physical properties beyond the thermal blob scale 
($s \gg \gT$, $r \gg \xi$, $q \ll 1/\xi$) where the structure factor becomes constant, 
$S(q) \approx \gT(\overlap)$, one may renormalize 
all spatial distances by $\xi$ and all curvilinear distances by $\gT$ and in these terms the system 
should behave as an incompressible packing of thermal blobs \cite{DegennesBook}.

We emphasize that the perturbation results \cite{DoiEdwardsBook} 
Eq.~(\ref{eq_theo_sRPA}) and Eq.~(\ref{eq_theo_sRPA2}) are supposed to apply 
only as long as the compressibility $\gT$ is not too small and the screening length $\xi$ remains
a respectable length, in any case much larger than the lattice constant $\sigma$.\footnote{We
remind that from the thermodynamic point of view the fundamental property characterizing the solution
is the dimensionless compressibility $\gT$ and {\em not} the correlation length $\xi$.} Formally, this 
is expressed by the criterion \cite{DoiEdwardsBook,BJSOBW07,WBM07}
\begin{equation}
\Gi \equiv \frac{\gT}{\rho (b \gT^{1/2})^d} \approx \frac{(\veff \rho)^{d/2-1}}{\rho b^d} \ll 1,
\label{eq_Ginzburg}
\end{equation}
with the Ginzburg parameter $\Gi$ being the small parameter of the standard perturbation theory.
Note that Eq.~(\ref{eq_Ginzburg}) sets a lower bound to
the correlation length $\xi \gg b/(\rho b^d)^{1/(d-1)}$.

Please also note that Eq.~(\ref{eq_theo_xidef}), Eq.~(\ref{eq_theo_sRPA2}) and Eq.~(\ref{eq_Ginzburg})
are consistent with relations given by Edwards \cite{DoiEdwardsBook}. 
(For instance, Eq.~(\ref{eq_Ginzburg}) corresponds to Eq.~(5.46) of Ref.~\cite{DoiEdwardsBook}.)
The only difference is that following \cite{ANS05a,WBM07} we have replaced the second virial coefficient 
$\vvir$ of the mono\-mers by the effective bulk modulus $\veff$ and the bond length 
of the unperturbed chain by the effective bond length $b$.

\begin{figure}[t]
\centerline{\resizebox{1.0\columnwidth}{!}{\includegraphics*{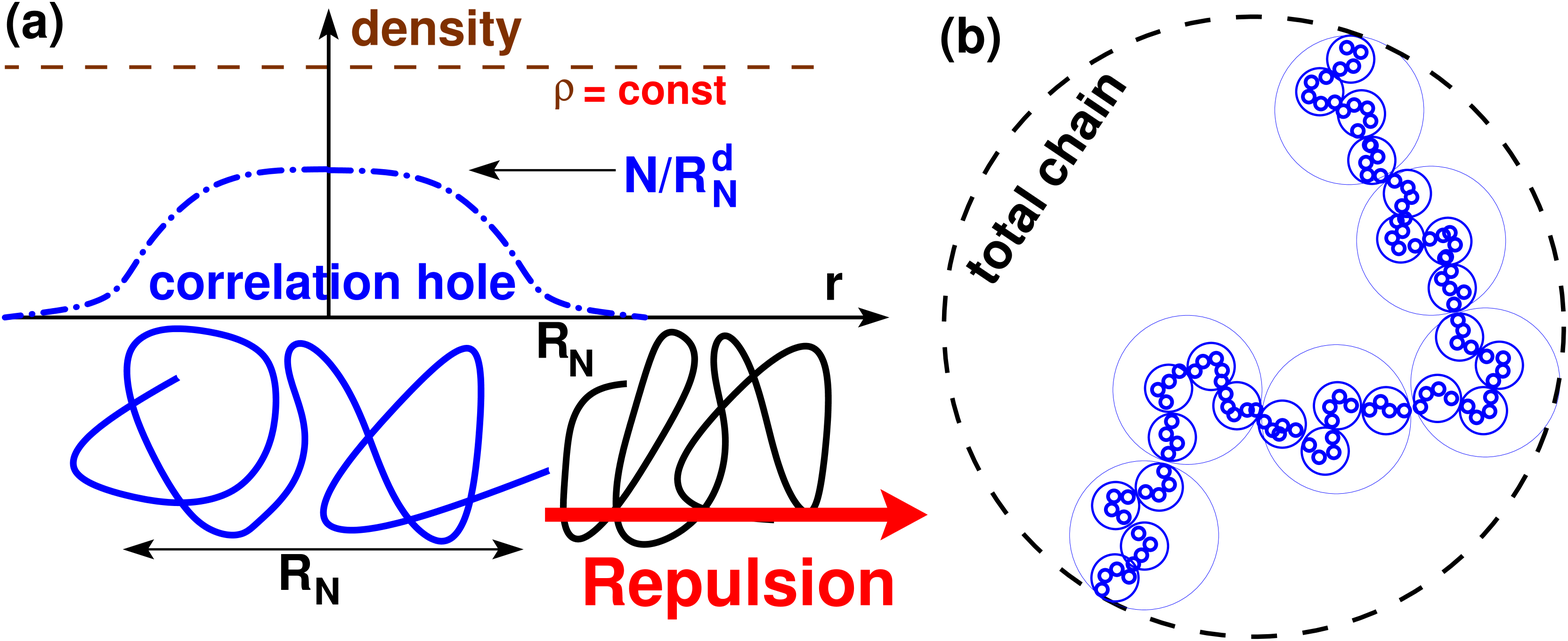}}}
\caption{Effective interactions due to the incompressibility constraint:
{\bf (a)}
Since the total density $\rho$ can barely fluctuate, chains of length $N$ are known to repel each other 
due to an entropic penalty $\UstarN \approx \rhostarN/\rho$ set by the self-density 
$\rhostarN \approx N/\RN^d$.
{\bf (b)}
Self-similar pattern of nested segmental correlation holes of decreasing strength
$\Ustars \approx s/\rho \Rs^d \approx \ce/\sqrt{s}$ in $d=3$ dimensions aligned along the backbone of a 
reference chain. 
The large dashed circle represents the classical correlation hole of the total chain ($s \approx N$).
This is the input of some recent approaches to model polymer chains as soft spheres 
\cite{EM01,Likos01,Guenza02}. 
We argue instead that the incompressibility constraint on {\em all} length scales $s$ matters
--- and not just for $s\approx N$ ---
leading to a short distance repulsion $\Ustars$ of their ``subchain correlation holes"
which increases with decreasing $s$.
\label{fig_theo_constraint}
}
\end{figure}
\subsubsection{Correlation hole effects}
\label{theo_corhole}

Let us for the clarity of the presentation return to monodisperse chains in the large-$\overlap$ limit, 
i.e. let us assume that $\gT \ll 1$ and that thus the total monomer density $\rho$ 
does not fluctuate. On the other hand, composition fluctuations of labeled chains or
subchains may certainly occur, however, subject to the total density constraint.
Composition fluctuations are therefore coupled and chains and subchains must feel an 
{\em entropic} penalty when the distance $r$ between their CM becomes comparable to their 
typical size \cite{ANS03,WBJSOMB07}.

\begin{figure}[tb]
\centerline{\resizebox{0.9\columnwidth}{!}{\includegraphics*{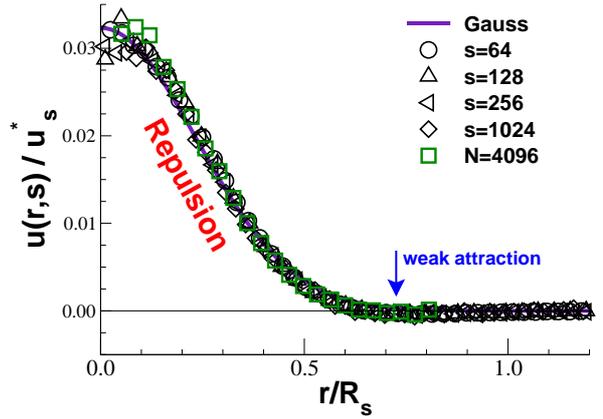}}}
\caption{Potential of mean force $u(r,s)$ for 3D polymer melts obtained using the classical BFM algorithm 
without monomer overlap ($\overlap=\infty$) described in Sec.~\ref{bfm_equil} for monodisperse chains of 
length $N=4096$ and volume fraction $\phi=0.5$ of occupied lattice sites \cite{WBCHCKM07}. 
Data for different $s \le N$ are successfully scaled tracing $u(r,s)/\Ustars$ as a function of 
$r/\Rs$ with $\Ustars \equiv s/\rho \Rs^3$.
The potential is positive and roughly Gaussian (bold line) for small $r/\Rs$ 
where the (sub)chains repel each other. It becomes weakly attractive 
for distances $r/\Rs \approx 0.7$ (vertical arrow).
\label{fig_theo_Ustar} 
}
\end{figure}

As sketched in Fig.~\ref{fig_theo_constraint}(a), let us first remind the well-known ``correlation
hole" effect for two test chains of length $N$ in the bath \cite{DoiEdwardsBook,ANS03}. 
The scaling of their effective interaction under the incompressibility constraint 
is obtained from the potential of mean force $u(r,N) \equiv - \ln(p(r,N)/p(\infty,N))$ \cite{ChandlerBook} 
with $p(r,N)$ being the pair correlation function of the chains, i.e. 
the probability distribution to find the CM of the second chain at a distance $r$ 
assuming the CM of the first chain at the origin ($r=0$).
Since the correlation hole is shallow for large $N$, expansion of the logarithm leads to
\begin{equation}
u(r,N) \approx 1 - \frac{p(r,N)}{p(\infty,N)} \approx 1 - \frac{\rho-\rhoN(r)}{\rho} 
= \frac{\rhoN(r)}{\rho}
\label{eq_theo_urN_expand}
\end{equation}
with $\rhoN(r)$ being the density distribution of the reference chain around its CM.
This distribution scales as 
\begin{equation}
\rhoN(r) \approx \rhostarN f(r/\RN)
\label{eq_theo_rhoNr_scal}
\end{equation}
with $\rhostarN \approx N/\RN^d$ being
the chain self-density, Eq.~(\ref{eq_intro_rhostarN}), and $f(x)$ a universal
function which becomes constant for $x \ll 1$ and decays rapidly for $x \gg 1$.\footnote{For
(to leading order) Gaussian chains $f(x)$ must also be Gaussian \cite{DegennesBook}
as indicated by the bold line in Fig.~\ref{fig_theo_Ustar}.}
The interaction penalty for two chains at $r/\RN \ll 1$ is thus given by 
\begin{equation}
u(0,N) \approx \UstarN \equiv \rhostarN/\rho \equiv N/\rho \RN^d \sim N^{1-d/2}
\label{eq_theo_UstarN}
\end{equation}
which decreases as $\UstarN \sim 1/\sqrt{N}$ in $d=3$ \cite{ANS03,WBJSOMB07}.
Hence, although the incompressibility constraint couples the chains, the correlations become
rapidly negligible with increasing chain length $N$ in agreement with our discussion 
in Sec.~\ref{intro_context}.

Interestingly, the above scaling argument does not only hold for chains ($s=N-1$) but also for 
the potential of mean force $u(r,s) \equiv - \ln(p(r,s)/p(\infty,s))$ obtained in a similar manner 
from the pair correlation function $p(r,s)$ of the center-of masses of subchains of arc-length 
$s \le N-1$ \cite{WBJSOMB07,WBM07}. 
Since 
\begin{equation}
\frac{\Ustars}{\UstarN} = (N/s)^{d/2-1} = \sqrt{N/s} \gg 1 \mbox{ for } s \ll N,
\label{eq_theo_UstarsUstarN}
\end{equation}
the correlation hole effect strongly increases with decreasing $s$,
albeit it remains always perturbative in $d=3$.  
Note that the subchain correlation hole potential $\Ustars$ does not depend explicitly on the bulk 
compression modulus $\veff$. It is dimensionless and independent of the definition of the monomer unit, 
i.e. it does not change if $\lambda$ monomers are regrouped to form an effective monomer
($\rho\to\rho/\lambda$, $s\to s/\lambda$) while keeping $\Rs$ fixed.
That the effective correlation hole potential for chains and subchains is more than a heuristic
scaling argument can be seen from Fig.~\ref{fig_theo_Ustar}. We present here the rescaled potential 
of mean force $u(r,s)$ for chains of length $N=4096$ obtained using the BFM algorithm described 
in Sec.~\ref{bfm_equil}.
The correlation hole potential for the total chains is indicated by the squares.
The predicted scaling is confirmed by the perfect data collapse of $u(r,s)/\Ustars$ 
plotted as function of $r/\Rs$. Please note that, strictly speaking, $u(r,s)$ for subchains
accounts also for the attractive interaction between subchains on the same chain and 
therefore differs slightly from the effective interaction potential of two {\em independent} 
chains of length $N=s+1$. This leads to the (very weak) attractive contribution to $u(r,s)$ for subchains 
(barely) visible in Fig.~\ref{fig_theo_Ustar}. However, this additional effect does not affect 
the scaling on short distances, $r \ll \Rs$, which matters here.

\subsubsection{Connectivity and swelling}
\label{theo_corhole_swelling}
To connect two test chains of length $N$ to form a chain of length $2N$ an effective free energy
$\UstarN$ has to be paid and this repulsion will push the two halves apart from each other \cite{ANS03}.
We consider next a subchain of length $s$ in the middle of a very long chain.
All interactions between the test subchain and the rest of the chain are first
switched off but we keep all other interactions, especially within the subchain 
and between the subchain monomers and monomers of surrounding chains.
The typical size $\Rs$ of the test subchain remains essentially unchanged
from the size of an independent chain of the same strand length.
If we now switch on the interactions between the tagged subchain and mono\-mers on
{\em adjacent} subchains of same length $s$,
this corresponds to an effective interaction of order $\Ustars$ as before.
(The effect of switching on the interaction to all other monomers of the chain
is inessential at scaling level, since these other monomers are more distant.)
Since this repels the respective subchains from each other, the corresponding subchain
is swollen compared to a Gaussian chain of non-interacting subchains.
It is this effect we want to characterize.

\subsubsection{Perturbation approach in three dimensions}
\label{theo_corhole_perturbation}
Let us return to systems of finite dimensionless compressibility $\gT$ 
but let us focus on subchains of length $s$ which are larger than the number 
of monomers $\gT$ contained in the thermal blob, i.e. we focus on scales
where the incompressibility constraint matters.
Interestingly, when taken at $s=\gT$ the subchain correlation hole potential becomes
\begin{equation}
\Ustar_{s=\gT} \approx \frac{\gT}{\rho (b \gT^{1/2})^d}
= \Gi
\label{eq_theo_UstarsGi1}
\end{equation}
with $\Gi$ being the standard Ginzburg parameter already defined in Eq.~(\ref{eq_Ginzburg}).
Hence, it follows for the subchain correlation hole potential that
\begin{equation}
\Ustars \approx \Gi (\gT/s)^{d/2-1} \ll \Gi \mbox{ for } d > 2 \mbox{ and }  s \gg g.
\label{eq_theo_UstarsGi2}
\end{equation}
Although for real polymer melts as for computational systems large values of $\Gi \approx 1$
may sometimes be found, 
$\Ustars \sim 1/\sqrt{s}$ decreases rapidly with $s$ in three dimensions,
as illustrated in Fig.~\ref{fig_theo_constraint}(b), and standard perturbation calculations
can be successfully performed.

As sketched in Sec.~\ref{theo_perturb} these calculations consider dimensionless quantities $\Ccal[\Ustar]$
which are defined such that they vanish ($\Ccal[\Ustar=0]=0$) if the perturbation
potential $\Ustars$ is switched off and are then shown to scale, to leading order,
linearly with $\Ustar$.
For instance, for the quantity $\Kp(s)$, defined in Eq.~(\ref{eq_theo_Kps_gauss}),
characterizing the deviation of the subchain size from Flory's hypothesis
one thus expects the scaling
\begin{equation}
\Kp[\Ustars] \approx + \Ustars \approx + \frac{s}{\rho \Rs^d}.
\label{eq_RsUs}
\end{equation}
The $+$-sign indicated marks the fact that the prefactor has to be positive to be consistent
with the expected swelling of the chains. Consequently, the rescaled mean-squared subchain size,
$\Rs^2/b^2 s \approx 1 - \Ustars$, must approach the asymptotic limit for large $s$
from below.
For 3D melts Eq.~(\ref{eq_RsUs}) implies that $\Kp(s)$ should
vanish rapidly as $1/(\rho b^3 \sqrt{s})$.\footnote{This is different in thin films where 
$\Ustars \approx \Gi$ decays only logarithmically \cite{ANS03}.}
Taking apart the prefactors --- which require a full calculation ---
this corresponds exactly to Eq.~(\ref{eq_intro_Rs}) with a
swelling coefficient $\ce \approx 1/\rho b^3$.
Note also that the predicted deviations are inversely proportional to $b^3$, i.e.
the more flexible the chains, the more pronounced the effect.
Similar relations $\Ccal[\Ustars] \sim \Ustars$ may also be formulated for other quantities
and will be tested numerically in Sec.~\ref{conf}. 

\subsection{Perturbation calculation}
\label{theo_perturb}

\subsubsection{General approach}
\label{theo_perturb_general}
We remind that the first-order perturbation calculation of an observable $\Acal$ under a 
dimensionless perturbation potential $U \ll 1$ (defined in units of $T$) generally reads \cite{DoiEdwardsBook}
\begin{equation}
\la \Acal \ra \approx \frac{\la \Acal (1 -U)\ra_0}{\la 1 - U \ra_0} \approx 
\la \Acal \ra_0 + \la U \ra_0 \la \Acal \ra_0 -\la U \Acal \ra_0
\label{eq_theo_Acalav}
\end{equation}
where averages performed over an unperturbed reference system of Gaussian chains are denoted 
$\la \ldots \ra_0$. Obviously, $\la \Acal \ra = \la \Acal \ra_0$ for $U \equiv 0$.
For the reduced quantity $\Bcal \equiv \Acal - \la \Acal \ra_0$ Eq.~(\ref{eq_theo_Acalav}) simplifies to 
\begin{equation}
\la \Bcal \ra \approx -\la U \Bcal \ra_0.
\label{eq_theo_Bcalav}
\end{equation}
For $\la \Acal \ra_0 \ne 0$ one may introduce the dimensionless observable $\Ccal = -\Bcal /\la \Acal \ra_0$ 
which, of course, also obeys Eq.~(\ref{eq_theo_Bcalav}). If $\la \Ccal U \ra_0 \approx \la U \ra_0$,
as is the case if self-energies and ultra-violet divergencies can be dumped into the reference of $U$, 
this is consistent with Eq.~(\ref{eq_RsUs}).
The observable $\Acal$ stands, for instance, for the $2p$-th moment $\Acal=\rvec_{nm}^{2p}$ of the vector 
$\rvec_{nm}=\rvec_m-\rvec_n$ between two monomers $n$ and $m=n+s$ on the tagged chain as shown in 
Fig.~\ref{fig_intro_lattice} \cite{WMBJOMMS04,WBM07} or 
for the scalar product $\Acal = \lvec_n \cdot \lvec_m / l^2$ of two bond vectors \cite{WJO10,ANS10a}.
If, as in the latter case, we have $\la \Acal \ra_0 = 0$ for linear chains by construction, 
one only has to compute $\la \Acal \ra = \la \Bcal \ra \approx -\la U \Bcal \ra_0$.

In this subsection we use $\bref= \sqrt{2d} \aref$ for the bond length of the unperturbed Gaussian 
reference chain which may {\em a priori} be different from the effective bond length $b =\sqrt{2d} a$ 
defined and measured according to Eq.~(\ref{eq_theo_bdef}). The reference bond length $\bref$ is a 
parameter which may be suitably chosen or adjusted in a Hartree-Fock iteration scheme \cite{ThijssenBook} 
(as shown in Sec.~\ref{conf_be}) for the specific problem and observable considered. If one is, 
for instance, interested in predicting the effective bond length $b$ for a weakly 
interacting system, a good trial value for $\bref$ should be the bond length $l$ of non-interacting 
chains on the lattice \cite{DoiEdwardsBook}. To be consistent the perturbation potential $U$ 
must in this case vanish if the interactions are switched off ($\veff \equiv 0$).
If on the other side the aim is to characterize the deviations from Flory's ideality hypothesis in 
incompressible melts ($\gT \ll 1$) one may naturally set $\bref=b$, i.e. one uses as reference the 
Gaussian chain which fits the chain on large scales (Sec.~\ref{theo_perturb_brenorm}). 
Obviously, in this case the perturbation potential $U$ must vanish in this low-wavevector limit such 
that $\la \Ccal \ra \approx - \la \Ccal U \ra_0 \to 0$. The last choice for $\bref$ turns out to be the 
best one in {\em all} cases where one does not need to {\em predict} the effective bond length $b$ but 
where it can be determined independently by {\em fitting} some large scale intrachain property such as 
the typical chain end-to-end distance $\RN$. 

The perturbation energy for a tagged test chain of length $\Ntest$ is given by the potential
\begin{equation}
U = \int_0^{\Ntest} {\rm d}k \int_0^k {\rm d}l \ \vpot(r_{kl}).
\label{eq_theo_Udef}
\end{equation}
As further discussed in Sec.~\ref{theo_perturb_vpot}, the effective interaction $\vpot(r)$ 
between two monomers of the test chain arises due to the presence of the bath of surrounding chains 
which screens the direct excluded volume interaction $\veff \delta(r)$.
The calculations are most readily performed in Fourier-Laplace space using definitions given in 
Sec.~\ref{theo_connectivity_Gqs}. 
%
See for instance the calculation presented in 
Appendix~\ref{app_stat_Grs} for the non-Gaussian contribution $\delta G(r,s) = G(r,s) - G_0(r,s)$ to the subchain size distribution
or Appendix~\ref{app_stat_Ps} for the bond-bond correlation function $\Pone(s)$.

\subsubsection{Effective interaction potential}
\label{theo_perturb_vpot}
We have still to specify the effective mono\-mer interaction $\vpot(q) = \Fcal[\vpot(r)]$
in reciprocal space with $q$ being the wavevector conjugated to the distance between two monomers $n$ and $m$ 
of the tagged chain.
Note that in general the test chain length $\Ntest$ and the mean chain length $\Nav$ of the bath may differ. 
Within linear response the effective pair interaction reads \cite{DoiEdwardsBook,BJSOBW07}
\begin{equation}
\frac{1}{\vpot(q) \rho} = \frac{1}{\veff \rho} + F_0(q).
\label{eq_theo_vpot_def}
\end{equation}
The first term stands here for the bare excluded volume interaction $\veff$ between monomers.
As we have already stressed above, Eq.~(\ref{eq_theo_gTdef}), thermodynamic consistency requires that 
$\veff$ is set by the excess contribution to the isothermal compressibility of the solution: 
$\veff \equiv 1/\gT \rho$ \cite{ANS05a,BJSOBW07,WCK09,GHF95}.\footnote{The bulk modulus $\veff$ only dominates 
$\vpot(q)$ for all $q$ in extremely compressible systems where $\gT \gg F_0(0) = \la N^2 \ra/\Nav$.}
$F_0(q)$ stands for the ideal chain intramolecular form factor of the bath of chains surrounding
the reference chain. According to Eq.~(\ref{eq_theo_Fq_poly}) the effective interaction $\vpot(q)$ 
depends thus in general on the length distribution $\pchain$ of the bath. 
We remind that for Flory-distributed melts the form factor is given by Eq.~(\ref{eq_theo_Fq_Flory}).
Replacing $\mu$ by $2/N$ this corresponds to the Pad\'e approximation, Eq.~(\ref{eq_theo_Fq_Pade}),
of the awkward Debye function for monodisperse chains.
%
%

Let us first assume that $F_0(q) \gg S(q) \approx \gT$, i.e. 
we assume $q \ll 1/\sigma$ in incompressible solutions ($\gT \ll 1$) 
and $q \ll 1/\xi$ for systems with a well-defined thermal blob ($\gT \gg 1$).
The effective interaction is then given by 
\begin{equation}
\vpot(q) \rho \approx \frac{1}{F_0(q)} \mbox{ for } q \ll 1/\xi \mbox{ and } q \ll 1/\sigma,
\label{eq_theo_vpot_lowq}
\end{equation} 
i.e. the effective interaction is given alone by the inverse structure factor of the bath
and does not depend {\em explicitly} on the compressibility $\gT$ of the solution.\footnote{Since
we shall set $\bref = b$ at the end of calculation and since the effective bond length $b$
depends on $\gT$, the effective potential $\vpot(q)$ depends {\em implicitly} on $\gT$.}
According to Eq.~(\ref{eq_theo_Fq_Guinier}) the effective potential becomes in the low-wave\-vec\-tor limit
\begin{equation}
\vpot_0 \equiv \vpot(q\to 0) = \frac{\la N \ra}{\la N^2 \ra \rho} \mbox{ for }  q \ll 1/\Rgyrz,
\label{eq_theo_vpotq0}
\end{equation}
i.e. $\vpot_0 = 2\mu/\rho$ for Flory-distributed melts and $\vpot_0 = 1/\rho N$ for mono\-dis\-perse melts. 
Long test chains are ruled by $\vpot_0$ which acts as a weak repulsive pseudo-potential with associated 
Fixman parameter $z \sim \vpot_0 \sqrt{\Ntest}$ \cite{DoiEdwardsBook}.\footnote{Characterizing 
the excluded volume interaction free energy of a chain with itself the Fixman parameter 
of a chain of length $N$ of excluded volume $\veff$ may be defined more generally as 
$$z(N) \equiv \veff (N/\RN^d)^2 \RN^d \approx (\veff/b^d) N^{2-d/2}.$$ 
Note that for a monodisperse melt of chain length $N = \gT$ and dimensionless compressibility
$\gT = 1/\veff \rho$ the Fixman parameter and the Ginzburg parameter become identical, $z(N=\gT) = \Gi$.} 
It follows that
\begin{equation}
z \gg 1 \ \mbox{ for } \Ntest \gg (\Nav^2/\Nav)^2 \approx \Nav^2
\label{eq_theo_zlarge}
\end{equation}
and the chains thus must swell and obey excluded volume statistics \cite{DegennesBook,SchaferBook}.\footnote{For 
this reason an upper bound is indicated, e.g., in Eq.~(\ref{eq_muEP_gauss}).}
We note that for a Flory-distributed bath Eq.~(\ref{eq_theo_vpot_lowq}) becomes
\begin{equation}
\vpot(q) \rho \approx \frac{1}{2} ((\aref q)^2 + \mu)
\mbox{ for } q \ll 1/\xi \mbox{ and } q \ll 1/\sigma
\label{eq_theo_vpot_Flory}
\end{equation}
which can also be used within the Pad\'e approximation ($\mu\equiv 2/N$)
for the calculation of monodisperse systems \cite{WMBJOMMS04,WBM07}.
More importantly, the effective potential becomes for intermediate wavevectors 
\begin{equation}
\vpot(q) \rho \approx \frac{(\aref q)^2}{2}
\mbox{ for } 1/\Rgyrz \ll q \ll 1/\xi
\label{eq_theo_vpot_q2}
\end{equation}
and this irrespective of the length distribution $\pchain$ of the bath.
Eq.~(\ref{eq_theo_vpot_q2}) lies at the heart of the announced power-law swell\-ing of (sub)chains,
Eq.~(\ref{eq_intro_keystat}) or Eq.~(\ref{eq_RsUs}) \cite{WMBJOMMS04,BJSOBW07,WBM07,WJO10}.

For later reference in Sec.~\ref{app_stat_Ps} we note that for a Flory-distributed bath of finite
compressibility the pair potential reads
\begin{equation}
\vpot(q) = \veff \frac{(\aref q)^2 + \mu}{(\aref q)^2 + (\aref/\xi)^2} \ \mbox{ for } q \ll 1/\sigma
\mbox{ and } \mu \gT \ll 1
\label{eq_theo_vpot_xi_Flory}
\end{equation}
where $\xi^2 \equiv \aref^2 \gT/2 =\aref^2/2\veff \rho$ following Eq.~(\ref{eq_theo_xidef}).
Allowing to characterize wavevectors below and above $1/\xi$, Eq.~(\ref{eq_theo_vpot_xi_Flory}) reduces to 
Eq.~(\ref{eq_theo_vpotq0}) for very low wavevectors and to Eq.~(\ref{eq_theo_vpot_q2}) in the intermediate 
wavevector range. 
In the limit of asymptotically long chains ($\mu \to 0$) Eq.~(\ref{eq_theo_vpot_xi_Flory}) becomes \cite{DoiEdwardsBook}
\begin{equation}
\vpot(q) = \veff \frac{q^2}{q^2 + \xi^{-2}} \mbox{ for } 1/\Rgyrz \ll q \ll 1/\sigma
\label{eq_theo_vpot_xi}
\end{equation}
which corresponds in real space to \cite{DoiEdwardsBook}
\begin{equation}
\vpot(r) = \veff \left(\delta(r) - \frac{\exp(-r/\xi)}{4\pi \xi^2 r} \right),
\label{eq_theo_vpot_xi_r}
\end{equation}
i.e. the effective potential consists of a strongly repulsive part $\veff \delta(r)$
of very short range ($r \approx \sigma$), and an attractive part of range $\xi$ 
stemming from the compression of the reference chain by the bath chains.\footnote{See 
Fig. II.1 of de Gennes' book \cite{DegennesBook}.}
Using Eq.~(\ref{eq_theo_vpot_xi}) it follows that 
\begin{equation}
\left. \int {\rm d}\rvec \ \vpot(\rvec) = \vpot(q) \right|_{q=0} = 0
\label{eq_theo_vpot_int}
\end{equation}
which is commonly taken as a proof that ``there is no excluded volume interaction among the
segments whose mean separation is larger than $\xi$" \cite{DoiEdwardsBook}. Unfortunately,
Eq.~(\ref{eq_theo_vpot_int}) does not imply mathematically that all other moments, 
say the integral over $\vpot(\rvec) \rvec^{2p}$, should also rigorously vanish.
It is thus incorrect to state that {\em all} possible correlation functions must be short-ranged. 
However, it remains relevant that $\vpot(q) \sim q^2$ vanishes with decreasing wavevector
and the same applies for the total perturbation $U(q)$ to the Gaussian reference.\footnote{Hence, 
$\la \Ccal \ra \to 0$ in the large-scale limit for an observable $\Ccal$ which suggests 
$\bref \equiv b$ for the Gaussian chain reference bond length.} 

\subsubsection{Free energy for high compressibilities}
\label{theo_FE}
For later use in Sec.~\ref{TD} we reformulate here a perturbation calculation result obtained long 
ago by Edwards \cite{DoiEdwardsBook} using Eq.~(\ref{eq_theo_vpot_xi}) which allows to predict  
thermodynamic properties of melts with sufficiently large compressibilities $\gT$.
Integrating twice with respect to the density $\rho$ the osmotic pressure given by Eq.~(5.45) or Eq.~(5.II.5) 
of \cite{DoiEdwardsBook} one obtains for monodisperse melts the free energy per monomer 
\begin{eqnarray}
\beta f(\beta)
& = &
\beta \eself + \frac{1}{N} \log(\rho \sigma^3/N) + \frac{1}{2} \vvir \rho
\nonumber \\
& - &
\underline{\frac{1}{12\pi}
\frac{1}{\xi^3\rho}} \mbox{ with } \xi^2 \equiv \frac{l^2}{12 \vvir \rho}
\label{eq_FE_Thigh}
\end{eqnarray}
and $\beta=1/T$ being the inverse temperature, $\bref=l$ the bond length of the Gaussian 
reference chains and
$\vvir(\beta)$ the second virial coefficient of a solution of unconnected mono\-mers. 
The first term $\beta \eself$ is due to the (essentially constant) intrachain self-energy which shall be 
discussed in Sec.~\ref{TD_energy}. It is due to the reference energy chosen in our numerical model Hamiltonian and 
it is normally not accessible experimentally. A similar intrachain energy contribution to the free energy arises 
also from Eq.~(5.43) of \cite{DoiEdwardsBook} if an upper cutoff $q_\mathrm{max}$ is introduced for the wavevectors $q$ 
to avoid the ultra-violet divergence. Such an upper cutoff is justified by the discreteness of 
the monomers of real polymers. This leads necessarily to a {\em non-universal} free energy contribution
which can be seen as an integration constant with respect to the integration of a {\em measurable} property
such as the osmotic pressure or the compressibility. 
The second term in Eq.~(\ref{eq_FE_Thigh}) represents the translational invariance of monodisperse 
chains of length $N$ (van't Hoff's law). Due to this contribution the compressibilities depend in general on $N$
as will be discussed in Sec.~\ref{TD_compress}.\footnote{For general polydisperse melts of given partial
densities $\rhochain = \rho \pchain/\Nav$ the ideal gas contribution becomes $\sum_N \rhochain \log(\rhochain) / \rho$.}
%
The (bare) excluded volume interaction between the mono\-mers is accounted for by the third term. 
The underlined term in the second line represents the leading correction to the previous term due to the fact
that the monomers are connected by bonds summing over the density fluctuations to quadratic order.
As one expects \cite{DegennesBook}, the corresponding correlations of the density fluctuations
reduce the free energy by about one $\kBT$ per thermal blob of volume $\xi^3$.
Interestingly, according to Edwards \cite{DoiEdwardsBook} the chain connectivity,
i.e. the presence of attractive forces between bonded monomers, does not change
the excluded volume $\vvir$ --- as one would expect naively --- but rather gives
rise to an additional term scaling differently with density.\footnote{A free energy contribution 
$\approx 1/\RN^3\rho \sim 1/N^{3/2}$ may be added to Eq.~(\ref{eq_FE_Thigh})
if one insists on taking as reference for the connectivity contribution to the free energy 
the limit $\gT \to N$, i.e. $\xi \to \RN$, where the chain connectivity becomes irrelevant.}
Various thermodynamic pro\-per\-ties are readily obtained from the quoted free energy and will be 
compared with our numerical results in Sec.~\ref{TD}. The underlined density fluctuation contribution to 
the free energy will be demonstrated numerically from the scaling of the specific heat $\cV$ (Sec.~\ref{TD_cV}).

As the reader might have noticed we have written the free energy in Eq.~(\ref{eq_FE_Thigh}) following Edwards 
assuming $\veff \equiv \vvir$ for the bare monomer interaction and $\bref \equiv l$ for the bond length of the 
Gaussian reference chain. As already alluded to above (Sec.~\ref{theo_perturb_general}), one would nowadays 
rather set $\veff \equiv 1/\gT \rho$ and $\bref \equiv b$ using the imposed or measured 
dimensionless compressibility $\gT$ and the measured effective bond length $b$. However, the choice
of Edwards has a clear advantage: $\gT$ and $b$ may not be known with sufficient 
precision while the second virial coefficient $\vvir$ and the bond length $l$ can always be calculated from the 
given model Hamiltonian. 
According to Eq.~(5.46) of Ref.~\cite{DoiEdwardsBook} the stated free energy is supposed to hold in the limit 
where $\Gi \ll 1$ with $\Gi \sim 1/\sqrt{\gT}$ being the Ginzburg parameter.
As we shall see in Sec.~\ref{TD}, this restricts the validity of the related predictions to rather weak values
of the (reduced) Lagrange multiplier $\overlap/T$ applied to control the compressibility.
In the range of validity of Eq.~(\ref{eq_FE_Thigh}) it turns out that 
$\vvir \approx 1/\gT \rho$ and $l \approx b$, i.e. the difference between both parameter choices
correspond to irrelevant higher order corrections.

\begin{figure}[t]
\centerline{\resizebox{1.0\columnwidth}{!}{\includegraphics{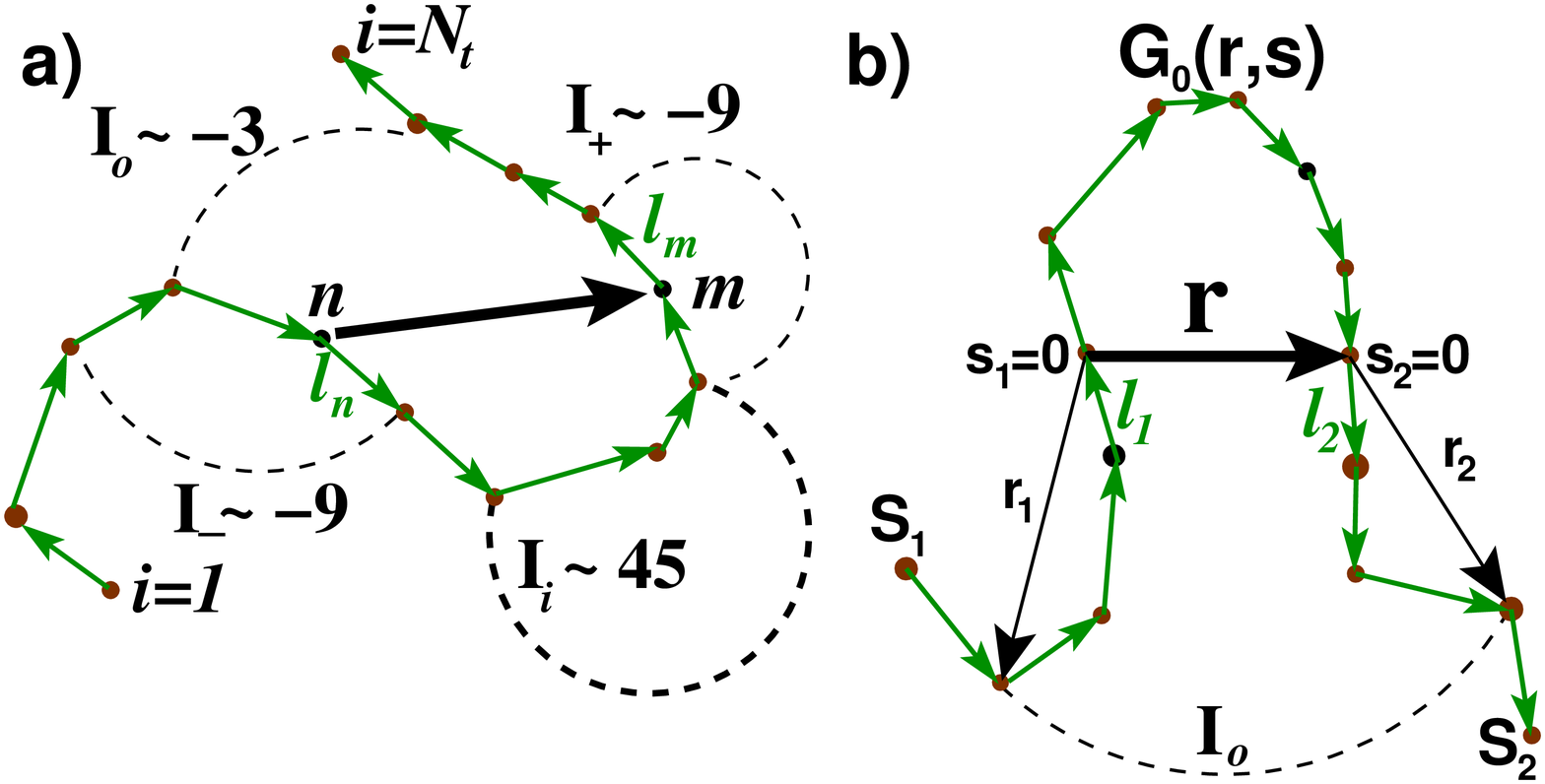}}}
\caption{Sketch of relevant interaction diagrams for the perturbation calculation of a long test chain 
of length $\Ntest$.
The dashed lines indicate the effective monomer interaction $\vpot(r)$ between two monomers $k$ and $l$
on the chain whose distance is weighted using the Gaussian propagator $G_0(r,|k-l|)$.
{\bf (a)} The calculation of the subchain size $\Rs^2$ requires the computation 
of four graphs, the dominant contribution stemming from the interactions of monomers within the subchain.
The numerical factors indicate the relative weights contributing to the $1/\sqrt{s}$ 
predicted by Eq.~(\ref{eq_intro_Rs}) \cite{WMBJOMMS04,WBM07}.
{\bf (b)} The direct calculation of the bond-bond correlation function $\Pone(s)$ is simplified by choosing 
two bonds $\lonevec$ and $\ltwovec$ {\em outside} the $s$-segment. For symmetry reasons only the graph $\Iouter$ 
for the effective interactions between the monomers in the two dangling tails of lengths $S_1$ and $S_2$ 
gives a non-vanishing contribution.
\label{fig_theo_graphD3}
}
\end{figure}
\subsubsection{Subchain size distribution and its moments}
\label{theo_perturb_Grs}

We turn now to the perturbation calculation predictions of intrachain conformational properties 
which will be com\-par\-ed with our numerical data in Sec.~\ref{conf}. We focus first on the scale-free
wavevector regime for arbitrarily long chains described by the effective interaction potential
Eq.~(\ref{eq_theo_vpot_xi}), i.e. effects related to the chain length distribution $\pchain$ are irrelevant.
For the observable $\Acal = \rvec_{nm}^2$ with $1 \ll n < m = n+s \ll \Ntest$ we indicate in 
Fig.~\ref{fig_theo_graphD3}(a) the different interaction graphs one may compute in real space \cite{WMBJOMMS04}
\begin{eqnarray}
\Iinner & = & \frac{12}{\pi} \frac{\veff \xi}{\bref^4} \bref^2 s
+                   \frac{-45}{\sqrt{24 \pi^3} \bref^3 \rho} \bref^2 \sqrt{s} \nonumber \\
\Iplus = \Iminus & = &   \frac{9}{\sqrt{24 \pi^3} \bref^3 \rho} \bref^2 \sqrt{s} \nonumber  \\
\Iouter & = &           \frac{3}{\sqrt{24 \pi^3} \bref^3 \rho} \bref^2 \sqrt{s}.
\end{eqnarray}
The diagram $\Iinner$ corresponds to the standard graph computed by Edwards for the total chain
($s=\Ntest-1$) \cite{DoiEdwardsBook}. Consistent with Edwards its leading Gaussian contribution 
describes how the effective bond length is increased from $\bref$ to $b$ under the
influence of a {\em small} excluded volume interaction inside the subchain between $n$ and $m$.
Note that all other contributions proportional to $\sqrt{s}$ correspond to the leading
non-Gaussian corrections predicted in Eq.~(\ref{eq_intro_Rs}). They only depend on
$\bref$ and $\rho$ but, more importantly, not on $\veff$ 
in agreement with the scaling discussed in Sec.~\ref{theo_corhole_perturbation}.
The relative weights of these four contributions are indicated in Fig.~\ref{fig_theo_graphD3}(a)
in units of $-\bref^2 \sqrt{s} / (\sqrt{24\pi^3} \bref^3 \rho)$.
The dominant correction stems from the interaction $\Iinner$ within the subchain.
The diagrams $\Iplus$ and $\Iminus$ are obviously identical in the scale-free limit.\footnote{Using 
Eq.~(\ref{eq_intro_PonesRsrelated}) it follows that the interactions described by the
strongest graph $\Iinner$ align the bonds $\lvec_n$ and $\lvec_m$ while the others tend to
reduce the effect \cite{WMBJOMMS04}. As shown in Fig.~\ref{fig_theo_graphD3}(b),
it is better to place the bond pair outside the subchain if one computes $\Pone(s)$ directly 
(Sec.~\ref{theo_perturb_bondbond} and Sec.~\ref{app_stat_Ps}).
For symmetry reasons only the interaction graph $\Iouter$ between the dangling ends matters
for the alignment of the bond pair. Both pictures are consistent and lead to the same result.} 
Summing over all contributions this yields
\begin{eqnarray}
\la \rvec_{nm}^2 \ra 
& = & \nonumber
\bref^2 s + \Iinner + \Iplus + \Iminus + \Iouter \\
& = &
b^2 s \left( 1 - \frac{\ce}{\sqrt{s}} \left( \frac{b}{\bref} \right) \right).
\label{eq_theo_R2pert}
\end{eqnarray}
where in the second line we have used the definition $\ce \equiv \sqrt{24/\pi^3}/\rho b^3$
already mentioned in Sec.~\ref{intro_key} and have set
\begin{equation}
b^2 \equiv \bref^2 \left(1 + \frac{12}{\pi} \frac{\veff \xi}{\bref^4} \right)
= \bref^2 \left(1 + \frac{\sqrt{12}}{\pi} \Giref \right)
\label{eq_theo_bepert}
\end{equation}
with $\Giref \equiv \sqrt{\veff\rho} / \bref^3 \rho$. 

For higher moments of the distribution $G(r,s)$ it is convenient to calculate first the perturbation
deviations of the Fourier-Laplace transformation $\delta \Ghat(q,t) = \Lcal[\Fcal[\delta G(r,s)]]$ 
with $\delta G(r,s) = G(r,s) - G_0(r,s)$ and to obtain the moments from the coefficients 
of the expansion of this ``generating function" in terms of the squared wavevector $q^2$.
As explained in detail in the Appendix~\ref{app_stat_Grs}, this leads to a deviation
\begin{equation}
\delta G(r,s) =  
\left( \frac{3}{2\pi \bref^2 s} \right)^{3/2} \exp\left(- \frac{3}{2} \frac{r^2}{\bref^2s} \right)
\frac{\ce}{\sqrt{s}} \left(\frac{b}{\bref} \right)^3 f(n)
\label{eq_theo_Grspert}
\end{equation}
with $n=r/\bref\sqrt{s}$ and the universal function
\begin{equation}
f(n) = \sqrt{\frac{3\pi}{32}} \left(- \frac{2}{n} + 9n - \frac{9}{2} n^3 \right)
\label{eq_theo_fn}
\end{equation}
which allows to specify all moments of $G(r,s)$ \cite{WBM07}.

\subsubsection{Adjusting the bond length of the reference chain}
\label{theo_perturb_brenorm}
The above perturbation result Eq.~(\ref{eq_theo_bepert}) is of relevance to describe the effect of a 
{\em weak} excluded volume $\veff$ on a reference system of ideal polymer melts with bond length $\bref=l$ 
where {\em all} interactions have been switched off ($\veff = 0$). It is expected to give a good estimation 
for the effective bond length $b$ only for a small Ginzburg parameter $\Gi \approx \Giref \ll 1$. 
For the dense incompressible melts we want to describe the latter condition does not hold and 
one cannot hope to find a good quantitative agreement with Eq.~(\ref{eq_theo_bepert}). 
Note also that large wavevectors contribute strongly to the leading Gaussian term. 
The effective bond length $b$ is, hence, strongly influenced by local and non-universal effects and is very 
difficult to predict in general (Sec.~\ref{conf_RNRs}).

Our more modest goal is to predict the coefficient of the $1/\sqrt{s}$-perturbation
and to express it in terms of a suitable variational reference Hamiltonian characterized
by a conveniently chosen $\bref$ and the {\em measured} effective bond length $b$
(instead of Eq.~(\ref{eq_theo_bepert})).
Following Refs.~\cite{Edwards82,WBM07} we argue that for dense melts
$\bref$ should be renormalized to $b$ to take into account higher order graphs.\footnote{The 
general scaling argument discussed in Sec.~\ref{theo_corhole_perturbation} states
that we have only {\em one} relevant length scale in this problem, the typical subchain size 
$\Rs \approx b \sqrt{s}$ itself. The incompressibility constraint cannot generate an additional scale.
It is this size $\Rs$ which sets the strength of the effective interaction which then in turn feeds back 
to the deviations of $\Rs$ from Gaussianity. Having a bond length $\bref$ in addition to the effective 
bond length $b$ associated with $\Rs$ would imply a {\em second} length scale $\bref \sqrt{s}$.}
Restating thus Eq.~(\ref{eq_theo_Grspert}) with $\bref \equiv b$ the subchain size distribution 
may be rewritten
\begin{equation}
\frac{\delta G(r,s)}{G_0(r,s)} = \frac{\ce}{\sqrt{s}} f(n)
\label{eq_theo_Grs_perturb}
\end{equation}
and for the $2p$-th moment of distribution this yields
\begin{equation}
\la \rvec^{2p}_{nm} \ra = 
\frac{(2p+1)!}{6^pp!} (b^2 s)^p 
\left( 1 - \frac{3 (2^p p! p)^2}{2 (2p+1)!} \frac{\ce}{\sqrt{s}} \right)
\label{eq_theo_Rps_perturb}
\end{equation}
which reduces for $p=1$ to Eq.~(\ref{eq_intro_Rs}) as stated in the Introduction.
As a consequence the non-Gaussianity parameters $\Kp(s)$ and $\alphap(s)$ defined in Sec.~\ref{theo_connectivity_Gqs} 
become
\begin{equation}
\Kp(s) = \frac{3(2^pp!p)^2}{2(2p+1)!} \frac{\ce}{\sqrt{s}}
\label{eq_theo_Kps_perturb}
\end{equation}
and
\begin{equation}
\alphap(s) = \left( \frac{3(2^pp!p)^2}{2(2p+1)!} - p \right) \frac{\ce}{\sqrt{s}}.
\label{eq_theo_alphaps_perturb}
\end{equation}
Eq.~(\ref{eq_theo_alphaps_perturb}) can be obtained from Eq.~(\ref{eq_theo_Kps_perturb}) by expanding
the second moment ($p=1$) in the denominator of the definition Eq.~(\ref{eq_theo_alphaps_gauss}).

\subsubsection{Bond-bond correlation function}
\label{theo_perturb_bondbond}
The bond-bond correlation function $\Pone(s)$ is a central property since it allows to probe directly 
the colored forces acting on the reference chain due to the incompressibility constraint, 
Eq.~(\ref{eq_app_displ5}). 
Using Eq.~(\ref{eq_intro_PonesRsrelated}) $\Pone(s)$ may be obtained by differentiating the
second moment $\Rs^2$ of the subchain size distribution $G(r,s)$ with respect to the arc-length $s$.
For arbitrarily large chains and $s \gg \gT$ this yields $\Pone(s) = \cP/s^{3/2}$ with
$\cP = \cinf \ce /8$ as announced in the Introduction.
%


It is also possible to obtain $\Pone(s)$ directly by averaging the observable $\Acal = \lvec_{n} \cdot \lvec_{m} / l^2$. 
Since for linear chains $\la \Acal \ra_0 =0$ by construction, the task is to compute Eq.~(\ref{eq_theo_Bcalav}).\footnote{We 
remind that for closed cycles the ring closure implies long-range angular correlations even for Gaussian chains, 
hence for rings $\la \Acal \ra \approx \la \Acal \ra_0 \ne 0$ to leading order \cite{WJO10}.}
Changing slightly the notations as indicated in Fig.~\ref{fig_theo_graphD3}(b), we consider
two bonds $\lonevec$ and $\ltwovec$ outside the $s$-segment. The lengths of the two tails of the
chains are denoted $S_1$ and $S_2$. One of the tails, say $S_2= \Ntest-s-S_1-1$, may be fixed
by the total length of the test chain. 
Placing the head of the first bond $\lonevec$ at the origin we consider the correlation function
$C_1(r) = \la \lonevec(0) \cdot \ltwovec(r) \ra/l^2$ between two bonds separated by the
distance $r=|\rvec|$. Interestingly, it can be seen by symmetry considerations that $C_1(r)$
does only depend on the effective interaction of the monomers in the first tail with the
monomers in the second tail and not on the monomers in the intermediate strand of length $s$.
Hence, we need only to calculate {\em one} interaction graph as opposed to the four graphs
required by the calculation of $\Pone(s)$ throught $\Rs^2$. It is for this reason we have chosen the
indicated positions of heads and tails of the bond vectors. 
With $B(\lvec)$ denoting the normalized distribution of the bond vector $\lvec$ of the polymer model considered,
the interaction graph in real space may be written  
\begin{eqnarray}
C_1(r) & = & - \int {\rm d}\lonevec {\rm d}\ltwovec B(\lonevec) B(\ltwovec) \ 
\Acal(\lonevec,\ltwovec)
\nonumber \\
& \times& \sum_{s_1=0}^{S_1} \sum_{s_2=0}^{S_2} \int {\rm d}\ronevec {\rm d}\rtwovec
G_0(\ronevec+\lonevec,s_1) G_0(\rtwovec-\ltwovec,s_2)
\nonumber \\
& \times& \vpot(\rtwovec+\rvec-\ronevec)
\label{eq_theo_Cone_r}
\end{eqnarray}
where $\ronevec$ points from the head of the bond $\lonevec$ to the monomer $s_1$ in the first tail 
and $\rtwovec$ from the tail of the bond $\ltwovec$ to the monomer $s_2$ in the second dangling chain end.
As the $s$-segment is not implied in the perturbation of $C_1(r)$, the constraint which consists in 
putting the two points on the same chain and putting a $s$-strand between them introduces, to lowest order,
the Gaussian propagator $G_0(r,s)$. Using Parseval's theorem the bond-bond correlation function reads
\begin{equation}
\Pone(s) = \int {\rm d}\rvec \ C_1(r) G_0(r,s) = \int \frac{{\rm d}\qvec}{(2\pi)^d}
C_1(q) G(q,s)
\label{eq_theo_Cone2Pone}
\end{equation}
with $C_1(q)$ being the Fourier transform of $C_1(r)$.\footnote{It can be shown that for infinite chains
$C_1(q) = 4 \vpot(q)/(bq)^2$ and, hence, $$C_1(r)= \frac{\veff}{\pi b^2 r} \exp(-r/\xi) \mbox{ in } d=3,$$
i.e. $C_1(r) \to 0$ for $\xi \to 0$ at fixed distance $r$.
\label{foot_theo_Cone}
Note that $C_1(r)$ is a mere technical intermediate quantity which should not be confused with the 
bond-bond correlation function $\Pone(r)$ discussed in Sec.~\ref{conf_Pr} and Sec.~\ref{app_stat_Pr}.} 
To simplify the notations we set from the start $\bref=b$, i.e. we take the effective bond length
as the bond length of the Gaussian reference chain. Although this is not strictly necessary, the calculation 
in reciprocal space may be strongly simplified by assuming the bond vector distribution to be a Gaussian
$B(\lvec) \equiv G_0(\lvec,s=1)$ with $b^2 \equiv 2d a^2 \equiv l^2 \equiv \la \lvec^2 \ra$.
This implies that the chain is perfectly flexible, i.e. $\cinf = (b/l)^2 = 1$.\footnote{The complete formula 
for systems of general rigidity can be recovered by multiplying the final perturbation calculation result with 
$\cinf$ as may be seen by scaling considerations \cite{WJO10,ANS10a} or by simply comparing the result with 
the bond-bond correlations obtained using $\Rs^2$.}
Under these premises a bond vector $\lvec$ may be represented in reciprocal space as
\begin{equation}
\Fcal[\lvec B(\lvec)] = {\rm i} \partial_q B(q) \approx {\rm i} a^2 \  2 \qvec
= \frac{{\rm i}}{d} \ b^2 \ \qvec
\label{eq_theo_bondFourier}
\end{equation}
with $B(\qvec) = \Fcal[B(\lvec)]= \exp(-(aq)^2)$ being the Fourier transformed bond vector distribution.\footnote{For
a general bond vector distribution one may expand $B(\qvec)$ at low momentum as indicated in Sec.~\ref{app_stat_moments}.}
Let us denote the wave\-vec\-tors conjugated to the bonds $\lonevec$ and $\ltwovec$ by $\qonevec$ and $\qtwovec$,
respectively. The Fourier transform of the observable $\Acal(\lonevec,\ltwovec)$ thus reads
\begin{equation}
\Acal(\qonevec,\qtwovec) = - \frac{b^2}{d^2} \qonevec \cdot \qtwovec.
\label{eq_theo_AcalFourier}
\end{equation}
Using Eq.~(\ref{eq_theo_AcalFourier}) for the observable and Eq.~(\ref{eq_theo_vpot_xi}) for the
interaction potential for Flory-distributed systems of given compressibility $\gT$ one may integrate 
Eq.~(\ref{eq_theo_Cone2Pone}) in reciprocal space as shown in Appendix~\ref{app_stat_Ps}. 
In the limit of very long chains ($\mu \to \infty$) one obtains in $d=3$ dimensions \cite{WCK09} 
\begin{equation}
\Pone(s) = \frac{\cP}{\gT^{3/2}} \left(\frac{4}{\sqrt{u}} - 4 \sqrt{2\pi} e^{2u} \mbox{erfc}(\sqrt{2u})
\right)
\label{eq_theo_Ponesoft}
\end{equation}
as a function of the reduced arc-length $u=s/\gT$ with $\mbox{erfc}(x)$ being the complementary
error function \cite{abramowitz}. As one expects, Eq.~(\ref{eq_theo_Ponesoft})
reduces to Eq.~(\ref{eq_intro_keystat}) for large $u \gg 1$, i.e. irrespective of the
compressibility $\gT$ the bond-bond correlation function behaves as in the incompressible limit. 
In the opposite limit where the structure within the thermal blob is probed Eq.~(\ref{eq_theo_Ponesoft})
corresponds to the weaker decay
\begin{equation}
\Pone(s) \approx \frac{\cP}{\gT^{3/2}} \frac{4}{\sqrt{u}}. 
\label{eq_theo_Ponefixm}
\end{equation}
This regime is consistent with the classical expansion result of the chain size in terms
of the Fixman parameter $z(s) \approx \veff \sqrt{s}/b^3$ \cite{WCK09}.\footnote{Omitting all
prefactors we remind \cite{DoiEdwardsBook} that to leading order $\Rs^2 \approx b^2s (1 + z(s) \ldots)$.
Using Eq.~(\ref{eq_intro_PonesRsrelated}) it follows that $\Pone(s) \sim 1/\gT\sqrt{s}$.} 
We therefore refer to this limit as the ``Fixman regime".

\subsubsection{Finite chain size effects}
\label{theo_perturb_Neffects}
To describe properly finite chain size corrections, Eq.~(\ref{eq_theo_vpot_q2}) must be
replaced by the general formula Eq.~(\ref{eq_theo_vpot_def}).
For mono\-disperse chains ($N=\Nav=\Ntest$) the form factor $F_0(q)$ is given by Debye's function Eq.~(\ref{eq_theo_Fq_Debye}).
This approximation allows in principle to compute, e.g., the mean-squared
total chain end-to-end distance, $\Acal = (\rvec_N - \rvec_1 )^2$.
One verifies readily (see \cite{DoiEdwardsBook}, Eq.~(5.III.9)) that the effect of
the perturbation may be expressed as
\begin{eqnarray}
& & \la \Acal \ra_0 \la U \ra_0 - \la \Acal U \ra_0
= \int \frac{{\rm d}\qvec}{(2\pi)^3} \
\vpot(q) \ 4 (\aref q)^2 \aref^2
\nonumber \\
& & 
\times \int_0^N {\rm d}s \ s^2(N-s)\exp\left(-(\aref q)^2\right).
\label{eq_theo_Rendcorrect1}
\end{eqnarray}
We take now first the integral over $s$. In the remaining integral over $q$ small $q$ wavevectors 
contribute to the $\sqrt{N}$-swelling while large $q$ renormalize the effective bond length of the 
dominant Gaussian behavior linear in $N$ (as discussed above). Since we wish to determine
the non-Gaussian corrections, we focus on small wavevectors $q \ll 1/\xi$, i.e.
the effective interaction potential is given by Eq.~(\ref{eq_theo_vpot_lowq}).
We thus continue the calculation using $\vpot(N,x)\rho = 1/(N\rho \fDebye(x))$
with $\fDebye(x)$ being Debye's function and $x=(\Rgyr(N) q)^2=(\aref q)^2 N$.
This allows us to express the swelling as
\begin{equation}
1 - \frac{\la (\rvec_N - \rvec_1)^2 \ra }{b^2 N} =
\frac{\ce}{\sqrt{N}} \ I(x_u).
\label{eq_theo_Rendcorrect2}
\end{equation}
We have set here $\bref=b$ in agreement with the renormalization of the
reference bond length discussed above.
The numerical integral $I(x_u) = \int_0^{x_u} dx \ldots$
over $x$ is slowly convergent at infinity. As a consequence the estimate
$I(\infty)=1.59$ may be too large for moderate
chain lengths. In practice, convergence is not achieved for values
$x_u(N) \approx (b/\xi)^2 N$ corresponding to the screening length $\xi$.

We remark finally that for various properties numerical integration can be avoided replacing
the Debye function by the Pad\'e approximation, Eq.~(\ref{eq_theo_vpot_Flory}).
This has been done for instance for the calculation of finite chain size effects
for the bond-bond correlation function $\Pone(s,N)$ discussed in 
Sec.~\ref{conf_Ps}.\footnote{It is interesting to compare the numerical value
$I(\infty) \approx 1.59$ obtained for the  {\em r.h.s.} of Eq.~(\ref{eq_theo_Rendcorrect2})
with the coefficients one would obtain by computing Eq.~(\ref{eq_theo_Rendcorrect1}) either
with the effective potential $\vpot(q)$ for infinite chains given by Eq.~(\ref{eq_theo_vpot_xi})
or with the Pad\'e approximation, Eq.~(\ref{eq_theo_vpot_Flory}). Within these approximations of
the full linear response formula, Eq.~(\ref{eq_theo_vpot_def}), the coefficients can be obtained
directly without numerical integration yielding overall similar values. In the first case
we obtain $15/8 \approx 1.87$ and in the second $11/8 \approx 1.37$ \cite{WBM07}.
While the first value is clearly not compatible with the measured end-to-end distances,
the second yields a reasonable fit, especially for small $N < 1000$.
}

\section{Bond-fluctuation model}
\label{bfm}

\subsection{Introduction}
\label{bfm_intro}

\begin{figure}[t]
\centerline{\resizebox{1.00\columnwidth}{!}{\includegraphics*{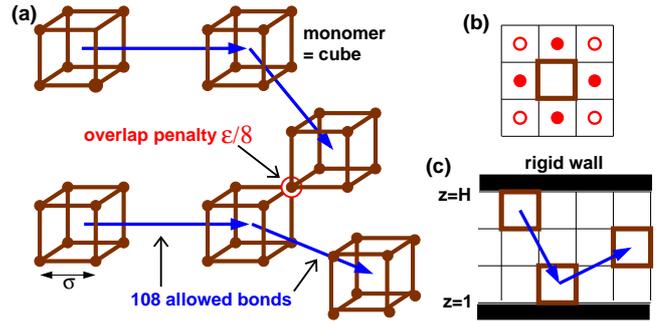}}}
\caption{
The BFM is an efficient lattice MC algorithm for coarse-grained polymer chains 
where monomers are represented by cubes on a simple cubic lattice
(of lattice constant $\sigma$) connected by a set of allowed bond vectors.
{\bf (a)}
The classical BFM assumes that lattice sites are at most occupied once.
The panel shows the recently proposed variant with {\em finite} excluded volume penalty \cite{WCK09}.
An energy $\overlap$ has to be paid if two cubes totally overlap. 
A corresponding fraction is associated with the partial monomer overlap, 
as sketched for two cube corners.
{\bf (b)}
Using {\em local} MC jump attempts to the next (filled circles) and next-nearest (open circles)
neighbors we investigate in Sec.~\ref{dyna} the influence of the incompressibility constraint
on the dynamics of overdamped polymer melts {\em without} topological interactions.
{\bf (c)} 
Using hard and structureless walls systems of reduced effective dimension $d < 3$ 
may be investigated as outlined in Sec.~\ref{dyna_beads} and Sec.~\ref{conc_outlook_slit}.
\label{fig_bfm_algo}
}
\end{figure}

The theoretical predictions sketched above should hold in {\em any} dense homopolymer solution assuming 
that the chains are asymptotically long, i.e. at least $N/\gT \gg 10^2$ and even better $N/\gT \gg 10^3$. 
The computational challenge is to equilibrate and to sample such configurations using an as simple as 
possible coarse-grained model for polymer melts \cite{VanderzandeBook,BWM04}. 
In this study we use the BFM, an efficient lattice MC algorithm proposed 
as an alternative to single-site self-avoiding walk models 
by Carmesin and Kremer in 1988 \cite{BFM}. 
As illustrated in Fig.~\ref{fig_bfm_algo}, the key idea of the model is to increase the size of the monomers 
which now occupy whole unit cells on a simple cubic lattice connected by a specified set of allowed bond vectors.
While the multitude of possible bond lengths and angles allows a better representation of the continuous-space behavior 
of real polymer solutions and melts, the model remains sufficiently simple retaining thus the 
computational efficiency of lattice models without being plagued by ergodicity problems \cite{BWM04}.
The BFM algorithm has been used for a huge range of problems addressing the generic behavior of 
long polymer chains of very different molecular architectures and geometries:
statics and dynamics of linear 
\cite{Paul91a,Paul91b,WPB92,WPB94,Shaffer94,Shaffer95,MWBBL03,MM00,WMBJOMMS04,WBJSOMB07,BJSOBW07,WBM07,MWK08,WCK09,Wittmann90,DeutschDickman90,Deutsch,MP94,WM94,Freire02} 
and cyclic \cite{MWC96,MWC00,MWB00} homopolymer melts,
polymer blends \cite{MarcusT3D95,Blends99,Cavallo03,Cavallo05,Kumar03},
gels and networks \cite{Sommer05},
glass transition \cite{OWBK97,BBP03,Mateo05,Freire08,Mateo09},
poly\-mers and co\-poly\-mers at surfaces \cite{CopolySurface99,Paul08},
brushes \cite{WJJB94,KBWB96,WCJT96},
thin films \cite{combpoly00,MBB01,MM02,MBDB02,CMWJB05},
equilibrium polymers \cite{Milchev95,WMC98a,WMC98b,papEPslit} and
other problems related to monomer and chain self-assembly \cite{Micelles06,Cavallo08}.
For recent reviews on the BFM algorithm 
see Refs.~\cite{BWM04,BFMreview05}.

Throughout this paper all lengths and densities are given in units of the lattice constant $\sigma$,
time scales are given in units of the Monte Carlo Step (MCS) and Boltzmann's constant $\kB$ is set to unity. 
Taking apart some paragraphs in Sec.~\ref{TD} we assume a temperature $T=1$. If not specified otherwise
the chains are monodisperse of length $N$.

We define first the classical BFM variant without mono\-mer overlap ($\overlap=\infty$)
and explain then how dense configurations may be obtained using a mix of local and global MC moves 
(Sec.~\ref{bfm_equil}). 
The generalization of the BFM Hamiltonian to finite monomer overlap penalties $\overlap$ 
is presented in Sec.~\ref{bfm_soft}. Finally, we turn in Sec.~\ref{bfm_EP} to polydisperse 
equilibrium polymer systems with annealed size distribution.

\subsection{Classical BFM without monomer overlap}
\label{bfm_classic}
The classical implementations of the BFM idea do not permit monomer overlap, i.e. each monomer occupies 
{\em exclusively} a unit cell of $2^d$ lattice sites on a $d$-dimensional simple cubic lattice 
\cite{BFM,BWM04,BFMreview05}.
The fraction $\phi$ of occupied lattice sites is thus $\phi = 2^d \rho$ with $\rho$ being the 
$d$-dimensional monomer number density.
A widely used choice for the allowed bond vectors for the 3D variant ($d=3$) of the BFM 
introduced by the Mainz condensed matter theory group around K. Binder 
\cite{Paul91a,Paul91b,WPB92,WPB94,Wittmann90,DeutschDickman90,Deutsch,MP94,WM94}
is given by
\begin{equation}
\Poper\left(\begin{array}{c}2\\0\\0\end{array}\right), 
\Poper\left(\begin{array}{c}2\\1\\0\end{array}\right), 
\Poper\left(\begin{array}{c}2\\1\\1\end{array}\right), 
\Poper\left(\begin{array}{c}2\\2\\1\end{array}\right), 
\Poper\left(\begin{array}{c}3\\0\\0\end{array}\right), 
\Poper\left(\begin{array}{c}3\\1\\0\end{array}\right) 
\label{eq_bfm_bondset}
\end{equation}
where $\Poper$ stands for all the possible permutations and sign combinations of a lattice vector.
This corresponds to 108 different bond vectors $\lvec$ of 5 possible bond lengths 
($2$, $\sqrt{5}$, $\sqrt{6}$, $3$, $\sqrt{10}$) and $100$ angles between consecutive bonds.
The smallest $13$ angles do not appear for the classical BFM because excluded volume forbids the
sharp backfolding of bonds. If only {\em local} hopping moves to the $2d=6$ nearest neighbor sites are 
performed --- called ``L06 moves" \cite{WBM07} --- this set of vectors ensures automatically that 
polymer chains cannot cross. (The corresponding ``L04 moves" for the 2D variant of the BFM are represented in 
Fig.~\ref{fig_bfm_algo}(b) by filled circles.) Topological constraints, e.g. 
in ring polymers \cite{MWC96,MWC00,MWB00} or polymer gels \cite{Sommer05}, 
hence are conserved.\footnote{Following \cite{Paul91a,Paul91b} we keep lists of the monomer 
positions in absolute space, their corresponding lattice positions and of the indices $1 \le i \le 108$
of the bond vectors connecting the monomers of the chains. Since the bond vector index can be encoded 
as a byte, this allows a rather compact storage of the configurations. 
Predefined tables allow the rapid verification of the excluded volume condition on 
the periodic lattice. Following M\"uller \cite{MarcusT3D95,BFMreview05} we use a Wigner-Seitz 
representation of the cubic lattice where a cube is not represented by 8 entries on the lattice 
but just by {\em one} variable in the cube center. This variable can be a boolean if we are only
interested in homopolymers or an integer if we deal with a mixture of different monomer types.
Although the Wigner-Seitz representation of the BFM algorithm is about a factor 3 slower than the 
original implementation, it has the advantage that the code becomes more compact and can be more
readily adapted to the various polymer architectures or interaction potentials of interest.}

\begin{figure}[t]
\centerline{
\resizebox{0.9\columnwidth}{!}{\includegraphics*{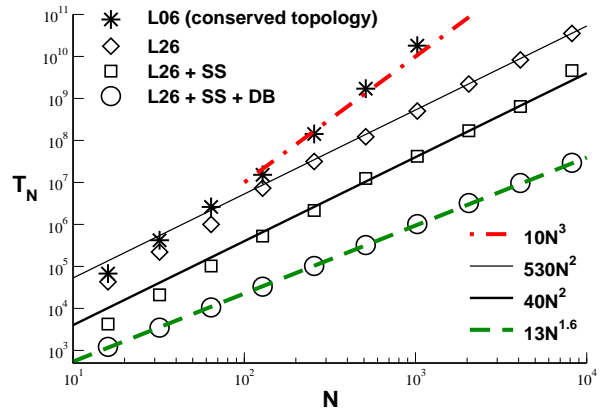}}
}
\caption{Diffusion time $\TN \equiv \RN^2/6\DN$ 
{\em vs.} chain length $N$ for different versions of the BFM without monomer overlap. 
All data are for our standard volume fraction $\phi=0.5$.
The BFM version with topology conserving local ``L06 moves" is represented by stars.
All other data sets use topology violating local ``L26 moves" \cite{WBM07}.
The L26 dynamics (diamonds) is essentially Rouse-like which allows the 
determination of $\DN$ although the monomers have not yet moved over $\RN$ 
for the largest chains considered. 
Additional ``slithering snake" (SS) moves increase the efficiency of the algorithm 
by approximately an order of magnitude (squares, bold line). 
A power-law exponent $1.6 \pm 0.1$ (dashed line) is found if ``double bridging" (DB) moves are included.
\label{fig_bfm_TN}
}
\end{figure}

Consequently, several authors report reptation-type dynamics for chain lengths $N \gg \Ne \approx 10^2$
at a standard ``melt" volume fraction $\phi=8 \rho = 0.5$ \cite{Paul91a,Paul91b,KBMB01,HT03}.\footnote{The 
BFM version by Shaffer \cite{Shaffer94,Shaffer95} assumes 
a different bond vector set which leads to a higher chain stiffness $\cinf$ and, 
hence, to smaller values of the entanglement length $\Ne$. Note that if one applies a local stiffness 
potential, as discussed Sec.~\ref{theo_connectivity}, one finds quite generally that topological effects 
become more pronounced \cite{WPB92,MWC00,MWB00}.} 
As may be seen from the stars indicated in Fig.~\ref{fig_bfm_TN}, the relaxation time $\TN$ obtained 
using L06 moves \cite{WBM07} becomes similar to the reptation theory prediction 
$\TN \sim N^3$ (dash-dotted line). We have used here --- as elsewhere if not stated otherwise ---
periodic simulation boxes of linear dimension $L=256$ containing $\nmon = \rho L^3 = 2^{20} \approx 10^6$ monomers. 
This large system size allows to eliminate finite-size effects even for the longest chain lengths studied.
The relaxation time $\TN \equiv \RN^2/6\DN$ has been estimated here (for historical reasons) 
using the self-diffusion coefficient $\DN$ obtained either from the monomer MSD 
$\MSDmon(t)$ or the chain CM MSD $\MSDcmN(t)$.
Other operational definitions of $\TN$ exist \cite{Paul91b,KBMB01} which lead to similar, vertically
slightly shifted results. %
The  last data point for $N=1024$ has to be taken with care as usual in computational as in 
experimental studies \cite{FeynmanJoking}. Being obtained by extrapolation using the expected 
shape of $\MSDcmN(t)/t$ in log-linear coordinates, it corresponds to a {\em lower bound} for $\TN$. 
Note that the relaxation time appears thus to increase even more strongly with $N$ than the standard 
reptation theory predicts \cite{DoiEdwardsBook}. We do not pursue this issue here (which has also been 
observed in MD simulations) the important point being merely that local 
topology conserving moves are too inefficient to equilibrate and sample large-$N$ polymer melts.

Note that the classical BFM without monomer overlap and using L06 moves is strictly speaking not ergodic, 
since some 
configurations may be easily constructed which are not accessible 
starting from an initial configuration of stretched linear chains.
Although topology conservation is irrelevant for the present work (taking apart the preliminary results
presented in Sec.~\ref{dyna_repta}) 
we keep the set of allowed bond vectors, Eq.~(\ref{eq_bfm_bondset}), for consistency with previous work.

\subsection{Local and global topology violating MC moves}
\label{bfm_equil}

\begin{figure}[t]
\centerline{
\resizebox{0.48\columnwidth}{!}{\includegraphics{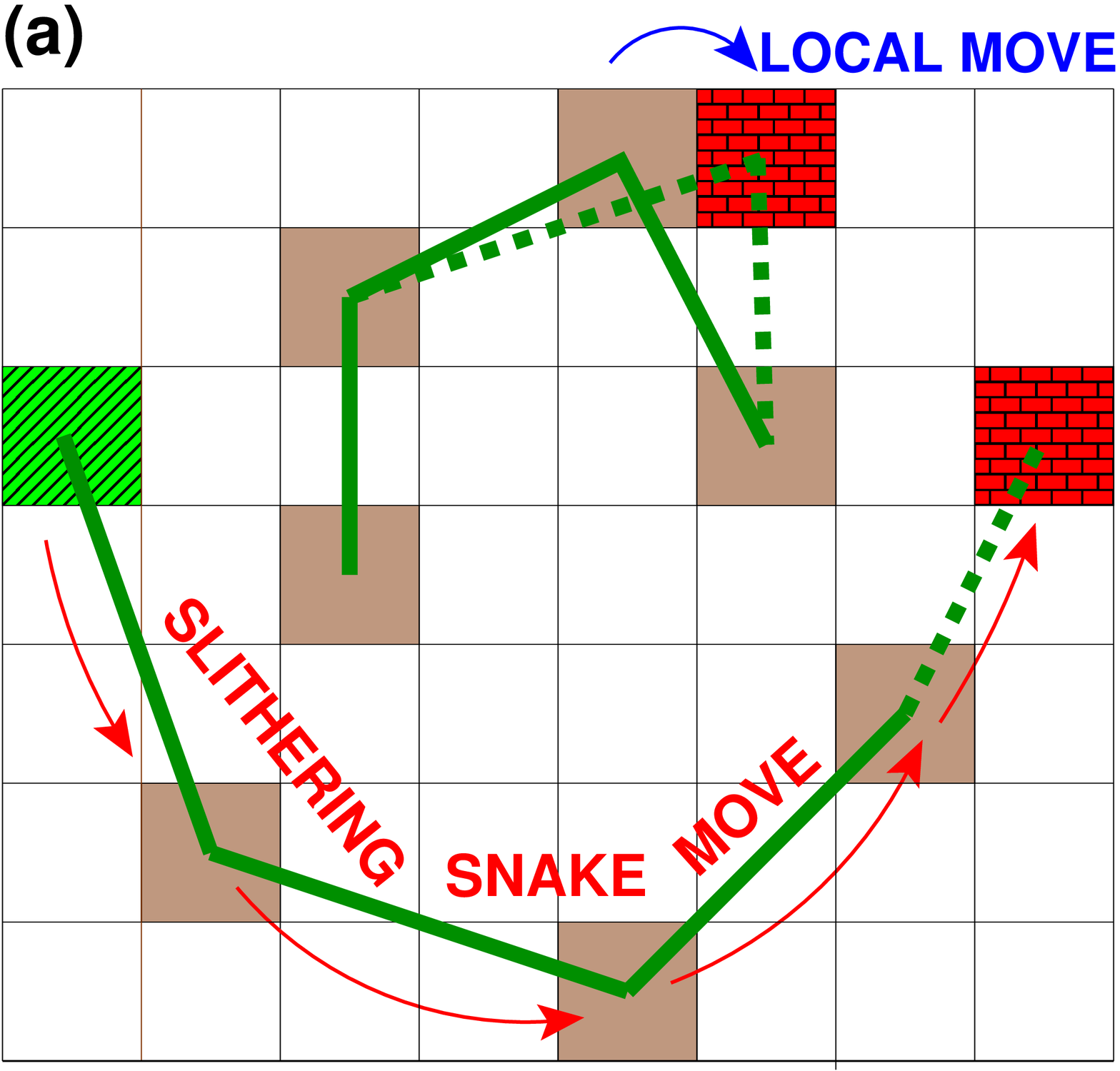}}
\hspace*{0.1cm}
\resizebox{0.48\columnwidth}{!}{\includegraphics{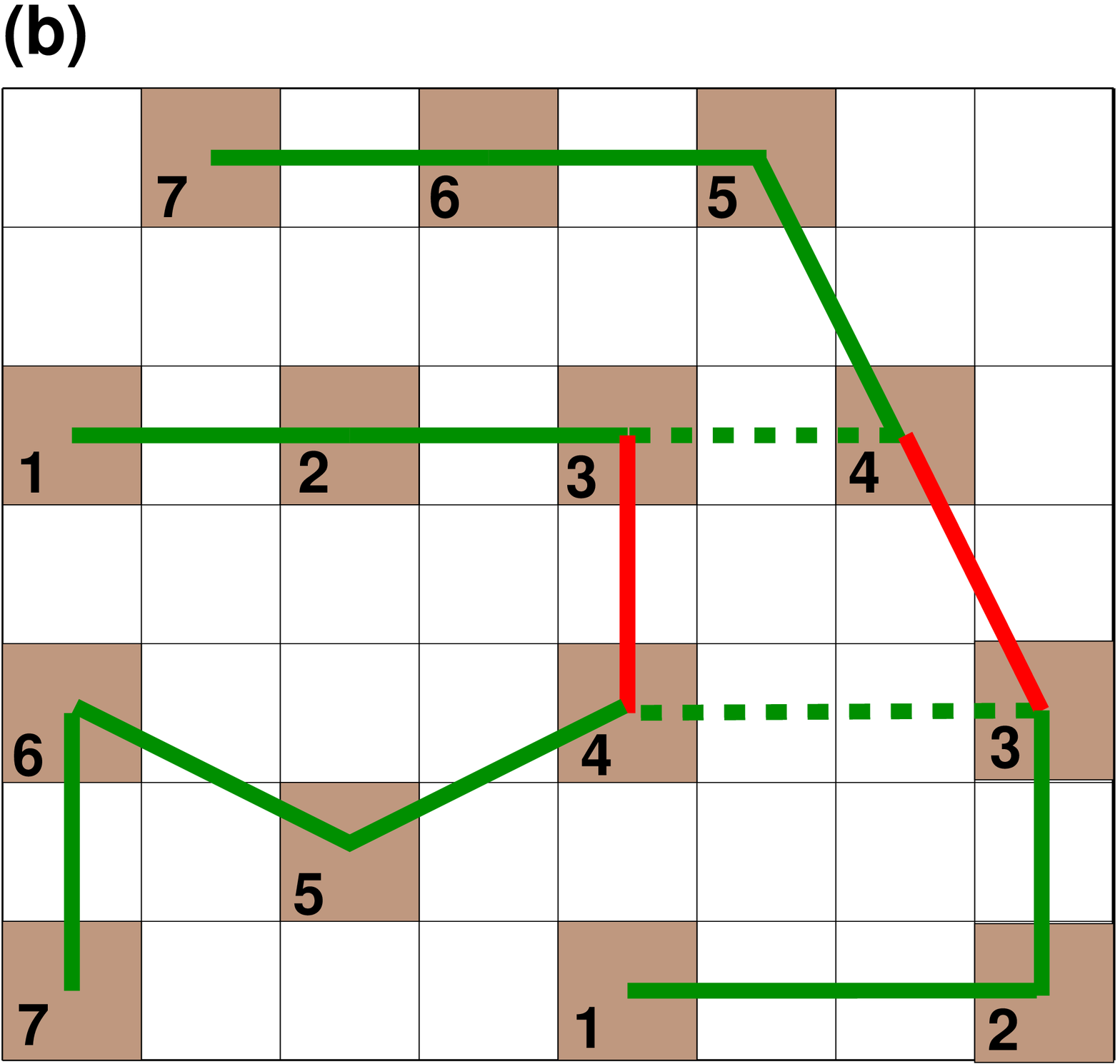}}
}
\caption{Two-dimensional sketch of the global MC moves used:
{\bf (a)} Monomers move collectively along the chain if a {\em slithering snake} move is performed.
Effectively, this amounts to removing a monomer (the striped one on the left),
connecting it to the other end of the chain and leaving the middle monomers unchanged.
Therefore, density fluctuations and constraint release only occurs at the chain end.
At higher volume fractions it is thus necessary to add local hopping moves, 
as shown at the top of the panel to maintain the computational efficiency \cite{MWBBL03}.  
{\bf (b)} Connectivity altering {\em double bridging} (DB) moves are very useful at 
high densities allowing us to extend the accessible molecular mass up to $N=8192$ \cite{WBM07}. 
Since density fluctuations do not couple to DB moves, local moves again must be added. 
\label{fig_bfm_equil}
}
\end{figure}

To equilibrate and sample BFM melts such as the ones presented in Fig.~\ref{fig_bfm_TN},
 we have replaced the realistic but very slow L06 moves by a mix of local topology violating so-called 
``L26 moves" \cite{WBM07} with global ``slithering snake" \cite{Kron65,Wall75,MWBBL03} and 
``double bridging" \cite{Theodorou02,dePablo03,Auhl03,BWM04} MC moves as shown in Fig.~\ref{fig_bfm_equil}.
%

\subsubsection{Local L26 moves}
\label{bfm_equil_L26}
Already the use of local moves to the $3^d-1=26$ next and next-nearest lattice sites surrounding 
the current mono\-mer position [Fig.~\ref{fig_bfm_algo}(b)] dramatically speeds up the relaxation dynamics, 
especially for $N > 512$, as can be seen from the diamonds indicated in Fig.~\ref{fig_bfm_TN}. 
Since the dynamics is to leading order of Rouse-type, as further discussed in Sec.~\ref{dyna}, 
the diffusion coefficient $\DN$ can readily be estimated from the CM MSD $\MSDcmN(t)$ 
even for our largest chains with $N=8192$. 
As shown in Fig.~\ref{fig_bfm_TN}, we find $\TN \sim 530 N^2$ for L26 dynamics. 
This is, of course, still prohibitive for the longest chains we aim to 
characterize.

L26 moves yield configurations {\em not} accessible with L06 moves. 
Concerning the static properties both system classes are practically equivalent. 
This has been confirmed by counting the number of monomers which become ``blocked" (in absolute 
space or with respect to an initial group of neighbor monomers) once one returns to the L06 scheme.
Typically, we find about 10 blocked monomers for a system of $2^{20}$ monomers \cite{WBM07}. 
While the few blocked monomers are irrelevant for static properties they obviously matter if
dynamical properties are probed.\footnote{The same problems arise if slithering snake or double 
bridging moves are used to equilibrate the configurations.}

\subsubsection{Slithering snake moves}
\label{bfm_equil_SS}
In addition to these local moves {\em one} slithering snake move per chain is attempted on 
average per MCS corresponding to the displacement of $N$ monomers along the chain backbone,
as sketched in Fig.~\ref{fig_bfm_equil}(a). Note that in our units {\em two} displacement attempts
per MCS are performed on average per monomer, one for a local move and one for a snake move.\footnote{It 
is computationally more efficient for large $N$ to take off a monomer at one chain end and to paste it at 
the other leaving all other monomers unaltered. Before static or dynamical measurements are performed the 
original order of the monomers must then be restored.} 
Interestingly, a significantly larger slithering snake attempt frequency would not be useful
since the relaxation time of snakes without or only few local moves increases
exponentially with mass as shown in \cite{Deutsch85,MWBBL03} due to the correlated motion
of snakes expected in analogy to the activated reptation limit for real polymer melts 
mentioned in Sec.~\ref{intro_Rouse} \cite{ANS97,ANS98}.
In order to obtain an efficient free snake diffusion [with an $N$-independent curvilinear
diffusion coefficient $\Dcurv(N) \sim N^0$ and $\TN \approx N^2/\Dcurv(N) \sim N^2$ \cite{MWBBL03,Wall75}
it is important to relax density fluctuations rapidly by local dynamical pathways. 
As shown in Fig.~\ref{fig_bfm_TN} (squares), we find a much reduced relaxation time $\TN \approx 40N^2$
which is, however, still inconveniently large for our longest chains. Note that most of the CPU time
is  used by the local moves and the computational load per MCS remains $N$-independent.

\subsubsection{Double bridging moves}
\label{bfm_equil_DB}
Double bridging (DB) moves are found to be very useful at high densities and help us to extend the 
accessible molecular masses close to $N\approx 10^4$. As for slithering snake moves we use all $108$ 
bond vectors to switch chain segments between two different chains. Only chain segments of equal length
are swapped to conserve monodispersity. Topological constraints are again deliberately 
violated. Since more than one swap partner is possible for a selected first monomer, delicate detailed
balance questions arise \cite{BWM04}. This is particularly important for short chains. To avoid the
computation of weights \cite{BWM04} one simple solution to this problem is to refuse all moves
with more than one swap partner.\footnote{Since we have a 
finite number of possible BFM bonds an even simpler option is to select randomly {\em one} bond vector $\lvec$ 
for a given first monomer at a lattice position $\rvec$ and to check whether a suitable monomer of another chain 
exists at $\rvec + \lvec$. Obviously, using only one instead of 108 bonds reduces the number of DB moves performed, 
but since correlated moves forth and back in topology space are pointless, this is not of disadvantage.}
The configurations are screened with a frequency $\fDB$ for possible DB moves where we scan
in random order over the monomers. The frequency should not be too large to avoid (more or less)
immediate back swaps and monomers should move at least over a couple of lattice constants between 
subsequent DB moves. 
In the example presented in Fig.~\ref{fig_bfm_TN} (spheres) a frequency $\fDB=0.1$ is used.
Empirically it is found that $\TN \approx 13 N^{1.6 \pm 0.1}$ using the diffusion coefficient $\DN$ obtained
from the monomer MSD $\MSDmon(t)$. For $N=8192$ this corresponds to $3 \times 10^7$ MCS. This
allowed us even for the largest chains to observe monomer diffusion over several $\RN$
within the $10^8$ MCS which were feasible in 2007 on our XEON-PC processor cluster \cite{WBM07}.
The influence of $\fDB$ on the performance has yet not been 
explored systematically, but preliminary results suggest smaller DB frequencies for future studies.
The power-law exponent $\approx 3/2$ remains robust as is also confirmed by
MD simulations of a bead-spring model coupled to DB moves \cite{WBM07}. 
This finding  begs for a systematic theoretical investigation. 

\begin{table}[t]
\begin{center}
\begin{tabular}{|c||c|c|c||c|c|c|}
\hline
$\overlap/T$&   $\gT$& $l$  &$ b $ & $A$   &  $W$  & $N \DN$  \\ \hline
0.0        & $\infty$&2.718 & 2.72 & 0.2109& 0.032 &0.065          \\
0.01       & 209     &2.718 & 2.80 & 0.2109& 0.030 &0.062           \\
0.03       & 67      &2.718 & 2.85 & 0.2099& 0.028 &0.060          \\
0.1        & 22      &2.719 & 2.92 & 0.2067& 0.024 &0.058          \\
0.3        & 7.1     &2.720 & 3.01 & 0.1992& 0.021 &0.050          \\
1          & 2.4     &2.721 & 3.13 & 0.1796& 0.015 &0.040       \\
3          & 0.85    &2.721 & 3.21 & 0.1432& 0.008 &0.024          \\
10         & 0.32    &2.670 & 3.24 & 8.8E-02& 0.003 &0.009          \\
30         & 0.25    &2.638 & 3.24 & 7.2E-02& 0.0013&0.004          \\ 
100        & 0.25    &2.636 & 3.24 & 6.9E-02& 0.0010&0.003          \\ 
$\infty$   & 0.25    &2.636 & 3.24 & 6.9E-02& 0.0010&0.003          \\ 
\hline
\end{tabular}
\vspace*{0.5cm}
\caption[]{Various properties for monodisperse BFM melts at volume fraction $\phi = 0.5$:
the dimensionless compressibility $\gT$,
the root-mean-square bond length $l$, 
the effective bond length $b$,
the acceptance rate $A$,
the local monomer mobility $W$ and 
the self-diffusion coefficient $\DN$. 
The dynamical data refer to local L26 moves to the nearest and next-nearest lattice sites 
which yields essentially Rouse-like dynamics.
The values of the dimensionless compressibility $\gT$ and the effective bond length $b$ 
for asymptotically long chains have been obtained using extrapolation schemes discussed
in Sec.~\ref{TD} and Sec.~\ref{conf}, respectively. 
\label{tab_overlap}}
\end{center}
\end{table}

\subsubsection{Summary of static properties}
\label{bfm_equil_static}
Some static properties obtained at our reference volume fraction $\phi=0.5$
assuming no monomer overlap ($\overlap=\infty$) are indicated in Table~\ref{tab_overlap}.
Averages are performed over all chains and typically 1000 configurations.
Taking apart the systems for $N=8192$, chains are always much smaller than the linear box size $L=256$. 
In the large $N$-limit we obtain an average bond length $\la |\lvec|\ra \approx 2.604$, 
a root-mean-squared bond length $l \approx 2.636$
and an effective bond length $b \approx 3.24$ as will be shown below in Sec.~\ref{conf_RNRs}.
This corresponds to a ratio $\cinf = (b/l)^2 \approx 1.52$ and, hence, to a persistence length
$\lpersist = l (\cinf + 1)/2 \approx 3.32$. The swelling coefficient $\ce =\sqrt{24/\pi^3}/b^3\rho$ 
defined in Sec.~\ref{intro_key} is thus $\ce \approx 0.41$. Especially, we find from the zero wavevector limit
of the total structure factor $S(q)$ a dimensionless compressibility $\gT \approx 0.246$
[Eq.~(\ref{eq_theo_gTdef})]
which compares well with real experimental melts. From the measured bulk compression modulus
$\veff \equiv 1/\gT(\rho)\rho \approx 66$ and the effective bond length $b$ one estimates a Ginzburg
parameter $\Gi = \sqrt{\veff \rho}/b^3\rho \approx 0.96$. 
Following Ref.~\cite{ANS05a} the interaction parameter $\veff$ is supposed here to be given
by the full inverse compressibility and not just by the second virial coefficient $\vvir=27$
of the BFM monomers \cite{DeutschDickman90}.

\subsection{BFM with finite excluded volume penalty}
\label{bfm_soft}

\subsubsection{Definition of Hamiltonian}
\label{bfm_soft_hamiltonian}
Fig.~\ref{fig_bfm_algo} shows how finite energy penalties may be introduced in the BFM algorithm \cite{WCK09}.
The overlap of two cube corners on one lattice site ($\Nov=1$) corresponds to an
energy cost of $\overlap/8$, the full overlap of two monomers ($\Nov=8$) to an energy $\overlap$.
More generally, with $\Nov$ being the total number of interacting cube corners
the total interaction energy of a configuration is
\begin{equation}
E = \frac{\overlap}{8} \Nov.
\label{eq_bfm_Hamiltonian}
\end{equation}
With the energies of the final ($E_\mathrm{f}$) and the initial configurations ($E_\mathrm{i}$) 
we accept a proposed MC move according to the Metropolis criterion 
with a probability $\mbox{min}(1,\exp[-(E_\mathrm{f} - E_\mathrm{i})/T])$
\cite{LandauBinderBook,KrauthBook}.
If the overlap penalty is the only energy scale as in the studies presented here,
one may, of course, either vary the overlap parameter $\overlap$ or the temperature $T$.
For the presentation of thermodynamic properties in Sec.~\ref{TD} it will be more
naturally to use $T$ as the control parameter and to fix arbitrarily $\overlap=1$.
The inverse temperature $\beta=1/T$ and the dimensionless overlap strength $x=\overlap/T$ 
are thus numerically equal. (Both notations are kept for dimensional reasons and for future
generalization to models with more than one energy scale.) In other parts of this review,
especially Sec.~\ref{conf} and Sec.~\ref{dyna}, it will be more natural to set 
temperature to unity, $T = 1$, using the overlap strength $\overlap$ as the control parameter.


\subsubsection{Second virial coefficient}
\label{bfm_soft_virial}
To illustrate this finite excluded volume interaction we indicate the second virial of an 
imperfect gas of unconnected monomers,
$\vvir = \int d\deltavec (1-e^{-E(\deltavec)/T})$,
which is shown below to be useful for roughly characterizing the effective strength of the potential.
$\deltavec$ stands for a possible lattice vector between the centers of two interacting cubes.
It is easy to see that there are 
8 vectors corresponding to $\Nov =1$ as shown in Fig.~\ref{fig_bfm_algo}(a),
12 to $\Nov=2$ (overlap of two cube corners),
6  to $\Nov = 4$ (overlap of two faces),
and 1 to $\Nov=8$ (full overlap).
Setting $x=\overlap/T$ this leads to a second virial
\begin{eqnarray}
\vvir(x) & = & 8 \times  (1 - \exp(-x /8)) \nonumber \\
  & + & 12 \times (1 - \exp(-x /4))        \nonumber \\
  & + & 6  \times (1 - \exp(-x /2))        \nonumber \\
  & + & 1  \times (1 - \exp(-x ))
\label{eq_bfm_vvir}
\end{eqnarray}
given in units of the lattice cube volume $\sigma^3$.
We note that the second virial becomes constant,
$\vvir = 27$, for large $x \gg 1$ as expected \cite{DeutschDickman90}.
This second virial coefficient is about half the effective bulk modulus $\veff \approx 66$ 
indicated above.  In the opposite limit we have
\begin{equation}
\vvir(x) \approx 8 x - \frac{27}{16} x^2 \mbox{ for } x \ll 1.
\label{eq_bfm_vThigh}
\end{equation}
We shall see in Sec.~\ref{TD_compress} that $\vvir \approx \veff$ in this limit.

\subsubsection{Implementation}
\label{bfm_soft_implementation}
%
%
%
%
Since a lattice site may be occupied now by more than one monomer, it is not possible to use a compact 
boolean occupation lattice 
as for the classical BFM.
Instead we have mapped Eq.~(\ref{eq_bfm_Hamiltonian}) onto a Potts spin model \cite{LandauBinderBook}
\begin{equation}
E = \frac{1}{2}
\sum_{\rvec} S(\rvec) \sum_{\deltavec}
J(\deltavec) S(\rvec+\deltavec)
- \frac{1}{2} \overlap \nmon
\label{eq_bfm_spinmodel}
\end{equation}
with constant monomer number $\nmon=\sum_{\rvec} S(\rvec) \stackrel{!}{=} L^3 \rho$.
We use the Wigner-Seitz representation of the BFM on the cubic lattice \cite{MarcusT3D95,BFMreview05}
where an integer spin variable $S(\rvec)$ counts the number of BFM monomers
($S=0,1,2,\ldots$) with cubes centered at a Wigner-Seitz lattice position $\rvec$.
Since we have now to compute the interaction between cube centers
instead of cube corners, the coupling constant $J$ characterizing the interaction between
two spins depends only on the relative distance $\deltavec$:
\begin{equation}
J(\deltavec) = \overlap \left\{
\begin{array}{ll}
1/8       & \mbox{if $\deltavec=\Poper(1,1,1)$ for cube corners,} \\
1/4       & \mbox{if $\deltavec=\Poper(1,1,0)$ for cube edges,} \\
1/2       & \mbox{if $\deltavec=\Poper(1,0,0)$ for cube faces,} \\
1         & \mbox{if $\deltavec=\Poper(0,0,0)$ for full overlap,} \\
0         & \mbox{otherwise.}
\end{array}
\right.
\label{eq_bfm_J}
\end{equation}
Since the interaction is still short-ranged and the values of $J$ are readily tabulated,
this remains an efficient rendering of the monomer interactions.
Note that the first term on the {\em r.h.s.} of Eq.~(\ref{eq_bfm_spinmodel})
contains a constant self-interaction contribution of the $\nmon$ monomers
with themselves for $\deltavec=\underline{0}$, which is subtracted by the second 
term.\footnote{Attractive interactions similar to the ones used in 
\cite{MarcusT3D95,Blends99,Cavallo03,Cavallo05}
may be easily added to the Potts spin formulation of the soft BFM. The simulation of
polymer blends requires additional Potts spin lattices as the two lattices 
used to obtain the chemical potential in Sec.~\ref{TD_muMD}.}

\begin{table}[t]
\begin{tabular}{|c||c|c|c||c|c|c|c|c|c|c|c|}
\hline
$\phi$   &   $\gT$ & $l$  &$ b $ & $A$    &  $W$  & $N \DN$ \\ \hline
0.5      & 0.32    &2.670 & 3.24 & 8.8E-02& 0.003 & 0.009   \\
0.25     & 1.1     &2.709 & 3.65 & 0.1455 & 0.004 & 0.018   \\
0.125    & 3.3     &2.725 & 3.95 & 0.1664 & 0.004 & 0.023   \\
0.0625   & 9.5     &2.731 & 4.38 & 0.1729 & 0.004 & 0.027   \\
0.03125  & 26      &2.733 & 4.77 & 0.1749 & 0.004 & 0.035   \\
\hline
\end{tabular}
\vspace*{0.5cm}
\caption[]{BFM solutions with overlap penalty $\overlap=10$ for different volume fractions $\phi = 8 \rho$.
The indicated dynamical properties ---
acceptance rate $A$, local mobility $W$ and self-diffusion constant $\DN$ ---
have been obtained using local L26-moves for chains of length $N=1024$.
\label{tab_density}}
\end{table}
\subsubsection{Equilibration and system properties}
\label{bfm_soft_equilibration}
As start configurations we have used the equilibrated BFM configurations without monomer overlap ($\overlap=\infty$) 
described in Sec.~\ref{bfm_classic} \cite{WBM07}.
As one may expect, the configurational properties are found essentially unchanged for $x \gg 5$
(Table~\ref{tab_overlap}).
Local L26-moves need to be added to the snake moves for $x \ge 1$.
Otherwise the slithering snake motion will become ineffective \cite{MWBBL03}.
Simple slithering snakes without local moves are sufficient, however, for smaller penalties.
We have crosschecked our results in this regime for $N=2048$ and $N=8192$ using boxes of linear size
$L=512$  by starting our simulations with Gaussian chains at $x=0$ and increasing then the penalty.
Table~\ref{tab_overlap} present some system properties obtained for our reference 
volume fraction $\phi=0.5$ such as dimensionless compressibility $\gT$ or the
effective bond length $b$.
Averages are performed over all chains and at least 100 configurations.
The chain lengths $N=64$, $N=1024$ and $N=2048$ have been studied with particular care.
%
Density effects have been studied more briefly. As will be discussed in  Sec.~\ref{TD_energy},
we have sampled weak overlap penalties ($x \ll 1$) for $N=8192$ to investigate the intrachain 
contributions to the mean energy. We have also probed various densities for $N=1024$ and 
$\overlap=10$ as summarized in Table ~\ref{tab_density}. This was done to
check for density effects on the deviations to Rouse dynamics as discussed in Sec.~\ref{dyna_robust}.

\subsection{BFM with annealed mass distribution}
\label{bfm_EP}

\subsubsection{Motivation and context}
\label{bfm_EP_motivation}
\begin{figure}[t]
\centerline{\resizebox{1.0\columnwidth}{!}{\includegraphics*{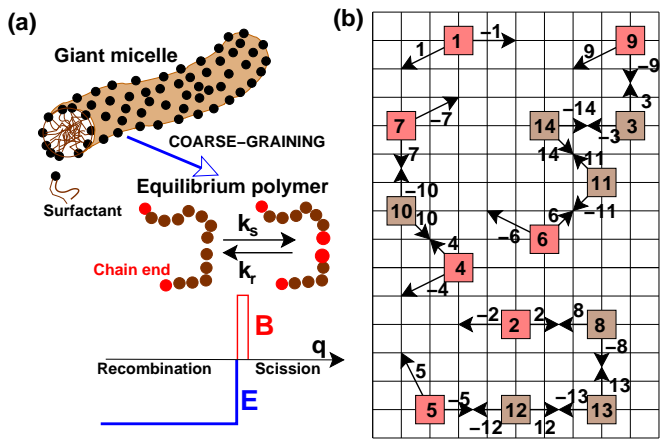}}}
\caption{Self-assembled linear EP: 
{\bf (a)} Bead-spring representation of a worm-like micelle
assuming a finite scission energy $E$ and a finite barrier height $B$ \cite{CC90}.
The scission energy determines the static properties and fixes the ratio 
of the scission and recombination rates, 
$\ks \sim \exp(-(E+B)/\kBT)$ and $\kr \sim \exp(-B/\kBT)$. 
Both energy scales are sketched {\em vs.} a generic reaction coordinate $q$ \cite{ChandlerBook}. 
The formation of closed rings and the branching of chains are not allowed in the presented studies. 
{\bf (b)}
Two-dimensional projection of EP modeled using the classical BFM without monomer overlap \cite{WMC98b}.
Chains consists of symmetrically connected lists of bonds. The pointers of end-bonds point to {\em NIL}.
The breaking of a saturated bond {\em ibond} requires to set the pointers of
the two connected bonds {\em ibond} and {\em jbond = pointer(ibond)} to {\em NIL}.
Setting {\em pointer(-2)=4} and {\em pointer(4)=-2} connects the two
end-monomers {\em imon=2} and {\em jmon=4}. 
\label{fig_bfm_EP}
}
\end{figure}

As discussed above (Fig.~\ref{fig_bfm_TN}), the equilibration and sampling of strictly monodisperse
polymer melts is a delicate issue. An elegant way to test the computed conformational
properties is given by associating a finite ``scission energy" $E \ge 0$ to each bond which has to be paid 
if a bond is broken \cite{CC90,Milchev95,WMC98a,WMC98b,Milchev00,HXCWR06,papEPslit}. 
Since we are only interested here in linear polymer melts, the formation of closed cycles and the branching 
of the chains is not permitted.\footnote{In systems of experimental relevance closed cycles are suppressed 
by the non-negligible rigidity of the chains \cite{SW99,MWL00b,WMSB00}.}
As sketched in Fig.~\ref{fig_bfm_EP}, we relax thus the constraint that bonds can never break,
i.e. that the connectivity matrix (defining which monomers are connected by BFM bonds) is quenched, 
and allow the polymerization of the chains and 
their respective monomers to take place under condition of chemical equilibrium. 
Such systems of self-assembled EP are not only useful for computational
purposes but are also of high experimental relevance.\footnote{In 
the surfactant literature \cite{CC90} EP are often referred to as ``living polymers" (LP) although this 
is potentially confusing since they are distinct from systems that polymerize stepwise, in the presence 
of a fixed number of initiators, for which this term has previously been reserved \cite{Szwarc56}. 
Since LP are held together by strong covalent carbon-to-carbon bonds, they do not break in the middle 
of the polymer chain.} 
An important example sketched in Fig.~\ref{fig_bfm_EP}(a) is that of some surfactant molecules forming 
long giant worm-like micelles which break and recombine constantly at random points along the sequence 
\cite{CC90,Israela}. Similar systems of EP are formed by liquid sulfur \cite{Scott65,Wheeler80}, 
selenium \cite{Faivre86} and some protein filaments \cite{OozawaBook}.
Although with respect to their static properties EP behave very much like quenched polymers, 
the constant reorganization of the chain connectivity offers an additional relaxation pathway
reducing strongly the relaxation times \cite{CC90,Faivre86,Milchev00,HXCWR06}. 
%
Obviously, EP are intrinsically polydisperse with an annealed length distribution $\pchain$ 
minimizing the free energy of the system (Sec.~\ref{theo_connectivity}). 
Since both the bonding energy per chain 
$-E (N-1)$ and all other free energy contributions to the chain chemical potential $\muchainN$ 
are extensive with repect to the chain length $N$ --- at least according to Flory's ideality hypothesis ---
one expects 
a Flory distribution decreasing exponentially with chain length $N$.
%
%

\subsubsection{Spatial monomer moves}
\label{bfm_EP_spatialmoves}
The EP systems presented in this study have been obtained with the classical BFM algorithm without
monomer overlap ($\overlap=\infty$, $T=1$) at the standard melt volume fraction $\phi=8\rho=0.5$. 
Using again the Wigner-Seitz representation we mark (only) the index of 
each monomer on the periodic lattice of linear size $L=256$. Thus the indices of the monomers in the 
neighborhood of a reference monomer are readily obtained which is helpful for 
the recombination of bonds described below.
Only local L06 or L26 monomer moves have been used since the breaking and recombination of the chains 
reduce the relaxation times dramatically compared to monodisperse systems \cite{HXCWR06}. 
Additional global MC moves as described in Sec.~\ref{bfm_equil} may be added, however, in future studies.

\subsubsection{Connectivity pointer list}
\label{bfm_EP_connectivity}
Self-assembled EP are only transient objects and it is thus inefficient to base the data structure on 
the chains \cite{Milchev95}, rather it should be based on the (saturated or unsaturated) 
{\em bonds} of each monomer \cite{WMC98b}. As sketched in Fig.~\ref{fig_bfm_EP}(b), this allows via a 
linear pointer list between the bonds to avoid all sorting procedures.
%
Using the assumption that no branching of chains is allowed, the two (possible) bonds of each monomer 
$imon$ are called $ibond=imon$ and $ibond=-imon$. No specific meaning (or direction) is attached to 
the sign: this is merely a convenience for finding the monomer from the bond list: $imon=|ibond|$.
Pointers are taken to couple independently of sign and the bonds are coupled by means of a pointer list 
in a completely transitive fashion. 
Only two simple operations are thus required for breaking or recombining bonds.
Unsaturated bonds at chain ends point to \mbox{NIL}. Only these bonds may recombine. 
%
A minor caveat attached to this data structure arises if the ends of a given chain are not 
allowed to bind together as in the presented studies.
Since there is no direct chain information in the data structure we have to check this constraint
before every recombination by working up the pointer list which only adds four lines to the source code.\footnote{For higher $E$ the simulation becomes actually
{\em faster} per MCS since the number of recombinations goes down like the squared 
density of unsaturated end monomers $(\rho/\Nav)^2 \sim \exp(-E)$ \cite{WMC98b}.}

\subsubsection{Connectivity altering moves}
\label{bfm_EP_connectivityaltering}
As sketched in Fig.~\ref{fig_bfm_EP}(a), EP systems are not only characterized by the mono\-mer density $\rho$ 
and the finite scission energy $E$ which determine the static properties but also by a barrier height $B \ge 0$ 
which only influences the scission and recombination rates. This barrier is taken into account by setting an 
attempt frequency $\freqB =\exp(-B/T)$ for choosing randomly one bond {\em ibond} out of the $2 \nmon$ bonds 
of the system. This frequency is a convenient tool for testing the dynamics of EP at different 
lifetimes of the chains \cite{Milchev00,HXCWR06,Oshaug95} although for the static properties discussed 
the choice of $B$ is irrelevant. 
The bond $ibond$ corresponds to a monomer $imon = |ibond|$ at a position $\rvec$.
Depending on whether the bond $ibond$ is saturated or unsaturated we try to break it or to connect it 
to a suitable nearby unsaturated monomer.
A delicate detailed balance problem arises \cite{HXCWR06} if $n_u > 1$ unsaturated monomers 
are available for recombination (the no-closed-cycle condition having been verified). 
If one chooses now one of these monomers at random, a weight $1/n_u$ has to be 
taken into account for the reverse breaking process. 
Choosing in addition to the reference bond $ibond$ a trial bond vector $\lvec$ allows to avoid these weights.
If $ibond$ is unsaturated one searches for an unsaturated monomer only at the position $\rvec + \lvec$ 
(taking into account the periodicity of the lattice). 
Since monomer overlap is forbidden there is at most one unsaturated monomer at this position. 
If this is case and if no closed cycle is formed, the recombination is accepted since the energy change is $-E \le 0$. 
To satisfy detailed balance a saturated bond can therefore only be broken if its bond vector is identical 
to the trial bond $\lvec$. Applying the Metropolis algorithm \cite{LandauBinderBook} a scission is performed 
whenever the value of a random number between 0 and 1 is smaller than $\exp(-E/T)$. 
The fact that we only probe one lattice vector $\lvec$ for possible recombinations and not all possible $108$ 
obviously strongly reduces the number of recombination and scission events. However, since we are interested in 
uncorrelated changes of the connectivity list, a broken monomer must anyway move over a certain distance,
say $10 \sigma$, before a new recombination attempt is made. Otherwise there is a strong chance that the newly 
created chain end monomer recombines with its previous partner \cite{Milchev00,HXCWR06,Oshaug95}.

\begin{table}[t]
\begin{center}
\begin{tabular}{|l||c|c|c|c|c|}                                                                                
\hline                                                                                                         
$E$  & \Nav & $F(0)$ & $l$   & $\Rendz$ & $\Rgyrz$ \\ \hline                                                 
1    & 6.4  & 11.9   & 2.632 & 12.6   & 5.2     \\                                                          
2    & 10.4 & 19.7   & 2.633 & 16.5   & 6.8     \\                                                          
3    & 16.8 & 32.4   & 2.633 & 21.6   & 8.8     \\                                                          
4    & 27.5 & 53.4   & 2.634 & 28.1   & 11.4    \\
5    & 44.9 & 87.9   & 2.634 & 36.3   & 14.8    \\                                                          
6    & 73.7 & 145    & 2.634 & 46.9   & 19.1    \\                                                          
7    & 121  & 239    & 2.634 & 60.7   & 24.7    \\                                                          
8    & 199  & 394    & 2.634 & 77.9   & 31.8    \\                                                          
9    & 328  & 650    & 2.634 & 102    & 41.4    \\                                                          
10   & 538  & 1075   & 2.634 & 129    & 52.7    \\
11   & 887  & 1766   & 2.634 & 165    & 67.7    \\
12   & 1453 & 4747   & 2.634 & 217    & 88.1    \\                                                          
13   & 2390 & 4747   & 2.634 & 270    & 110     \\
14   & 3911 & 7868   & 2.634 & 348    & 143     \\
15   & 6183 & 12272  & 2.634 & 426    & 184     \\
\hline
\end{tabular}
\vspace*{0.5cm}
\caption[]{Various properties of EP obtained by means of the 3D BFM
algorithm without monomer overlap ($\overlap=\infty$) at volume fraction $\phi = 8\rho = 0.5$:
imposed scission energy $E$,
the mean chain length $\Nav$, 
the ratio $F(0) = \la N^2 \ra/\Nav$ comparing the first and the
second moment of the number distribution,
the root-mean-squared bond length $l$, 
the $z$-averaged end-to-end distance $\Rendz$ and radius of gyration $\Rgyrz$
obtained using Eq.~(\ref{eq_theo_R2PN}) with $p=2$.
For all scission energies we have used periodic simulation boxes of 
linear size $L=256$ containing $n_{mon} = 2^{20}$ monomers.
\label{tab_bfm_EP}}
\end{center}
\end{table}
\begin{figure}[t]
\centerline{\resizebox{0.9\columnwidth}{!}{\includegraphics*{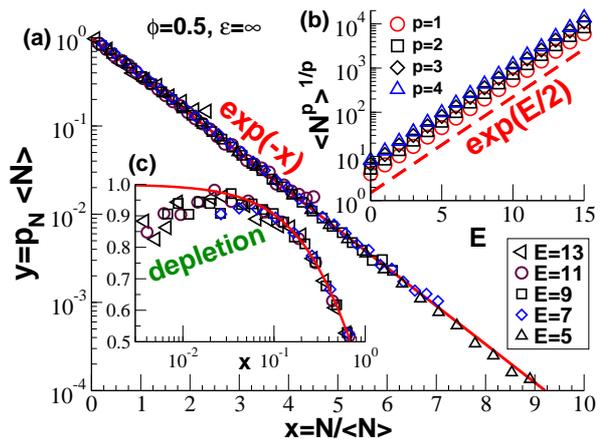}}}
\caption{Normalized chain length distribution $\pchain$ of linear EP 
for different scission energies $E$ obtained
using the classical BFM algorithm without monomer overlap:
{\bf (a)} The main panel demonstrates the collapse of the rescaled distribution $y=\pchain \Nav$
as a function of $x=N/\Nav$.
The exponential decay (solid line) implied by Flory's ideality hypothesis is
(to first order) consistent with our data.
{\bf (b)} First four moments of the distribution {\em vs.} $E$.
{\bf (c)} Replot of the data of panel (a) in log-linear coordinates focusing on short chains.
The data points are systematically {\em below} the exponential decay (solid line) for $x \ll 1$.
\label{fig_bfm_EP_PN}
}
\end{figure}

\subsubsection{Some computational results}
\label{bfm_EP_someresults}
As summarized in Table~\ref{tab_bfm_EP}, we have sampled EP systems with scission energies up to $E=15$, 
the largest energy corresponding to a mean chain length $\Nav\approx 6183$. 
As for the athermal classical monodisperse BFM systems we obtain a dimensionless compressibility $\gT=0.24$, 
an effective bond length $b=3.244$ and a swelling coefficient $\ce = 0.41$. 
%
As discussed in Sec.~\ref{theo_connectivity_poly} one expects EP melts to be Flory distributed
[Eq.~(\ref{eq_theo_PN_Flory})]
if the chemical potential of the chains is extensive with respect to their mass, $\muchainN \sim N$.
The main panel of Fig.~\ref{fig_bfm_EP_PN} pre\-sents the normalized length distribution $\pchain$
for different scission energies $E$ as indicated. A nice data collapse is apparently
obtained if $\pchain \Nav$ is plotted as a function of the reduced chain length $x =N/\Nav$
using the measured mean chain length $\Nav$. At first sight, there is {\em no} sign of
deviation from the exponential decay indicated by the solid line.
The mean chain length itself is given in panel (b) as a function of $E$ together with
some higher moments $\Npav= \sum_N N^p \pchain$ of the distribution.
As indicated by the dash\-ed line, we find $\Npav \sim \exp(p E/2)$ as expected
from Eq.~(\ref{eq_theo_Nav}) \cite{CC90,WMC98b}.
The data presented in the first two panels of Fig.~\ref{fig_bfm_EP_PN} is thus fully consistent
with older computational work \cite{Milchev95,WMC98a,WMC98b,Milchev00,HXCWR06,Kroeger96,Padding04} 
which has let us to believe (incorrectly) that Flory's ideality hypothesis must hold rigorously.
Closer inspection of the histograms reveals, however, deviations for small $x \ll 1$.
As can be seen from panel (c), the probability for short chains is reduced with
respect to the Flory distribution (solid line).
We shall further investigate this depletion in Sec.~\ref{TD_muEP}.

We note finally that all EP systems presented here have been sampled within 4 months while the sample of 
monodisperse configurations for $N=8192$ alone required about 3 years on a similar XEON processor. 
EP are therefore very interesting from the computational point of view, allowing for an efficient test 
of theoretical predictions.

\section{Thermodynamic properties of BFM melts}
\label{TD}

\subsection{Introduction}
\label{TD_intro}
To characterize the soft BFM model introduced in Sec.~\ref{bfm_soft} we will first investigate thermodynamic 
pro\-per\-ties such as the mean overlap energy per monomer $e$, the specific heat $\cV$, the dimensionless 
compressibility $\gT$ or the excess chain chemical potential $\muchainN$ as functions of the reduced 
overlap strength $x = \overlap/T = \overlap \beta$. For the small-$x$ limit these properties have 
been calculated long ago by Edwards \cite{DoiEdwardsBook} as summarized in Eq.~(\ref{eq_FE_Thigh}).
Various thermodynamic pro\-per\-ties obtained from the quoted free energy will be 
compared with our numerical results \cite{WCK09}. 
%
To demonstrate that deviations from Flory's ideality hypothesis are also present in thermodynamic 
properties we will investigate in detail in Sec.~\ref{TD_muEP} the chemical potential in systems of annealed 
EP using the classical BFM algorithm without monomer overlap. 

\begin{figure}[t]
\centerline{\resizebox{0.9\columnwidth}{!}{\includegraphics*{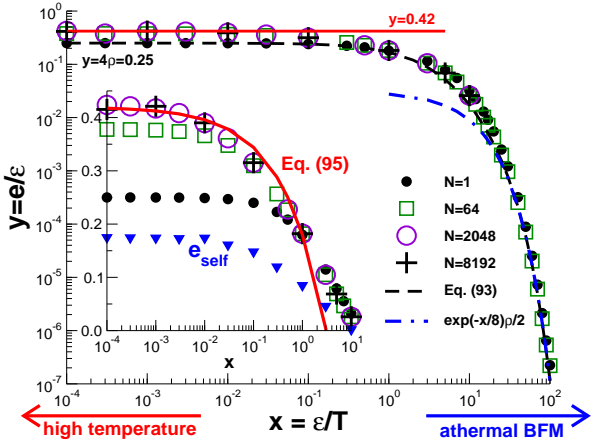}}}
\caption{Mean overlap energy per monomer $y=e/\overlap$ {\em vs.} $x = \overlap/T$ 
for several chain lengths $N$ as indicated. The energy decreases monotonously with increasing $x$. 
The decay becomes Arrhenius-like for $x \gg 10$ (dash-dotted line).
The dashed line indicates the energy predicted from the second virial of soft BFM beads, 
Eq.~(\ref{eq_TD_evirial}).
The main figure demonstrates the weak chain length dependence on logarithmic scales,
especially for strong excluded volume interactions ($x > 1$).
Inset: Same data plotted with linear vertical axis emphasizing the higher mean energy
for long polymers ($N > 64$) for $x \ll 1$ caused by a self-energy contribution
$\eself/\overlap \approx 0.18$. The self-energies are indicated by the filled triangles.
Eq.~(\ref{eq_TD_eThigh}) is represented by the bold line.
\label{fig_TD_energy}
}
\end{figure}
\subsection{Mean overlap energy}
\label{TD_energy}
%
%
From the numerical point of view the simplest thermodynamic property to be investigated here
is the mean interaction energy per monomer, $e = \la E \ra/\nmon$, due to the Hamiltonian,
Eq.~(\ref{eq_bfm_Hamiltonian}). Fig.~\ref{fig_TD_energy} presents the dimensionless energy 
$y = e/\overlap$ for BFM melts ($\phi=0.5$) for different chain length $N$.
Decreasing the overlap strength $x$ starting from configurations obtained using the classical BFM 
(Sec.~\ref{bfm_equil}), the interaction energy increases first exponentially for large $x$ and 
levels off for $x \ll 1$ where the monomers freely overlap.
The data for unconnected beads ($N=1$) represented by the filled spheres and polymer chains
($N \gg 1$) are broadly speaking similar, especially for large overlap penalties, $x > 1$.
Interestingly, the mean energy of polymer melts increases more strongly for $x \ll 1$
as can be seen better from the log-linear data representation chosen in the inset of 
Fig.~\ref{fig_TD_energy}.
Also shown in the inset is the mean intrachain self-energy per mono\-mer $\eself$
(filled triangles) obtained for the largest chain length available for a given $x$.
In fact about half of the energy of polymer melts
for all $x$ is due to the self-interactions of the chains \cite{WCK09}.
For $x \ll 1$ the self-energy becomes $\eself/\overlap \approx 0.18$
which is exactly the observed energy difference between polymer and bead systems.

%
Before addressing this point let us consider the energy of soft BFM beads ($N=1$)
for which the second virial coefficient $\vvir(x)$ has been given in Eq.~(\ref{eq_bfm_vvir}).
Since $e = \partial_{\beta} (\beta f(\beta))$ the mean energy becomes to
leading order \cite{McQuarrieBook}
\begin{eqnarray}
y(x) & \approx & \frac{1}{2} \rho \frac{\partial \vvir(x)}{\partial x}  \nonumber \\ 
     & = & \frac{\rho}{2}
\left(e^{-x/8} + 3 e^{-x/4} + 3 e^{-x/2} + e^{-x} \right)
\label{eq_TD_evirial}
\end{eqnarray}
corresponding to the first term in the third line of Eq.~(\ref{eq_FE_Thigh}).
Eq.~(\ref{eq_TD_evirial}) is represented by the dashed line in Fig.~\ref{fig_TD_energy}.
It corresponds to an Arrhenius behavior with
$y \approx \rho \exp(-x/8)/2$ for $x \gg 1$ (dash-dotted line) 
and to $y \rightarrow \frac{1}{2} 8\rho = 4\rho$ for $x \ll 1$.
This simple formula predicts well the bead data over the entire range of $x$
and also yields a remarkable fit for polymer chains with larger overlap penalties.

\begin{figure}[t]
\centerline{\resizebox{0.9\columnwidth}{!}{\includegraphics*{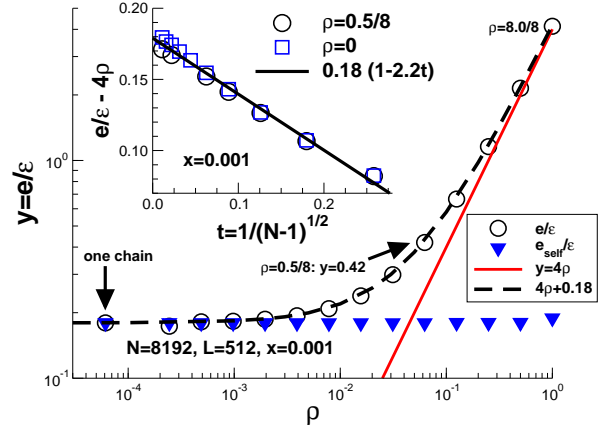}}}
\caption{Reduced mean energy $e/\overlap$ (spheres)
and self-energy $\eself/\overlap$ (triangles) as functions of
the number density $\rho$ for $N=8192$, $L=512$ and $x = 0.001$.
As shown by the dashed line,
$e(\rho)$ is a superposition of the mean field energy $4\rho$ and
the (essentially) constant self-energy $\eself/\overlap \approx 0.18 N^0 x^0 \rho^0$.
Inset: $e/\overlap - 4\rho$ as a function of chain length $1/\sqrt{N-1}$
for our reference density $\rho=0.5/8$ and for a single chain ($\rho=0$).
The linear slope (bold line) is expected from the return probability of
Gaussian chains.
\label{fig_TD_energy_phi}
}
\end{figure}
The energy difference between polymer chains and beads for $x \ll 1$ is
accounted for by the first free energy contribution indicated in Eq.~(\ref{eq_FE_Thigh}).
This contribution is further investigated in Fig.~\ref{fig_TD_energy_phi}
presenting data for such small $x$ that the entropy dominates
all conformational properties. The self-energy of a chain is thus given by the probability 
$p(s,\deltavec)$ that a random walk of $s$ BFM bonds returns to a relative position 
$\deltavec$ with respect to a reference monomer at $\rvec$.  Hence,
\begin{equation}
\eself = \frac{2}{N} \sum_{\deltavec} \sum_{s=2}^{N-1} (N-s) J(\deltavec) p(s,\deltavec)
\label{eq_TD_enoninter}
\end{equation}
where the first sum runs over all positions with non-vanish\-ing coupling constant
$J(\deltavec)$ as defined in Eq.~(\ref{eq_bfm_J}).
The probability $p(s,\deltavec)$ and the weights $J(\deltavec) p(s,\deltavec)$ can be
tabulated in principle for small $s$. Since the return probability decreases strongly with $s$,
these  model-specific small-$s$ values dominate the integral, Eq.~(\ref{eq_TD_enoninter}).
As can be seen from the inset of Fig.~\ref{fig_TD_energy_phi} for single chains 
($\rho=0$), $\eself \approx 0.18 \overlap$ for large $N$.
The weak $N$-dependence visible in the panel stems from the upper integration
boundary over the Gaussian return probability which leads to a chain length correction
linear in $t \equiv 1/\sqrt{N-1}$ (bold line). 
Also shown in the panel are energies for our reference density $\rho=0.5/8$.
They are shifted vertically by the mean field energy $4 \rho$
assuming that density fluctuations of different chains do not couple.
The main panel presents the mean energy $e$ and the mean self-energy $\eself$
as functions of the density $\rho$ for chains of length $N=8192$.
The self-energy (triangles) stays essentially $\rho$-independent.
The total interaction energy sums over the self-energy and mean-field energy contributions
as shown by the dashed line. The self-energy contribution can only be neglected
for volume fractions larger than unity.

%
Summarizing Eqs.~(\ref{eq_FE_Thigh}) and (\ref{eq_bfm_vThigh}) 
the energy should scale to leading order in $x$ as
\begin{equation}
y \approx 0.18 + 4\rho
- \
\underline{ \frac{24^{3/2}}{\pi} \frac{\sqrt{x \rho}}{l^3 \rho}}
+ \ldots
\mbox{ for } x \ll 1
\label{eq_TD_eThigh}
\end{equation}
where the two $x$-independent contributions have already been discussed above.
The underlined term stems from the density fluctuation contribution
in Eq.~(\ref{eq_FE_Thigh}).
%
Eq.~(\ref{eq_TD_eThigh}) is indicated by the bold line in the inset of Fig.~\ref{fig_TD_energy}.
It yields a reasonable description for small $x$.
Since the energy is dominated by the two constant contributions to Eq.~(\ref{eq_TD_eThigh})
for $x \le 0.001$ and since higher expansion terms become relevant for $x  > 0.1$,
the predicted $\sqrt{x}$-decay corresponds unfortunately only to the small-$x$ regime.
To show that it is indeed the density fluctuation term which dominates the
temperature dependence for $x \ll 1$ we will consider now the
specific heat $\cV$, i.e. the second derivative of the free energy with respect to $\beta$.

\subsection{Energy fluctuations}
\label{TD_cV}
\begin{figure}[t]
\centerline{\resizebox{0.9\columnwidth}{!}{\includegraphics*{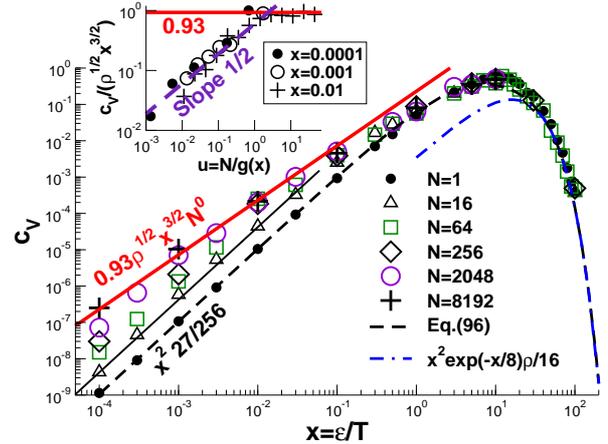}}}
\caption{Specific heat per bead $\cV$ {\em vs.} $x$.
The dashed line indicates the energy fluctuations predicted from the second virial,
Eq.~(\ref{eq_TD_cVvirial}), which fits the data of soft BFM beads ($N=1)$ over six decades.
While the chain length does not matter for strong excluded volume interactions,
the energy fluctuations are found to increase strongly with $N$ for $x \ll 1$.
For short chains we observe $\cV \sim \rho N^{1/2} x^2$ as can be seen for $N=16$ (thin solid line).
The chain length effect drops out for large $N$ where $\cV \approx \rho^{1/2} x^{3/2} N^0$ (bold line)
as suggested by Eq.~(\ref{eq_TD_cVThigh}).
Inset: $\cV /(\rho^{1/2} x^{3/2})$ as a function of the reduced chain length $u=N/\gT(x)$.
\label{fig_TD_cV}
}
\end{figure}

The fluctuations of the interaction energy are addressed in Fig.~\ref{fig_TD_cV} displaying
the enthalpic contribution to the specific heat per monomer,
$\cV = -\beta^2 \partial^2_{\beta} (\beta f(\beta))$
\cite{McQuarrieBook}.
Using the second virial of soft BFM beads, Eq.~(\ref{eq_bfm_vvir}), one obtains
\begin{equation}
\cV = \frac{\rho}{2} x^2  \left(
\frac{1}{8} e^{-x/8} 
+ \frac{3}{4} e^{-x/4} 
+ \frac{3}{2} e^{-x/2} 
+ e^{-x} \right)
\label{eq_TD_cVvirial}
\end{equation}
as represented by the dashed line.
In the large-$x$ limit this yields an exponential decay $\cV \approx \rho x^2 \exp(-x/8)/16$ (dash-dotted line)
while for $x \ll 1$ a power-law limiting behavior is obtained: $\cV \approx \frac{27}{16} \rho x^2 \sim x^2$.
Eq.~(\ref{eq_TD_cVvirial}) predicts the energy fluctuations of BFM beads for essentially all $x$, 
slightly underestimating the maximum of $\cV$ at $x \approx 10$.
Chain length effects are small for large $x$ where Eq.~(\ref{eq_TD_cVvirial}) can be used to fit the 
specific heats of polymer melts.
%
%
Strong $N$-effects are, however, visible for $x \ll 1$ where $\cV$ increases mono\-ton\-ously with $N$.
This can better be seen from the inset where the specific heat is plotted
as a function of the reduced chain length $u=N/\gT$ with $\gT$ being the dimensionless
compressibility determined in Sec.~\ref{TD_compress}.\footnote{Since $e$ and 
$\cV$ correspond to different derivatives of the free energy $f$
with respect to $\beta$, there is obviously no inconsistency in the finding that $\cV$
reveals larger chain length effects than $e$.}
For large chains with $u \gg 1$ this increase levels off at an $N$-independent envelope
\begin{equation}
\cV \approx
\frac{24\sqrt{6}}{\pi} \frac{\rho^{1/2}}{l^3} x^{3/2} N^0
+ \ldots
\label{eq_TD_cVThigh}
\end{equation}
due to the density fluctuation contribution in Eq.~(\ref{eq_FE_Thigh}).
In contrast to Eq.~(\ref{eq_TD_eThigh}) for the mean energy the density fluctuation term does
now correspond to the leading contribution to the numerically measured property.
This increases the range where the density fluctuation contribution can be demonstrated
to over three decades in $x$.
Eq.~(\ref{eq_TD_cVThigh}) is indicated by the bold lines in the main panel and the inset
of Fig.~\ref{fig_TD_cV}. 

\subsection{Compressibility}
\label{TD_compress}
\begin{figure}[t]
\centerline{\resizebox{0.9\columnwidth}{!}{\includegraphics*{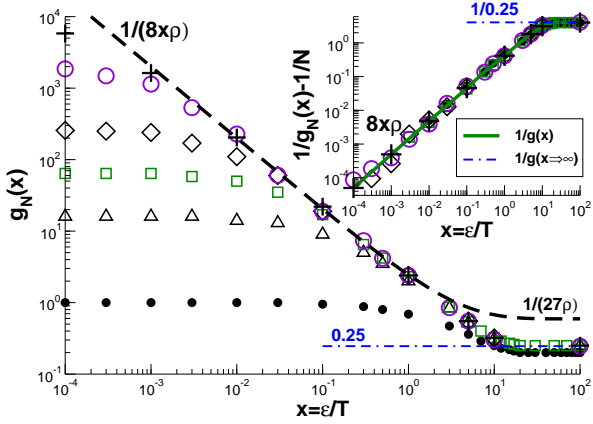}}}
\caption{Dimensionless compressibility $\gN(x) \equiv \lim_{q\to\infty} S(q,N)$ {\em vs.} $x$ for
different $N$ using the same symbols as in Fig.~\ref{fig_TD_cV}.
Main panel: Raw data as obtained from the low-wavevector limit of the structure factor.
Chain length effects become irrelevant for $x \ge 0.1$ if $N \ge 64$
and for $x > 0.001$ if $N \ge 2048$.
The data are compared to the simple second virial approximation $1/\vvir(x) \rho$ (dashed line)
which reduces to $1/(8x\rho)$ for $x \ll 1$.
As one expects, the compressibility levels off for large $x$ and becomes identical to the value
$\gT \approx 0.25$, known for the classical BFM \cite{WBM07} (dash-dotted line).
Inset: As suggested by Eq.~(\ref{eq_TD_gNeffect}) the excess part of the inverse compressibility
$1/\gN(x)-1/N$ becomes chain length independent, i.e. the data points for all $N$ collapse.
The bold line indicates $\gT(x) = \lim_{N\to\infty} \gN(x)$ from Table~\ref{tab_overlap}.
\label{fig_TD_compress}
}
\end{figure}

The key control property characterizing the decree of coupling of the polymer chains due to 
the imposed penalty $\overlap$ is the dimensionless compressibility 
$\gT(x) \equiv \lim_{N\to \infty} \gN(x)$ of asymptotically long chains. 
As suggested by Eq.~(\ref{eq_theo_gTdef}), we compute first the dimensionless compressibility
$\gN(x) \equiv \lim_{q\to 0} S(q,N)$ from the low-$q$ limit of the total monomer structure factor
for different overlap penalties $x$ and chain lengths $N$  (see below for details).
These raw data are presented in Fig.~\ref{fig_TD_compress} as a function of $x$.
As one expects, $\gN(x)$ decreases mono\-ton\-ously with $x$.
%
%
Note that the structure factor $S(q,t)$ measures the complete compressibility, 
not just its excess contribution.
As can be seen, e.g., from Eq.~(\ref{eq_FE_Thigh}) or from the virial expansion of
polymer solutions \cite{DegennesBook}, the compressibility can be written in general as
\begin{equation}
\frac{1}{\gN(x)} =
\rho \ \frac{\partial^2(\beta f(\beta) \rho)}{\partial \rho^2}
=
\frac{1}{N} + \frac{1}{\gex(x,N)}
\label{eq_TD_gNeffect}
\end{equation}
for all $x$ with $\gex(x,N)$ being the excess contribution to the compressibility
which may, at least in principle, depend on $N$.\footnote{Small corrections may arise as they 
do arise for the chemical potential as shown below in Sec.~\ref{TD_muEP}.} 
As can be seen from the inset of Fig.~\ref{fig_TD_compress}, {\em all} rescaled compressibilities 
collapse, however, on {\em one} $N$-independent master curve if one plots 
$1/\gex(x,N) \equiv 1/\gN(x)-1/N$ as a function of $x$,
even the compressibilities obtained for unconnected beads ($N=1$).
Within numerical accuracy the $N$-dependence observed for $\gN(x)$ can therefore be attributed
to the trivial osmotic contribution and the excess compressibility $\gex \sim N^0$ is thus
identical to the compressibility $\gT(x)$ of asymptotically long chains.
The bold line indicated in the inset presents the best values of $\gT(x)$
summarized in Table~\ref{tab_overlap}.
These values have been obtained from the excess compressibilities for the largest chain length
available for $x \ge 0.001$. A {\em precise} numerical determination of $\gex(x)$ becomes impossible
for even smaller overlap penalties. We thus have used for the smallest $x$-values sampled
the theoretical prediction
\begin{equation}
\frac{1}{\gT(x)}
\approx \vvir(x) \rho
\left(1 - \
\underline{\frac{3\sqrt{3}}{2\pi} \frac{(\vvir(x)\rho)^{1/2}}{b^3(x)\rho}}
\ldots \right)
\label{eq_TD_gThigh}
\end{equation}
for $x \ll 1$ due to the postulated free energy, Eq.~(\ref{eq_FE_Thigh}).
The prefactor $\vvir(x)\rho$ representing the bare monomer interaction is indicated
by the dashed line in the main panel of Fig.~\ref{fig_TD_compress}.
Hence, $\gT(x) \approx 1/(8x\rho) = 2/x$ for weak interactions.
The underlined term is the leading correction due to the density fluctuation contribution
to the free energy. It implies that the excess compressibilities for polymer melts and
unconnected beads cannot be completely identical. However, 
the difference is far too small to be measurable in the limit where Eq.~(\ref{eq_TD_gThigh}) applies.
Although this result is unfortunate from the theoretical point of view,
the data collapse observed in the inset suggests that it is acceptable to numerically estimate
the long chain compressibility $\gT(x)$ by computing the structure factors of rather short chains.

\begin{figure}[t]
\centerline{\resizebox{0.9\columnwidth}{!}{\includegraphics*{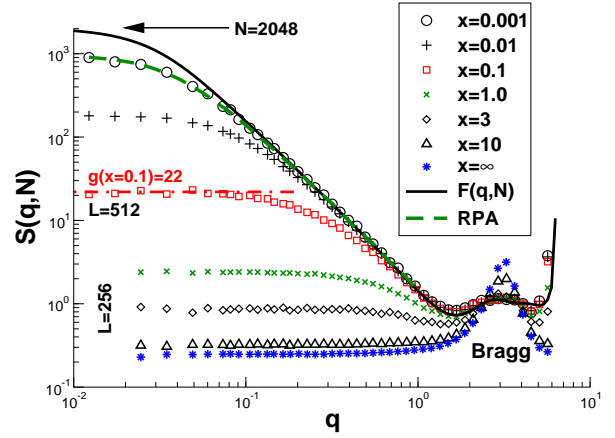}}}
\caption{Total structure factor $S(q)$ as a function of wavevector $q$ for $N=2048$
for different overlap penalties $x=\overlap/T$ as indicated.
For comparison, we have also included the single chain form factor $F(q)$ for $x=0.001$.
The low-wavevector limit of the structure factor is used to determine the dimensionless
compressibility $\gN(x)$. 
Only for $x \le 3$ does the structure factor decay monotonously with $q$
as suggested by the RPA formula, Eq.~(\ref{eq_TD_RPA}).
$G(q)$ becomes essentially constant for smaller temperatures
except for wavevectors corresponding to the first sharp diffraction peak.
The box size $L=256$ allows only a direct and fair determination of $\gN(x)$ for $x > 0.1$.
We have been forced to increase the box size to $L=512$ for smaller $x$ as may be seen
for an example with $x=0.1$ (dash-dotted line).
As shown by the bold dashed line, the RPA formula is used to improve the estimation of $\gN(x)$
for small $x$.
\label{fig_TD_Sq}
}
\end{figure}
\begin{figure}[t]
\centerline{\resizebox{0.9\columnwidth}{!}{\includegraphics*{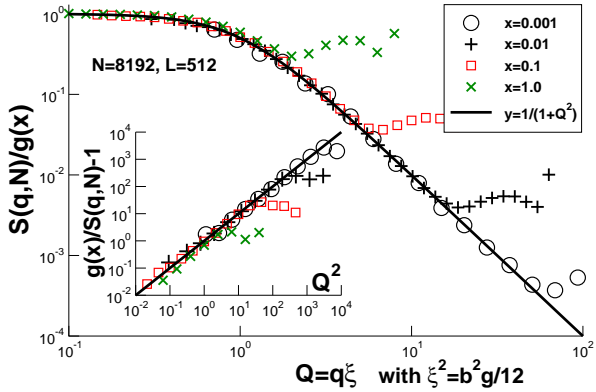}}}
\caption{Rescaled total structure factor $S(q)/\gT(x)$
as a function of the reduced wavevector $Q \equiv q\xi$
for chain length $N=8192$ and several $x \le 1$ as indicated.
The screening length $\xi$ of the thermal blob is obtained according to Eq.~(\ref{eq_theo_xidef})
using $\gT(x)$ and $b(x)$ from Table~\ref{tab_overlap}. The bold line compares the data with 
Eq.~(\ref{eq_TD_RPAapprox}). If replotted as indicated in the inset
the data collapse on the bisection line. Deviations from the RPA formula
become visible for larger $x$ as shown for $x=1$ (crosses).
\label{fig_TD_Sqxi}
}
\end{figure}

%
We now turn to the total structure factor $S(q,N)$ shown in Fig.~\ref{fig_TD_Sq} to explain how the 
compressibilities have been obtained. Only chains of length $N=2048$ are presented for clarity. 
Since the wavevectors $q$ used for computing $S(q)$ must be commensurate with the cubic simulation box 
of linear dimension $L$, i.e. the smallest possible wavevector is $2\pi/L$, it thus is important to 
have a sufficiently large box for a reasonable determination of $\gN(x)$.
Note that around and above $q \approx 2$ monomer structure and lattice effects become important.
Being interested in universal physical behavior we focus on wavevectors $q \ll 1$.
For comparison, we have also included the single chain form factor $F(q)$
for $x=0.001$ (bold line).
Note that the qualitative shape of $F(q)$
--- decaying monotonously with $q$ from its maximum value $F(q=0,N)=N$ ---
depends very little on the overlap penalty $x$ (not shown).
We remind that the ``random phase approximation" (RPA) formula
\cite{DegennesBook,DoiEdwardsBook}
\begin{equation}
\frac{1}{S(q,N)} = \frac{1}{F(q,N)} + \frac{1}{\gex(x,N)}
\label{eq_TD_RPA}
\end{equation}
relates the total structure factor to the measured form factor.
Eq.~(\ref{eq_TD_RPA}) is of course consistent with Eq.~(\ref{eq_TD_gNeffect}) in the $q\to 0$ limit.
It allows to directly fit for the excess compressibility $\gex(x,N) \approx \gT(x)$
using the measured structure factor $S(q,N)$ and form factor $F(q,N)$,
at least in the $x$-range where the RPA approximation applies.
As may be seen from the figure, $S(q,N)$ indeed decreases systematically with $x$,
i.e. with decreasing $\gT(x)$. For $x \le 3$ it also decays monotonously
with $q$, again in agreement with Eq.~(\ref{eq_TD_RPA}).
Interestingly, this becomes qualitatively different for larger excluded volume interactions
($x > 3$) where the total structure factor is essentially constant (in double-lo\-gar\-ith\-mic coordinates),
very weakly {\em increasing} monotonously with $q$.
The RPA formula apparently does not apply in this limit in agreement with Eq.~(\ref{eq_Ginzburg}).
Fortunately, this is of no concern for our main purpose --- to compute $\gT(x)$ ---
since in precisely this limit the compressibility is readily obtained from a broad plateau
(even for much smaller boxes) which in addition becomes chain length independent,
as we have already seen from the inset of Fiq.~(\ref{fig_TD_compress}).
Using boxes with $L=256$ it is possible to directly measure the plateau values for $x \le 0.3$.
For smaller $x$ we have simulated boxes with $L=512$ containing $\nmon \approx 8.4 \cdot 10^6$ monomers
and corresponding to a smallest wavevector $q \approx 0.01$.
This box size becomes again insufficient for the smallest reduced overlap penalties $x$ we have simulated,
as shown in Fig.~\ref{fig_TD_Sq} for $x=0.001$ (dashed line). It is for these values where the
RPA formula, Eq.~(\ref{eq_TD_RPA}), allowing to fit the {\em deviation}
from the (barely visible) plateau, has been particulary useful.

%
As already stated in Sec.~\ref{theo_incompress}, for intermediate wavevectors
(where $q$ corresponds to distances much smaller than the radius of gyration
and much larger than the monomer size) the general RPA Eq.~(\ref{eq_TD_RPA}) may be rewritten as
\begin{equation}
\frac{1}{S(q)} = \frac{1}{\gT(x)} + \frac{1}{2} a^2(x) q^2 = \frac{1}{\gT(x)} \left(1 + (q \xi)^2 \right)
\label{eq_TD_RPAapprox}
\end{equation}
where we have used that the form factor becomes $F(q) \approx 2/(a q)^2$ \cite{DoiEdwardsBook}.
This assumes that corrections to Gaussian chain statistics may be ignored \cite{WBJSOMB07,BJSOBW07}
and that finite chain size effects are negligible.
From the numerical point of view the approximated RPA Eq.~(\ref{eq_TD_RPAapprox})
has the disadvantage that the effective bond length $b(x)=\sqrt{6} a(x)$ needs to be determined first.
As shown in Fig.~\ref{fig_TD_Sqxi}, it has the advantage that it allows for an additional
test of the values of $\gT(x)$ and $b(x)$ indicated in Table~\ref{tab_overlap}.
The main panel presents the rescaled structure factor $S(q)/\gT(x)$ for chains of length $N=8192$
as a function of $Q \equiv q\xi$ with $\xi$ being obtained from $\gT(x)$ using Eq.~(\ref{eq_theo_xidef}).
All data collapse on the master curve $1/(1+Q^2)$ indicated by the bold line provided that the
wavevector $q$ remains sufficiently small and no local physics is probed.
That the used compressibilities are accurate is emphasized further in the inset
where $\gT(x)/S(q)-1$ is plotted as a function of $Q^2$ using only sufficiently small wavevectors $q$.
According to Eq.~(\ref{eq_TD_RPAapprox})
all data should collapse on the bisection line in double-logarithmic coordinates
if the correct compressibilities are used. This is indeed the case.
%
Please note the weak deviations visible for $x=1$ which is due to the breakdown of the RPA formula
for large $x$ mentioned above.

\subsection{Chemical potential: Gaussian contribution}
\label{TD_muMD}

\begin{figure}[t]
\centerline{\resizebox{0.9\columnwidth}{!}{\includegraphics*{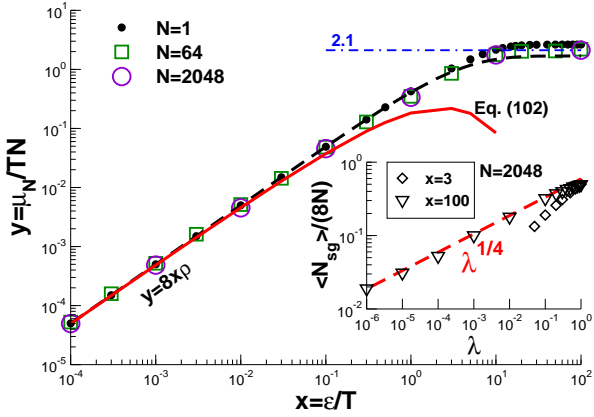}}}
\caption{Reduced chemical potential $y=\muchainN/TN$ measuring the reversible work for bringing a 
test chain of length $N$ into a bath of chains of equal length $N$ at volume fraction $\phi=0.5$.
Increasing linearly (dashed line) for small $x$ it levels off for large $x \gg 1$ (dash-dotted line).
The dashed line shows the second virial approximation $y \approx \vvir(x) \rho$
for unconnected beads, fitting successfully the data below $x \approx 1$. The bold line 
indicates Eq.~(\ref{eq_muMD_Thigh}).
Inset:
The chemical potential has been obtained by thermodynamic integration over the 
excluded volume interaction of an inserted ghost chain \cite{MP94}. 
\label{fig_TD_muMD}
}
\end{figure}

%
%
According to Flory's ideality hypothesis the chemical potential $\muchainN$ of 
polymer melts is expected to be extensive with respect to their mass \cite{DegennesBook}.
Fig.~\ref{fig_TD_muMD} presents the reduced excess contribution to the chemical potential, 
$y \equiv \muchainN/TN$, obtained using thermodynamic integration (as explained below) 
for three chain lengths $N=1$, $64$, and $2048$ as functions of $x=\overlap/T$. 
As one expects, $y(x)$ increases first linearly with $x$ and then levels off. 
Chain length effects are again small on the logarithmic scale chosen in the plot.
For large $x$ the chemical potential becomes slightly larger for beads ($y\approx 2.64$) 
than for long chains where $y \approx 2.1$ as shown by the dash-dotted line. 
%
That the chemical potential of polymer chains is reduced compared to unconnected 
beads is expected due to the (effectively) attractive bond potential.
For $x \ll 1$ this reduction should be described by the density fluctuation 
contribution to the free energy [Eq.~(\ref{eq_FE_Thigh})] 
which corresponds to an excess chemical potential
\begin{eqnarray}
y(x) & = & \frac{\partial (\beta f(\beta) \rho)}{\partial \rho}
\nonumber \\ 
& \approx & \vvir(x) \rho \left( 1 - \
 \underline{ \frac{3\sqrt{3}}{\pi} \frac{(\vvir(x)\rho)^{1/2}}{b^3(x)\rho}}
+ \ \ldots \right)
\label{eq_muMD_Thigh}
\end{eqnarray}
for $x \ll 1$ with $\vvir(x)$ being the second virial of unconnected beads.
The dashed line in Fig.~\ref{fig_TD_muMD} presents the leading contribution 
$\vvir(x) \rho$ for unconnected beads,
the bold line in addition the underlined connectivity contribution in Eq.~(\ref{eq_muMD_Thigh}).
Surprisingly, it turns out that the simple second virial approximation provides a better fit 
over the entire $x$-range than the full prediction. 
%
That the density fluctuation contribution overestimates the reduction of the chemical potential 
for $x > 1$ is in agreement with Eq.~(\ref{eq_Ginzburg}).
For $x \ll 1$ where Eq.~(\ref{eq_muMD_Thigh}) applies in principle the relative correction, 
scaling as $\sqrt{x/\rho}$, becomes unfortunately too small to allow a fair test of the theory
using the measured chemical potential. 

%
We now explain how the chemical potential presented in Fig.~\ref{fig_TD_muMD} has been obtained.
Obviously, the simple insertion method due to Widom \cite{FrenkelSmitBook} becomes rapidly 
inefficient with increasing $x$. Generalizing the method suggested in \cite{MP94,WM94} we have 
performed the thermodynamic integration \cite{FrenkelSmitBook}
\begin{equation}
\beta \muchainN = \int_{\lambda(\overlap)}^{1} d\lambda \frac{\la \Novghost \ra}{\lambda}
\label{eq_muMD_thermoint}
\end{equation}
over discrete values of the affinity $\lambda = \exp(-\overlapghost \beta/8)$ 
characterizing the excluded volume interaction of a ghost (g) chain that is inserted into 
an equilibrated system (s). 
$\la \Novghost \ra$ refers to the mean number of lattice sites where system and ghost monomer cube corners overlap
at a given interaction $\lambda$.
Generalizing the Potts spin mapping, Eq.~(\ref{eq_bfm_spinmodel}),
of the excluded volume interactions for homopolymers presented above, we use now {\em two} spin lattices,
$S_\mathrm{s}(\rvec)$ describing (as before) the interaction of the system monomers and
$S_\mathrm{g}(\rvec)$ the ghost chain. The spin lattices are kept at the same temperature $T$
and are both characterized by the same penalty $\overlap=1$ which has to be paid 
for a complete overlap of two system monomers or two ghost monomers.
The interaction of both spins is described by
\begin{equation}
\Delta E_{sg} = \sum_{\rvec} S_\mathrm{s}(\rvec) \sum_{\deltavec} J_\mathrm{sg}(\deltavec) 
S_\mathrm{g}(\rvec+\deltavec)
\label{eq_muMD_ghostinter}
\end{equation}
with coupling constants $J_\mathrm{sg}(\deltavec) \sim \overlapghost$ defined as in Eq.~(\ref{eq_bfm_J}) 
taken apart the energy parameter $\overlap$ which is replaced by the tunable interaction energy 
$\overlapghost$.
Starting with decoupled system and ghost configurations at $\overlapghost=0$, i.e. $\lambda=1$, 
we increase the interaction parameter 
up to $\overlapghost=\overlap$, i.e.~ $\lambda(\overlap) = \exp(-\overlap \beta/8)$, 
always keeping the coupled system at equilibrium. 
Monitoring the distribution of the number $\Novghost$ of overlaps between system and ghost cube corners 
we use multihistogram methods as described in \cite{MP94,WM94} to improve the precision of the integral.
The mean overlap number $\la \Novghost \ra$ is shown in the inset of 
Fig.~\ref{fig_TD_muMD} as a function of $\lambda$ for $N=2048$ and two inverse temperatures 
$x=3$ and $x=100$. Starting from $\lambda=1$ the overlap number decreases mono\-tonous\-ly 
with increasing coupling between system and ghost mono\-mers. 
Interestingly, a power-law behavior 
\begin{equation}
\la \Novghost \ra/N \approx \lambda^{1/4}
\label{eq_TD_lambdapower}
\end{equation}
is found empirically for large $x \gg 10$ (dashed line). Fitting this power law and integrating 
then analytically Eq.~(\ref{eq_muMD_thermoint}) provides a useful crosscheck of the numerical 
integration using the multihistogram analysis. This is a technically important finding, 
since the multihistogram analysis requires overlapping distributions of $\Novghost$ and 
hence much more equilibrated intermediate values $\lambda$ as indicated for $x=100$. 

\subsection{Chemical potential: Non-extensive corrections}
\label{TD_muEP}

\subsubsection{Theoretical predictions}
\label{TD_muEP_theo}
\begin{figure}[t]
\centerline{\resizebox{0.9\columnwidth}{!}{\includegraphics*{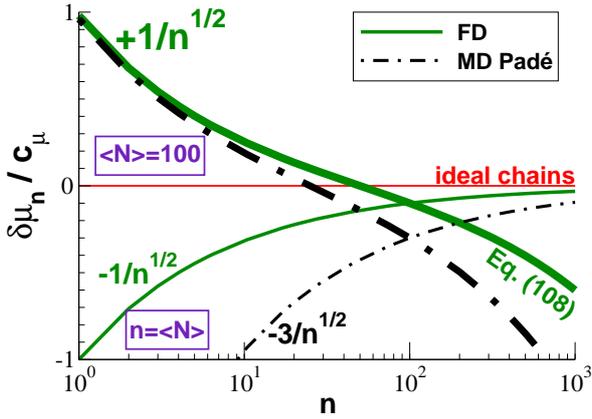}}}
\caption{Non-extensive deviation of the chemical potential $\dmuchainn/\cep$ 
as a function of the test chain length $n$ as predicted by perturbation calculation \cite{WJC10}. 
The reference for the chemical potential $\muchainn$ is set here by the chemical potential
of Gaussian chains of same effective bond length $b$.
Flory-distributed (FD) melts are indicated by solid lines, monodisperse (MD) melts
by dash-dotted lines (Pad\'e approximation).
The bold lines refer to test chains in melts of constant (mean) chain length with $\Nav=100$.
The deviation changes sign at $n \approx \Nav$.
The thinner lines refer to test chains of same length as the typical melt chain, $n \equiv \Nav$,
where the asymptotic Gaussian behavior ($\dmuchainn=0$) is approached systematically from below.
\label{fig_muEP_sketch}
}
\end{figure}
We have seen in Sec.~\ref{TD_muMD} that the chemical potential $\muchainN$ of a test chain
of length $N$ plugged into a melt of chemically identical mono\-di\-sperse polymers of the same length
increases essentially linearly with $N$. 
Focusing now on classical BFM systems where $\overlap = \infty$ and setting temperature to unity ($T=1$),
we show how small non-linear deviations may be captured numerically \cite{WJC10}.
More generally, the challenge is to characterize the chemical potential $\muchainn$ of a test chain of 
length $n$ immersed into a bath of $N$-chains of an arbitrary (normalized) length distribution $\pchain$. 
We remind that according to Flory's ideality hypothesis one expects \cite{DegennesBook},
\begin{equation}
\muchainn = \mu \ n  \mbox { for } \gT \ll n \ll \Nav^2,
\label{eq_muEP_gauss}
\end{equation}
with $\mu > 0$ being the effective chemical potential {\em per} mo\-no\-mer.
The upper boundary $\Nav^2$ indicated in Eq.~(\ref{eq_muEP_gauss})  
is due to the well-known swelling of extremely large test chains where the bath acts as a good solution 
\cite{DegennesBook,SchaferBook}.
As we have pointed out in Sec.~\ref{intro_key}, Flory's hypothesis has been challenged 
by the discovery of long-range correlations imposed by the incompressibility constraint.
These correlations lead to the systematic swelling of chain segments as further discussed in
Sec.~\ref{conf} \cite{WMBJOMMS04,WCK09}.
We question here the validity of Flory's hypothesis for an important {\em thermodynamic} property, 
the chemical potential $\muchainn$ of a test chain inserted into a three-dimensional melt.
One expects that the correlation hole potential (Fig.~\ref{fig_theo_constraint}) leads to a deviation 
\begin{equation}
\dmuchainn \equiv \muchainn - \mu n \approx \Ustar(n) \sim +1/\sqrt{n}
\mbox{ for } n \ll \Nav
\label{eq_muEP_claim0}
\end{equation}
that is non-extensive in chain length and this irrespective of the distribution $\pchain$ of the bath. 
Assuming a {\em quenched} Flory-size distribution, Eq.~(\ref{eq_theo_PN_Flory}), it can be demonstrated 
as shown in Appendix~\ref{app_stat_muEP} that
\begin{equation}
\dmuchainn \approx \frac{\cep}{\sqrt{n}} \left(1 - 2 \mu n \right)
\mbox{ for } g \ll n \ll \Nav^{2}
\label{eq_muEP_claim1}
\end{equation}
where we have set $\cep = 3 \ce /8$ using the swelling coefficient $\ce = \sqrt{24/\pi^3} / \rho b^3$ 
defined in the Sec.~\ref{intro_key}. 
Eq.~(\ref{eq_muEP_claim1}) is represented by the bold solid line in Fig.~\ref{fig_muEP_sketch}.
(A corresponding prediction for a monodisperse bath \cite{WJC10} is indicated by the dash-dotted line.)
As anticipated by Eq.~(\ref{eq_muEP_claim0}), the first term in Eq.~(\ref{eq_muEP_claim1}) dominates 
for short test chains. 
The second term dominates for large
test chains with $n > 1/(2\mu)$ becoming non-perturbative for $n \gg 1/\mu^2$. Both contributions
to $\dmuchainn$ {\em decrease} with increasing $n$.\footnote{This corresponds to an effective enhancement
factor of the partition function quite similar to the $\dmuchainn = -(\gamma-1) \log(n)$ in
the standard excluded volume statistics with $\gamma \approx 1.16 > 1$ being the self-avoiding
walk susceptibility exponent \cite{DegennesBook}.}
Interestingly, while $\dmuchainn$ decreases at fixed $\Nav$, it {\em increases} as
$\dmuchainn/\cep = -1/\sqrt{n}$ (thin solid line) for a test chain with $n \equiv \Nav$. 
The chemical potential of typical chains of the bath approaches thus the ideal chain limit from below.

Flory-distributed polymer melts are obtained naturally in systems of self-assembled linear
EP where branching and the formation of closed rings are forbidden \cite{CC90}.
Since the suggested correction, Eq.~(\ref{eq_muEP_claim1}), to the ideal chain chemical potential
is weak, the system must remain to leading order Flory distributed and Eq.~(\ref{eq_muEP_claim1})
should thus hold.\footnote{The chemical potential of a test chain {\em does} depend on the length 
distribution $\pchain$ of the bath. However, for an infinite macroscopically homogeneous systems it is 
{\em independent} on whether this distribution is annealed or quenched, i.e. if it is allowed to 
fluctuate or not. This follows from the well-known behavior of fluctuations of
extensive parameters in macroscopic systems: the relative fluctuations vanish as $1/\sqrt{V}$
as the total volume $V\to \infty$. The latter limit is taken first in our calculations, 
i.e. we consider an infinite number of (annealed or quenched) chains. The large-$N$ limit is 
then taken {\em afterwards}.}
This correction implies for the {\em annealed} length distribution of linear EP 
that (to leading order)
\begin{eqnarray}
\pchain & \approx&  \mu e^{-\mu N - \dmuchainN}
\label{eq_muEP_claim2}
\\
&\approx& \mu e^{-\mu N} \left( 1 - \frac{\cep}{\sqrt{N}} (1-2\mu N) \right)
\label{eq_muEP_claim3}
\end{eqnarray}
where both the lower ($\gT \ll N$) and the upper limit ($N \ll \Nav^2$) of validity
are irrelevant in the large-$\Nav$ limit.
Note that Eq.~(\ref{eq_muEP_claim3}) is properly normalized,
i.e. the prefactor $\mu$ of the distribution remains exact
if $\dmuchainN$ is given by Eq.~(\ref{eq_muEP_claim1}). 
At given $\mu$ the first moment increases slightly 
\begin{equation}
\Nav = \mu^{-1} \left(1+ \cep \sqrt{\mu \pi} \right).
\label{eq_muEP_mu2Nav}
\end{equation}
and, more generally, one expects for the  $p$th moment 
\begin{equation}
\frac{\mu^p \Npav}{p!} - 1 = 
\frac{\cep \sqrt{\mu}}{p!} \left[ 2\Gamma(p+3/2) - \Gamma(p+1/2)\right]
\label{eq_muEP_pmoment}
\end{equation}
with $\Gamma(x)$ being the Gamma function \cite{abramowitz}.
The non-ex\-po\-nen\-tial\-ity parameter $\nonexp \equiv 1 - \Npav / p! \Nav^p$ thus scales as
\begin{eqnarray}
\nonexp & = & \wep \cep \sqrt{\mu} \mbox{ with } \nonumber \\
\wep & \equiv & (\Gamma(p+1/2) + \sqrt{\pi} p p! -2 \Gamma(p+3/2))/p!
\label{eq_muEP_nonexp}
\end{eqnarray}
being a $p$-dependent geometrical factor.\footnote{The polydispersity index $I$, 
i.e. the ratio of weight average and number average, becomes 
$I = \la N^2 \ra/\la N \ra^2 = 2 (1 - w_2 \cep \sqrt{\mu}) < 2$. 
In this sense, the distribution becomes narrower just as for dilute good solvent 
EP where $I = 1  + 1/\gamma < 2$ \cite{WMC98b,Yeruk77}.}
%

\begin{figure}[t]
\centerline{\resizebox{0.9\columnwidth}{!}{\includegraphics*{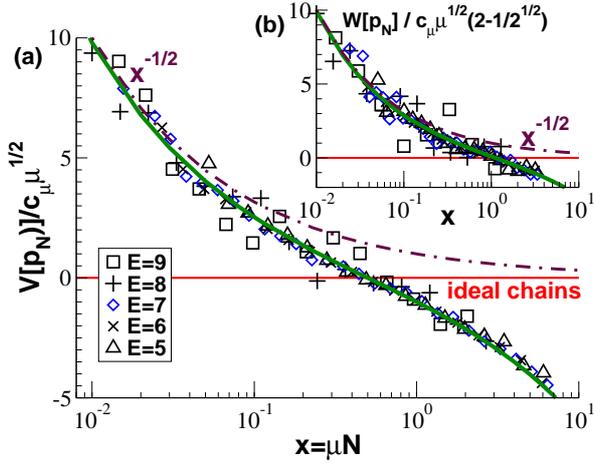}}}
\caption{Characterization of $\pchain$ 
using the functionals
{\bf (a)} $V[\pchain]  \approx \dmuchainN$ and
{\bf (b)} $W[\pchain] \equiv 2 V[\pchain] - V[\pchaintwo]$ 
which should both vanish for perfectly Flory-distributed ideal chains 
as indicated by the horizontal lines in the panels.
Data points for different scission energies $E$ collapse
if $V[\pchain]/\cep \sqrt{\mu}$ and $W[\pchain]/\cep \sqrt{\mu} (2-1/\sqrt{2})$
are plotted {\em vs.} the reduced chain length $x=\mu N$.
For small $x$ both functionals decay as $1/\sqrt{x}$ as shown by the dash-dotted lines.
The bold lines correspond to the full predictions Eq.~(\ref{eq_muEP_VN}) and Eq.~(\ref{eq_muEP_WN})
for $V[\pchain]$ and $W[\pchain]$, respectively.
\label{fig_muEP_WN}
}
\end{figure}
\subsubsection{Computational results for EP melts}
\label{TD_muEP_simu}

Eq.~(\ref{eq_muEP_claim3}) and Eq.~(\ref{eq_muEP_nonexp}) allow us to demonstrate numerically the prediction, 
Eq.~(\ref{eq_muEP_claim1}), from the observed non-ex\-ponen\-tiality of the length distribution of EP melts 
obtained as described in Sec.~\ref{bfm_EP}.
As demonstrated in the main panel of Fig.~\ref{fig_bfm_EP_PN} these EP melts are indeed essentially
Flory distributed, i.e. the length distribution $\pchain$ decays to leading order exponentially 
with the reduced chain length $N/\Nav$. However, deviations for $N/\Nav \ll 1$ are 
visible in Fig.~\ref{fig_bfm_EP_PN}(c) in qualitative agreement with the predicted positive 
deviation of the chemical potential, Eq.~(\ref{eq_muEP_claim0}). 
The curvature of $-\log(\pchain)$ is further analyzed in Fig.~\ref{fig_muEP_WN}.
Motivated by Eq.~(\ref{eq_muEP_claim2}), we present in panel (a) the functional
\begin{equation}
V[\pchain] \equiv - \log(\pchain) - \mu N + \log(\mu)
\label{eq_muEP_VNdef}
\end{equation}
where the second term takes off the ideal contribution to the chemical potential.
The last term is due to the normalization of $\pchain$ and eliminates a 
trivial vertical shift depending on the scission energy $E$.
Consistently with Eq.~(\ref{eq_muEP_mu2Nav}), the chemical potential per monomer $\mu$ has
been obtained from the measured mean chain length $\Nav$ using
\begin{equation}
\mu \equiv {\Nav}^{-1} \left(1+ \cep \sqrt{\pi} / \sqrt{\Nav} \right).
\label{eq_muEP_Nav2mu}
\end{equation}
Note that $\mu$ and $1/\Nav$ become numerically indis\-tin\-guish\-able for $E \ge 7$. 
If the Gaussian contribution to the chemical potential is properly subtracted
one expects to obtain directly the non-Gaussian deviation to the chemical potential, 
$\dmuchainN \approx V[\pchain]$.  
Due to Eq.~(\ref{eq_muEP_claim1}) the functional should thus scale as
\begin{equation}
V[\pchain] / \cep \sqrt{\mu} \approx (1 - 2 x)/\sqrt{x}
\label{eq_muEP_VN}
\end{equation}
with $x = \mu N$ as indicated by the bold line in the panel. 
This is well born out by the data collapse obtained up to $x \approx 5$.
Obviously, the statistics deteriorates for $x \gg 1$ for all energies 
due to the exponential cutoff of $\pchain$.
Unfortunately, the statistics of the length histograms decreases strongly with $E$
and becomes too low for a meaningful comparison for $E > 9$.
It is for this numerical reason that we use Eq.~(\ref{eq_muEP_Nav2mu}) rather 
than the large-$E$ limit $\mu = 1/\Nav$ since this allows us to add the two histograms 
for $E=5$ and $E=6$ for which high precision data is available. Otherwise these 
energies would deviate from Eq.~(\ref{eq_muEP_VN}) for large $x$ due to an insufficient
subtraction of the leading Gaussian contribution. 
%

%
Since the subtraction of the large linear Gaussian contribution is in any case
a delicate issue we present in panel (b) of Fig.~\ref{fig_muEP_WN} a second functional,
\begin{equation}
W[\pchain] \equiv 2 V[\pchain] - V[\pchaintwo] = \log\left[\frac{\pchaintwo \mu}{p_\mathrm{N}^2}\right],
\label{eq_muEP_WNdef}
\end{equation}
where by construction this contribution is eliminated
following a suggestion made recently by Semenov and Johner \cite{ANS03}.
The normalization factor $\mu$ appearing in Eq.~(\ref{eq_muEP_WNdef}) eliminates again 
a weak vertical scission energy dependence of the data. Obviously, $W[\pchain] \equiv 0$ 
for perfectly Flory-distributed chains. 
Following Eq.~(\ref{eq_muEP_claim1}) and Eq.~(\ref{eq_muEP_claim2}) one expects
\begin{equation}
\frac{W[\pchain]}{\cep \sqrt{\mu} (2-1/\sqrt{2})} \approx
\frac{1- 0.906 x}{\sqrt{x}}
\label{eq_muEP_WN}
\end{equation}
with $x=\mu N$.
Eq.~(\ref{eq_muEP_WN}) is indicated by the bold line 
which compares again rather well with the presented data.

\begin{figure}[t]
\centerline{\resizebox{0.9\columnwidth}{!}{\includegraphics*{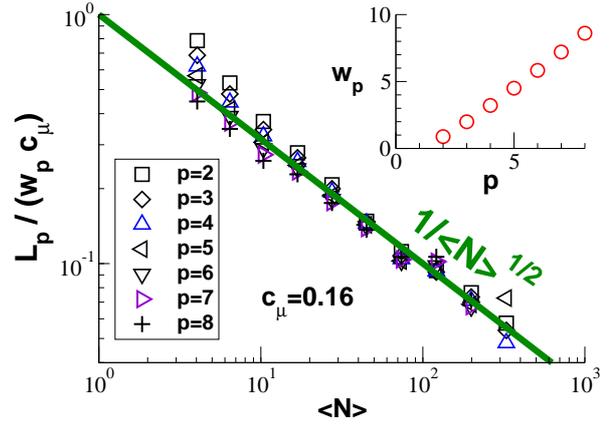}}}
\caption{Non-exponentiality parameter $\nonexp$ for different moments $p$
as a function of mean chain length $\Nav$.
$\nonexp$ is finite decreasing with chain length as suggested by Eq.~(\ref{eq_muEP_nonexp}).
The vertical axis has been rescaled successfully using the $p$-dependent weights $\wep$
indicated in the inset.
\label{fig_muEP_Lp}
}
\end{figure}
The functionals presented in Fig.~\ref{fig_muEP_WN} require high-accur\-acy histo\-grams. 
That $\pchain$ is only approximately Flory distributed can be more readily seen using the
``non-exponential\-ity parameter" $\nonexp \equiv 1 - \Npav / p! \Nav^p$.
Obviously, $\nonexp \equiv 0$ for rigorously Gaussian chains.
As stated in Eq.~(\ref{eq_muEP_nonexp}), we expect the non-exponentiality parameter
to decay as $\nonexp = \wep \cep \sqrt{\mu} \sim 1/\sqrt{\Nav}$, 
i.e. as the correlation hole potential of the typical melt chain. 
The main panel of Fig.~\ref{fig_muEP_Lp} presents 
$\nonexp / \wep \cep$ as a function of $\Nav \approx 1/\mu$.
The predicted power-law decay is clearly demonstrated by the data.
Note that the scaling of the vertical axis with the $p$-dependent geometrical factors 
$\wep$ allows to bring all moments on the same master curve. 
As can be seen from the inset of Fig.~\ref{fig_muEP_Lp},
this scaling is significant since $\wep$ varies over nearly a decade.
%
Deviations from the predicted scaling are visible, not surprisingly,
for small $\Nav < 10$.
Since the coefficient $\cep$ is known the data collapse on the theoretical prediction
(bold line) has been produced without any free adjustable parameter.

\subsection{Summary}
\label{TD_summary}
In this section we have discussed various thermodynamic properties of BFM melts at our 
reference volume fraction $\phi=8\rho=0.5$. 
For large overlap penalties $x =\overlap/T \gg 1$ all thermodynamic properties remain similar 
to the known values for athermal BFM systems ($x=\infty$) \cite{MP94,WM94,WBM07}. 
From the computational point of view this is of some interest since it suggests 
the equilibration and sampling of configurations at a reasonable cost at overlap penalties 
between $x=1$ and $x = 10$ where the slithering snake algorithm is still efficient.
Decreasing the monomer interaction further we have found, as one expects, good agreement 
with the prediction by Edwards, Eq.~(\ref{eq_FE_Thigh}), for small overlap penalties ($x \ll 1$). 
In this limit the thermodynamics is essentially determined by the second virial contribution,
Eq.~(\ref{eq_bfm_vvir}). Interestingly, being the second derivative of the free energy with 
respect to the inverse temperature the specific heat $\cV$ allows the verification of the
predicted scaling of the free energy contribution $-\kBT/\xi^3\rho$ due to the chain connectivity 
(Fig.~\ref{fig_TD_cV}). 
The central control parameter of this work,
the di\-men\-sion\-less compressibility $\gT \equiv \lim_{N\to\infty} ( \lim_{q\to 0} S(q,N))$,
is discussed in Sec.~\ref{TD_compress}.
While the determination of $\gT(x)$ is numerically trivial for large overlap penalties, 
more care is needed for $x \ll 1$ (Fig.~\ref{fig_TD_compress}). 
Note that the static RPA, Eq.~(\ref{eq_theo_sRPA}), breaks down in the opposite limit ($x \gg 1$) 
where the structure factors of beads and polymer melts become ultimately identical, $S(q) \approx \gT \ll 1$. 

For a test chain of length $n=N$ immersed into a bath of monodisperse chains of length $N$ we have
confirmed by thermodynamic integration in Sec.~\ref{TD_muMD} 
that the chemical potential $\muchainN$ is essentially extensive with respect to the chain length,
as expected from Flory's ideality hypothesis. 
In order to demonstrate that the small deviations predicted by the perturbation theory \cite{WJC10} 
exist, we have investigated in Sec.~\ref{TD_muEP} essentially Flory-distributed
(Fig.~\ref{fig_bfm_EP_PN}) self-assembled EP systems. 
The detailed analysis of the measured length distribution $\pchain$ (Fig.~\ref{fig_muEP_WN}) 
has allowed us to show the existence of a correction $\dmuchainn$, Eq.~(\ref{eq_muEP_claim1}), 
scaling essentially as the correlation hole penalty $\dmuchainn \approx \Ustar(n)$ for $n \ll \Nav$. 
The observed deviations for EP systems \cite{WJC10} beg for an improved numerical verification 
by means of the thermodynamic integration method (Sec.~\ref{TD_muMD}) for test chains immersed into 
mono\-disperse melts.\footnote{The presented results for dense polymer solutions
may also be of relevance to the chemical potential of dilute polymer chains at and around the
$\Theta$-point which has received attention recently \cite{Rubinstein08,SNM09}. The reason for
this connection is that (taken apart different prefactors) the {\em same} effective interaction
potential $\vpot(q) \sim q^2$ enters the perturbation calculation in the low wavevector limit.
A non-extensive correction $\dmuchainn \sim +1/\sqrt{n}$ in three dimensions
is thus to be expected.}
%

%

\section{Intramolecular conformational properties}
\label{conf}
\subsection{Introduction}
\label{conf_intro}
We turn now to the description of intrachain conformational properties of BFM melts
at volume fraction $\phi=0.5$ and temperature $T=1$ comparing our numerical results to the theoretical
predictions announced in Sec.~\ref{intro_key} and Sec.~\ref{theo}.
We discuss first the properties on monomeric level (Sec.~\ref{conf_bonds}) and
show then how the effective bond length $b$ for asymptotically long chains
can be obtained using the theoretical input developed in Sec.~\ref{theo}.
We present then the bond-bond correlation function as a function
of curvilinear distance $s$ (Sec.~\ref{conf_Ps}) and as a function of
the spatial distance $r$ between both bonds (Sec.~\ref{conf_Pr}).
Higher moments of the distribution $G(r,s)$ and its deviations $\delta G(r,s)$
from Flory's ideality hypothesis will be analyzed in Sec.~\ref{conf_moments} and Sec.~\ref{conf_Grs}, respectively.
We turn finally to the characterization of an experimental relevant observable,
the intramolecular form factor $F(q)$.

\begin{figure}
\centerline{\resizebox{0.9\columnwidth}{!}{\includegraphics*{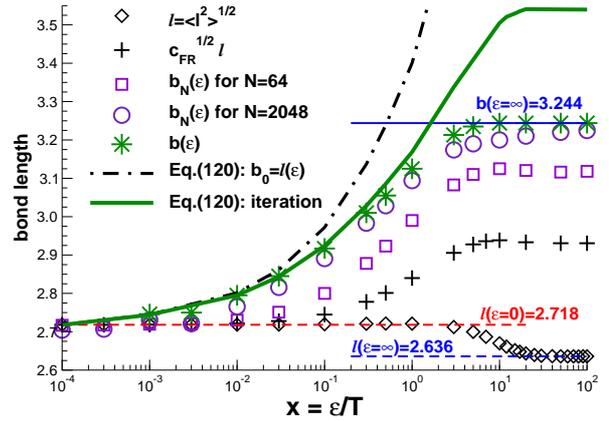}}}
\caption{Various bond properties {\em vs.} overlap penalty $\overlap$
for volume fraction $\phi=0.5$ and temperature $T=1$.
The data for the root-mean-square bond length $l(\overlap)$ and the effective bond length 
$b(\overlap) \equiv \lim_{N\to\infty} \bratio(\overlap)$ for asymptotically long chains are 
given in Table~\ref{tab_overlap}.
The dash-dotted line indicates the effective bond length
as predicted by Eq.~(\ref{eq_conf_bepredict}) assuming $\bref = l(\overlap)$ 
for the bond length of the reference chain.
The bold line shows the fix points obtained by iteration of Eq.~(\ref{eq_conf_bepredict}).
}
\label{fig_conf_bonds}
\end{figure}

\subsection{Bond properties}
\label{conf_bonds}

We begin by characterizing local-scale features of the algorithm described in Sec.~\ref{bfm}. 
By definition of our version of the BFM algorithm the bond length is allowed to fluctuate 
between $2$ and $\sqrt{10}$. One expects that switching on the overlap penalty
$\overlap$ will suppress large bonds due to the increasing pressure.
The mean bond length is characterized by the root-mean-square length $l$.
(Other moments yield similar results.) The mean bond length rapidly becomes ($N > 20$)
chain length independent \cite{Paul91a}. As can be seen from Fig.~\ref{fig_conf_bonds},
$l$ shows a monotonous decay between $\overlap \approx 3$ and $\overlap \approx 20$.
As other local properties, the bond length becomes constant in the small-$\overlap$ and 
large-$\overlap$ limits (dashed lines).  
The value $l(\overlap=0)=2.718$ gives the lower bound for the effective bond length 
$b(\overlap)$ of asymptotically long chains (stars) obtained below.

Defining the bond angle $\theta$ between two subsequent bonds by the scalar product 
$\cos(\theta) = \ehat_n \cdot \ehat_{n+1}$ of the normalized bond vectors
$\ehat_i = \lvec_i/|\lvec_i|$, the {\em local} chain rigidity may be characterized by
$\la \theta \ra$ and $\la \cos(\theta) \ra$. 
Note that $\la \theta \ra$ and  $\la \cos(\theta) \ra$ can be regarded as chain length independent, 
just as the mean bond length. The local rigidity is negligible for $\overlap \ll 1$, i.e. 
$\la \theta \ra \approx 90^{\circ}$ and $\la \cos(\theta)) \ra \approx 0$ due to the symmetry of the distribution $p(\theta)$ 
with respect to $90^{\circ}$. The rigidity then increases around $\overlap \approx 1$ and becomes constant again for large 
$\overlap$ where $\la \theta \ra \approx 82.2^{\circ}$ and $\la \cos(\theta) \ra \approx 0.106$ \cite{WCK09}. 
The increase of the local rigidity for larger excluded volume interactions is of course expected 
due to the suppression of immediate backfoldings corresponding to bond angles $\theta > 143^{\circ}$ \cite{WPB92}. 
The distribution $p(\theta)$ therefore becomes lopsided towards smaller $\theta$ (not shown).
According to Eq.~(\ref{eq_theo_cinf_FR}) the effective bond length of a ``freely rotating" (FR) chain
is given by $b(\overlap) = l(\overlap) \sqrt{c_\mathrm{FR}}$ with 
$c_\mathrm{FR} = (1+\la \cos(\theta) \ra)/(1-\la \cos(\theta) \ra)$. 
This simple model, indicated by the crosses in Fig.~\ref{fig_conf_bonds}, yields a qualitatively reasonable
trend (monotonous increase of the effective bond length at $\overlap \approx 1$) but fails 
to fit the directly measured effective bond lengths quantitatively \cite{WCK09}.

\subsection{Mean-squared chain and subchain size}
\label{conf_RNRs}
\subsubsection{Total chain size $\RN$}
\label{conf_RN}

One way to characterize the total chain size $\RN$ is to measure the second moment of the 
chain end-to-end distance $\RN^2 \equiv \la (\bm{r}_N-\bm{r}_1)^2 \ra$. 
We consider the ratio $\bratio(\overlap) \equiv \RN(\overlap)/\sqrt{N-1}$ to compare 
the measured chain size with the ideal chain behavior which is commonly taken as granted
\cite{DegennesBook,DoiEdwardsBook,Flory49} and which is the basis of our perturbation calculation
(Sec.~\ref{theo_perturb}). The task is to extrapolate for the effective bond length 
$b(\overlap)\equiv \lim_{N\to\infty} \bratio(\overlap)$ 
of asymptotically long chains.
The ratio $\bratio(\overlap)$ for $N=64$ and $N=2048$ and the asymptotic limit $b(\overlap)$ 
--- obtained by extrapolation as described below --- are presented in Fig.~\ref{fig_conf_bonds}. 
Obviously, $\bratio(\overlap) \to l(\overlap=0)$ for all $N$ in the small-$\overlap$ limit. 
$\bratio(\overlap)$ increases then in the intermediate $\overlap$-window before it levels off at 
$\overlap \approx 10$.
The swelling due to the excluded volume interaction is the stronger the larger the chain length,
i.e. $\bratio(\overlap)$ increases monotonously with $N$. This swelling thus cannot be attributed
to a {\em local} rigidity as described, e.g., by the freely-rotating chain model.

\begin{figure}[t]
\centerline{\resizebox{0.9\columnwidth}{!}{\includegraphics*{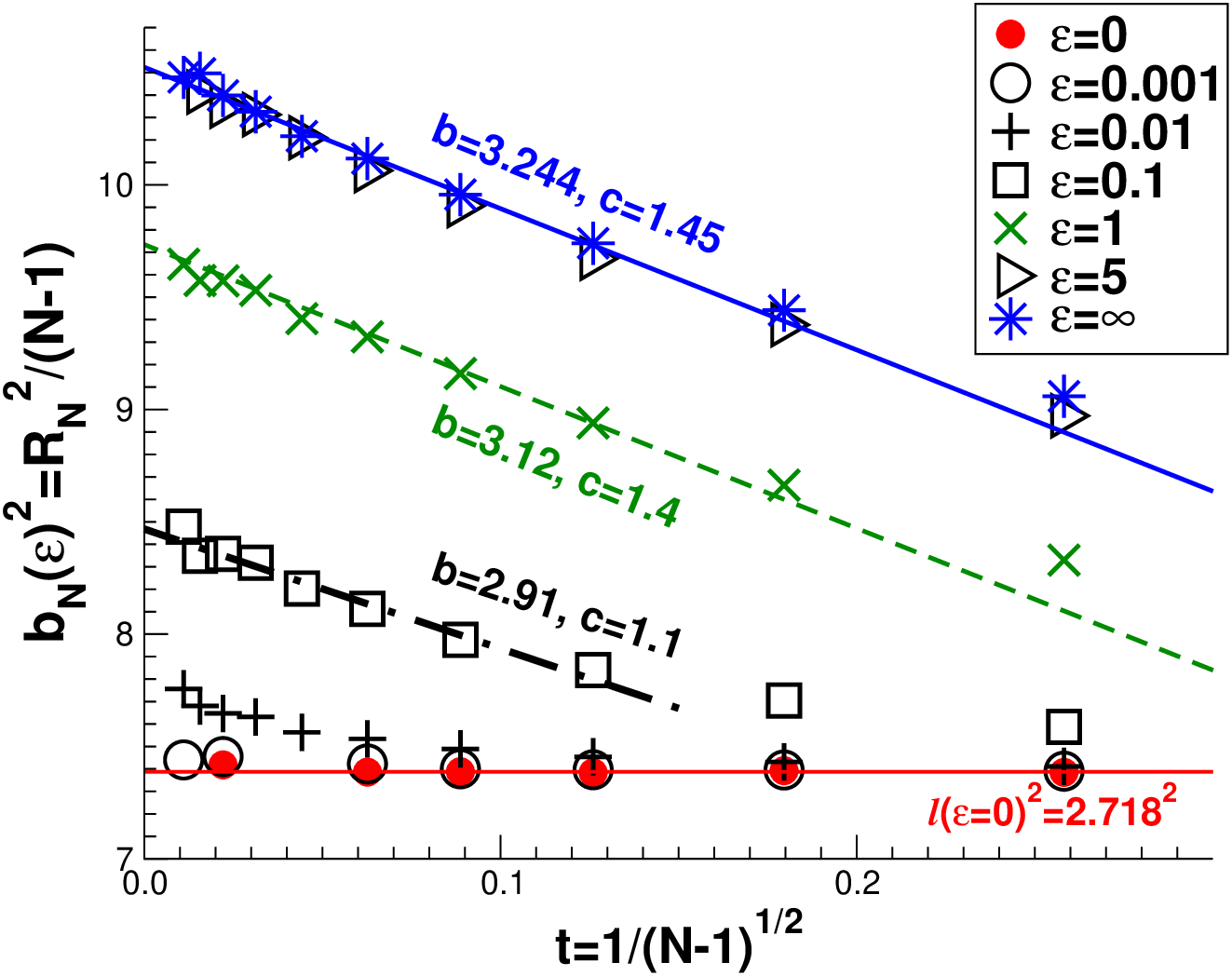}}}
\caption{Rescaled end-to-end distance $\bratio^2(\overlap) \equiv \RN^2(\overlap)/(N-1)$
as a function of $t=1/\sqrt{N-1}$ for different $\overlap$.
The chains only remain Gaussian on all scales and all $N$ for extremely small $\overlap$.
For $\overlap \ge 0.1$ one observes $\bratio^2(\overlap)$ to decay linearly  
in agreement with Eq.~(\ref{eq_conf_RNfit}).
This can be used for a simple two-parameter fit for the effective bond length 
$b(\overlap)$ as indicated for $\overlap=0.1$, $1.0$ and $\infty$.
}
\label{fig_conf_RN}
\end{figure}
The $N$-effect can be seen better in Fig.~\ref{fig_conf_RN} where we have plotted
$\bratio$ for several penalties $\overlap$ as a function of $t=1/\sqrt{N-1}$.
The choice of the horizontal axis is motivated by Eq.~(\ref{eq_intro_Rs}) suggesting 
the linear relation
\begin{equation}
\bratio^2(\overlap) \approx b^2(\overlap) \ \left( 1 - c(\overlap) \ce(\overlap) t \right) 
\mbox{ for } N/\gT \gg 1
\label{eq_conf_RNfit}
\end{equation}
with $\ce \equiv \sqrt{24/\pi^3}/\rho b^3$ being the swelling coefficient 
defined in Sec.~\ref{intro_key} and $c(\overlap)$ an additional numerical prefactor of order unity.
This prefactor has been introduced in agreement with Eq.~(\ref{eq_theo_Rendcorrect2}). 
The reason for this coefficient is that the corrections to Gaussian behavior differ slightly 
for internal chain segments [as described by Eq.~(\ref{eq_intro_Rs})] 
and the total chain size which is characterized in Fig.~\ref{fig_conf_RN}. 
We remind that $c \to I(\infty) \approx 1.59$ for $N \to \infty$.
However, since this value corresponds to the limit of a very slowly converging integral \cite{WBM07} 
it is better to use Eq.~(\ref{eq_conf_RNfit}) as a two-parameter fit for $b(\overlap)$ and $c(\overlap)$
and to crosscheck then whether the fitted $c$ is of order unity.  
As shown in the figure for three overlap penalties, this method can be used reasonably for overlap penalties 
as low as $\overlap \approx 0.1$, albeit with decreasing $\overlap$ it systematically underestimates
the ``true" $b(\overlap)$-values indicated in Table~\ref{tab_overlap}. 
Please note that $N/\gT \approx 400$ for $\overlap=0.1$ and $N=8192$. 
Chains with $N \gg 8192$ would be required to use this method for even smaller $\overlap$.
In this limit it is better to use as a first step the value $\bratio(\overlap)$ of the 
largest chain length simulated as a (rather reasonable) lower bound for $b(\overlap)$.

\subsubsection{Subchain size $\Rs$}
\label{conf_Rs}
\begin{figure}[t]
\centerline{\resizebox{0.9\columnwidth}{!}{\includegraphics*{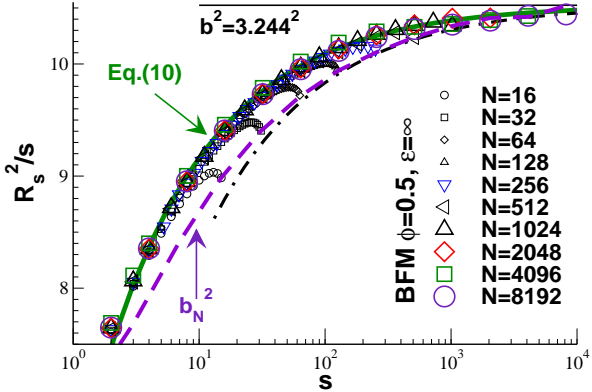}}}
\caption{
\label{fig_conf_Rslog}
Mean-squared subchain size $\Rs^2/s$ {\em vs.} $s$ for different chain length $N$.
Log-linear coordinates are used to emphasize the power law swelling over several 
orders of magnitude of $s$. The data approach the asymptotic limit (horizontal line) 
from below, i.e. the chains are {\em swollen}. This behavior is well fitted by 
Eq.~(\ref{eq_intro_Rs}) for $1 \ll s \ll N$ (bold line). 
Non-monotonous behavior is found for $s \rightarrow N$, especially for small $N$ \cite{Auhl03}. 
The dashed line indicates the measured total chain end-to-end distances $\bratio$
showing even more pronounced deviations.
The dash-dotted line compares this data with Eq.~(\ref{eq_theo_Rendcorrect2}).
}
\end{figure}

\begin{figure}[t]
\centerline{\resizebox{0.9\columnwidth}{!}{\includegraphics*{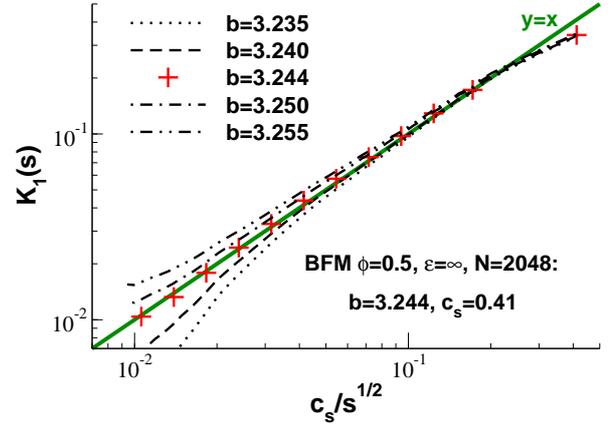}}}
\caption{
\label{fig_conf_K1s}
Replot of the mean-squared subchain size as $K_1(s) = 1 - \Rs^2 /b^2s$ 
{\em vs.} $\ce/\sqrt{s}$ with $\ce \equiv \sqrt{24/\pi^3}/\rho b^3$, 
as suggested by Eq.~(\ref{eq_intro_Rs}), for different trial effective bond lengths $b$ 
as indicated.
Only chains of length $N=2048$ and $\overlap=\infty$ are considered for clarity. 
This procedure is very sensitive to the value chosen and allows a precise determination.
}
\end{figure}

The mean-squared subchain size $\Rs^2$ is presented in Figs.~\ref{fig_conf_Rslog} and \ref{fig_conf_K1s}
where we focus on melts without monomer overlap ($\overlap = \infty$) \cite{WBM07}. We show how the 
effective bond length $b$ for asymptotically long chains may be extrapolated from the measured $\Rs^2$ 
for finite arc-length $s$ and finite chain length $N$ using the perturbation prediction Eq.~(\ref{eq_intro_Rs}). 

As already remarked by Auhl {\em et al.} \cite{Auhl03}, Fig.~\ref{fig_conf_Rslog} shows clearly that
the chains are swollen, i.e. $\Rs^2/s$ increases systematically and this up to very large arc-length $s$.
In agreement with Eq.~(\ref{eq_intro_Rs}), the Gaussian behavior (horizontal line) is approached 
from below and the deviation decays as $\Ustars \sim 1/\sqrt{s}$.
The indicated bold line corresponds to $b = 3.244$ and $\ce \approx 0.41$
which fits nicely the data over several decades in $s$ ---
provided that chain end effects can be neglected ($s \ll N$).
Note that a systematic {\em underestimation} of the true effective bond length
would be obtained by taking the largest $\Rs^2/s \approx 3.23^2$ value
available, say, for monodisperse chains of length $N=2048$.
Interestingly, $\Rs^2/s$ does not approach the asymptotic limit monotonically \cite{Auhl03,WBM07}.
Especially for short chains one finds a {\em non-monotonic} behavior for $s \rightarrow N$.
This means that the total chain end-to-end distance $\RN$ shows even more pronounced deviations 
from the asymptotic limit.
%
We emphasize that the non-monotonicity of $\Rs^2/s$  becomes weaker with increasing $N$
and that, as one expects, the inner distances, as well as the total chain size,
are characterized by the {\em same} effective bond length $b$ for large $s$ or $N$.\footnote{We
note {\em en passant} that this is not the case for the compact chain conformations adopted by
chains in strictly 2D melts where the effective bond length associated to the chain ends is smaller 
than the one for subchains even for $N \to \infty$ \cite{MWK10}.}
The non-monotonic behavior may be qualitatively understood by the reduced self-interactions at the chain ends 
which lessens the swelling on these scales. Our perturbation prediction Eq.~(\ref{eq_theo_Rendcorrect2})
is indicated in Fig.~\ref{fig_conf_Rslog} by the dash-dotted line. As already remarked, 
this prediction contains a slowly converging numerical integral of order unity. The dash-dotted line
corresponds to the value $c=I(\infty) \approx 1.59$ predicted for asymptotically long chains,
the dashed line to the slightly smaller value $c=1.45$ fitted in Fig.~\ref{fig_conf_RN}. 
In summary, it is clear that one should use the subchain size $\Rs$ rather
than the total chain size $\RN$ to obtain in a computational study a reliable 
fit of the effective bond length $b$.

The representation chosen in Fig.~\ref{fig_conf_Rslog} is not the most convenient one
for an accurate determination of $b$ and $\ce$. How precise coefficients may be obtained 
according to Eq.~(\ref{eq_intro_Rs}) is addressed in the 
Fig.~\ref{fig_conf_K1s} for chains of length $N=2048$. In a first step $\Rs^2/s$ should 
be plotted as a function of $1/\sqrt{s}$ just as in Fig.~\ref{fig_conf_RN}. 
This allows for a first rough estimation of $b$.
Since data for large $s$ are less visible in this representation, we recommend for the 
fine-tuning of $b$ to switch then to logarithmic coordinates with a vertical axis 
$K_1(s) = 1 - \Rs^2/ b^2 s$ for different trial values of $b$.
The correct value of $b$ is found by adjusting the axes such that $K_1(s)$ extrapolates 
{\em linearly} as a function of $\ce/\sqrt{s}$ to zero for large $s$.\footnote{For 
the bead-spring model presented in Ref.~\cite{WBM07} the empirical swelling
coefficient $\ce$ is shown to differ for yet unknown reasons by a few percents
from the predicted value $\sqrt{24/\pi^3}/\rho b^3$.}
We assume for the fine-tuning that higher order perturbation corrections may be neglected,
i.e. we take Eq.~(\ref{eq_intro_Rs}) literally. 
The plot shows that this method is very sensitive, yielding a best value $b=3.244$ 
for $\overlap=\infty$ that agrees with the theory over more than one order of magnitude 
{\em without} curvature.
Using this method we have determined the effective bond lengths $b(\overlap)$ for a broad range of 
overlap penalties $\overlap$ as indicated in Table~\ref{tab_overlap} and shown by the stars in 
Fig.~\ref{fig_conf_bonds} \cite{WCK09}.

\subsubsection{Predicting the effective bond length}
\label{conf_be}

Up to now, we have used the theoretical results to {\em fit} the effective bond length $b(\overlap)$,
rather than to predict it from the known thermodynamic properties and local model features
such as the bond length $l(\overlap)$.
As reminded in Eq.~(\ref{eq_theo_bepert}) the increase of the effective bond length for weakly interacting 
and asymptotically long polymer melts has been in fact calculated long ago by Edwards \cite{DoiEdwardsBook}.
Following a suggestion made by Muthukumar and Edwards \cite{Edwards82}
this perturbation result may be rewritten as a recursion relation
\begin{equation}
b_{i+1}^2 = l^2 \left( 1 + \frac{\sqrt{12}}{\pi} \Giite \right)
\mbox{ with } \Giite \equiv \frac{1}{\sqrt{\gT} b_{i}^3\rho}
\label{eq_conf_bepredict}
\end{equation}
with $b_i$ being the bond length after the $i$-th iteration step, $\gT$ the measured dimensionless compressibility
and $\Giite(b_i,g)$ the relevant Ginzburg parameter quantifying the strength of the interaction acting on a chain 
segment of length $s=\gT$. Since $\Gi \approx \Giite$ becomes small for large compressibilities $\gT(\overlap)$, 
one expects good agreement with our data for small $\overlap$.
%
If we set $\bref=l(\overlap)$ this yields after one iteration ($i=1$)
the dash-dotted line indicated in Fig.~\ref{fig_conf_bonds}, i.e. a reasonable prediction is only achieved
up to $\overlap \approx 0.01$. 
The predictive power of Eq.~(\ref{eq_conf_bepredict}) can be considerably improved over 
nearly two decades up to $\overlap \approx 1$ if one applies the formula iteratively 
using the effective bond length $b_i$ obtained at step $i$ as input for the Ginzburg parameter
for computing $b_{i+1}$. This recursion converges rapidly as shown by the
bold line indicated in Fig.~\ref{fig_conf_bonds} obtained after 20 iterations.\footnote{Essentially 
the same result is obtained up to $\overlap \approx 1$ if one sets directly $b_0=b$ using the {\em measured} 
effective bond length as start value of the iteration.} 
Note that $\Gi < 0.34$ for $\overlap < 1$ where Eq.~(\ref{eq_conf_bepredict}) fits our data nicely.
The fix-point solution of Eq.~(\ref{eq_conf_bepredict}) does not capture correctly the leveling off 
of $b(\overlap)$ setting in above $\overlap \approx 1$. Since the Ginzburg parameter becomes there 
of order one, this is to be expected. 
In summary, we have shown that the iteration of Eq.~(\ref{eq_conf_bepredict}) allows a good 
prediction for $b(\overlap)$ for $\Gi(\overlap) \ll 1$. If reliable values for compressibilities $\gT(\overlap) \gg 1$ 
are available, this is the method of choice if one cannot afford to simulate very long chains.

\subsection{Bond-bond correlation function}
\label{conf_bondbond}

\subsubsection{Bond-bond correlation function $\Pone(s)$}
\label{conf_Ps}
%
%
As we have seen in Fig.~\ref{fig_conf_K1s}, to demonstrate the deviations from Flory's ideality hypothesis
starting from the subchain size $\Rs^2$ requires to subtract a large Gaussian contribution $b^2s$.
Unfortunately, this requires as a first step the precise determination of the effective bond length $b(\overlap)$ 
for asymptotically long chains which might not be available.
Indeed we have used in the preceding Sec.~\ref{conf_Rs} the fact that the scaling of $K_1(s)$ critically depends 
on this accurate value to {\em improve} the estimation of $b(\overlap)$. 
Hence, it would be nice to demonstrate directly the scaling implied by our key prediction {\em without} any tunable parameter. 
The trick to achieve this is similar to our demonstration of the density fluctuation 
contributions to the free energy, Eq.~(\ref{eq_FE_Thigh}), presented in Sec.~\ref{TD_cV}:
We consider the {\em curvature} of $\Rs^2$, i.e. its second derivative with respect to $s$.
Using Eq.~(\ref{eq_intro_PonesRsrelated}) this second derivative is obtained directly from the bond-bond 
correlation function $\Pone(s)$ computed by averaging over all pairs of monomers $(n,m=n+s)$.
We remind that $\Pone(s)$ is generally believed to decrease exponentially as in Eq.~(\ref{eq_theo_Pone_spersist}).
This textbook belief is based on the assumption that all long range interactions are negligible
on distances larger than $\xi$. Hence, only correlations along the backbone of the chains
should matter and it is then straightforward to work out that an exponential cutoff
is inevitable due to the multiplicative loss of any information transferred recursively
along the chain \cite{FloryBook}.

%
%
\begin{figure}[t]
\centerline{\resizebox{0.9\columnwidth}{!}{\includegraphics*{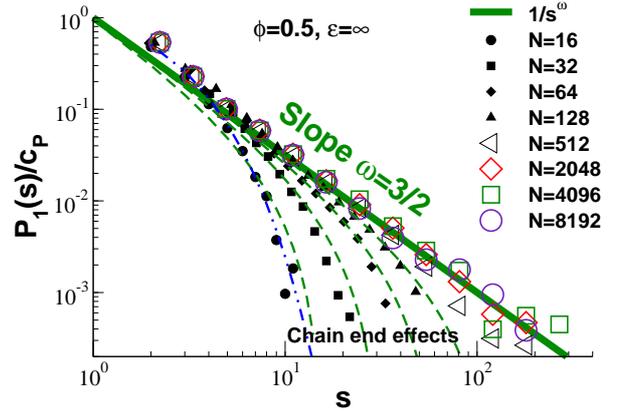}}}
\caption{The bond-bond correlation function $\Pone(s)/\cP$ as a function of the
curvilinear distance $s$ for a broad range of chain lengths $N$ \cite{WMBJOMMS04,WBM07}.
Provided that $1 \ll s \ll N$, all data sets collapse on the power law slope
with exponent $\omega=3/2$ (bold line) as predicted by Eq.~(\ref{eq_intro_keystat}).
The dash-dotted curve $\Pone(s) \approx \exp(-s/1.5)$ shows that exponential behavior is
only compatible with very small chain lengths.
The dashed lines correspond to the theoretical prediction, Eq.~(\ref{eq_conf_P1sN}),
for short chains with $N=16,32,64$ and $128$ (from left to right).
}
\label{fig_conf_Ps}
\end{figure}

The bond-bond correlation function $\Pone(s)$ obtained for monodisperse chains
at $\phi=0.5$ and $\overlap=\infty$ is presented in Fig.~\ref{fig_conf_Ps}.
The power-law decay with exponent $\omega=3/2$ predicted by our key perturbation result
Eq.~(\ref{eq_intro_keystat}) is perfectly confirmed by the larger chains ($N > 256$).
As can be seen for $N=16$, exponentials are compatible with the data of
short chains, however. This might explain why the power-law scaling has been overlooked
in older numerical studies, since good statistics for large chains ($N > 1000$)
has only become available recently. However, it is clearly shown that
$\Pone(s)$ approaches systematically the scale-free asymptote with increasing $N$.
The departure from this limit is fully accounted for by the theory if chain
end effects are carefully considered (dashed lines). Generalizing Eq.~(\ref{eq_intro_keystat})
using the Pad\'e approximation, Eq.~(\ref{eq_theo_vpot_Flory}), perturbation theory yields
\begin{equation}
\Pone(s) = \frac{\cP}{s^{3/2}} \frac{1+3u + 5u^2}{1+u} (1-u)^2
\label{eq_conf_P1sN}
\end{equation}
where we have set $u=\sqrt{s/N}$ \cite{WBM07}. 
For $u\ll 1$ this is consistent with Eq.~(\ref{eq_intro_keystat}).
In the limit of large $s \rightarrow N$, the correlation functions vanish rigorously
as $\Pone(s) \propto (1-u)^2$.
Considering that non-universal features cannot be neglected for short chain properties
and that the theory does not allow for any free fitting parameter, the agreement found
in Fig.~\ref{fig_conf_Ps} is rather satisfactory.

%
\begin{figure}[t]
\centerline{\resizebox{0.9\columnwidth}{!}{\includegraphics*{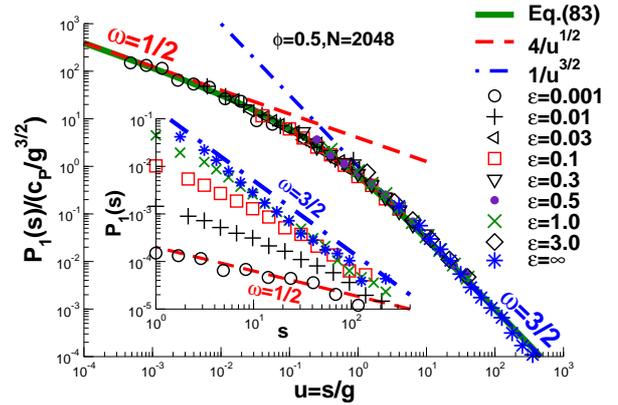}}}
\caption{$\Pone(s)$ for different overlap penalties $\overlap$ \cite{WCK09}.
Inset: $\Pone(s)$ in log-log coordinates. 
The data approaches a power law behavior, $\Pone(s) \sim 1/s^{\omega}$, 
with exponent $\omega =1/2$ for small $\overlap$ (dashed line) and 
$\omega=3/2$ for $\overlap \ge 1$ (dash-dotted line).
Main panel: $\Pone(s)/\left[\cP(/\gT^{3/2}\right]$ {\em vs.} $u=s/\gT$ as suggested by Eq.~(\ref{eq_theo_Ponesoft}). 
For large $u$, where an incompressible packing of thermal blobs is probed, all data collapse onto 
the dash-dotted line as predicted by Eq.~(\ref{eq_intro_keystat}), i.e. $\Pone(s)$ becomes independent 
of the compressibility $\gT(\overlap)$.
}
\label{fig_conf_Ps_eps}
\end{figure}
The bond-bond correlation function $\Pone(s)$ for different overlap penalties $\overlap$ is presented
in Fig.~\ref{fig_conf_Ps_eps} for chains of length $N=2048$ \cite{WCK09}. As can be seen from the unscaled data
shown in the inset, $\Pone(s)$ approaches a power law with exponent $\omega=1/2$ (dashed line)
in the limit of weak overlap penalties in agreement with Eq.~(\ref{eq_theo_Ponefixm}). 
For $\overlap \ge 1$ our data is compatible with an exponent $\omega=3/2$ (dash-dotted line)
as suggested by Eq.~(\ref{eq_intro_keystat}). Hence, we have demonstrated without any tunable parameter 
that Flory's ideality hypothesis is systematically violated for all segment lengths $s$ and 
all overlap penalties $\overlap$.
As suggested by Eq.~(\ref{eq_theo_Ponesoft}), the main panel of Fig.~\ref{fig_conf_Ps_eps} presents
$\Pone(s)/(\cP/\gT^{3/2})$ as a function of the reduced arc-length $u=s/\gT$
using the dimensionless compressibilities $\gT(\overlap)$ and effective bond lengths $b(\overlap)$ from
Table~\ref{tab_overlap}. The data collapse is remarkable as long as $1 \ll s \ll N$. 
The relation Eq.~(\ref{eq_theo_Ponesoft}) is indicated by the bold line; 
it is in perfect agreement with the simulation data.\footnote{Eq.~(\ref{eq_theo_Ponesoft}) has also been applied 
successfully to the $\Pone(s)$ obtained from single-chain-in-mean-field simulation \cite{MMuller06}.}
The asymptotical behavior with $\omega=1/2$ for $u \ll 1$ and $\omega=3/2$ for $u \gg 1$ is shown by the dashed and 
dash-dotted lines, respectively. As predicted by Eq.~(\ref{eq_intro_keystat}), one recovers the power law 
$\Pone(s) = \cP/s^{3/2}$ irrespective of the blob size $g$. 
This suggests that the exponent $\omega=3/2$ is not due to {\em local} physics on the monomer scale,
since for $s \gg g \gg 1$ distances much larger than the monomer or even the thermal blob are probed.

\subsubsection{Distance dependence of angular correlations}
\label{conf_Pr}

%
We have seen in the previous paragraph that $\Pone(s)$ 
decays for $s/\gT \gg 1$ as a power law with an exponent $\omega=d \nu = 3/2$.  
Since the decay of $\Pone(s)$ resembles the return probability of
a random walk $G_0(r \to 0,s)$, it is tempting \cite{Rubinstein08}
to attribute the observed effect to ``local self-kicks" at a distance $r \approx \sigma$
involving the bonds $\lvec_n$ and $\lvec_m$ themselves (or their immediate neighbors). 
Accordingly, the bond-bond correlation function should reveal a $\delta(r)$-correlation
if sampled as a function of the distance $r = |\rvec|$
between bond pairs.
This interpretation turns out to be incorrect, however, and we will show that
the power law in $s$ simply translates as \cite{WJO10}
\begin{equation}
s^{-\omega} \Leftrightarrow (r/b)^{-\omega/\nu} = (r/b)^{-d}.
\label{eq_conf_sr_scaling}
\end{equation}
As demonstrated analytically in Sec.~\ref{app_stat_Pr}, one expects indeed for incompressible solutions
of infinite chains that
\begin{equation}
\Pone(r) \approx \Poneasym(r) \equiv  \frac{\cinf}{12\pi \rho r^3}
\mbox{ for } \xi \ll r \ll \rstar
\label{eq_conf_P1rkey}
\end{equation}
as suggested by Eq.~(\ref{eq_conf_sr_scaling}),
i.e.~the angular correlations are genuinely long-ranged.
%
%
As discussed in Sec.~\ref{app_stat_Pr_finiteN} the upper cutoff $\rstar$ arises due to the enhanced 
weight of stretched chain segments which align bond pairs for distances $r \gg \rstar$. Unfortunately, 
this cutoff increases rather slowly with chain length \cite{WJO10}
\begin{equation}
\rstar \approx b \Nav^{1/d} \ll \RN \approx b \Nav^{\nu}.
\label{eq_conf_rstar}
\end{equation}
The simulation of computationally challenging chain lengths thus is required
to demonstrate numerically the predicted power-law decay of $\Pone(r)$.
Generalizing Eq.~(\ref{eq_conf_P1rkey}) for Flory-distributed EP one obtains according to
Eq.~(\ref{eq_app_Pr_EPcorr}) that 
\begin{equation}
\Pone(r) = \Poneasym(r) h(x) + \ctwo \mu 
\label{eq_conf_Pr_EPcorr}
\end{equation}
with $h(x) = (1+2x)^2 \exp(-2x)$ being a scaling function of $x= \sqrt{\mu}r/2a$ and $\ctwo$
a phenomenological constant set (in practice) by the finite local chain rigidity. The angular 
correlation function $\Pone(r)$ for EP systems is thus predicted to level off for large reduced 
distances.\footnote{For monodisperse systems one finds instead $\Pone(r) \approx (r/N)^2$ for 
very large distances $r$ with $\RN \ll r \le N l$ due to the increasing contribution of stretched bonds \cite{WJO10}.}

\begin{figure}[t]
\centerline{\resizebox{1.0\columnwidth}{!}{\includegraphics*{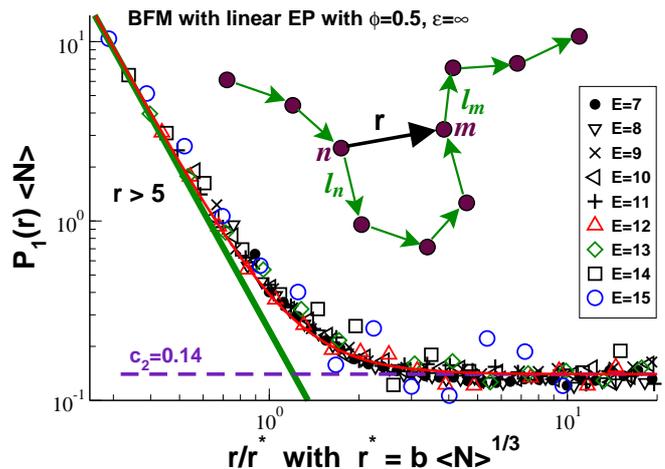}}}
\caption{The angular correlations are characterized by the bond-bond correlation function
$\Pone(r) = \la \lvec_n \cdot \lvec_m\ra /l^2$ averaging over all pairs of bonds of a chain
of same distance $r=|\rvec|$. 
The data shown has been obtained for EP of a broad range of scission energies $E$.
If $\Pone(r) \Nav$ is traced as a function of $r/\rstar$ all data points collapse.
Confirming the general scaling idea Eq.~(\ref{eq_conf_sr_scaling}) the bold line indicates the predicted power-law asymptote 
Eq.~(\ref{eq_conf_P1rkey}) for $r/\rstar \ll 1$, i.e. the bond-bond correlations are truely long-ranged.
The plateau $\Pone(r) \Nav = \ctwo= 0.14$ seen for $r/\rstar \gg 1$ is due the the enhanced weight
of stretched chain segments \cite{WJO10}. 
The complete perturbation prediction Eq.~(\ref{eq_conf_Pr_EPcorr}) given by the thin line interpolates perfectly 
between the power-law asymptote (bold line) and the plateau (dashed line).
}
\label{fig_conf_Pr_EP}
\end{figure}
We present in Fig.~\ref{fig_conf_Pr_EP} numerical results obtained for systems containing Flory-distributed EP.
As indicated in the sketch, $\Pone(r)$ is obtained by averaging of all intrachain bond pairs of same distance $r$. 
To avoid trivial correlations the second bond $\lvec_m$ is outside the subchain between the
monomer $n$ and $m$ defining the vector $\rvec$ \cite{WJO10,ANS10a}.
Model-depending physics not taken into account by theory obviously becomes relevant for short
distances corresponding to sub\-chains of a couple of monomers. For clarity, we have thus omitted data 
points with $r \le 5$. For small distance $r \ll \rstar$ the (unscaled) bond-bond correlation function 
$\Pone(r)$ is found to decay strongly as expected from Eq.~(\ref{eq_conf_Pr_EPcorr}). Note that in this
limit $\Pone(r)$ does neither depend on the mean chain length nor the chain length distribution, 
i.e. the same behavior is observed for monodisperse chains. See Ref.~\cite{WJO10} for details. 
To scale away the $\Nav$-dependence in the large-$r$ limit, we trace in Fig.~\ref{fig_conf_Pr_EP} 
the rescaled bond-bond correlation function $\Pone(r) \Nav$ as a function of $r/\rstar$ with $\rstar \equiv b \Nav^{1/3}$
using the mean chain lengths $\Nav$ indicated in Table~\ref{tab_bfm_EP}. 
Note that the error bars (not shown) become clearly much larger 
than the symbol size for large bond energies $E > 12$. 
It is fair to state, however, that all data points collapse nicely on the {\em one} master curve indicated
by the thin line predicted by Eq.~(\ref{eq_conf_Pr_EPcorr}). 
The asymptotic power law Eq.~(\ref{eq_conf_P1rkey}) is indicated by the bold line.
That $\Pone(r)$ for EP becomes constant for $r/\rstar \gg 1$ confirms a non-trivial prediction of the theory.\footnote{The 
clearly visible plateau can be used to determine the coefficient $\ctwo$.
The same coefficient can then be used for fitting the large-$r$ behavior of monodisperse systems.
Note that the best fit value $\ctwo=0.14$ is close to
$\Pone(s=1)= \la \lvec_n \cdot \lvec_{n+1} \ra/l^2 \approx 0.10$,
the independently determined bond-bond correlation between adjacent bond vectors.}

\subsection{Higher moments}
\label{conf_moments}

\begin{figure}
\centerline{\resizebox{0.9\columnwidth}{!}{\includegraphics*{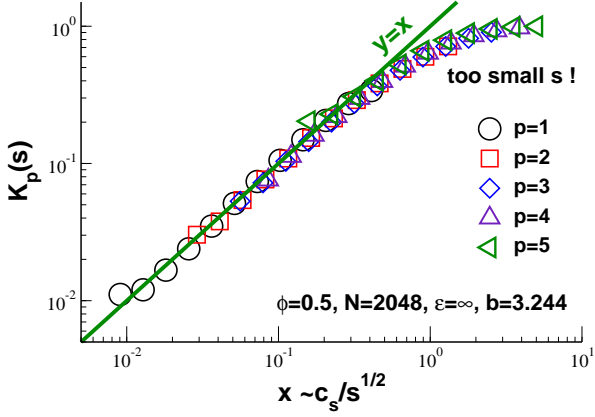}}}
\caption{Test of Eq.~(\ref{eq_theo_Kps_perturb}) where the rescaled moments $\Kp(s)$
are plotted {\em vs.} $x \sim \ce/\sqrt{s}$. Setting $b=3.244$ for all moments,
{\em all} data sets extrapolate linearly to zero for small $x$.
The saturation at large $x$ is due to the finite extensibility of subchains.
Since this effect becomes more marked for larger moments, the fit of $b$ is
best performed for $p=1$.
}
\label{fig_conf_Kps}
\end{figure}
The preceding discussion focused on the {\em second} moment of the subchain size
distribution $G(r,s)$ and its derivatives with respect to $s$.
We have also computed 
higher moments $\la \rvec^{2p} \ra$ with $p \le 5$ \cite{WBM07}.
(For clarity we focus here on monodisperse chains.)
%
%
We compare these moments in Fig.~\ref{fig_conf_Kps} with Eq.~(\ref{eq_theo_Kps_perturb})
tracing the cummulant $\Kp(s)$ defined in Eq.~(\ref{eq_theo_Kps_gauss}) as a function of 
\begin{equation}
x=\frac{3 (2^p p! p)^2}{2 (2p+1)!} \ \frac{\ce}{\sqrt{s}}. 
\label{eq_conf_Kps_xdef}
\end{equation}
All data sets collapse
nicely on the prediction (bold line) for small $x$.\footnote{The curvature of the data at small $s$ 
is due to the finite extensibility of the subchains which becomes more marked for higher moments.}
It is important that the {\em same} effective bond length $b$ is obtained from
the analysis of all functions $\Kp(s)$ as illustrated in Fig.~\ref{fig_conf_Kps}.
Otherwise we would regard equilibration and statistics as insufficient.

\begin{figure}
\centerline{\resizebox{0.9\columnwidth}{!}{\includegraphics*{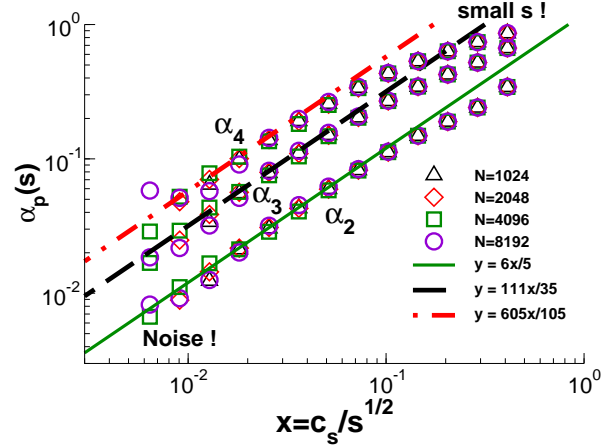}}}
\caption{Non-Gaussianity parameter $\alphap(s)$ computed {\em vs.} $\ce/\sqrt{s}$.
Perfect data collapse for all chain lengths $N$ indicated is obtained for each $p$.
A linear relationship over nearly two orders of magnitude is found as theoretically expected,
Eq.~(\ref{eq_theo_alphaps_perturb}).
The data curvature for small $s$ becomes more pronounced for larger $p$.
}
\label{fig_conf_alphas}
\end{figure}
The failure of Flory's hypothesis can also be demonstrated by means of the 
non-Gaussianity parameter $\alphap(s)$ defined in Eq.~(\ref{eq_theo_alphaps_gauss})
which compares the $2p$-th moment with the second moment ($p=1$).
In contrast to the related function $\Kp(s)$ this has the advantage
that here two {\em measured} properties are compared without any tunable parameter,
such as $b$.
Fig.~\ref{fig_conf_alphas} presents $\alphap(s)$ {\em vs.} $\ce/\sqrt{s}$
for three moments. For each $p$ we find perfect data collapse
for all $N$ which confirms the expected linear relationship
$\alphap(s) \approx \Ustar(s)$.\footnote{If one plots $\alphap(s)$ as a function 
of the {\em r.h.s.} of Eq.~(\ref{eq_theo_alphaps_perturb}) all data points for all moments and 
even for too small $s$ collapse on one master curve just as in Fig.~(\ref{fig_conf_Kps}).}
The lines indicate the theoretical prediction Eq.~(\ref{eq_theo_alphaps_perturb}). 
The prefactors $6/5$, $111/35$ and $604/105$ for $p=2$, $3$ and $4$ respectively
are nicely confirmed. They increase strongly with $p$, i.e. the non-Gaussianity
becomes more pronounced with increasing $p$.
Hence, $b$ should be best fitted by the second moment where the non-Gaussian behavior is the weakest.
%

\begin{figure}
\centerline{\resizebox{0.9\columnwidth}{!}{\includegraphics*{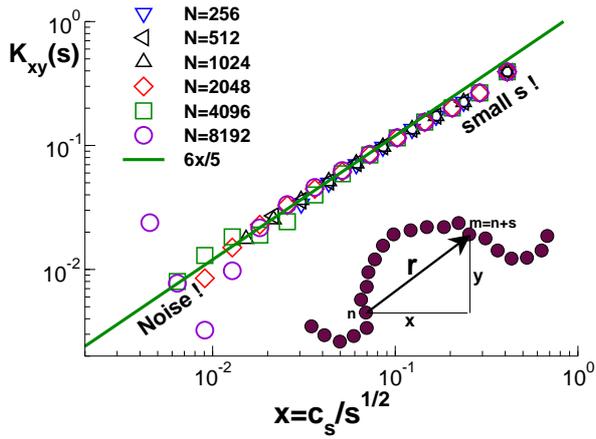}}}
\caption{Plot of $K_{xy}(s) = 1 - \la x^2 y^2 \ra / \la x^2 \ra \la y^2 \ra$
averaged over all pairs of monomers ($n,m=n+s$) and three different direction pairs
as a function of $\ce/\sqrt{s}$.
As indicated by the sketch at the bottom of the figure, $K_{xy}(s)$ measures the
correlation of the components of the subchain vector $\rvec$.
All data points collapse and show again a linear relationship $K_{xy} \approx \Ustars$.
Different directions are therefore coupled!
No curvature is observed over two orders of magnitude confirming that
higher order perturbation corrections are negligible.
Noise cannot be neglected for large $s > 100$ and finite subchain-size effects
are visible for $s \approx 1$.
}
\label{fig_conf_Kxys}
\end{figure}
Figure~\ref{fig_conf_Kxys} presents a similar correlation function which measures
the non-Gaussian correlations of different spatial directions. It is defined by
$K_{xy}(s) \equiv 1 - \la x^2 \ y^2 \ra/\la x^2 \ra \la y^2 \ra$
for the two spatial components $x$ and $y$ of the vector $\rvec$
as illustrated by the sketch given at the bottom of Fig.~\ref{fig_conf_Kxys}.
Symmetry allows to average over the three pairs of directions $(x,y)$, $(x,z)$ and $(x,z)$.
Following the general scaling argument given in Sec.~\ref{theo} we expect
$K_{xy}(s) \approx \Ustar(s) \approx \ce/\sqrt{s}$ which is confirmed by the
perturbation result 
\begin{equation}
K_{xy}(s) = K_2(s) = \frac{6}{5} \frac{\ce}{\sqrt{s}}.
\label{eq_conf_Kxy}
\end{equation}
This is nicely confirmed by the linear relationship found (bold line)
on which all data from both simulation models collapse perfectly.
The different directions of subchains are therefore coupled.
As explained at the end of Appendix~\ref{app_stat_moments}, $K_{xy}(s)$ and
$\alpha_2(s)$ must be identical if the Fourier transformed subchain size
distribution $G(q,s)$ can be expanded in terms of $q^2$ and this irrespective
of the values the expansion coefficients take.
Fig.~\ref{fig_conf_Kxys} confirms, hence, that our computational systems are perfectly
isotropic and tests the validity of the general analytical expansion.
The correlation function $K_{xy}$ is of particular interest since the
zero-shear viscosity should be proportional to $\la \sigma_{xy}^2 \ra
\sim \la x^2 y^2 \ra = \la x^2 \ra \la y^2 \ra (1 - K_{xy}(s))$ 
where we assume following Edwards \cite{DoiEdwardsBook} that only intrachain stresses
contribute to the shear stress $\sigma_{xy}$.
Hence, our results suggest that the classical calculations \cite{DoiEdwardsBook}
--- assuming incorrectly $K_{xy}=0$ --- should be revisited.

\subsection{Corrections to the subchain size distribution}
\label{conf_Grs}
\begin{figure}[t]
\centerline{\resizebox{0.9\columnwidth}{!}{\includegraphics*{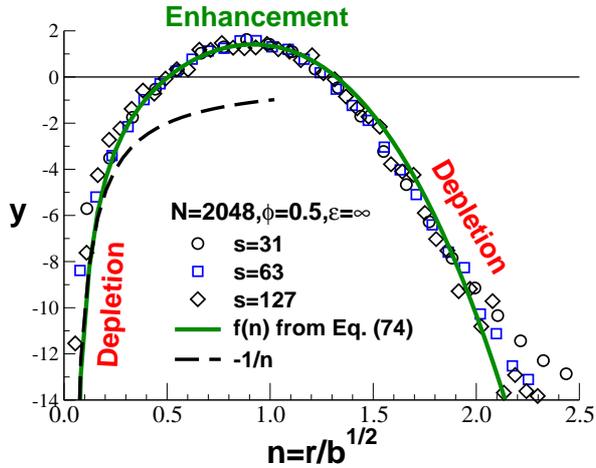}}}
\caption{Deviation $\delta G(r,s) = G(r,s) - G_0(r,s)$
of the measured subchain size distribution
from the Gaussian behavior $G_0(r,s)$ for several $s \ll N$ and $N=2048$.
As suggested by Eq.~(\ref{eq_theo_Grs_perturb}), we have plotted
$y= (\delta G(r,s)/G_0(r,s))/(\ce/\sqrt{s})$ as a function of $n=r/b \sqrt{s}$.
The data collapse confirms that the deviation scales linearly with $\Ustars \approx \ce/\sqrt{s}$.
The bold line indicates the universal function $f(n)$ predicted by Eq.~(\ref{eq_theo_fn}).
\label{fig_conf_dGrsbe}
}
\end{figure}
We turn finally to the subchain size distribution $G(r,s)$ itself which is presented in Fig.~\ref{fig_conf_dGrsbe}.
From the theoretical point of view $G(r,s)$ is the most fundamental property from which all others can be derived. 
The normalized histograms $G(r,s)$ are computed by counting the number of subchain vectors between $r-dr/2$ and
$r + dr/2$ with $dr$ being the width of the bin and one divides then by the spherical bin volume. Since the BFM model 
is a lattice model, this volume is not $4\pi r^2 dr$ but given by the number of lattice sites the
subchain vector can actually point to for being allocated to the bin.
Incorrect histograms are obtained for small $r$ if this is not taken into account.
Averages are taken over all (sub)chains, just as before.
Clearly, non-universal physics must show up for small vector length $r$ and small
curvilinear distance $s$ and we concentrate therefore on values $r \gg \sigma$ and $s \ge 31$.
When plotted in linear coordinates as in Fig.~11 of \cite{WBM07}, $G(r,s)$ compares roughly
with the Gaussian prediction $G_0(r,s)$ given by Eq.~(\ref{eq_intro_Grs_gauss}), but presents a
distinct depletion for small subchain vectors with $n \equiv r/ b \sqrt{s} \ll 1$
and an enhanced regime for $n \approx 1$. 
%
To analyse the data it is better to consider instead of $G(r,s)$ the relative deviation
$\delta G(r,s) / G_0(r,s) = G(r,s)/G_0(r,s) -1$ which should further be divided by the
strength of the subchain correlation hole, $\ce/\sqrt{s}$. As presented in Fig.~\ref{fig_conf_dGrsbe}
this yields a direct test of the relation Eq.~(\ref{eq_theo_Grs_perturb}) derived in Appendix~\ref{app_stat_Grs}.
The figure demonstrates nicely the scaling of the data for all $s$.
It shows further a good collapse of the data close to the universal function $f(n)$
predicted by theory (bold line).
Note that the depletion scales as $1/n$ for small subchain vectors (dashed line).
The agreement of simulation and theory is by all standards remarkable.
Obviously, error bars increase strongly for $n \gg 1$ where $G_0(r,s)$ decreases strongly.
The regime for very large $n$ where the finite extensibility of subchain matters has been omitted for clarity.
We emphasize that this scaling plot depends very strongly on the value $b$
which is used to calculate the Gaussian reference distribution.\footnote{As shown in Ref.~\cite{WBM07},
a similar plot can be achieved which does not require a precise value of $b$ if the reduced deviation $y$ 
is plotted as a function of $r/\Rs$ using the {\em measured} subchain size $\Rs$.}

\subsection{Intramolecular form factor $F(q)$}
\label{conf_Fq}
The form factor $F(q)$ defined in Sec.~\ref{theo_connectivity_Fq} is an 
important property since it allows to make a connection between theory and simulation on the one hand and 
experiments of real systems on the other hand \cite{BenoitBook}. 
\begin{figure}[t]
\centerline{\resizebox{0.9\columnwidth}{!}{\includegraphics*{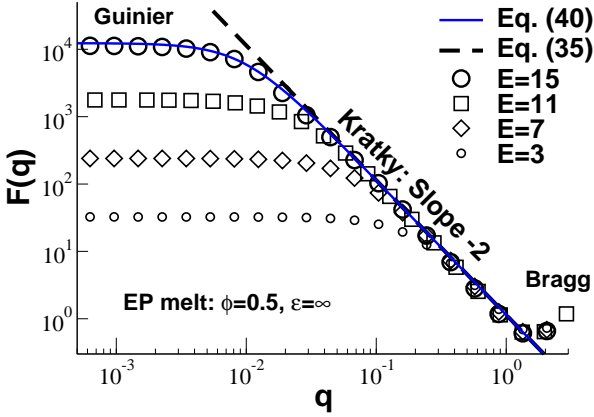}}}
\caption{Intramolecular form factor $F(q)$ {\em vs.} wavevector $q$ for EP of various scission energies $E$ \cite{BJSOBW07}.
The ideal chain form factor for Flory-distributed polymers, Eq.~(\ref{eq_theo_Fq_Flory}), is indicated by the solid line.
In the Kratky regime between the total chain and monomer sizes the form factor expresses the fractal dimension of the 
Gaussian coil, Eq.~(\ref{eq_theo_Fq_qlarge}), as shown by the dashed line. Experimentally, this is the most important 
regime since it is, e.g., not affected by the (a priori unknown) polydispersity.
The computational data reveal an additional regime at large wavevectors corresponding to the monomer structure
(``Bragg regime") which is not treated by the theory.
\label{fig_conf_Fq_Flory}
}
\end{figure}
Figure~\ref{fig_conf_Fq_Flory} presents the (unscaled) form factors obtained for four different scission energies $E$
for our EP model at $\phi=0.5$. The three different $q$-regimes are indicated. Details of the length distribution $\pchain$
matter in the Guinier regime which probes the total coil size. Non-universal contributions to the form factor arise 
in the ``Bragg regime" at large wave\-vectors. Obviously, the larger $E$ the wider the intermediate Kratky regime where chain length, 
polydispersity and local physics do not contribute much to the deviations of the form factor from ideality.
A very similar plot has been obtained for monodisperse polymers (not shown). Not surprisingly, it demonstrates that the form 
factors of both system classes become indistinguishable for large wavevectors.
Note that the $F(q)$ obtained for EP
of different $E$ can be brought to collapse by tracing $F(q)/F(0)$ as a function of $Q \equiv q \Rgyrz$ with 
$F(0)=\la N^2 \ra/\Nav$ and $\Rgyrz$ the measured $z$-averaged radius of gyration as indicated in Table~\ref{tab_bfm_EP}
\cite{BJSOBW07}.
A similar plot can again be obtained for monodisperse chains using $F(0)=N$ and $Q =q \Rgyr(N)$. 

\begin{figure}[t]
\centerline{\resizebox{0.9\columnwidth}{!}{\includegraphics*{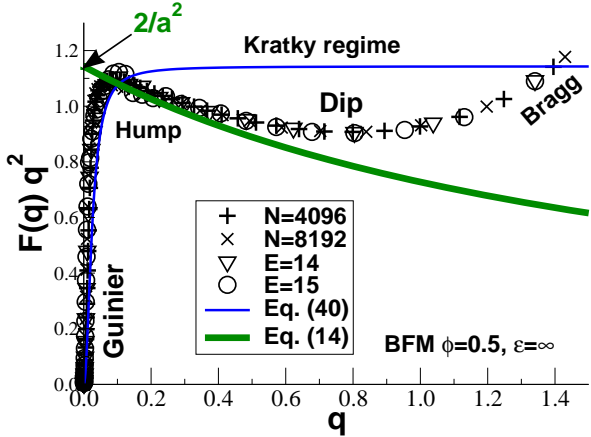}}}
\caption{Kratky representation of $F(q) q^2$ {\em vs.} $q$ for monodisperse quenched polymers
(crosses) and equilibrium polymers (open symbols). The {\em non-monotonous} behavior predicted by the 
theory is clearly demonstrated.
The ideal chain form factor (thin line), overpredicts the dip of the form factor at $q\approx 0.7$ by about $20\%$.
The bold line indicates the prediction for infinite chains, Eq.~(\ref{eq_intro_Fq}).
The data is fitted assuming the effective bond length $b \approx 3.244$.
\label{fig_conf_Fq_Kratky}
}
\end{figure}
Interestingly, a careful inspection of Fig.~\ref{fig_conf_Fq_Flory} reveals that Eq.~(\ref{eq_theo_Fq_Flory}) overestimates 
systematically the data in the Kratky regime. 
This can be seen more clearly in the Kratky representation given in Fig.~\ref{fig_conf_Fq_Kratky} in linear 
coordinates.\footnote{Such ``Kratky plots" are commonly used to represent neutron scattering experiments and
to test Flory's ideality hypothesis. In applications, the existence of a Kratky plateau appears to be elusive
\cite{BenoitBook}. Possible causes for deviations are effects of chain stiffness and finite chain thickness.
In special cases, the scattering signals from chain stiffness and thickness may compensate one another,
leading fortuitously to an extended Kratky plateau \cite{Rawiso87,Gagliardi05}. Apparently, ``Kratky plots
have to be interpreted with care" \cite{BenoitBook}.}
We present here the systems with the longest masses currently available for both monodisperse ($N=4096$ and $N=8192$) and 
EP systems ($E=14$ and $E=15$) \cite{BJSOBW07}. The  non-monotonous behavior is in striking conflict with Flory's hypothesis. 
The difference between the ideal Gaussian behavior (thin line) and the data becomes up to 20\%.
%
This difference is qualitatively expected from the perturbation theory result,  Eq.~(\ref{eq_intro_Fq}), derived in
Appendix~\ref{app_stat_Fq}. The predicted scaling can be easily understood by means of a simple scaling relation 
following the discussion in Sec.~\ref{theo_corhole_perturbation}. As the arc-length $s$ is related to the 
wavevector $q$ and the ideal form factor $F_0(q)$ by $s(q) \sim 1/|q|^2 \sim F_0(q)$, it
follows for the difference of measured and ideal form factors that \cite{WBJSOMB07}
\begin{equation}
\delta \left( \frac{1}{F(q)} \right) = \frac{1}{F_0(q)} \times \left( \frac{F_0(q)}{F(q)} -1 \right)
\approx \frac{1}{s} \times \Ustars \approx \frac{|q|^d}{\rho}
\label{eq_conf_Fq_scal}
\end{equation}
which agrees with Eq.~(\ref{eq_intro_Fq}).\footnote{This is the same scaling dependence which leads to
$\Pone(r) \approx 1/\rho |r|^d \approx 1/\rho (b s^{1/2})^d \approx \Pone(s)$ for the bond-bond correlation
function as a function of $r$ or $s$ as investigated in Sec.~\ref{conf_Ps}.} 
The important feature of this $|q|^3$-correction is
that it depends neither on the strength of the excluded volume interaction nor on the effective bond length $b$.
Hence, it must be generally valid, even for semidilute solutions.\footnote{It applies then for $q \ll 1/\xi$
with $\xi$ being the semidilute blob length \cite{DegennesBook}. We have checked that the
result from the renormalization group theory for semidilute solutions \cite{SchaferBook,MM00} takes the same
form as Eq.~(\ref{eq_intro_Fq}) with an amplitude $0.03124$ that is within $0.03\%$ of our $1/32$ \cite{BJSOBW07}.}
The prediction for infinite $N$, Eq.~(\ref{eq_intro_Fq}), cannot capture the decrease of the form factor
for small $q$ leading to the Guinier regime where $F(q)$ is determined by the finite size of the simulated chains.
A clearer evidence for the theory should thus be obtained by a different comparison between theory and simulation,
which accounts for the finite-$N$ effects.

\begin{figure}[t]
\centerline{\resizebox{0.9\columnwidth}{!}{\includegraphics*{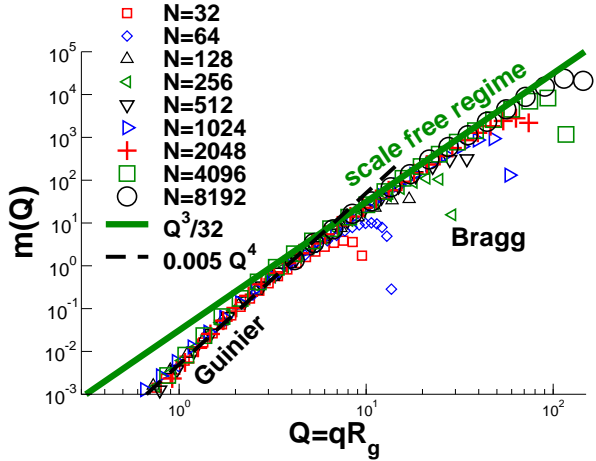}}}
\caption{Scaling attempt of the non-Gaussian deviations for monodisperse polymers 
in terms of the {\em measured} radius of gyration $\Rgyr(N)$. As suggested by Eq.~(\ref{eq_intro_Fq}),
the difference $1/F(q)-1/F_0(q)$ of the measured and the ideal
chain Debye form factor has been rescaled by the factor $N \rho/\rhostarN$
and plotted as a function of $Q= q \Rgyr$.
We obtain perfect data collapse for all chain lengths included.
(Obviously, data points in the Bragg limit $q\approx 1$ do not scale.)
Note that the power law slope, $m(Q) = Q^3/32$, predicted by Eq.~(\ref{eq_intro_Fq}),
can be seen over more then one order of magnitude. In the Guinier regime,
the difference increases more rapidly, $m(Q) \sim Q^4$ (dashed line),
as one expects from a standard analytic expansion in $Q^2$.
\label{fig_conf_Fq_mQ}
}
\end{figure}

This is achieved in Fig.~\ref{fig_conf_Fq_mQ} which focuses on deviations 
$\delta (1/F(q)) = 1/F(q) - 1/F_0(q)$ for monodisperse systems with $F_0(q)$ being the Debye formula. 
As above in our discussion of the subchain distribution $G(r,s)$ and its moments one should avoid to use as Gaussian reference
the ideal chain form factor $F_0(q)$ expressed in terms of the effective bond length $b$ since
the latter property might not be sufficiently accurate. A variation of a few percents breaks 
the scaling and leads to qualitatively different curves \cite{BJSOBW07} as in Fig.~\ref{fig_conf_K1s}
for the moment $K_1(s)$. Since such a precision is normally not available (neither in simulation nor
in experiment) it is of interest to seek for a more robust representation of the form factor deviations
which does not rely on $b$. The reference chain size is thus set in Fig.~\ref{fig_conf_Fq_mQ} by the 
{\em measured} radius of gyration $\Rgyr(N)$ which is used for rescaling the axis and, more importantly, 
to compute $F_0(q)$. (A virtually indistinguishable plot has been obtained for EP.)
The scaling of the vertical axis is suggested by Eq.~(\ref{eq_intro_Fq}) which predicts the difference of the 
inverse form factors to be proportional to $N^0 q^3$. Without additional parameters ($\Rgyr$ is known to high precision)
we confirm the scaling of
\begin{equation}
\label{eq_Fqinvscal}
m(Q) \equiv \left( \frac{N}{F(q)} - \frac{N}{F^{(0)}(q)} \right) \frac{\rho}{\rhostarN}
\label{eq_conf_Fq_mQ}
\end{equation}
as a function of $Q=q \Rgyr(N)$ with $\rhostarN \equiv N/\Rgyr^3$ being the self-density.
Importantly, our simulations allow us to verify for $Q \gg 5$ the fundamentally novel $Q^3$ behavior of the master curve
predicted by Eq.~(\ref{eq_intro_Fq}) and this over more than an order of magnitude!
In this representation we do {\em not} find a change of sign for the form
factor difference ($\delta F(q)$ is always negative) and all regimes can
be given on the same plot in logarithmic coordinates.
In the Guinier regime we find now $m(Q) \propto Q^4$ which is readily explained in terms of a standard expansion in 
$Q^2$ since the first two terms in $Q^0$ and $Q^2$ must vanish by construction because of the definition of $\Rgyr(N)$,
Eq.~(\ref{eq_theo_Fq_Guinier}).

\subsection{Summary}
\label{conf_summary}
In this section we have investigated various intrachain static properties of polymer melts
comparing our numerical data to the predicted deviations with respect to Flory's ideality hypothesis.
From the computational side the most important point is discussed in Sec.~\ref{conf_Rs} where
we show how the effective bond length $b$ for asymptotically long chains should be extrapolated
from subchains of finite arc-length with $\gT \ll s \ll N$. That the deviations are indeed due 
to long-range interactions has been demonstrated from the scaling of the bond-bond correlation function 
$\Pone(r) \sim 1/r^3$ in Sec.~\ref{conf_Pr}.

The most important finding from the experimental side concerns the scale-free deviations demonstrated
for the form factor $F(q)$, Eq.~(\ref{eq_intro_Fq}). We have shown that the Kratky plot does not exhibit 
the plateau expected for Gaussian chains in the intermediate wavevector range. These deviations should be measurable 
by neutron scattering experiments of {\em flexible} chains. Unfortunately, finite persistence length 
effects may mask the predicted behavior if the chains are not sufficiently long. An experimental verification ---
e.g. following the lines of the the promising recent study \cite{Gagliardi05} --- would be of great 
fundamental interest; it could also delineate the conditions where the predicted conformational
corrections to ideality are relevant in real polymer systems and must be considered in understanding
their structure, phase behavior and equilibrium dynamics. 
It is to the latter point we turn now our attention.

\begin{figure}[t]
\centerline{\resizebox{.95\columnwidth}{!}{\includegraphics*{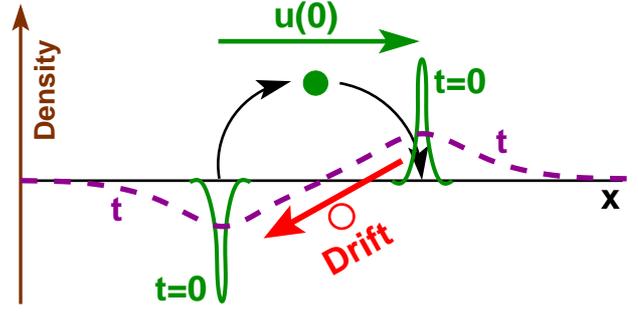}}}
\caption{The displacement auto-correlation function $C(t)$ of dense overdamped colloids 
is known to reveal a negative algebraic long-time tail due to long-range dynamical interactions 
caused by the weak compressibility of the solution \cite{DhontBook}: 
If a particle (filled sphere) is displaced at time $t=0$ by a distance $u(0)$
along the $x$-axis, this creates a density dipole (bold peaks) decaying in time
by cooperative diffusion (dashed line).
The gradient of the chemical potential associated to the decaying field
generates a drag pulling the particle (open sphere) back to its original position. 
We argue \cite{WPC11} that a related mechanism causes scale-free correlations 
of the displacements of (sub)chains in dense polymer solutions without topological constraints.
\label{fig_dipole}
}
\end{figure}
\section{Scale-free dynamical correlations in polymer melts}
\label{dyna}

\subsection{Introduction}
\label{dyna_intro}

\subsubsection{Prelude: Overdamped colloids}
\label{dyna_intro_colloid}
Dense, essentially incompressible simple liquids with conserved momentum  are known to 
exhibit long-range correlations of the particle displacement field \cite{DhontBook,HansenBook}. 
As first shown in the MD simulations by Alder and Wainwright \cite{HansenBook},
the coupling of displacement and momentum fields manifests itself by an algebraic decay of the VCF, 
\begin{equation}
C(t) \equiv \la \vvec(t) \cdot \vvec(0) \ra 
\sim + 1/\xidyn(t)^{d} \sim +1/t^{\alpha d},
\label{eq_dyna_Alder}
\end{equation}
with $\vvec(t)$ being the particle velocity at time $t$
and $\xidyn(t) \sim t^{\alpha}$ the typical particle displacement 
with exponent $\alpha = 1/2$.\footnote{The discussion is strongly simplified. 
          Strictly speaking,
          it is not the particle diffusion which sets the dynamical length scale
          but the diffusive propagation of the transverse momentum \cite{HansenBook}.}
Interestingly, even if the momentum conservation is dropped,
as justified for overdamped dense colloidal suspensions \cite{DhontBook},
scale-free albeit much weaker correlations are to be expected
due to the incompressibility constraint \cite{DhontBook,Fuchs02}. 
As illustrated in Fig.~\ref{fig_dipole}, the motion of a tagged colloid is coupled to the 
collective density dipole field,\footnote{Prefactors are omitted for simplicity. 
          Especially, we do not distinguish between the {\em self} diffusion of the particle 
          and the {\em collective} diffusion of the field characterized by rather
          different coefficients \cite{DhontBook}.}
\begin{equation}
\delta \rho(\rvec,t) \approx \frac{1}{\xidyn^d(t)} \frac{\pjump \cdot \rvec}{\xidyn^2(t)} \ \exp\left[-(\rvec/\xidyn(t))^2\right] 
\mbox{ for } t > 0,
\label{eq_dyna_dipolefield}
\end{equation}
created by the colloid's own displacement $\pjump$ at $t=0$.
After averaging over the typical displacements of the test particle and assuming a 
Cahn-Hilliard response proportional to the gradient $\nabla \mu(\rvec)$ of the chemical potential $\mu(\rvec)$
of the density field \cite{ChaikinBook}, this leads to a {\em negative} algebraic decay 
\begin{equation}
C(t) \sim -1/\xidyn(t)^{d+2} \sim - 1/t^{\omega}
\mbox{ with } \omega = (d+2) \ \alpha,
\label{eq_dyna_VCFcolloid}
\end{equation}
e.g., $\omega=5/2$ in $d=3$ dimensions.
This phenomenological scaling picture agrees with more systematic
mode-coupling calculations \cite{DhontBook,Fuchs02,hagen97}.
It has been confirmed computationally by means of 
Lattice-Boltzmann simulations \cite{hagen97},
MD simulations \cite{williams06} 
and even MC simulations with local moves \cite{WPC11} 
as demonstrated in Sec.~\ref{dyna_beads}.\footnote{Interestingly, 
the same power-law exponent $\omega=(d+2)/2$ is also seen
in Brownian dynamics simulations of the Lorentz model \cite{Franosch10}.}

\subsubsection{Polymer melts without topological constraints}
\label{dyna_intro_poly}
%
The finding of long-range static correlations in dense polymer solutions discussed in Sec.~\ref{conf} 
begs the question of whe\-ther a similar interplay of chain connectivity and incompressibility leads
to similar scale-free and $\gT$-independent {\em dynamical} correlations. To avoid additional physics 
we focus on model systems where hydrodynamic \cite{ANS11a} and topological constraints may be considered 
to be negligible, as in the pioneering work by Paul {\em et al.} \cite{Paul91a,Paul91b}, or are 
deliberately switched off \cite{Shaffer94,Shaffer95,WBM07}. 
As we have summarized in Sec.~\ref{intro_key}, deviations from Rouse-type dynamics
have been reported for such systems in various numerical and experimental studies.
The effective exponent $\beta \approx 0.8$ characterizing the short-time CM diffusion,
Eq.~(\ref{eq_intro_betadef}), is of course inconsistent with the key assumption of the Rouse model
that the random forces acting on the chains are uncorrelated.
We attempt here to clarify this problem using the BFM variant with topology non-conserving local L26-moves 
described in Sec.~\ref{bfm}. Obviously, it would not be possible to equilibrate BFM configurations with chain 
lengths up to $N=8192$ using local moves. Taking thus advantage of the configurations obtained using global MC moves 
we demonstrate that the VCF $\CN(t)$ associated to the chain CM displacements 
does {\em not} vanish, but instead decays as \cite{WPC11}
\begin{equation}
\CN(t) \approx - \left(\frac{\RN}{\TN}\right)^2 \frac{\rhostarN}{\rho} \ f(t/\TN)
\label{eq_dyna_key} 
\end{equation} 
with $f(x)$ being a universal scaling function. 
Note that the postulated Eq.~(\ref{eq_dyna_key}) does not depend explicitly on the 
compressibility of the solution.
The squared characteristic ``velocity" $(\RN/\TN)^2$ arises for dimensional reasons.
The second prefactor $\rhostarN/\rho$ stems from the interaction of chains and
subchains imposed by the incompressibility constraint.\footnote{That the effect
is proportional to the correlation hole penalty, $\UstarN \approx \rhostarN/\rho$,
may be guessed from the static correlations we have discussed in Sec.~\ref{conf}.
The fact that the polymers behave as weakly interacting colloids for large times $t/\TN \gg 1$
confirms this scaling.} 
As one expects from Eq.~(\ref{eq_dyna_VCFcolloid}), the scaling function decays for $x \gg 1$ as 
$f(x) \sim 1/x^{\omega}$ with an exponent $\omega = (d+2)/2$.
More importantly, it will be shown that this long-time behavior is preceded for $x \ll 1$ 
by a much weaker algebraic decay with
\begin{equation}
\omega= (d+2) \alpha = 
5/4 \mbox{ for } d=3
\label{eq_dyna_keyomega}
\end{equation}
due to the much slower relaxation ($\alpha = 1/4$) of the collective dipole field of subchains
which was generated by the initial displacement of the tagged subchain at $t=0$. 
The gradient of the chemical potential $\nabla \mu(\rvec)$ pulling the reference subchain back to 
its original position is of course not only due to the density fluctuation of the subchain density 
field but also to the tensional forces along the chains caused by the displacement. However, 
since subchain density fluctuations and tensions are coupled, their free energy contributions 
cause chemical potential gradients of the same order $\kBT s^0 N^0$ and are thus identical from
the scaling point of view \cite{WPC11,papFarago}.

\subsubsection{Outline}
\label{dyna_intro_outline}
%
Reminding first how a VCF may be defined and determined in a Monte Carlo simulation
\cite{papBigblurb}, we confirm in Sec.~\ref{dyna_beads} that Eq.~(\ref{eq_dyna_VCFcolloid}) 
holds for dense BFM beads. 
Section~\ref{dyna_jean} focuses then on polymer melts demonstrating
the dynamical coupling of (sub)chains.
Excluded volume and density effects are discussed in Sec.~\ref{dyna_robust}. 
Preliminary simulation data for melts {\em with} topological constraints presented in 
Sec.~\ref{dyna_repta} indicate that the early-time behavior, $\CN(t) \sim - N^{-1}t^{-(d+2)/4}$ for $t \ll \TN$,
is preserved before entanglement effects set in.
A perturbation cal\-cu\-la\-tion prediction for this time window is outlined in Sec.~\ref{dyna_perturb}. 
In demonstrating Eqs.~(\ref{eq_dyna_key}) and (\ref{eq_dyna_keyomega}) our study focuses on {\em one} 
mechanism explaining the striking deviations from the Rouse behavior observed in the literature.
See Refs. \cite{Schweizer89,Guenza02,Chong07,ANS11a,ANS11b} for related theoretical studies.
%
We stress that the presented MC simulations \cite{WPC11} are necessarily incomplete since important 
additional dynamical correlations arise due to the incomplete screening of hydrodynamic interactions 
mentioned briefly in Sec.~\ref{conc_outlook_offlattice} \cite{ANS11a,ANS11b}.

\begin{figure}[tb]
\centerline{\resizebox{0.9\columnwidth}{!}{\includegraphics*{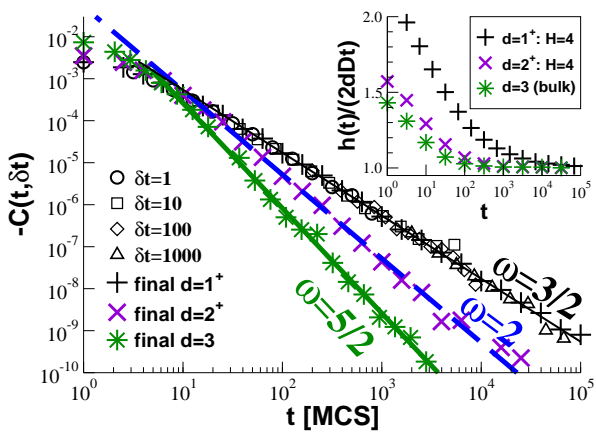}}}
\caption{Diffusion of BFM beads ($N=1$) without monomer overlap ($\overlap=\infty$) 
in $d=1^+$, $d=2^+$ and $d=3$ dimensions obtained using L26-moves at the reference 
``melt density" ($\phi=0.5$) \cite{papBigblurb}.
Inset: Although the bead diffusion is essentially free, small deviations are visible
if $\MSDall(t)/(2d D t)$ is plotted in log-linear coordinates.
Main panel: Collapse of displacement correlation functions 
$C(t,\deltat)$ for $d=1^+$ and various time increments $\deltat$ (open symbols). 
The ``final" function $C(t)$ is obtained by adding the first decade of data for each $\deltat$
and logarithmic averaging. The cummulants for each effective dimension agree nicely with the 
predicted power-law exponent $\omega=(d+2)/2$, Eq.~(\ref{eq_dyna_VCFcolloid}).
\label{fig_BFMbeads}
}
\end{figure}

\subsection{Diffusion of dense BFM beads}
\label{dyna_beads}

Since in MC simulations there is no ``mono\-mer mass",
no (conserved or non-conserved) ``monomer momentum" and not even an instantaneous velocity,
it might at first sight appear surprising that a well-posed ``velocity correlation function"
(VCF) can be defined and measured. To illustrate that this is indeed the case is the
first purpose of this subsection. The second is to verify that the negative analytic decay 
of the VCF expected for overdamped colloids, Eq.~(\ref{eq_dyna_VCFcolloid}),
is also of relevance for dense BFM beads ($N=1$) diffusing through configuration space 
by means of {\em local} hopping moves on the lattice as shown in Fig.~\ref{fig_bfm_algo}(b).
The systems presented in Fig.~\ref{fig_BFMbeads} correspond to three different effective dimensions $d$.
The effectively 1D systems ($d=1^+$) have been obtained by confining the beads to a thin
capillary of square cross-section, the effectively 2D systems ($d=2^+$) by confining the
beads to a thin slit as shown in Fig.~\ref{fig_bfm_algo}(c). 
The distance $H=4$ between parallel walls was chosen to allow the free crossing of the beads. 
(For the 1D case this is important, of course, since ordering the beads along the capillary 
alters dramatically the dynamics.)
The data has been obtained for beads without monomer overlap ($\overlap=\infty$)
by means of L26-moves on a 3D cubic lattice with half of the available lattice sites being occupied ($\phi=0.5$).
We average over the $2^{20}$ beads contained in each configuration.
One standard measure characterizing the monomer displacements is the mean-square displacement (MSD)
$\MSDall(t) \equiv \la (\rvec(t)-\rvec(0))^2 \ra$ displayed in the inset of 
Fig.~\ref{fig_BFMbeads} (with $\rvec(t)$ being the particle position at time $t$).
As one expects, the monomer displacements become uncorrelated for $t \gg 1$, 
i.e. $\MSDall(t) \approx 2d D t$ with $D$ being the monomer self-diffusion constant:
$D=0.0187$ for $d=1^+$, $D=0.0154$ for $d=2^+$, $D=0.0382$ for $d=3$.
The typical displacement $\xidyn(t) \equiv \MSDall^{1/2}(t)$ thus scales as 
$\xidyn(t) \sim t^{\alpha}$ with  $\alpha=1/2$.
However, small deviations are clearly visible for short times if $\MSDall(t)/(2d D t)$ is plotted 
in log-linear coordinates. The deviations are particular strong for $d=1^+$.
The effective random forces acting on the beads are thus not completely white.
Obviously, one might try to characterize these deviations by fitting various polynomials to the
measured MSD. However, since one needs to subtract the rather large free diffusion contribution from
the measured signal to characterize tiny deviations this is a numerically difficult if not impossible route.

In analogy to the bond-bond correlation function $\Pone(s) \sim \partial_s^2 \Rs^2$ 
discussed in Sec.~\ref{conf_bondbond} allowing to make manifest deviations from the Gaussian 
chain assumption, it is numerically much better to {\em directly} compute the second derivative 
of $\MSDall(t)$ with respect to time to avoid this large, but trivial contribution. How this can be done 
is illustrated in the main panel of Fig.~\ref{fig_BFMbeads}.
We sample equidistant series of configurations at time intervals $\deltat = 1, 10, 100, \ldots$ 
as indicated by the open symbols.
Each time series contains $10^4$ configurations. Averaging over all possible pairs of configurations
$(t_0,t_0+t)$ we compute 
$C(t,\deltat) \equiv \la \uvec(t_0+t) \cdot \uvec(t_0) \ra_{t_0}/\deltat^2$,
i.e. a four-point correlation function of the monomer trajectories
with $\uvec(t) = \rvec(t+\deltat) - \rvec(t)$ being the monomer displacement vector 
at time $t$ in a time interval $\deltat$.
By construction $C(t,\deltat) \equiv 0$ if both displacement vectors are uncorrelated.
%
According to Eq.~(\ref{eq_app_ht2Ct}) one expects 
\begin{equation}
C(t,\deltat) 
 \approx \frac{1}{2} \partial_t^2 \MSDall(t) \ \deltat^0 \mbox{ for } t \gg \deltat.
\label{eq_dyna_Ctht}
\end{equation}
%
As can be seen for BFM beads in $d=1^+$ dimensions, the $\deltat$-dependence indeed drops 
out for $t/\deltat > 1$ and we thus avoid the second index $\deltat$ writing $C(t)$ for the 
displacement (or velocity) correlation function.
Obviously, the statistics deteriorates for very large $t/\deltat$ where fewer 
configuration pairs contribute to the average (taking apart that the signal itself decays). 
It is for this reason that a hierarchy of time series of different $\deltat$ is needed.
Taking for each $\deltat$ only the first decade of data ($2 \le t/\deltat < 20$),
these data sets are pasted together and averaged logarithmically.
The exponents $\omega=3/2$ (thin line), $\omega=2$ (dashed line) and $\omega=5/2$ (bold line)
predicted by Eq.~(\ref{eq_dyna_VCFcolloid}) for $d=1$, $d=2$ and $d=3$, respectively,
compare well with our data over several orders of magnitude in time,
especially for $d=1^+$. 
Note that if one is satisfied with less orders of magnitude it is sufficient to check the exponents 
using just a time window $\deltat=1$ 
as may be seen from the open spheres.
The superposition of data from different time series is just a numerical trick
which reduces the number of configurations to be stored and the number of configuration pairs
to be computed for a given time $t$.

\subsection{Polymer melts without topological constraints}
\label{dyna_jean}
\begin{figure}[tb]
\centerline{\resizebox{0.9\columnwidth}{!}{\includegraphics*{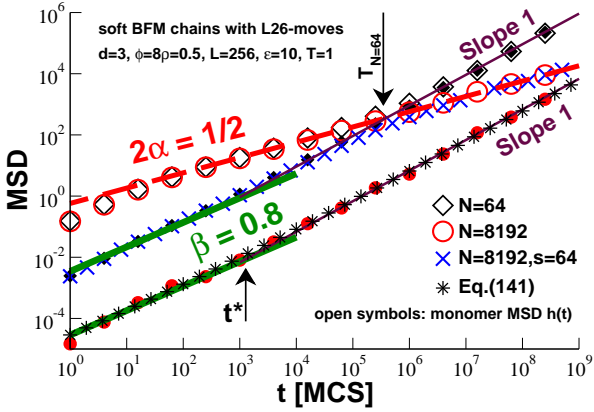}}}
\caption{
Various MSD for chains of lengths $N=64$ and $N=8192$ \cite{WPC11}.
The open symbols refer to the monomer MSD $\MSDmon(t)$,
the filled symbols to the MSD $\MSDcmN(t)$ of the CM of chains,
the crosses to the MSD $\MSDcms(t)$ of the CM of subchains of arc-length $s=64$ 
of total chains of length $N=8192$. 
The dashed line indicates the monomer MSD expected for Rouse chains for $t \ll \TN$,
the thin solid lines the free diffusion limit. 
As emphasized by the bold lines corresponding to the exponent $\beta=0.8$
suggested by Eq.~(\ref{eq_intro_betadef}), correlations are visible for the 
short-time behavior of the CM motion.
The stars indicate Eq.~(\ref{eq_dyna_Ct2MSD}) using $c = 1$,
i.e.  the time window $t \ll \tstar\approx 10^3$ is described by an exponent $\beta=3/4$. 
\label{fig_MSD_jean}
}
\end{figure}

\begin{figure}[tb]
\centerline{\resizebox{0.9\columnwidth}{!}{\includegraphics*{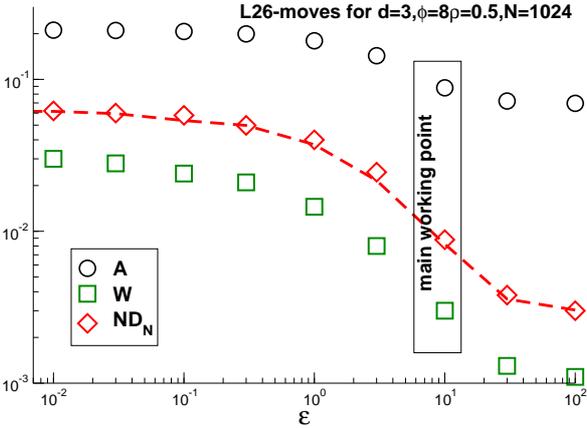}}}
\caption{Acceptance rate $A$, effective local mobility $W$ and self-diffusion coefficient $N \DN$
{\em vs.} the monomer overlap penalty $\overlap$. The data has been obtained using L26-moves at 
volume fraction $\phi=0.5$ \cite{WPC11}. 
All dynamical properties decrease monotonously with the interaction penalty but
become essentially constant above our main working point at $\overlap = 10$. 
The dashed line corresponds to the diffusion coefficient according to  
Eq.~(\ref{eq_dyna_W2DN}) and assuming the indicated mobilities $W$.
\label{fig_AW}
}
\end{figure}

\subsubsection{Mean-square displacements}
\label{dyna_jean_MSD}
Having shown that scale-free dynamical correlations exist for dense BFM beads as expected for 
overdamped colloids (Fig.~\ref{fig_dipole}), we turn now our attention to 3D melts of long and 
flexible homopolymers. As de\-scribed in Sec.~\ref{bfm_soft}, we focus on systems with finite 
overlap penalty $\overlap=10$ and volume fraction $\phi=0.5$ which we sample using local topology 
non-conserving L26-moves. 
As can be seen in Fig.~\ref{fig_MSD_jean} for chains of length $N=64$ and $N=8192$,
these systems are essentially of Rouse-type.
The monomer MSD $\MSDmon(t)$ is indicated by the open symbols.\footnote{For the large $N$ sampled 
here it is inessential whether this average is performed over all mono\-mers 
or only over a few monomers in the center of chains at $i \approx N/2$ as in \cite{Paul91a,Paul91b,KBMB01}.}
As expected from Rouse dynamics we obtain the $N$-independent short-time asymptotics
\cite{DoiEdwardsBook}
\begin{equation}
\MSDmon(t) = b^2 (W t)^{1/2} N^0 \mbox{ for }  10 \ll t \ll \TN \approx N^2/W
\label{eq_dyna_MSDmonshort}
\end{equation}
indicated by the dashed line ($2\alpha=1/2$). 
As can be seen for $N=64$, the monomers diffuse again freely with a power-law slope $1$ (thin lines) 
for times larger than the Rouse time $\TN$. We remind that it was neither computationally feasible 
nor our goal to sample for our larger chains ($N > 1000$) over the huge times needed to make this 
free diffusion regime accessible.
Following Paul {\em et al.} \cite{Paul91b}, the short-time power law, 
Eq.~(\ref{eq_dyna_MSDmonshort}), can be used to determine the effective mono\-mer mobility. 
We obtain $W(\overlap=10)=0.003$ for our main working point. 
Mobilities for other penalties are listed in Table~\ref{tab_overlap}.
As may be seen from Fig.~\ref{fig_AW}, $W(\overlap)$ decays with increasing excluded volume
just as the acceptance rate $A(\overlap)$ (spheres) of the Monte Carlo attempts, but the decay 
is even more pronounced for $W(\overlap)$, i.e. increasingly more accepted monomer moves do not 
contribute to the effective motion \cite{Paul91b}. 
%

The full symbols displayed in Fig.~\ref{fig_MSD_jean} refer to the MSD 
$\MSDcmN(t) = \la (\rN(t)-\rN(0))^2 \ra$ of the CM $\rN(t)$ of chains of length $N$.
As one expects for Rouse chains, the amplitude of $\MSDcmN(t)$ decreases inversely with $N$
and the diffusion appears to be Fickian ($\MSDcmN(t) \sim t$) at least for times 
$t \gg \tstar \approx 10^3 N^0$ as indicated by the vertical arrow.
Obviously, $\MSDcmN(t)$ and $\MSDmon(t)$ merge for times beyond the Rouse time ($t \gg \TN$). 
Fortunately, since $\MSDcmN(t)$ becomes linear for $\tstar \ll \TN$, it is possible for all $N$
to measure the self-diffusion coefficient $\DN$ by plotting 
$N \MSDcmN(t)/ 6 t$ {\em vs.} time $t$. 
For our main working point we obtain $N \DN(\overlap=10) \approx 0.009$.
Values for other $\overlap$ are again given in Table~\ref{tab_overlap} and 
are represented in Fig.~\ref{fig_AW} (diamonds). 
These values compare nicely with the Rouse model prediction \cite{DoiEdwardsBook}
\begin{equation}
N \DN  = \frac{\pi}{2} a^2 \ W
\label{eq_dyna_W2DN}
\end{equation}
shown by the dashed line in Fig.~\ref{fig_AW}.
Similarly, it is possible (at least for our shorter chains)
to measure the longest Rouse relaxation time $\TN$ by an analysis of the Rouse modes and to compare it with
the Rouse model prediction \cite{DoiEdwardsBook}
$\TN  =  4N^2 / \pi^3 W$.
We obtain again a nice agreement between directly and indirectly computed relaxation times (not shown).

Up to now we have insisted on the fact that our systems are to leading order of Rouse type
and we have characterized them accordingly. However, deviations from the Rouse picture are clearly
revealed for short times, especially for $\MSDcmN(t)$.\footnote{That the monomer MSD $\MSDmon(t)$ also 
deviates for very short times is due to the trivial lower cut-off associated to the discretization.}
In agreement with Eq.~(\ref{eq_intro_betadef}) the short-time CM motion may be characterized by 
a power-law exponent $\beta \approx 0.8$ (bold lines). 
Since in our BFM version topological constraints are irrelevant, this shows that the deviations 
obtained for the classical BFM algorithm with topological constraints \cite{Paul91b} 
cannot be attributed alone to precursor effects to reptational dynamics 
(which indeed exist as shown in Fig.~\ref{fig_dyna_repta}). 

Before we turn to the more precise numerical characterization of these deviations by means of the
associated displacement correlation function, let us ask whether the observed colored forces acting 
for short times on the CM of the entire chain are also relevant on the scale of subchains of 
arc-length $s$ ($1 \ll s \le N$). To answer this question we compute the MSD 
$\MSDcms(t) = \la (\rs(t)-\rs(0))^2 \ra$ associated to the subchain CM $\rs(t)$ 
as shown in Fig.~\ref{fig_MSD_jean} for subchains of length $s=64$ in the middle of total chains of 
length $N=8192$ (crosses).
Since for short times the subchain does not ``know" that it is connected to the rest of the chain,
one expects it to behave as a total chain of the same length ($s=N$). This is indeed borne out
by our data which are well described by
\begin{equation}
s \MSDcms(t) \approx N \MSDcmN(t) \mbox{ for } t \ll \Ts \approx s^2/W
\label{eq_dyna_MSDlocality}
\end{equation} 
for all chain length $N$ and subchain length $s$ studied. The subchains reveal thus for
sufficiently short times the {\em same} colored forces as the total chain
as can be clearly seen from the example given in Fig.~\ref{fig_MSD_jean}.
For larger times the subchain becomes ``aware" that it is connected to the rest of the chain
and gets enslaved by the monomer MSD, i.e. as expected from the Rouse model we observe 
\begin{equation}
\MSDcms(t) \approx \MSDmon(t) \approx b^2 (W t)^{1/2} s^0 N^0
\mbox{ for }  \Ts \ll t \ll \TN
\label{eq_dyna_MSDcmsintermediate}
\end{equation}
and, obviously, $\MSDcms(t) \approx \MSDcmN(t) \approx \MSDmon(t) \approx 6 \DN t$ for even larger times $t \gg \TN$.
%
%
%

\subsubsection{Locality and relevant exponent $\alpha$}
\label{comp_jean_comments}
Two comments are in order here. First, it should be noticed that Eq.~(\ref{eq_dyna_MSDlocality}) expresses 
the fact that the effective forces acting on the $N/s$ subchains of length $s$ in a chain of total 
length $N$ add up {\em independently} to the forces acting on the total chain. In this sense 
Eq.~(\ref{eq_dyna_MSDlocality}) states that the deviations from the Rouse picture must be {\em local}.
We will explicitly verify this below (Fig.~\ref{fig_VCF_jeansub}).\footnote{The ``locality" of the 
          correlations described by Eq.~(\ref{eq_dyna_MSDlocality})
          or Eq.~(\ref{eq_dyna_VCFlocality}) does not imply that the displacements of subchains
          around the reference subchain displaced at $t=0$ are $\delta(r)$-correlated.}
Second, if one chooses following Eq.~(\ref{eq_dyna_Ctht}) 
an arbitrary time window $\deltat$ to characterize the displacement correlations,
this corresponds to dynamical blobs containing $s \approx \sqrt{W \deltat}$ adjacent monomers,
which must move together due to the chain connectivity. Eq.~(\ref{eq_dyna_MSDcmsintermediate}) implies now that
the dipole field\footnote{We remind that, strictly speaking, 
          it is not the dipole field associated to the {\em density}
          of the subchain center-of-masses, but to their {\em chemical potential}. The use of the imprecise 
          notion ``density dipole field" might be excused by the fact that both fields are supposed to decay 
          similarly with time $t$, chain length $N$ and subchain length $s$.} 
associated with the CM of these $s$-subchains and 
created at $t=0$ by a tagged $s$-subchain must decay according to a typical displacement 
$\xidyn(t) \sim t^{\alpha}$ with $\alpha=1/4$.
It is this exponent $\alpha$ which is mentioned in Eq.~(\ref{eq_dyna_keyomega}).
Since this exponent is smaller than for colloids ($\alpha=1/2$), 
the subchain field must decay more slowly and one thus expects
a much weaker decay of the associated VCF. 

\begin{figure}[tb]
\centerline{\resizebox{0.9\columnwidth}{!}{\includegraphics*{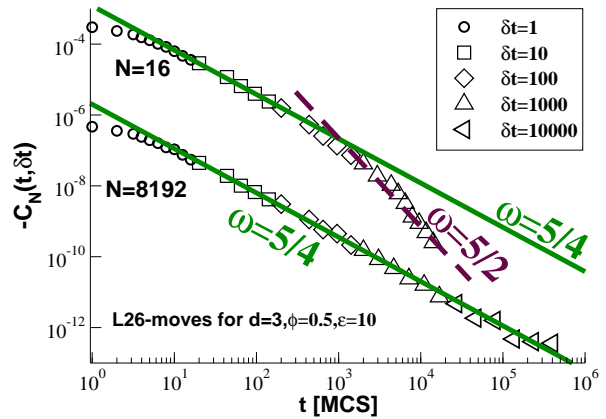}}}
\caption{VCF $\CN(t,\deltat)$ for two chain lengths, $N=16$ (top) and $N=8192$ (bottom), 
obtained using L26-moves \cite{WPC11}. We only indicate for each $\deltat$ the data window 
which is used to construct (by logarithmic averaging) the final VCF $\CN(t)$. 
The bold lines represent the exponent $\omega=5/4$
which is generally observed for times $10 \ll t \ll \TN$.
For $N=16$ we also indicate the exponent $\omega=5/2$ (dashed line) 
expected for larger times where the polymer coils behave according to 
Eq.~(\ref{eq_dyna_VCFcolloid}).
\label{fig_VCF_jeanconstruct}
}
\end{figure}

\begin{figure}[tb]
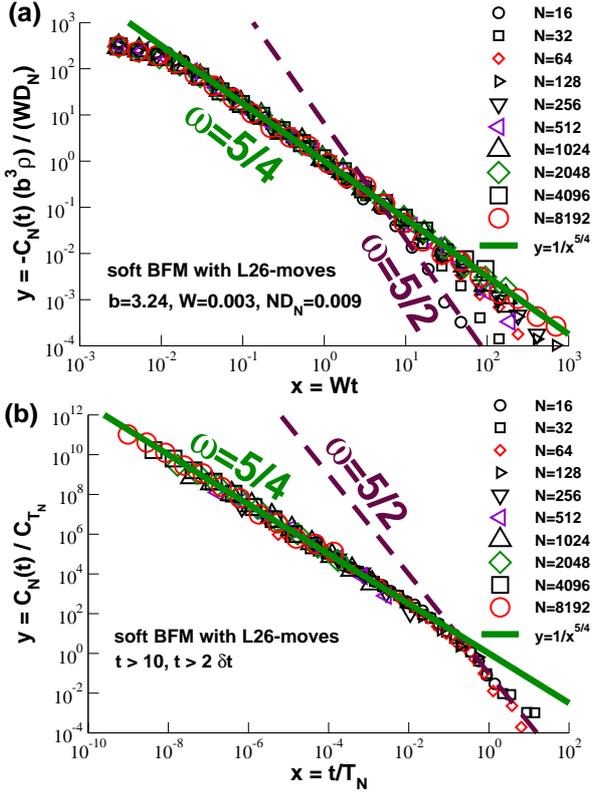

\centerline{\resizebox{0.9\columnwidth}{!}{\includegraphics*{fig40a}}}
\centerline{\resizebox{0.9\columnwidth}{!}{\includegraphics*{fig40b}}}
\caption{Scaling of VCF $\CN(t)$ for different $N$ \cite{WPC11}:
{\bf (a)}
Rescaled VCF $y = -\CN(t) (b^3 \rho) /(W \DN)$ {\em vs.} reduced time $x=W t$.
For times $t \ll \TN$ the VCF thus clearly scales as $\CN(t) \sim 1/N$,
i.e. the correlations must be due to $\sim N$ local independent events.
The exponent $\omega=5/4$ (bold line) is observed over up to five orders of magnitude in time.
{\bf (b)}
Collapse of $y=\CN(t)/\CNstar$ with $\CNstar \sim 1/N^{7/2}$ 
as a function of $x=t/\TN$ confirming Eq.~(\ref{eq_dyna_key}). 
The large time behavior ($x \gg 1$) is described by an exponent $\omega = 5/2$ (dashed line).
The exponent $\omega = 5/4$ (bold line) for $x \ll 1$ is demonstrated over eight orders of magnitude.
\label{fig_VCF_jean}
}
\end{figure}

\begin{figure}[tb]
\centerline{\resizebox{0.9\columnwidth}{!}{\includegraphics*{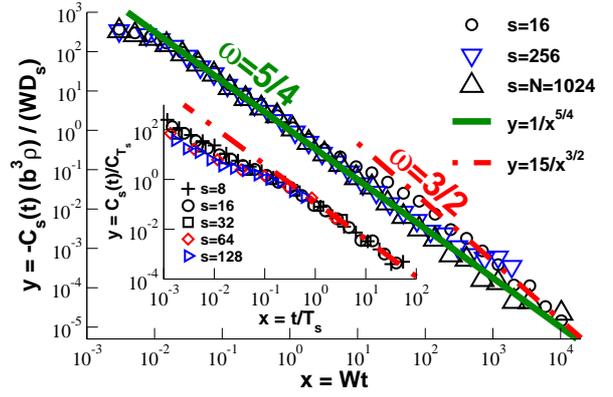}}}
\caption{$\Cs(t)$ for subchains obtained for chains of length $N=1024$ 
using an overlap penalty $\overlap=10$ and L26-moves \cite{WPC11}.
Main panel:
For $t \ll \Ts$ the subchains scale as the total chains in 
agreement with Eq.~(\ref{eq_intro_keydyna}) as shown by the bold line.
For intermediate times $\Ts \ll t \ll \TN$
the subchain displacements ($1 \ll s\ll N$) follow the monomer MSD and thus
$\Cs(t) \sim s^0 (b W)^2 (Wt)^{1/2-2}$ (dash-dotted line). 
Inset: The latter regime can be better seen by
setting $\Ts=s^2/W$ and $\Csstar = - (b W)^2 s^{-3}$ and plotting
$y=\Cs(t)/\Csstar$ as a function of $x=t/\Ts$.
The dashed slope corresponds to $y= 1/8x^{3/2}$ for $x \gg 1$.
\label{fig_VCF_jeansub}
}
\end{figure}

\subsubsection{Center-of-mass velocity correlation function}
\label{comp_jean_VCF}
Following the numerical strategy presented in Sec.~\ref{dyna_beads},
we characterize now more precisely the correlations seen for the 
chain MSD $\MSDcmN(t)$ and the subchain MSD $\MSDcms(t)$ by computing directly
their second derivative with respect to time $t$, i.e. the 
associated correlation functions $\CN(t) \approx \partial_t^2 \MSDcmN(t)/2$ 
and $\Cs(t) \approx \partial_t^2 \MSDcms(t)/2$. 

The VCF $\CN(t,\deltat) = \la \uvec(t+t_0) \cdot \uvec(t_0) \ra_{t_0}/\deltat^2$ 
for the displacement vector  $\uvec(t) = \rN(t+\deltat)-\rN(t)$ of the chain CM $\rN(t)$ is shown 
in Fig.~\ref{fig_VCF_jeanconstruct} for two chain lengths, $N=16$ (top) and $N=8192$ (bottom). 
Averages are again performed over all configuration pairs $(t_0,t+t_0)$ in the set of $10^4$
configurations sampled for each $\deltat$.
As in Sec.~\ref{dyna_beads} we find that $\CN(t,\deltat) \sim \deltat^0$ for $t \gg \deltat$.
For clarity, only the data subset is indicated for each $\deltat$ which is used to construct the 
final VCF $\CN(t)$ (as shown below in Fig.~\ref{fig_VCF_jean}). 
The bold lines represent the predicted short-time exponent $\omega=5/4$
which can be observed for $N=8192$ over nearly five orders of magnitude.
For $N=16$ we also indicate the exponent $\omega=5/2$ (dashed line) for $t \gg \TN$ 
where the polymers should behave as colloids according to Eq.~(\ref{eq_dyna_VCFcolloid}).
Note that the magnitude of the signal decreases strongly with $N$, 
which together with the fact that fewer chains per box are available, makes the determination of 
$\CN(t)$ increasingly more delicate.

The $N$-dependence of the VCF $\CN(t)$ is further analyzed in Fig.~\ref{fig_VCF_jean}.
The rescaled VCF $y=-\CN(t) (b^3 \rho) / (W \DN)$ is traced in panel (a) as function of 
the reduced time $x=Wt$ using the monomer mobility $W = 0.003$ and 
the diffusion coefficient $\DN=0.009/N$ determined above (Table~\ref{tab_overlap}). 
This scaling makes the axes dimensionless and re\-scales the vertical axis by a factor $N$.
As shown by the successful data collapse for chain lengths ranging from $N=16$ up to $N=8192$
on the slope indicated by the bold line, the VCF scales exactly as $\CN(t) \sim 1/N$ 
for $t \ll \TN$. This confirms the already stated ``locality" of the correlations, 
Eq.~(\ref{eq_dyna_MSDlocality}), i.e. the forces acting on subchains add up 
independently to the forces acting on the entire chain.
We have still to motivate the precise scaling used for the axes.
According to Eq.~(\ref{eq_dyna_key}) the VCF scales as a function of the reduced time $t/\TN$.
Substituting the typical chain size $\RN \approx b N^{1/2}$ and 
relaxation time $\TN \approx \RN^2/\DN \approx N^2/W$ this reduces for $\omega=5/4$ to
\begin{equation}
y \equiv - \CN(t) (b^3 \rho)/ (W \DN) = c \ (W t)^{-5/4} N^0
\mbox{ for } t \ll \TN
\label{eq_dyna_VCF_Wt}
\end{equation}
with $c$ being a dimensionless constant. 
The bold slope indicated in the plot corresponds to a value $c = 1$.
Interestingly, since $\CN(t) \approx \partial_t^2 \MSDcmN(t)/2$,
it follows from Eq.~(\ref{eq_dyna_VCF_Wt}) that 
\begin{equation}
\MSDcmN(t) = 6 \DN t \left( 1 + \frac{16 c}{9 b^3\rho} (Wt)^{-1/4} \right).
\label{eq_dyna_Ct2MSD}
\end{equation}
As may be seen in Fig.~\ref{fig_MSD_jean} (stars) for $N=8192$, Eq.~(\ref{eq_dyna_Ct2MSD}) with $c=1$
provides an excellent fit of the measured $\MSDcmN(t)$. 
%
%
We also note that the second term in Eq.~(\ref{eq_dyna_Ct2MSD}) 
dominates the short-time dynamics for $t \ll \tstar \equiv W^{-1} (16 c / 9 b^3\rho)^4 \ N^0 \approx 10^3$.
This is indicated by the vertical arrow in Fig.~\ref{fig_MSD_jean}. Hence, for $t \ll \tstar$
the stars correspond to a power-law slope with exponent $\beta = 2 - \omega = (6-d)/4 =  3/4$. 
This is close to the phenomenological exponent $\beta = 0.8$ from the literature.
The central advantage of computing the VCF $\CN(t)$ lies in the fact that it allows us thus to make 
manifest that (negative algebraic) deviations from the Rouse behavior exist for {\em all} times and 
not just for $t \ll \tstar$. 

Returning to our discussion of Fig.~\ref{fig_VCF_jean}(a) we emphasize that the VCF of shorter chains 
decay more rapidly for large times following roughly the exponent $\omega=5/2$ expected for effective colloids. 
We verify explicitly in Fig.~\ref{fig_VCF_jean}(b) that the bending down of the data
is consistent with the announced scaling in terms of a reduced time $x=t/\TN$ 
and a vertical axis $y= \CN(t)/\CNstar$ using the amplitude
$\CNstar \equiv \CN(t=\TN) \approx - (\RN/\TN)^2 \rhostarN/\rho \sim -1/N^{7/2}$ stated in 
Eq.~(\ref{eq_dyna_key}).
The successful data collapse confirms that the only relevant time scale in this problem
is the relaxation time $\TN$ for which the deviations for short ($\omega=5/4$) and long times 
($\omega=5/2$) match.
It is worthwhile to emphasize that the general scaling Eq.~(\ref{eq_dyna_key}) together
with the locality of the deviations, $\CN(t) \sim 1/N$, immediately
imply the exponent $\omega$. This can be seen by counting the powers
of the chain length $N$,
\begin{equation}
-1 \stackrel{!}{=} (1/2-2) 2 + (1 - d/2) + 2 \omega, 
\label{eq_dyna_Ncounting}
\end{equation}
thus $\omega = (d+ 2)/4 = 5/4$ in agreement with Eq.~(\ref{eq_dyna_keyomega})
and the numerically observed time dependence.
Assuming that Eq.~(\ref{eq_dyna_key}) holds, the exponents for $N$ and $t$ 
thus contain the same information.

The VCF $\Cs(t)$ for subchains of arc-length $s \le N$ is presented in Fig.~\ref{fig_VCF_jeansub}.
The subchain VCF $\Cs(t)$ is obtained as the total chain VCF $\CN(t)$, the only difference being 
that the CM $\rs(t)$ of the subchain defines now the displacement vector $\uvec(t) = \rs(t+\deltat)-\rs(t)$.
The data presented in the main panel is rescaled as in Fig.~\ref{fig_VCF_jean}(a)
with $\Ds = 0.009/s$ setting now the relevant diffusion constant used for the vertical axis. 
At short times the data for all $s$ collapses on the same power-law slope with exponent 
$\omega=5/4$ (bold line) as in Eq.~(\ref{eq_dyna_VCF_Wt}). Confirming Eq.~(\ref{eq_intro_keydyna}) 
and Eq.~(\ref{eq_dyna_MSDlocality})
we obtain the scaling
\begin{equation}
\frac{\Cs(t)}{W\Ds} \approx \frac{\CN(t)}{W\DN} \sim -(W t)^{-(d+2)/4} \mbox{ for } t \ll \Ts,
\label{eq_dyna_VCFlocality}
\end{equation}
i.e. the colored forces acting on subchains add up independently to the colored 
forces acting on the total chain.

\begin{figure}[tb]
\centerline{\resizebox{0.9\columnwidth}{!}{\includegraphics*{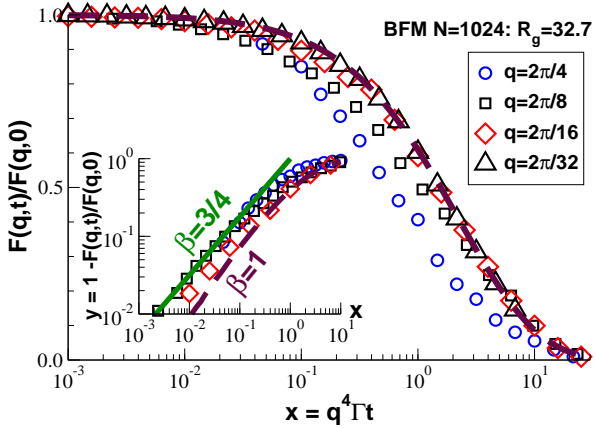}}}
\caption{Reduced dynamical coherent form factor $F(q,t)/F(q,0)$ as a function 
of reduced time $x = q^4 \Gamma t$ for several wavevectors $q$ as indicated.
The data has been obtained for BFM melts ($\phi=0.5$, $\overlap=10$, $\Gamma=0.0072$)
of chain length $N=1024$ with radius of gyration $\Rgyr = 32.7$.
Main panel:
Although the Rouse model scaling Eq.~(\ref{eq_dyna_Fqt_scal}) holds for sufficiently small $q$,
deviations are clearly visible for large wavevectors  $q \ge 2\pi/8$.
The dashed line indicates the numerical solution of Eq.~(\ref{eq_dyna_Fqt_Rouse})
for the intermediate wavevector regime. 
Inset: The double-logarithmic representation of $y(x) \equiv 1 - F(q,0)/F(q,0)$
reveals for large wavevectors and small times ($x \ll 1$) an exponent $\beta = 3/4$ (bold line).
\label{fig_Fqt_jean}
}
\end{figure}

\subsubsection{Dynamic coherent form factor}
\label{dyna_jean_Fqt}
Neither the MSD nor the VCF of the (sub)chain CM can be measured directly in a real experiment.
We discuss now how the exponent $\beta=2-\omega=3/4$ may be tested experimentally by means of 
the dynamical coherent form factor
\cite{DoiEdwardsBook,BenoitBook}
\begin{eqnarray}
F(q,t) & = & N \la \rhotagged(\qvec,t) \rhotagged(\qvec,0) \ra \\
 & =  & \frac{1}{N} \sum_{n,m=1}^{N}\la \exp\left(i \bm{q} \cdot (\bm{r}_n(t)-\bm{r}_m(0)\right)\ra
\label{eq_dyna_Fqt}
\end{eqnarray}
where $\rhotagged(\qvec,t) = \sum_{n=1}^{N} \exp(- i \qvec \cdot \rvec_n)$ stands for the 
Fourier transform of the density of a labeled chain.
For $t=0$ the dynamical form factor $F(q,t)$ reduces to the static form factor $F(q)$ 
discussed in Sec.~\ref{conf_Fq}.
In the Guinier regime $F(q,t)$ probes the overall chain motion which
allows to determine quite generally $\MSDcmN(t)$ using the expansion \cite{DoiEdwardsBook}
\begin{equation}
y \equiv 1 - \frac{F(q,t)}{F(q,0)} \approx 1 - \exp\left(- \frac{\MSDcmN(t) q^2}{4} \right)
\approx  \frac{\MSDcmN(t) q^2}{4},
\label{eq_dyna_Fqt_Guinier}
\end{equation}
at least if sufficiently precise small angle data are available \cite{Smith00,PaulGlenn}.
Thus, $y \sim t^{\beta}$ for $t \ll \tstar$.
Obviously, Eq.~(\ref{eq_dyna_Fqt_Guinier}) also holds for our numerical data (not shown).

Larger wavevectors ($1/\RN \ll q \ll 1/\xi$) are experimentally more readily accessible \cite{BenoitBook}.
Since in this regime the dynamical form factor probes the MSD of subchains 
of arc-length $s \sim 1/q^2$ with $\MSDcms(t)$ replacing $\MSDcmN(t)$ in 
Eq.~(\ref{eq_dyna_Fqt_Guinier}) \cite{WMJ10},
it is of relevance to verify whether it is possible in practice to confirm the exponent $\beta=3/4$ using 
the numerically computed dynamical form factor.
We remind that since in this wavevector regime $F(q,t)$ becomes independent of $N$, 
the Rouse model implies the scaling 
\begin{equation}
\frac{F(q,t)}{F(q,0)} = \Fscal(x) \mbox{ with } x=q^4 \Gamma t
\label{eq_dyna_Fqt_scal}
\end{equation}
being the scaling variable and
\begin{equation}
\Gamma =  \frac{a^2}{2} \frac{T}{\zeta} = \frac{\pi}{4} a^4 W
\label{eq_dyna_Gamma}
\end{equation}
a rescaled monomer mobility \cite{DoiEdwardsBook}. Using the directly measured mobility $W$
we have $\Gamma= 0.0072$ for the BFM data 
presented in Fig.~\ref{fig_Fqt_jean}. 
As can be seen from the main panel, a satisfactory data collapse is indeed obtained 
if sufficiently large subchains are probed ($q \le 2\pi/8$).
This is consistent with the fact that our systems are to leading order
of Rouse-type, especially if large times are probed. The scaling function $\Fscal(x)$ indicated by the dashed line 
has been obtained by integrating numerically the Rouse model prediction \cite{DoiEdwardsBook}
\begin{eqnarray}
\Fscal(x) & = & \int_0^{\infty} {\rm d}u \exp\left[- u - x^{1/2} h\left( u x^{-1/2}\right) \right] \mbox{ with} \nonumber \\
h(u) & = & \frac{2}{\pi} \int_0^{\infty} dx \frac{\cos(x u)}{x^2} \left(1 -\exp(-x^2)\right).
\label{eq_dyna_Fqt_Rouse}
\end{eqnarray}
Considering that there is no adjustable parameter this fit is satisfactory.
Expansion of Eq.~(\ref{eq_dyna_Fqt_Rouse}) for small $x$ yields\footnote{Note that 
the approximation $\Fscal(x) = \exp(-2\sqrt{x/\pi})$ given in \cite{DoiEdwardsBook} is unfortunately wrong.
Effective mobilities $W$ determined using this formula are therefore also incorrect \cite{WPB92}. 
A better approximation taking into account the fluctuations around the saddle point is given by 
$\Fscal(x) \approx (\pi^3 x/4)^{1/4}\exp(-2\sqrt{x/\pi})$.}
\begin{equation}
\Fscal(x) = 1- x+x^{3/2}\frac{4\sqrt{2}}{3\sqrt{\pi}} + \ldots \mbox{ for } x \ll 1.
\label{eq_dyna_Fscal_smallx}
\end{equation}
Thus, according to the Rouse model one expects the rescaled dynamical form factor $y(x) = 1 - F(q,t)/F(q,0)$ 
traced in the inset of Fig.~\ref{fig_Fqt_jean} to increase (to leading order) 
as $y(x) \sim x^{\beta}$ with $\beta=1$ (dashed line). Instead we find for larger
wavevectors a power-law increase with exponent $\beta=3/4$ (bold line) in agreement with 
the scaling relation
\begin{equation}
y \approx q^2\MSDcms(t) \approx q^2 t^{3/4}/s \approx q^4 t^{3/4}
\label{eq_dyna_Fscal_beta}
\end{equation}
for $t \ll \tstar \ll \Ts$ [Eq.~(\ref{eq_dyna_Ct2MSD})].\footnote{That with decreasing wavevector the data 
approaches systematically the Rouse prediction $\beta=1$ is expected 
{\em (i)} since for the same value $x= q^4 \Gamma t$ a smaller wavevector $q$ corresponds to a larger time $t$ and 
{\em (ii)} since for $t \gg \tstar \sim s^0 N^0$ the colored noise becomes masked by the white forces, 
Eq.~(\ref{eq_dyna_Ct2MSD}).}
In other words, an experimental observation of a power law slope of the reduced dynamical structure factor 
with $\beta \approx 3/4$ would thus confirm scale-free CM displacement correlations with 
$\omega=2-\beta \approx 5/4$.

\begin{figure}[tb]
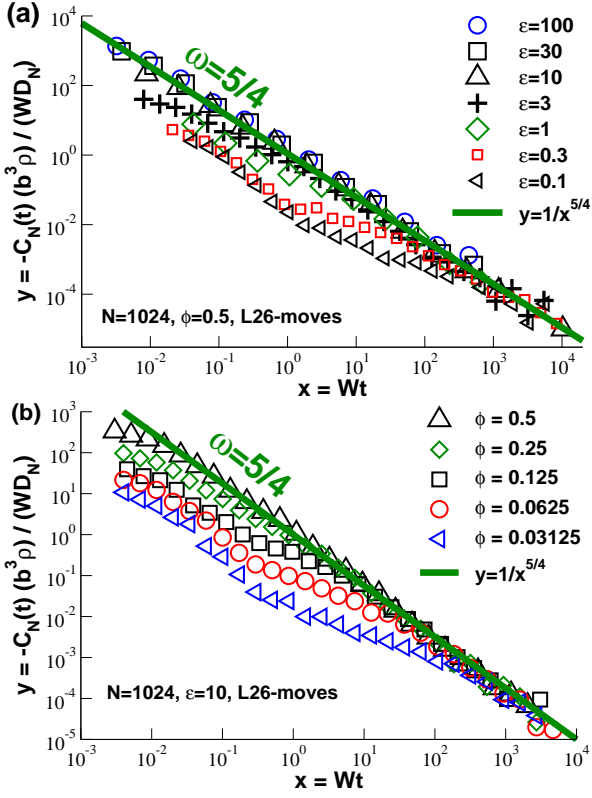

\centerline{\resizebox{0.9\columnwidth}{!}{\includegraphics*{fig43a}}}
\centerline{\resizebox{0.9\columnwidth}{!}{\includegraphics*{fig43b}}}
\caption{Robustness of the scaling of $\CN(t)$ \cite{WPC11}. 
Data obtained for $N=1024$ using L26-moves is shown for 
{\bf (a)} various overlap penalties $\overlap$ at volume fraction $\phi=0.5$ and
{\bf (b)} various volume fractions $\phi$ at overlap penalty $\overlap=10$. 
For sufficiently large times (but still $t \ll \TN$) all data appears to collapse
on the same asymptotic power-law exponent $\omega=5/4$ (bold line) for 
incompressible polymer solutions. 
The statistics deteriorates with decreasing $\overlap$ and $\rho$ and 
additional physics is visible at short times. 
\label{fig_dyna_robust}
}
\end{figure}

\subsection{Compressibility effects}
\label{dyna_robust}
Up to now we have focused on one working point at overlap penalty $\overlap=10$ and volume fraction $\phi=0.5$, 
i.e. all data corresponds to the same static properties, especially to the same dimensionless compressibility 
$\gT= 0.32$. It is natural to ask how the observed scaling changes with the compressibility of the solution. 
Our key scaling relation Eq.~(\ref{eq_dyna_key}) corresponding to an incompressible packing of blobs does in fact 
only depend {\em implicitly} on $\gT$. This suggest that Eq.~(\ref{eq_dyna_VCF_Wt}) for times $t \ll \TN$ 
should also remain valid if one uses for rescaling of the axes the effective bond length $b$, the mobility $W$ and 
the diffusion constant $\DN$ measured independently for the given operational parameters $\overlap$ and $\phi$.
This is in fact borne out for the asymptotic behavior of the rescaled VCFs displayed 
in Fig.~\ref{fig_dyna_robust}(a) for different overlap penalties $\overlap$ at $\phi=0.5$  and 
in Fig.~\ref{fig_dyna_robust}(b) for different volume fraction $\phi$ at $\overlap=10$. 
(We only display chains with $N = 1024$ to avoid the colloidal regime for $t \gg \TN$.)
The values used for the rescaling of the axes are listed in Table~\ref{tab_overlap} for the variation of the 
overlap penalty and in Table~\ref{tab_density} for different volume fractions. The scaling collapse for long 
times presented in panel (b) demonstrates explicitly that $\CN(t) \sim 1/b^3\rho$ 
if one scales out the additional variation of the monomer mobility $W(\phi)$. We remind that this density 
dependence stems originally from the factor $\rhostarN/\rho$ in Eq.~(\ref{eq_dyna_key}) due to the correlation 
hole forces setting the repulsion between chains and subchains which drive the dynamical correlations.
That the data deviates for short time from Eq.~(\ref{eq_dyna_VCF_Wt}) is to be expected qualitatively since the 
incompressibility constraint is only felt by the chains if the dynamics is probed on a scale corresponding 
to the static screening length. As we have seen in Sec.~\ref{conf_bondbond}, 
a similar crossover is observed for static properties such as the angular correlation 
function \cite{WCK09}.
At present we are still lacking a detailed description for the short time dynamical behavior
and how to match it with the correlated motion of incompressible blobs at long times.

\begin{figure}[tb]
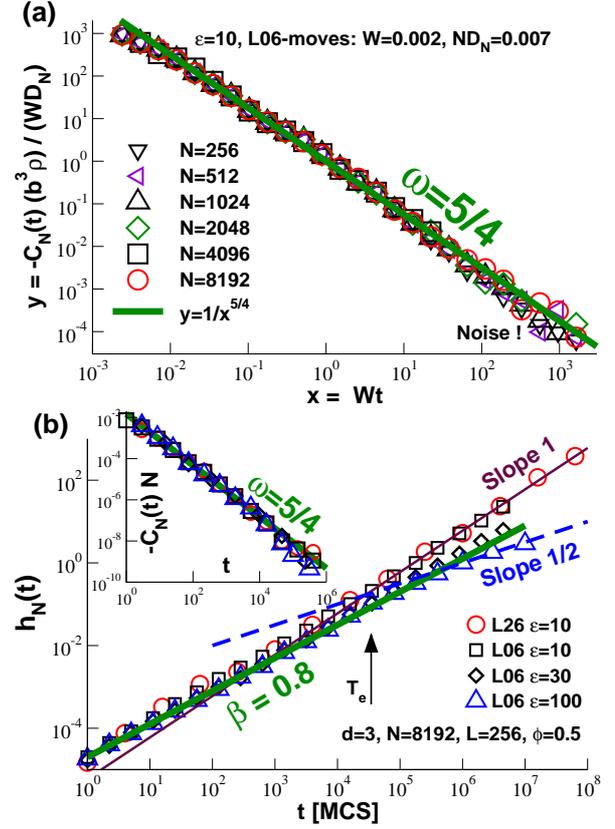

\centerline{\resizebox{0.9\columnwidth}{!}{\includegraphics*{fig44a}}}
\centerline{\resizebox{0.9\columnwidth}{!}{\includegraphics*{fig44b}}}
\caption{Dynamical correlations for 3D BFM melts at volume fraction $\phi=0.5$
obtained using L06-moves:
{\bf (a)}
VCF $\CN(t)$ for $\overlap=10$ focusing on larger chains with $N \ge 256$.
Using $W=0.002$ and $\DN=0.007/N$ we rescale the data as in Fig.~\ref{fig_VCF_jean}(a).
This allows to collapse the data on the same power-law slope (bold line)
as obtained using L26-moves, Eq.~(\ref{eq_dyna_VCF_Wt}).
{\bf (b)}
Topology induced additional correlations for chains of length $N=8192$.
Main panel: MSD $\MSDcmN(t)$ {\em vs.} time $t$ comparing L26-moves for $\overlap=10$ (spheres) 
to L06-moves for different overlap penalties $\overlap$.
Topological constraints become important for L06-moves with $\overlap \gg 10$
and, as expected from reptation theory for times larger than the entanglement time 
$\Te \approx 10^5$ \cite{DoiEdwardsBook},
the data bends towards a power-law exponent $1/2$ (dashed line).
Inset: The reduced VCF $-\CN(t) N$ confirms that for all system classes
the short-time dynamics is described by the power-law exponent $\omega=5/4$ (bold line).
\label{fig_dyna_repta}
}
\end{figure}

\subsection{L06-moves and effects of topological constraints}
\label{dyna_repta}
Up to now we have deliberately tuned our model to avoid topological constraints.
Obviously, these constraints are expected to matter for the dynamics of real 3D polymer melts 
\cite{DegennesBook,DoiEdwardsBook}. It is thus of interest to see 
how the presented picture changes if topology is again switch\-ed on using the topology conserving 
L06-moves of the classical BFM \cite{Paul91a,Paul91b,KBMB01}.

Obviously, for small penalties the dynamics remains of Rouse-type as can be seen from
Fig.~\ref{fig_dyna_repta}(a) for the same static conditions ($\phi=0.5$,$\overlap=10$)
as in Sec.~\ref{dyna_jean}. One determines readily a local mobility $W=0.002$ 
and a self-diffusion coefficient $\DN = 0.007/N$, i.e. the dynamics is slightly slower than for the 
larger L26-moves. If using these parameters $\CN(t)$ is rescaled as shown in Fig.~\ref{fig_VCF_jean}(a),
all data sets collapse perfectly on the same slope as for L26-moves (bold line).
This is consistent with the idea that the scaling Eq.~(\ref{eq_dyna_VCF_Wt}) only depends
implicitly on the specific MC moves used.

Preliminary data comparing L26-moves and L06-moves for chains of length $N=8192$
and higher overlap penalties is presented in Fig.~\ref{fig_dyna_repta}(b). 
With increasing $\overlap$ the crossing of the chains gets more improbable
for L06-moves and the topological constraints become more relevant. This can be seen for 
$\overlap=30$ and even more for $\overlap=100$. 
The latter data set approaches the slope $1/2$ (dashed line) 
expected from reptation theory for times larger than the entanglement time $\Te$ \cite{DoiEdwardsBook}.
The vertical arrow indicates a value for $\Te$ obtained from an analysis
of the monomer displacements $\MSDmon(t)$ \cite{KBMB01}. 
Apart from the much larger chain length $N=8192$ used, the data for $\overlap=100$ is 
consistent with the results obtained using the classical BFM  ($\overlap=\infty$) \cite{Paul91b,KBMB01}. 
As in Fig.~\ref{fig_MSD_jean} the bold line represents the exponent $\beta=0.8$ \cite{Paul91b,PaulGlenn}. 
Superficially, it does a better job for systems with conserved topology due to the broad crossover to the 
entangled regime.
However, that this empirical exponent is by no means deep as revealed by the VCF
$\CN(t)$ plotted in the inset. For short times {\em all} systems
with and without topology conservation are well-described by the {\em same}
exponent $\omega=5/4$ (bold line) in agreement with the proposed deviations 
from the Rouse model, Eq.~(\ref{eq_dyna_keyomega}). 
%
Note that the last points given for $\overlap=100$ decay slightly more rapidly.
Unfortunately, the length of the analyzed time series does currently not allow us 
to verify whether our data become consistent with the decay
\begin{equation}
\CN(t) \approx - \frac{1}{N} \left(\frac{\de}{\Te} \right)^2 \left( \frac{t}{\Te} \right)^{-3/2} 
\mbox{ for } t \gg \Te
\label{eq_dyna_CNentangled}
\end{equation}
expected for reptating chains with $\de \sim \Te^{1/2}$ being the tube diameter \cite{DegennesBook}. 
Much longer time series, just as for the L26-moves with $\overlap=10$ we have focused on,
are currently under production to clarify this issue.
The numerical demonstration is challenging, since the difference between
the exponents, $3/2-5/4=1/4$, is rather small and several orders of magnitude in time 
are needed to discriminate the power laws. In any case it is thus due to the dynamical correlations 
first seen in the BFM simulations of Paul {\em et al.} \cite{Paul91b,PaulGlenn} 
that the crossover between Rouse and reptation regimes becomes broader and
more difficult to describe than suggested by the standard Rouse-reptation theory 
\cite{DegennesBook,DoiEdwardsBook} 
which does not take into account the (static and dynamical) correlations 
imposed by the incompressibility constraint.

\subsection{Perturbation calculation predictions}
\label{dyna_perturb}

\subsubsection{Colored collective forces}
\label{dyna_colored}

Focusing now on the short-time correlations ($t \ll \TN$) in $d$-di\-men\-sion\-al 
polymer melts without topological constraints, we present now a linear response
calculation to describe the coupling between the degrees of freedom of a tagged test chain 
and the degrees of freedom of the collective bath characterized, respectively,
by the dynamical form factor $F(q,t)$ and the dynamical structure factor $S(q,t)$ at equilibrium. 
The underlying physics is that the displacement of a test chain at $t=0$ creates a free energy 
perturbation of the bath which decays by collective diffusion but survives at intermediate times. 
The collective force $\fcolvec(t)$ associated with the perturbation of the molecular field
pushes the test chain towards its original position at $t=0^{-}$ causing thus the anomalous 
diffusion of the CM motion demonstrated numerically above.
%
%
We remind first that the forces acting on the CM $\rN(t)$ of an {\em a priori} Rouse chain 
may be written \cite{DoiEdwardsBook}
\begin{equation}
N \zeta \frac{d \rN(t)}{dt} = \ftotvec(t) = \franvec(t) + \fcolvec(t),
\label{eq_dyna_forceonchain}
\end{equation}
i.e. in addition to the random white force $\franvec(t)$ of the standard Rouse model 
we have to account for the force $\fcolvec(t)$ from the molecular field.
The friction coefficient $\zeta$ may be obtained 
using Eq.~(\ref{eq_dyna_Gamma}) from the effective monomer mobility $W$.
According to the relation Eq.~(\ref{eq_app_displ5}) given in Appendix~\ref{app_ht2Ct}, 
the VCF $\CN(t)$ of the chain CM is given by
\begin{equation}
\CN(t) = - \frac{1}{(N\zeta)^2} \la \fcolvec(t) \cdot \fcolvec(0) \ra \mbox{ for } t > 0,
\label{eq_dyna_VCF_fcol}
\end{equation}
i.e. our task is to compute the correlation of the collective force $\fcolvec(t)$
due to the molecular field surrounding the reference chain.
The {\em negative} sign corresponds to the fact that the collective forces push the 
reference chain back towards its original position, just as for the overdamped 
colloids. 
\subsubsection{Dynamical random phase approximation}
\label{dyna_perturb_dRPA}
As already discussed in Sec.~\ref{dyna_jean_Fqt}, the degrees of freedom of the 
test chain are encoded by the dynamical form factor $F(q,t)$, Eq.~(\ref{eq_dyna_Fqt}). 
As seen from the main panel of Fig.~\ref{fig_Fqt_jean}, the scaling $F(q,t)=F(q,0) \Fscal(x)$ 
with $x = q^4 \Gamma t$ is nicely obeyed for not too large wavevectors, 
although deviations from the Rouse prediction are visible for small $x$. 
Since these deviations are small, 
it is justified in the spirit of a perturbation calculation to assume that 
Eq.~(\ref{eq_dyna_Fqt_Rouse}) holds for all wave\-vectors $q \gg 1/\RN$ and times $t \ll \TN$. 

\begin{figure}[tb]
\centerline{\resizebox{0.95\columnwidth}{!}{\includegraphics*{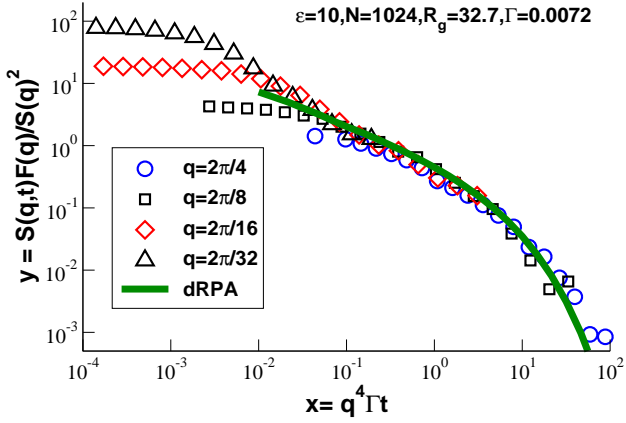}}}
\caption{Scaling of the dynamical collective structure factor $S(q,t)$ characterizing the
degrees of freedom of the bath surrounding the test chain. 
Data obtained from BFM simulations (open symbols) 
are compared to the theoretical prediction, Eq.~(\ref{eq_dyna_dRPAc}), 
indicated by the bold line \cite{papFarago}.
In agreement with theory we find a good scaling collapse if $y(x) = S(q,t) F(q)/S^2(q)$
is plotted as a function of the reduced time $x = q^4 \Gamma t$ with $\Gamma$ obtained
independently from the monomer mobility $W$, Eq.~(\ref{eq_dyna_Gamma}). 
Deviations from the Rouse model prediction are visible for small $x$, 
i.e. for short times $t \ll \tstar$, in agreement with Fig.~\ref{fig_Fqt_jean}.
\label{fig_Sqt_jean}
}
\end{figure}

The degrees of freedom of the bath are described by the dynamical collective structure factor 
\begin{equation}
S(q,t) = \la \rho(\qvec,t) \rho(\qvec,0) \ra / (\rho V)
\label{eq_dyna_Sqt}
\end{equation}
with $\rho(\qvec,t) = \sum_{n=1}^{\nmon} \exp\left(-i \qvec \cdot \rvec_n(t) \right)$ 
being the Fourier transform of the collective monomer density.
For $t=0$ Eq.~(\ref{eq_dyna_Sqt}) reduces to the static structure factor $S(q)$ analyzed in 
Sec.~\ref{TD_compress}.
The dynamical structure factor cannot be devised using the Rouse theory alone,
for $S(q,t)$ encodes the collective behavior of the chains, a cooperativity destroyed
by the assumptions underlying the Rouse approach. Fortunately, to deal with this question
the ``dynamical Random Phase Approximation" (dRPA) is a valuable tool which provides
sensible predictions for $S(q,t)$ \cite{ANS86,ANS97b,papFarago}. 
Similarly to the static RPA discussed in Sec.~\ref{TD_compress}, the dRPA is based on a
self-consistent closure of a mean-field theory: 
upon a perturbation by an external field, the response of a polymer is assumed to be the 
combination of a direct response to the field plus a response mediated by the environment of other
chains, which are also subjected to the external field.
This autocoherent reasoning leads to a prediction for 
$\Shat(q,z)$, the Laplace transform of $S(q,t)$, given by \cite{papFarago}
\begin{equation}
z \Shat(q,z) - S(q) = \frac{z \Fhat(q,z)-F(q)}{1+\rho \veff F(q) \left[F(q) -z \Fhat(q,z) \right]}
\label{eq_dyna_dRPAa}
\end{equation}
with $\veff = 1/\rho \gT$ being the effective monomer excluded volume 
and $\Fhat(q,z)$ the Laplace transform of the dynamical form factor $F(q,t)$.
For $z=0$, i.e. by summing over all times $t$, this equation yields of course the static RPA 
$S(q)^{-1}=F(q)^{-1}+ \veff \rho$ discussed in Sec.~\ref{TD_compress}.
Using this limit one may rewrite Eq.~(\ref{eq_dyna_dRPAa}) which yields the dRPA formula
\begin{equation}
\frac{\Shat(q,z)}{S(q)} = \frac{\Fhat(q,z)/F(q)}{1+\rho \veff F(q) \left[1 -z \Fhat(q,z)/F(q) \right]}.
\label{eq_dyna_dRPAb}
\end{equation}
Focusing on incompressible systems ($\gT \ll 1$), large wavevectors ($1/\RN \ll q \ll 1/\xi$) 
and short times $t\ll \TN$, Eq.~(\ref{eq_dyna_dRPAb}) can be further simplified.
As shown in \cite{papFarago} 
\begin{equation}
\frac{S(q,t)}{S(q)} \approx \frac{S(q)}{F(q)} \Sscal(x)
\mbox{ with } \Sscalhat(z) = \frac{\Fscalhat(z)}{1-z\Fscalhat(z)}
\label{eq_dyna_dRPAc}
\end{equation}
where $\Fscalhat(z)$ and $\Sscalhat(z)$ stand, respectively, for the
Laplace transform of the scaling functions $\Fscal(x)$ and $\Sscal(x)$.
Assuming for $\Fscal(x)$ the Rouse model prediction, Eq.~(\ref{eq_dyna_Fqt_Rouse}),
this yields for $y(x) \equiv S(q,t)F(q) / S^2(q)$
the bold line presented in Fig.~\ref{fig_Sqt_jean}.
%
Taking apart deviations for small $x$, i.e. for small times, we obtain a nice scaling
collapse for the BFM data presented in Fig.~\ref{fig_Sqt_jean}, especially considering
that there is no free fit parameter available.
We note finally that since for the wavevectors used $S(q) \approx \gT = 1/(\veff \rho)$,
it follows from Eq.~(\ref{eq_dyna_dRPAc}) that 
\begin{equation}
(\veff \rho)^2 S(q,t) F(q,t) = \frac{2}{(a q)^{2}} \Fscal(x) \Sscal(x)
\label{eq_dyna_dRPAd}
\end{equation}
for sufficiently low wavevectors at fixed compressibility or 
for sufficiently small compressibilities at fixed wavevector.

\subsubsection{Lowest-order perturbation}
\label{dyna_perturbation}
A collective density fluctuation $\delta\rho$ corresponds to a molecular field $v\delta\rho$ driving 
the mono\-mers of the test chain with a force $- T \veff \nabla \delta \rho$. 
%
The lowest order force correlation corresponds to the following mechanism: the test chain interacts 
with the collective field at time $0$ and again at time $t$. 
%
Using the densities $\rhotagged(q,t)$ and $\rho(q,t)$ of the test chain and the bath
in Fourier space defined above, the collective force $\fcolvec(t)$ may be rewritten as
\begin{equation}
\fcolvec(t) = T \veff \int{\frac{{\rm d}^d q}{(2\pi)^d} (i \qvec) \rhotagged(q,t) \rho(q,t)}.
\label{eq_dyna_fcol1}
\end{equation}
The correlation $\la \fcolvec(t) \cdot \fcolvec(0)\ra$ for $t > 0$ becomes thus
after lowest-order factorization
\begin{equation}
\frac{T}{V}\int{\frac{{\rm d}^d q}{(2\pi)^d}v^2q^2 \la\rhotagged(q,0)\rhotagged(q,t)\ra\la\rho(q,0)\rho(q,t)\ra}.
\label{eq_dyna_fcol2}
\end{equation}
Using the dynamical form and structure factors, Eq.~(\ref{eq_dyna_Fqt}) and Eq.~(\ref{eq_dyna_Sqt}), 
this yields
\begin{equation}
\la \fcolvec(t) \cdot \fcolvec(0)\ra = T^2 N\rho\int{\frac{{\rm d}^d q}{(2\pi)^d}v^2q^2 F(q,t) S(q,t)}.
\label{eq_dyna_fcol3}
\end{equation}
Using Eq.~(\ref{eq_dyna_VCF_fcol}) and Eq.~(\ref{eq_dyna_dRPAd}) the VCF becomes
\begin{equation}
\CN(t) = - \frac{T^2}{N\zeta^2\rho} \int{\frac{{\rm d}^d q}{(2\pi)^d}q^2 \Sscal(x) \Fscal(x)}.
\label{eq_dyna_theo3}
\end{equation}
Since $\Fscal(x)$ and $\Sscal(x)$ decrease rapidly for $x \gg 1$, the integral is 
well-behaved.
After substituting the friction coefficients $\zeta$ and $\Gamma$ by $W$, Eq.~(\ref{eq_dyna_Gamma}),
this may be rewritten as
\begin{equation}
\CN(t) = -c_d \frac{W \DN}{b^d \rho} \ (W t)^{-\omega} 
\mbox{ with } \omega = (d+2)/4.
\label{eq_dyna_theo4}
\end{equation}
Hence, Eq.~(\ref{eq_dyna_theo4}) confirms finally the scaling announced in Eq.~(\ref{eq_dyna_key}) 
and Eq.~(\ref{eq_dyna_keyomega}).
%
The coefficient $c_d$ 
is given by 
\begin{equation}
c_d = \sqrt{\pi} \left(4d/\sqrt{\pi} \right)^{d/2}
\int{\frac{{\rm d}^d x}{(2\pi)^d}x^2 \Fscal(x) \Sscal(x)}
\label{eq_dyna_cd}
\end{equation}
which may be computed numerically \cite{papFarago}.
For 3D melts ($d=3$) we find $c_3 \approx 0.42$.
This is of the same order but slightly lower as the empirical coefficient 
$c \approx 1$ estimated from Fig.~\ref{fig_VCF_jean}. 
Where this small difference comes from is currently still an unresolved question.
One possibility is that the single loop approximation presented here
--- especially assuming Eq.~(\ref{eq_dyna_Fqt_Rouse}) and Eq.~(\ref{eq_dyna_dRPAc}) for small $x$ ---
is not sufficient and higher order terms should be considered. 
Another (physically more appealing) option is that only about half of the free energy is stored 
in the longitudinal composition fluctuations and that the transverse stress due to the shearing 
of the bath at constant density generates an additional contribution to the correlations 
of same magnitude \cite{papFarago,ANS11b}.

\subsection{Summary}
\label{dyna_summary}
Focusing on a variant of the BFM algorithm with topology non-conserving moves we 
investigated the effective dynamical forces acting on the chains and subchains 
imposed by the incompressibility constraint. 
Sampling chain lengths up to $N=8192$ 
allowed us to carefully check the $N$-dependence of deviations with respect to the Rouse model.
Such deviations are visible from the short-time scaling of the CM MSD $\MSDcmN(t)$
in agreement with the literature, however, a more precise characterization can be achieved by means of the CM VCF
$\CN(t) \approx \partial^2_t \MSDcmN(t)/2$ which allows to probe directly the correlated random forces,
Eq.~(\ref{eq_dyna_VCF_fcol}). 
How such a VCF can be computed within a MC scheme has been first illustrated for dense BFM beads 
(Fig.~\ref{fig_BFMbeads}) confirming the negative algebraic decay expected for overdamped colloids, 
Eq.~(\ref{eq_dyna_VCFcolloid}).  The observed exponent $\omega=(d+2) \alpha$ with $\alpha=1/2$ can be 
understood by the coupling of a tagged colloid to the gradient of the collective density field decaying in time.
As shown in Fig.~\ref{fig_VCF_jean}(b), the same exponents are relevant for polymer chains 
for large times ($t \gg \TN$) where the chains behave as effective colloids.
More importantly, the short-time deviations for $\MSDcmN(t)$ have been traced back to the negative analytic decay 
$\CN(t) \sim - N^{-1} t^{-\omega}$ for $t \ll \TN$ with an exponent $\omega \approx 5/4$. 
That $\CN(t)$ decays inversely with mass shows that the process is {\em local}, i.e. the displacement correlations 
of subchains of arc-length $s$ add up independently (Fig.~\ref{fig_VCF_jeansub}).
Assuming according to the postulated scaling relation Eq.~(\ref{eq_dyna_key})
the chain relaxation time $\TN$ to be the only characteristic time scale, both 
asymptotic regimes can be brought to a successful data collapse.
The $\omega$-exponents for short times proposed in Eq.~(\ref{eq_dyna_keyomega}) are consistent
with the crossover scaling, Eq.~(\ref{eq_dyna_key}), and the locality of the correlations 
($\CN(t) \sim 1/N$) which implies $\omega=(d+2)/4$.
This scaling can be understood by the generalization of the above-mentioned correlation experienced 
by overdamped colloids to the displacement field of subchains of length $s \sim \deltat^{1/2}$ with 
$\deltat$ being the (arbitrary) time window used to define the displacements. Since subchains repel 
each other due to the incompressibility constraint,
a tagged subchain is pulled back to its original position by the subchain dipole field.
Since for times $\deltat \ll t \ll \TN$ the relevant dipole field decays much slower 
than for colloids ($\alpha =1/2 \to 1/4$), the correlations are much more pronounced. 
That our scaling relations for $t \ll \TN$ do not depend explicitly on the compressibility
of the solution, as stated by Eq.~(\ref{eq_dyna_key}), has been checked by the systematic
variation of excluded volume penalty $\overlap$ and volume fraction $\phi$ \cite{WPC11}. 
%
As shown for melts {\em with} topological constraints the early-time behavior, 
$\CN(t) \sim - N^{-1}t^{-5/4}$, is preserved before entanglement effects set in
which leads to a broad crossover between Rouse and reptation regimes.
%
As shown in Sec.~\ref{dyna_perturb}, our scaling approach is consistent with a standard 
linear response calculation. {\em En passant} we have verified the so-called 
``dynamical Random Phase Approximation" for the collective dynamical response
of the bath surrounding the reference chain (Fig.~\ref{fig_Sqt_jean}) \cite{papFarago}.

\section{Conclusion}
\label{conc}

\subsection{Summary}
\label{conc_summary}

Until very recently it has been generally assumed that {\em all} long-range static and
dynamical correlations are negligible in dense 3D solutions of flexible homopolymers beyond 
the excluded volume screening length $\xi \sim \gT^{1/2}$ which characterizes the decay of 
the density fluctuations \cite{FloryBook,DegennesBook,DoiEdwardsBook,RubinsteinBook}.
For static properties this general screening assumption leads to Flory's ideality hypothesis 
(Sec.~\ref{intro_Flory}) \cite{FloryBook,DoiEdwardsBook} stating that the chains must 
obey Gaussian chain statistics. For the equilibrium dynamics it implies Rouse model dynamics 
(Sec.~\ref{intro_Rouse}) if in addition momentum conservation (``hydrodynamic screening") 
and topological constraints \cite{DoiEdwardsBook} may be neglected or are deliberately 
switched off as one can readily do in a com\-put\-er experiment using local MC hopping moves
(Sec.~\ref{bfm_equil_L26}).

%
%
That some deviations from Flory's ideality hypothesis must exist 
has been suggested in earlier work by de Gennes \cite{DegennesBook,Brochard79}, Obukhov \cite{Obukhov81}, 
Duplantier \cite{Duplantier89}, Sch\"afer {\em et al.} \cite{SchaferBook,MM00}
or Semenov and Johner \cite{ANS03}.
For instance, it has been known for a long time \cite{DegennesBook} that due to the incompressibility 
constraint an entropic penalty $\UstarN \approx \rhostarN/\rho$ set by the correlation hole self-density 
$\rhostarN \approx N/\RN^d$ has to be paid for bringing two chains of length $N$ close to each other
(Fig.~\ref{fig_theo_constraint}). In the large-$N$ limit this effect must become negligible in $d=3$, however.
That the dynamics of polymer melts without relevant topological interactions may deviate from 
the Rouse model has been pointed out in various theoretical, computational and experimental
studies by Schweizer \cite{Schweizer89}, Guenza {\em et al.} \cite{Guenza02} or 
Paul {\em et al.} \cite{Paul91b,Paul98} as summarized in Ref.~\cite{PaulGlenn}.
%
However, since topological and hydrodynamical effects have not been systematically separated from the 
incompressibility con\-straint \cite{Guenza02} --- with the notable exception of the BFM simulations
of Shaffer \cite{Shaffer94,Shaffer95} --- 
it has been difficult to pin-point the precise physical origin of the key finding 
observed for the MSD $\MSDcmN(t)$ of the chain CM, Eq.~(\ref{eq_intro_betadef}).
%

Summarizing recent theoretical and computational results 
we have argued in the present contribution that due to the incompressibility constraint 
both static \cite{WMBJOMMS04,WBCHCKM07,WBJSOMB07,BJSOBW07,WBM07,MWK08,WCK09,WJO10,WJC10,ANS10a} 
and dynamical \cite{WPC11,papFarago} correlations of the composition fluctuations, 
i.e. of the density and displacement fields of marked chains and subchains, must arise for
distances $r \gg \xi$ and for corresponding arc-lengths $s \gg \gT$.
Since the applied constraint is scale-free, the resulting correlations are also scale-free
--- taken apart the usual upper and lower cutoffs set by the chain and the monomer size ---
and do not depend explicitly on the dimensionless compressibility $\gT$ of the solution 
(Sec.~\ref{theo_incompress}).
%
The central point stressed by us \cite{WBCHCKM07} is that not only chains
but also subchains of arbitrary length $s$ must repel each other due to the incompressibility
constraint and this with a penalty $\Ustars \approx s/\rho \Rs^d$ being for $s \ll N$ much 
larger than the penalty $\UstarN$ between the chains (Fig.~\ref{fig_theo_Ustar}).
Since subchains repel each other and this with a strength decreasing with $s$ in $d=3$, 
the chains become weakly swollen as discussed in Sec.~\ref{conf}. 
Our study shows that a polymer in dense solutions should not be viewed as {\em one} soft sphere
(or ellipsoid) \cite{EM01,Likos01,Guenza02}, but as a hierarchy of nested segmental correlation holes 
of all sizes aligned and correlated along the chain backbone 
[Fig.~\ref{fig_theo_constraint}(b)].\footnote{Similar deviations from Flory's ideality hypothesis
have been reported recently for polymer gels and networks \cite{Sommer05,Everaers05}.} 
The effective interaction between subchains has also dynamical consequences since a subchain 
displaced at $t=0$ causing thus a perturbation of the subchain density field (Fig.~\ref{fig_dipole})
will be slightly pushed back by the bath to its original position (Sec.~\ref{dyna}).

In this review we have focused on numerical results obtained by means of a BFM variant with finite
monomer interactions without topological con\-straints (Fig.~\ref{fig_bfm_algo}). Since this ``soft BFM" 
is fully ergodic (in contrast to the classical BFM) and very efficient due to its implementation as a 
Potts spin model \cite{WCK09}, it may be an interesting alternative to various popular coarse-grained 
simulation approaches with self-con\-sistently calculated effective pair interactions 
\cite{EM01,Likos01,Guenza02,MMuller05,MMuller06}.
Our study has not been limited to monodisperse polymers but we have also investigated
(essentially) Flory-di\-stri\-buted EP \cite{WBCHCKM07,BJSOBW07,WJO10,WJC10}. 
Since in these systems chains break and recombine constantly (Fig.~\ref{fig_bfm_EP}) 
equilibration and sampling become much faster than for their monodisperse counterparts. 
This is computationally of interest since the numerical demonstration of the various theoretical predictions  
requires the sampling of large chain lengths to avoid additional chain end effects.
%
%
%

The deviations from the screening assumption are indeed scale-free as made manifest 
by the numerical observation of the analytic decay of various correlation functions.
%
As announced by Eq.~(\ref{eq_intro_PsD}), it has been shown in Sec.~\ref{conf} that 
the bond-bond correlation function $\Pone(s) \sim \partial_s^2 \Rs^2$ decays as 
$\Pone(s) \sim +1/s^{d/2}$ for $s \ll N$ and $d=3$. That these correlations are 
long-ranged is made explicit by the power-law decay of the bond-bond correlation 
function $\Pone(r) \sim +1/r^d$ as a function of the monomer distance $r$ 
(Sec.~\ref{conf_Pr}).
As a second key result of this study, the displacement correlation function 
$\CN(t) \approx \partial_t^2 \MSDcmN(t)$ is shown to reveal for short times $t \ll \TN$ 
a negative algebraic decay according to $N \CN(t) \sim - 1/t^{(d+2)/4}$ as predicted by scaling 
arguments and perturbation theory (Sec.~\ref{dyna}). As stated by Eq.~(\ref{eq_intro_keydyna}),
this analytic decay holds more generally for the displacement correlation function $\Cs(t)$ 
of subchains of length $s \le N$ and times $t \ll \Ts \approx s^2/W$, Fig.~\ref{fig_VCF_jeansub},
which shows that the correlations of different subchains add up independently to the correlations
of the total chain (``locality").

Note that earlier computational studies have focused on (second and higher) moments of the generalized 
displacement field under consideration such as the mean-squared size  $\Rs^2$ of subchains of arc-length $s$ 
\cite{Auhl03} or the CM MSD $\MSDcmN(t)$ \cite{Paul91b} and have thus only probed indirectly the respective 
colored forces, Eq.~(\ref{eq_app_displ5}), corresponding to the deviations from the general screening 
assumption.\footnote{In line
with Appendix~\ref{app_ht2Ct} we view here the bond vectors $\lvec_i$ of a chain conformation as displacement 
vectors with the monomer index $i$ playing the role of time.} 
Computing numerically correlation functions such as $\Pone(s)$ or $\CN(t)$,
rather than twice their integral, allows to probe directly the colored forces, Eq.~(\ref{eq_app_displ5}),
without having to subtract first the white noise and the local physics 
which contribute both to $\Rs^2$ and $\MSDcmN(t)$.
As a consequence, this allows to demonstrate that the deviations from the screening assumption are 
not due to (non-universal) physics at the lower cutoff but, in fact, are present for all 
arc-length $s$ (Fig.~\ref{fig_conf_Ps})
and all times $t$ (Fig.~\ref{fig_VCF_jean}). 

\subsection{Related questions and outlook}
\label{conc_outlook}

\begin{figure}[tb]
\centerline{\resizebox{0.9\columnwidth}{!}{\includegraphics*{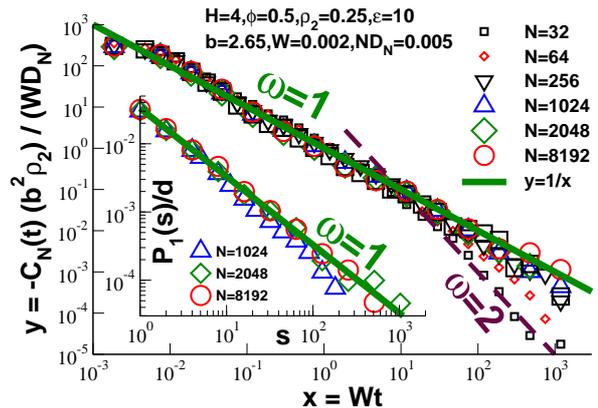}}}
\caption{Polymer melts confined to ultrathin films of width $H=4$ for
volume fraction $\phi=0.5$, projected 2D density $\rhotwo=H \rho =0.25$,
overlap penalty $\overlap=10$ and several chain lengths $N$ as indicated.
Inset: Bond-bond correlation function $\Pone(s)$ compared to 
the power-law exponent $\omega=d/2=1$ predicted by Eq.~(\ref{eq_intro_PsD}).
Main panel: VCF $\CN(t)$ using a similar representation as in Fig.~\ref{fig_VCF_jean}(a).
The exponent $\omega=(d+2)/2=2$ expected for large times, Eq.~(\ref{eq_dyna_VCFcolloid}), 
is represented by the dashed line, the short-time exponent $\omega=(d+2)/4=1$
predicted by Eq.~(\ref{eq_dyna_key}) by the bold line. 
A perfect data collapse is observed for $t \ll \TN$, i.e. $\CN(t) \sim 1/N$. 
\label{fig_conc_slit}
}
\end{figure}

\subsubsection{Confined polymer melts}
\label{conc_outlook_slit}
In the presented numerical work we have focused on 3D polymer bulks ($d=3$). 
It has been shown that even the deviations with respect to the general screening assumption are 
well-described by means of one-loop perturbation calculations (Sec.~\ref{theo}, Sec.~\ref{dyna_perturb}) 
and this for arbitrarily long chains and subchains. As already noted in Sec.~\ref{intro_context},
the success of the mean-field approach is expected due to the scaling of the (sub)chain self-density
$\rhostars/\rho \sim s^{1-d\nu}$,
i.e. the number of subchains a reference subchain interacts with increases as
$\rho/\rhostars \sim \sqrt{s} \gg 1$ in $d=3$.
Due to the increasing experimental interest on mechanical and rheological properties of nano\-scale 
systems in general \cite{NanoscienceBook} and on polymer melts confined to surfaces, 
thin slits ($d=2$) or capillaries ($d=1$) 
in particular \cite{Russel96,RLJones1999,mai99,Granick03,Russel04,McKenna2005,Deutsch05,Rubinstein07,Obukhov07}
one is naturally led to question theoretically such mean-field calculations for systems of
reduced (effective) dimension \cite{Brochard79,Duplantier89,ANS03,LFM11,EisenrieglerBook,Saleur03,Saleur07}. 
Especially, the perturbations to chain dynamics due to geometric constraints remain a challenge 
of significant technological relevance with opportunities ranging from tribology to biology
\cite{Russel96,RLJones1999,Granick03,McKenna2005}.
%

Note that for ideal chain systems with $d \le 1/\nu = 2$ the self-density $\rhostars$ ultimately 
exceeds the density $\rho$ at a characteristic chain length $\gstarup$ which depends on the
specific problem considered. 
For (sub)chains larger than this length $\gstarup$ one expects the chains to adopt compact and 
segregated conformations, i.e. Flory's exponent $\nu$ is given by the spatial dimension, $\nu = 1/d$ 
\cite{MandelbrotBook}.
Since these compact structures arise ultimately due to the scale-free incompressibility constraint,
their surfaces are expected not to be ruled by a finite surface tension (which would imply
a length scale) \cite{ANS03,MSZ11} but to be also scale-free and described by a fractal surface exponent 
$\ds \le d$ \cite{MandelbrotBook}. In most cases of interest the surface exponent becomes 
$\ds = d - \thetatwo$ \cite{MSZ11}, i.e. it is set by the well-known contact exponent $\thetatwo$ 
describing the scaling of the distribution $G(r,s) \sim r^{\thetatwo}$ for short distances $r \ll \Rs$ 
between two monomers $n$ and $m=n+s$ on a chain. 
(Obviously, $\thetatwo=0$ for Gaussian chains, Eq.~(\ref{eq_intro_Grs_gauss}).)
According to the generalized Porod law the intramolecular structure factor $F(q)$
of such compact objects scales as 
\begin{equation}
F(q)/N \approx 1/(\RN q)^{2d-\ds}
\label{eq_conc_Porod}
\end{equation}
in the intermediate range of the wavevector $q$ \cite{BenoitBook,bray88,MSZ11}. If $\ds < 2d-2$,
i.e. $\thetatwo > 2-d$, the standard Kratky representation reveals thus a much more pronounced 
non-monotonous behavior as for the 3D melts discussed in Sec.~\ref{conf_Fq}.\footnote{Incidentally, 
melts of non-concatenated 
\label{foot_ring}
rings in $d=3$ have been argued to become ``marginally compact" with $\nu=1/d=1/3$ and $\ds \to d=3$, 
i.e. the form factor is predicted to scale as $F(q) q^2 \sim 1/q$ \cite{MSZ11,KK11a} in the large-$N$ limit. 
That the rings should adopt compact configurations is expected due to the mutual repulsion caused by 
the topological constraints 
\cite{MWC96,MWC00,MWB00,Gagliardi05,Cates86,Obukhov94,winkler06,Vettorel09,suzuki09,KK11a,winkler11,MSZ11}.
Since there is no obvious reason for a finite surface tension, the surface must become fractal
which may be determined both in experiment as in a computer simulation by the generalized Porod law 
of the form factor. A {\em marginally} compact structure allows to keep all monomers evenly
exposed to the topological constraints imposed by other rings, i.e. all subsegments of the rings
are thus ruled in a self-similar manner by the same statistics. Such behavior is known for various 
biological systems, such as the lungs of mammals, attempting to maximize the surface at constant 
overall embedding volume \cite{WBE99}.}

Strictly 1D and 2D self-avoiding polymer melts ($\overlap = \infty$) are known to become compact and 
segregated with $\gstarup \approx 1$ and are thus {\em not} Gaussian beyond the monomer scale. 
While the 1D case is trivial, it should be noted that for strictly 2D melts compactness means 
$\nu=1/d=1/2$ \cite{DegennesBook}. This does of course not imply ideal chain behavior since other 
critical exponents characterizing the chain conformations differ from those of non-interacting Gaussian 
chains as shown in the pioneering work by Duplanier \cite{Duplantier89}. For instance, 
since $\thetatwo = 3/4$ \cite{Duplantier89} 
this implies that the compact 2D chains have a fractal 
perimeter of dimension $\ds = d - \thetatwo = 5/4$ \cite{ANS03}.
This theoretical prediction has been verified numerically \cite{MKA09,MWK10,MSZ11} 
from the scaling of the chain perimeter (determined by ``box counting" \cite{MandelbrotBook}) and 
of the form factor shown to be consistent with Eq.~(\ref{eq_conc_Porod}).

Self-avoiding melts in strictly one and two dimensions correspond to rather specific universality 
classes \cite{ANS03,Saleur03,Saleur07,LFM11}. Since systems of strictly one mono\-mer 
layer thickness at high volume fractions remain experimentally a challenge \cite{mai99,Deutsch05}, 
it is of 
interest to relax the non-cross\-ing constraint by either allowing some finite monomer overlap penalty 
$\overlap$ (Sec.~\ref{bfm_soft}) or by considering chains confined to capillaries or slits between 
parallel walls of finite distance $H$ as in Fig.~\ref{fig_bfm_algo}(c).
Note that $\gstarup \approx (a H^2 \rho)^2 \sim H^4$ for melts confined to a 1D capillary of finite width $H$
as already shown by Brochard and de Gennes in 1979 \cite{Brochard79} and $\gstarup$ is known to increase even 
exponentially with $H$ for essentially 2D ultrathin films \cite{ANS03}. 
Since $\gstarup$ becomes thus rapidly large, perturbation results, such as the scaling of the
bond-bond correlation function $\Pone(s)$ stated in Eq.~(\ref{eq_intro_PsD}), must become relevant for 
finite (sub)chain lengths. 
A lower limit of validity $\gstardo$ of the $d$-dimensional perturbation calculation is set 
for systems with finite overlap penalty by $\gstardo \approx \gT$
and (sub)chains confined to capillaries or slits of finite width should be larger than 
$\gstardo \approx H^2$, otherwise the chains do not ``feel" the wall constraint and behave 
as in the 3D bulk.

Let us focus now on effectively 2D films of width $H$
\cite{Cavallo03,Cavallo05,CMWJB05,thinfilmdyn} as shown in Fig.~\ref{fig_conc_slit}. 
Note that all properties considered here are the 2D projection of the 3D observables.
The relevant density is, e.g., the projected number density $\rhotwo = H \rho = 0.25$ and
the indicated bond length $b = 2.65$ is the projection of the effective bond length
of the 3D melt. Since the width $H=4$ used here is much smaller than the typical 
subchain size $\Rs$ considered, i.e. $s \gg \gstardo \approx 1$, these systems can be regarded as 
effectively $d=2^{+}$ dimensional for all $s$ and $N$. (We write $d=2^+$ to stress that
monomer overlap and chain crossings are allowed.) Since also $\gstarup \gg s$ one expects the 
perturbation result for the bond-bond correlation function $\Pone(s) \sim 1/s^{\omega}$ with 
$\omega = d/2 = 1$ to yield a reasonable fit to the measured data for all $s \ll N$. 
This is indeed borne out nicely for the data presented in the inset. 
According to Eq.~(\ref{eq_intro_PonesRsrelated}) this implies that $\Rs^2/s$ and $\RN^2/N$ must 
increase logarithmically with, respectively, $s$ and $N$. 
This can be checked directly as seen in Ref.~\cite{CMWJB05}. The same paper also investigates the 
form factor $F(q)$ for different $H$. It is shown that $q^2 F(q)$ becomes systematically more 
non-monotonous with decreasing width $H$, i.e. the more $\rhostarN/\rho$ increases.

We turn now to the dynamics of BFM melts confined to thin slits of width $H=4$ sampled using local 
topology non-conserving L26-moves as in Sec.~\ref{dyna_jean} \cite{thinfilmdyn}. It can be demonstrated 
that to leading order the dynamics remains consistent with the Rouse model, 
especially Eq.~(\ref{eq_intro_DNTN_Rouse}).\footnote{Since in thin films the ratio $\RN^2/N$ diverges 
logarithmically with $N$ \cite{ANS03,CMWJB05,papEPslit} and since $\TN \approx \RN^2/\DN$ 
still holds, there are some (rather weak) logarithmic corrections to Eq.~(\ref{eq_intro_DNTN_Rouse}) 
as discussed in \cite{thinfilmdyn}.}  
As before, we obtain from the monomer MSD $\MSDmon(t)$ a local mobility, $W \approx 0.002$.
That the self-diffusion coefficient scales as $\DN \approx 0.005/N$ can be seen by tracing 
$N \MSDcmN(t)/2d t$ in log-linear coordinates. 
(Note that $W$ and $\DN$ are consistent with Eq.~(\ref{eq_dyna_W2DN}).)
Systematic deviations are again revealed by the short-time behavior of the MSDs $\MSDcmN(t)$ and $\MSDcms(t)$ 
of chains and subchains (not shown). These deviations are, however, much more pronounced than in 3D and are 
visible over time scales up to the total chain relation time $\TN$. 
A more precise characterization of the colored forces acting on the chains is again achieved by means of the 
VCF $\CN(t)$ which is represented in the main panel of Fig.~\ref{fig_conc_slit}. The representation chosen is 
similar to Fig.~\ref{fig_VCF_jean}(a) for the bulk case; the factor $b^3 \rho$ being replaced by $b^2 \rhotwo$.
The faster decay observed for the shorter chains at $t \gg \TN$ is described by the exponent $\omega=(2+d)/4=2$ 
(dashed line) expected for 2D overdamped colloids, Eq.~(\ref{eq_dyna_VCFcolloid}).
As may be seen from the collapse for short times ($t \ll \TN$), the VCF scales again as 
$\CN(t) \sim 1/N$, i.e. the colored forces are again local.\footnote{Note that a different scaling, 
$\CN(t) \sim 1/\sqrt{N}$, has been reported in a study of strictly 2D 
polymer melts sampled by MD simulation using a standard Langevin thermostat \cite{WMJ10}.
The correlations of the CM motion are argued to become {\em non-local} in this special limit
due to the ``constant surface constraint" of the compact and segregated subchains.}
In agreement with Eq.~(\ref{eq_dyna_keyomega}) or Eq.~(\ref{eq_dyna_Ncounting})
the VCF is observed to decay algebraically with an exponent $\omega= (d+2)/4=1$ (bold line).
Since $\CN(t) \approx \partial^2_t \MSDcmN(t)/2$ it follows that $\MSDcmN(t)/t$ is not constant
but must increase logarithmically for all times $t \ll \TN$. In analogy to Eq.~(\ref{eq_dyna_Ct2MSD}) 
for the bulk case this thus explains the observed deviations from the Rouse model.

We note finally that qualitatively similar results consistent with Eq.~(\ref{eq_intro_PsD}) and 
Eq.~(\ref{eq_dyna_keyomega}) are also found in preliminary BFM simulations of polymer melts 
confined to thin capillaries with a square-section of width $H = 8 \ll \RN$ and a broad range
of overlap penalties $\overlap$.

\subsubsection{Interchain correlations and anti-Casimir forces}
\label{conc_outlook_casimir}
\begin{figure}[t]
\centerline{\resizebox{.95\columnwidth}{!}{\includegraphics*{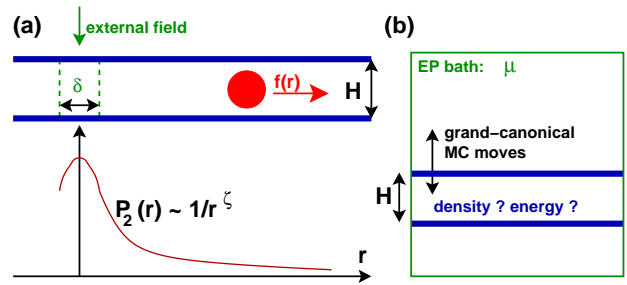}}}
\caption{The presented studies focused on {\em intrachain} properties which
are numerically more readily accessible than the long-range {\em interchain} correlations
associated to the ``anti-Casimir" forces \cite{ANS05a,ANS05b}.
The two panels indicate two possibilities to probe such correlations for EP melts
confined to thin slits:
{\bf (a)}
A weak external field conjugated to the bond direction is applied to the bonds in a thin slap 
of width $\delta$: The second Legendre polynomial $\Ptwo(r)$ characterizing the orientation of the bonds 
with respect to the direction perpendicular to the slap and the force $f(r)$ experienced by a colloid 
placed at a distance $r$ should decay analytically with not too large exponents.
{\bf (b)}
EP in grand-canonical contact with a large bath:
The difference of, e.g., the densities of the confined system and the bath 
decays analytically with $H$.
\label{fig_conc_casimir}
}
\end{figure}

The presented work has focused on {\em intrachain} properties such as the bond-bond correlation 
function $\Pone(s)$.
%
The main reason for this is that the power-law exponents associated to the intrachain static and
dynamical deviations due to the incompressibility constraint are not too large being thus
numerically still accessible with reasonable computational effort.
In fact, similar correlations have been predicted theoretically also for {\em interchain} 
properties \cite{ANS05a,ANS05b} which for macroscopic thermodynamic and mechanical 
properties may even be more relevant. Note that for these interchain correlations it is not
only the incompressibility constraint which matters but in addition the constraint that closed
loops are disallowed and must be eliminated from an extended grand-canonical ensemble containing 
both linear chains and rings, as shown by Obukhov and Semenov \cite{ANS05a,ANS05b}. 
The ``throw\-ing-away" of configurations is argued to generate additional entropic forces
{\em repelling}, e.g., two large colloids immersed in a linear-chain polymer melt.
Unfortunately, these ``anti-Casimir forces" correspond to such strong exponents 
as a function of distance $r$ or wavevector $q$ (similar to the standard van der Waals forces
\cite{Israela}) that, at present, it has turned out to be elusive to verify them numerically 
although a brave attempt has been made by means of off-lattice MC simulations \cite{Milchev02}. 
Following this recent work, we sketch in Fig.~\ref{fig_conc_casimir} two possibilities which 
might allow to probe correlations due to the no-loop constraint by means of EP melts confined 
to thin slits.
The use of EP instead of monodisperse chains should speed up the sampling of independent 
configurations while confining the monomers reduces the exponents of the 
various analytic scaling relations predicted. For instance, the second Legendre polynomial 
$\Ptwo(r)$, measuring the orientation of the bonds at a distance $r$ from a small region where
an external field weakly aligns the bond vectors, should decay as
$\Ptwo(x) \sim r^{-\zeta}$ with $\delta =2$ for $d=2^+$
as illustrated in panel (a) of Fig.~\ref{fig_conc_casimir}.
If successful these simulations may stimulate real micro-mechani\-cal
experiments using optical tweezers aligning bond vectors and repell\-ing thus nearby colloids.

\subsubsection{Dynamical properties beyond MC stochastics}
\label{conc_outlook_offlattice}
%
%
We have checked by means of MD simulations of a bead-spring model \cite{papFarago,ANS11a} that 
$\CN(t) \approx -N^{-1} t^{-5/4}$ for $t \ll \TN$ holds as long as a large friction 
constant $\gamma$ is used for the Langevin thermostat applied \cite{FrenkelSmitBook}. 
Concerning the power-law amplitude it should be noted, however, 
that the coefficient $c$ determined from the data using Eq.~(\ref{eq_dyna_VCF_Wt}) is 
much larger for these MD simulations ($c/c_3 \approx 20$) compared to our BFM result ($c/c_3 \approx 2$). 
As we have pointed out in Sec.~\ref{dyna_perturb}, small differences of order one 
to the predicted value $c_3 = 0.42$ are expected due to the approximations made
in the one-loop perturbation theory, especially since the forces associated to the shear 
stress may not be properly accounted for \cite{papFarago}. However, the much larger ratio $c/c_3$ 
found for our MD simulations points to additional physics related to the 
screening of the mono\-mer momentum in these strongly interacting viscoelastic systems 
\cite{ANS11a,ANS11b}.\footnote{The 
beads bounce into each other on time scales much shorter than $1/\gamma$. Kicks due to conservative or 
random forces can thus propagate through the system before the thermostat has time to do its job.}
Since our BFM and MD simulations correspond to different model systems, it is currently not 
possible to clarify this issue. 
In the future we plan to address this problem using a generic
(off-lattice) bead-spring Hamiltonian with soft beads and large springs such as, e.g.,
the model used by Spenley \cite{Spenley00} since we require a model which is 
(i) essentially incompressible ($\gT \ll 1$),
(ii) without having too much useless local structure (as for the Lennard-Jones 
beads used in \cite{papFarago,ANS11a,ANS11b}),
(iii) does not conserve topology and
(iv) can be used for different dynamical methods such that we can compare the
melt dynamics under the {\em same} static conditions. 
For such a model one should then compare the dynamics obtained by means of
\begin{itemize}
\item[$\bullet$]
MD simulation with a strong Langevin thermostat \cite{FrenkelSmitBook};
\item[$\bullet$]
local MC moves of the monomers \cite{LandauBinderBook};
\item[$\bullet$]
BD simulation in the sense of a ``position Langevin equation"
where the momentum variables are dropped 
from the equations of motion \cite{AllenTildesleyBook,Ermak75}. 
\end{itemize}
The BD method has the advantage that one needs not to inject strong random forces 
as for MD to get rid of the momentum to get closer to the overdamped motion assumption 
implicit to theory \cite{papFarago}. We expect to obtain for the MC and BD 
simulations similar ratios $c/c_3 \approx 2$ as for our BFM study.


Interestingly, much stronger dynamical correlations are revealed \cite{ANS11a} in MD simulations 
using a weak Langevin thermostat ($\gamma \to 0$) or a momentum conserving ``Dissipative Particle
Dynamics" (DPD) thermostat \cite{Warren97,FrenkelSmitBook}. 
As expected from the hydrodynamic screening assumption \cite{DoiEdwardsBook}, 
the chains are described to leading order by the Rouse model, 
i.e. Eqs.~(\ref{eq_dyna_MSDmonshort}), (\ref{eq_dyna_W2DN}) or (\ref{eq_dyna_Fqt_Rouse}) hold.
However, it is shown that some long-range correlations arise due to an intricate coupling of 
hydrodynamics (momentum conservation) and the (transverse) viscoelastic response of the bath following 
the displacement of a reference chain.\footnote{Note
the analogy to the static excluded volume screening discussed in Sec.~\ref{theo}. Although the chains are 
essentially described in $d=3$ by Gaussian chain statistics, not all long-range correlations vanish.} 
As a consequence, even the VCF $\Cs(t)$ of subchains for $t \ll \Ts$ depends now explicitly 
on the total chain length $N$, i.e. the correlations are {\em non-local} with $\CN(t) \sim 1/\sqrt{N}$
as suggested by the Zimm model. Since the scaling is still described by Eq.~(\ref{eq_dyna_key})
this implies an exponent $\omega=3/2$ for short times. Hence,
\begin{equation}
\CN(t) \approx - \left(\frac{\RN}{\TN}\right)^2 \frac{\rhostarN}{\rho} \ (\TN/t)^{\omega} \sim
\frac{1}{\sqrt{N t^3}}
\label{eq_conc_keySascha}
\end{equation}
which is the central formula real experimental data should be compared with.
Although the VCF decays now more rapidly in time, it is stronger than the VCF for perfectly
overdamped melts, since the $N$-dependence in Eq.~(\ref{eq_conc_keySascha}) is weaker. 
Note that at $t \approx \TN$ both mechanisms become of the same magnitude: $\CN(\TN) \sim 1/N^{7/2}$.
Using the same bead-spring model mentioned above it would now be of high interest to sample 
the dynamical response using
\begin{itemize}
\item[$\bullet$]
a DPD thermostat following the work by Spenley \cite{Spenley00}. 
\end{itemize}
Obviously, these issues are clearly outside the realm of 
standard MC based stochastics which shows their limitations. 


\begin{acknowledgements}
We thank A.N.~Semenov (Strasbourg), I. Erukhimovich (Moscow),
M.~M\"{u}ller (G\"{o}tt\-ing\-en) and T.~Kreer (Dresden) for helpful discussions.
A.C. acknowledges the MIUR (Italian Ministry of Research) for financial support within the
program ``Incentivazione alla mobilit\`a di studiosi stranieri e italiani residenti all'estero",
H.X. the CNRS for supporting her sabbathical stay in Strasbourg, 
J.E.Z. a grant by the ANR Blanc (FqSimPIB), P.P. a grant by the IRTG ``Soft Matter Science" and 
N.S. financial support by the R\'egion d'Alsace.
\end{acknowledgements}

\appendix
\appendix


\section{Displacement correlations}
\label{app_ht2Ct}

Let $x(n)$ denote a (real valued) stochastic variable in some vector space with $n$ 
being a discrete or continuous index characterizing this variable. Examples are the position $\rvec_i$ 
of the monomers of a given chain as a function of the monomer index $i$ 
(Fig.~\ref{fig_intro_lattice})\footnote{This monomer index $i$ may be either a discrete (as in our study) 
or a continuous variable as, e.g., in Refs.~\cite{MMuller05,MMuller06}.} 
or the center-of-mass (CM) position $\rN(t)$ of a chain of length $N$ as a function of a time $t$
measured by the discrete number of Monte Carlo Steps (MCS).
We assume {\em translational invariance} of ensemble averaged properties with respect to $n$, 
i.e. properties such as the ``mean-square displacement" (MSD)
\begin{equation}
h(n) \equiv h(n_1=n_0+n,n_0) \equiv \la (x(n_1) - x(n_0))^2 \ra
\label{eq_app_MSD}
\end{equation}
only depend on the difference $n=n_1-n_0 \ge 0$ and not on the absolute indices $n_0$ or $n_1$. 
Taking advantage of the translational invariance, the ensemble average $\la \ldots \ra$ is often 
sampled over all available pairs of indices $(n_1=n_0+n,n_0)$ for a given $n$.
Examples for MSDs obtained as ``gliding averages" are the typical segment size $\Rs^2$ presented 
in Sec.~\ref{conf_Rs} or the MSD $h(t)$ of BFM beads ($N=1$) discussed in Sec.~\ref{dyna_beads}.
In this review we consider essentially Fickian stochastic processes where the associated 
generalized MSDs are to leading order proportional to time $n$, Eq.~(\ref{eq_intro_Rs_gauss}) or 
Eq.~(\ref{eq_intro_hN_Rouse}). Since the deviations with respect to the Fickian behavior are small, 
it is important to sample directly correlation functions corresponding to higher derivatives of $h(n)$ 
with respect to time.
%
%

Since we have to deal with indices $n$ which may be either discrete or continuous, let us introduce the 
(forward) difference operator $\Delta_n$ \cite{abramowitz} defined by $\Delta_n f(n) = f(n+ \delta n) - f(n)$ 
with $f(n)$ being a function of our vector space and $\delta n$ an arbitrary (typically) small shift of $n$. 
Let us define the ``velocity vector" 
\begin{equation}
v(n) \equiv \frac{\Delta_n x(n)}{\delta n} \equiv \frac{u(n)}{\delta n} \equiv \frac{x(n+\delta n) - x(n)}{\delta n}
\label{eq_app_vel}
\end{equation}
associated to $x(n)$ and the ``velocity correlation function" (VCF)
\begin{equation}
C(n) \equiv C(n_1=n_0+n,n_0) \equiv \la v(n_1=n_0+n) \cdot v(n_0) \ra
\label{eq_app_DCF}
\end{equation}
which measures the correlations of two velocities $v(n_0)$ and $v(n_1)$.
$C(n)$ automatically vanishes if both vectors are uncorrelated.
Note that $C(n)$ is a four-point correlation function with respect to $n$
depending in general on the shift $\delta n$ of the index. If $\delta n > n$ 
both displacements become trivially correlated and we have
$\lim_{n\to 0} C(n) = h(\delta n)/ \delta n^2 > 0$.
To see how $h(t)$ and $C(n)$ are related for $n \gg \delta n$ 
let us apply twice the difference operator to Eq.~(\ref{eq_app_MSD}).
This yields
\begin{equation}
\Delta_{n_1} \Delta_{n_0} \la (x(n_1) - x(n_0))^2 \ra =
-2 \la v(n_1) \cdot v(n_0) \ra \delta n^2
\label{eq_app_displ1}
\end{equation}
where we have used that the difference operator and the averaging procedure commute.
Since the MSD $h(t)$ can be assumed to be mathematically well behaved
and using the translational invariance it follows on the other hand side that
\begin{equation}
\Delta_{n_1} \Delta_{n_0} h(n_1-n_0) \approx - \partial_n^2 h(n) \ \delta n^2
\mbox{ for } n \gg \delta n.
\label{eq_app_displ2}
\end{equation}
Altogether this demonstrates that
\begin{equation}
C(n) \approx \frac{1}{2} \partial_n^2 h(n) \mbox{ for } n \gg \delta n
\label{eq_app_ht2Ct}
\end{equation}
in agreement with Eq.~(\ref{eq_intro_PonesRsrelated}) for the bond-bond correlation function $\Pone(s)$
and with Eq.~(\ref{eq_intro_CNdef}) for the chain velocity correlation function. Note that since $h(n)$ 
does not depend on $\delta n$, $C(n)$ does not depend on $\delta n$ either in the limit $n \gg \delta n$.
%


The stochastic variables of interest in this review are described by a
position Langevin equation of form \cite{vanKampenBook}
\begin{equation}
\zeta \frac{d x(n)}{dn} \approx \zeta v(n) = 
\ftot(n) = \fran(n) + \fcol(n)
\label{eq_app_displ3}
\end{equation}
with $\zeta$ being the ``friction constant" and $\ftot(n)$ the total ``force" acting on $x(n)$
which may be decomposed in a random white force contribution $\fran(n)$ 
and a colored force $\fcol(n)$ stemming from the remaining (non-white) interactions of the  
degree of freedom under consideration with the bath and its constraints. Perfectly Fickian behavior
would be obtained if only the white force contribution were present ($\fcol(n)\equiv 0$).\footnote{While 
$\fran(n)$ and $\fcol(n)$ are both of zero mean, one may add in addition an external force $\fext$ 
which allows to determine the friction constant from the drift velocity:  $\zeta \la v(n) \ra = \fext$.} 
For $x(n)$ being the position of a monomer or the CM of a chain as a function of time $t$, 
Eq.~(\ref{eq_app_displ3}) corresponds to an overdamped motion as discussed in Sec.~\ref{dyna_beads} 
for MC beads or in Sec.~\ref{dyna_jean} for Rouse-like chains in the melt, Eq.~(\ref{eq_dyna_forceonchain}).
From Eq.~(\ref{eq_app_displ3}) it follows for the displacement correlation function that
\begin{equation}
C(n)  = \la v(n) \cdot v(0) \ra
 = \zeta^{-2} \la \ftot(n) \cdot \ftot(0) \ra.
\label{eq_app_displ4}
\end{equation} 
The total force correlation function $\la \ftot(n) \cdot \ftot(0) \ra$  at $n > 0$ decomposes now into
four contributions:
\begin{itemize}
\item[(a)] $\la \fran(n) \cdot \fran(0)\ra = 0$ since $\fran(n)$ is a white force,
\item[(b)] $\la \fran(n) \cdot \fcol(0)\ra = 0$ since future white forces cannot be anticipated,
\item[(c)] $\la \fcol(n) \cdot \fcol(0)\ra$ and
\item[(d)] $\la \fcol(n) \cdot \fran(0)\ra = -2 \la \fcol(n) \cdot \fcol(0)\ra$
due to the odd $n$-symmetry of the velocity $v(n) = -v(-n)$ \cite{ANS98b}.\footnote{Since
$v(n) = - v(-n)$ this implies $\ftot(n) = -\ftot(-n)$ for the total force which in turn yields 
$\la \ftot(n) \cdot \fcol(0) \ra = -\la \ftot(-n) \cdot \fcol(0) \ra$.
The identity (d) is then obtained by substituting
$\la \ftot(n) \cdot \fcol(0) \ra = \la \fcol(n) \cdot \fcol(0) \ra$ (due to (b)) and
$\la \ftot(-n) \cdot \fcol(0) \ra = \la \fcol(n) \cdot \ftot(0) \ra
= \la \fcol(n) \cdot \fran(0) \ra + \la \fcol(n) \cdot \fcol(0) \ra.$}
\end{itemize}
Summing up over all contributions and using Eq.~(\ref{eq_app_displ4}) it follows 
\begin{equation}
C(n) = - \zeta^{-2} \la \fcol(n) \cdot \fcol(0) \ra \mbox{ for } n > 0,
\label{eq_app_displ5}
\end{equation}
i.e. the displacement correlation function probes directly the correlations of the colored force $\fcol(n)$. 
This general result is used in Sec.~\ref{dyna_perturb} for the motion of the chain CM with time $t$
but does also apply for the monomer position $\rvec_i$ as a function of the monomer index $i$ 
where the bond-bond correlation function $\Pone(s)$ measures the colored forces acting on the chain.  
It should be stressed, however, that in this case the forces noted in Eq.~(\ref{eq_app_displ3}) 
do not correspond directly to the standard forces acting on the monomers in real time.


\section{Static properties}
\label{app_stat}

\subsection{Moments and generating function}
\label{app_stat_moments}

Higher moments of the segmental size distribution $G(r,s)$ can be systematically 
obtained from its Fourier transform $G(q,s) = \Fcal[G(r,s)]$
which is in this context sometimes called the ``generating function" \cite{vanKampenBook}.
For ideal Gaussian chains the generating function is given by $G_0(q,s)=\exp(-s (a q)^2)$
where we have used $a^2 = b^2/2d$ to simplify the notation. 
Moments of the size distribution are given by proper derivatives of $G(q,s)$ taken at $q=0$.
For example, 
\begin{equation}
\la \rvec^{2p}\ra= \left. (-1)^p\Delta^p G(q,s)\right|_{q=0}
\label{eq_app_r2p}
\end{equation}
with $\Delta$ being the Laplace operator with respect to the wavevector $\qvec$.
A moment of order $2p$ is, hence, linked to only {\em one} coefficient $A_{2p}$ in 
the systematic expansion, $G(q,s) = \sum_{p=0} A_{2p} q^{2p}$, of $G(q,s)$ around $q=0$.
For our example this implies 
\begin{equation}
\la \rvec^{2p} \ra = (-1)^p (2p+1)! \ A_{2p}
\label{eq_app_r2pA2p}
\end{equation}
in general and more specifically for a Gaussian distribution
${\la \rvec^{2p}\ra}_0=\frac{(2p+1)! }{p!}s^p a^{2p}$. 
The non-Gaussian parameters read, hence,
\begin{equation}
\alphap(s) \equiv 
1 - \frac{6^p p!}{(2p+1)!} \frac{\la \rvec^{2p} \ra}{\la \rvec^2 \ra^p}
= 1 - p! \ \frac{A_{2p}}{A_2^p},
\label{eq_app_alphapAp}
\end{equation}
which implies (by construction) $\alphap = 0$ for a Gaussian distribution.
As various moments of the same global order $2p$ are linked to the same $A_{2p}$ they differ 
by a multiplicative constant independent of the details of the (isotropic) distribution $G(q,s)$. 
For example, $\la \rvec^{2}\ra= 6 |A_2|$, 
$\la \rvec^{4}\ra=120 A_4$, 
$\la x^{2}\ra = \la y^{2}\ra = 2 |A_2|$, 
$\la x^{2}y^{2}\ra=8A_4$
with $x$ and $y$ denoting the spatial components of the segment vector $\rvec$.
Using Eq.~(\ref{eq_app_alphapAp}) for $p=2$ it follows that
\begin{equation}
K_{xy}(s) \equiv 1 - \frac{\la x^{2}y^{2}\ra}{\la x^{2}\ra\la y^{2}\ra}
= 1- 2 \frac{A_4}{A_2^2} = \alpha_2(s),
\label{eq_app_KxyKalpha2}
\end{equation}
i.e. the properties $\alpha_2(s)$ and $K_{xy}(s)$ discussed in Figs.~\ref{fig_conf_alphas} and 
\ref{fig_conf_Kxys} must be identical in general provided that 
$G(q,s)$ is isotropic and can be expanded in $q^2$.
%

\subsection{Deviations of the segmental size distribution}
\label{app_stat_Grs}

\begin{figure}[t]
\centerline{
\resizebox{0.9\columnwidth}{!}{\includegraphics{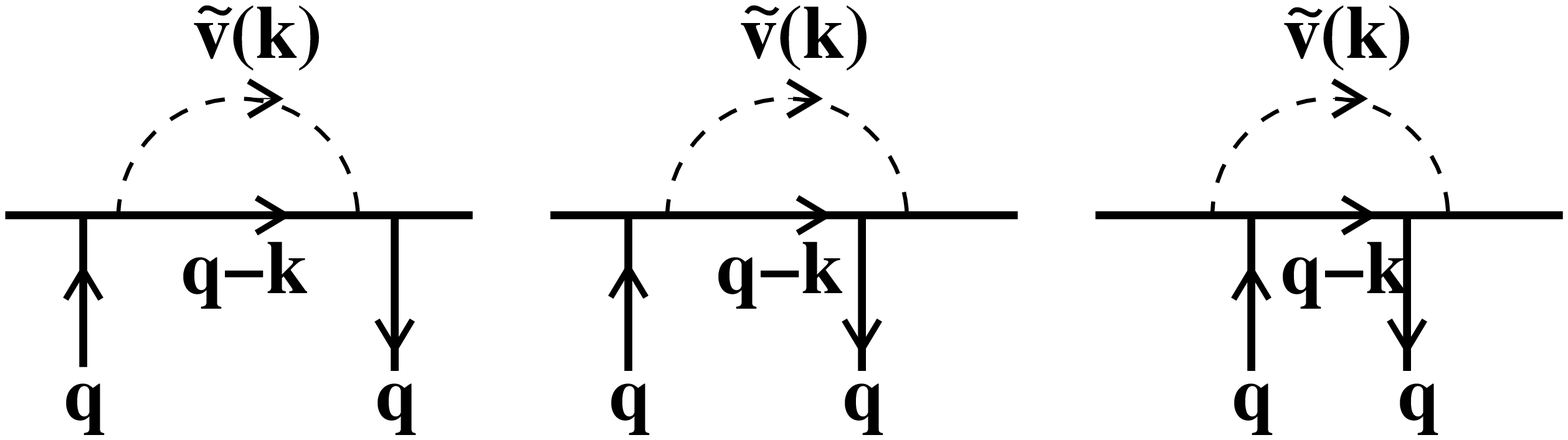}}
}
\caption{
Interaction diagrams used in reciprocal space for the calculation of $\delta G(q,t)$
in the scale free limit.
There exist three nonzero contributions to first-order perturbation,
the first involving two points inside the segment (first two lines of Eq.~(\ref{eq_app_deltaGqt})),
the second one point inside and one outside the segment
(third line of Eq.~(\ref{eq_app_deltaGqt})) and the third one point on either side
of the segment (last line of Eq.~(\ref{eq_app_deltaGqt})).
Momentum $q$ flows from one correlated point to the other.
Integrals are performed over the momentum $k$.
Dotted lines denote the effective interactions $\veff(k)$ given by Eq.~(\ref{eq_theo_vpot_xi}),
bold lines the propagators which carry each a momentum $q$ or $q-k$ as indicated.
\label{fig_app_diagGqt}
}
\end{figure}

We turn now to specific properties of $G(q,s)$ computed for formally {\em infinite} polymer chains 
in the melt. In practice, these results are also relevant for small segments in large chains, 
$N \gg s \gg 1$, and, especially, for segments located far from the chain ends.
These chains are nearly Gaussian and the generating function can be written as 
$G(q,s)=G_0(q,s)+\delta G(q,s)$ where according to Eq.~(\ref{eq_theo_Acalav}) we have
a small perturbation $\delta G(q,s) = - \la U G \ra_0 + \la U \ra_0 \la G \ra_0$ 
due to the effective interaction potential $\vpot(q)$ given by Eq.~(\ref{eq_theo_vpot_xi}). 
In this paragraph we use $\bref^2=6 \aref^2$ for the bond length of the Gaussian reference chain
of the perturbation calculation and $b^2$ for the measured effective bond length, 
Eq.~(\ref{eq_theo_bdef}).
To compute the different integrals it is more convenient to work in Fourier-Laplace space ($q,t$) 
with $t$ being the  Laplace variable conjugate to $s$:
\begin{equation}
\nonumber
\delta \Ghat(q,t) = \int_0^{\infty} ds \ \delta G(q,s) e^{-s t}.
\label{eq_app_Gqt}
\end{equation}
As illustrated in Fig.~\ref{fig_app_diagGqt}, there are three contributions to this perturbation: 
one due to interactions between two monomers inside the segment (left panel), 
one due to interactions between an internal monomer and an external one (middle panel) and
one due to interactions between two external monomers located on opposite sides (right panel). 
In analogy to the derivation of the form factor described in Ref.~\cite{BJSOBW07} this yields:
\begin{eqnarray}
\delta \Ghat(q,t) & = & \nonumber
-\frac{1}{Q^2}\frac{\veff}{4\pi \aref^3}\left(\sqrt{Q} - \sqrt{t}\right)  \\ \nonumber
& + & \frac{1}{Q^2}\frac{\veff}{4\pi q \aref^2\xi^2}
\left(\mbox {Arctan}\left[\frac{q \aref}{\aref / \xi +\sqrt{t}}\right] - 
\frac{q \aref}{\aref / \xi+\sqrt{Q}}\right)\\ \nonumber
&-&\frac{1}{Q}\frac{2\veff}{ 4\pi q \aref^4}
\left(\mbox{Arctan}\left[\frac{q \aref}{\aref / \xi + \sqrt{t}}\right] - 
\frac{q \aref}{\aref / \xi + \sqrt{Q}}\right)\\ 
& - &\frac{\veff \xi^2}{4\pi q \aref^6}
\left(\mbox{Arctan}\left[\frac{q \aref}{\sqrt{t}}\right] - \frac{q \aref}{\sqrt{Q}}\right)
\label{eq_app_deltaGqt}
\end{eqnarray}
where we have set $Q = (\aref q)^2 + t$.
The graph given in the left panel of Fig.~\ref{fig_app_diagGqt} corresponds to the first two lines,
the middle panel to the third line and the right panel to the last one.
Seeking for the moments we expand $\delta \Ghat(q,t)$ around $q=0$. 
Having in mind chain strands counting many monomers ($s\gg 1$), 
we need only to retain the most singular terms for $t \rightarrow 0$.
Defining the two dimensionless constants 
$c = (3 \pi^{3/2} \aref^3 \rho )^{-1} = \sqrt{24/\pi^3}/\bref^3 \rho$ and
$\dref = \veff \xi / 3 \pi \aref^4  = 12 \veff \xi / \pi \bref^4$ 
this expansion can be written as
\begin{eqnarray}
\label{eq_app_Gexpansion}
\delta \Ghat(q,t) = 
& - & \frac{1}{1!} \frac{\Gamma(2)}{t^2}  \ \dref         \ (\aref q)^2
  +   \frac{1}{1!} \frac{\Gamma(3/2)}{t^{3/2}} \  c   \ (\aref q)^2   + \ldots 
\\ \nonumber
& + & \frac{2}{2!} \frac{\Gamma(3)}{t^3} \  \dref         \ (\aref q)^4 
  -   \frac{1}{2!} \frac{16}{5} \frac{\Gamma(5/2)}{t^{5/2}}  \  c   \  (\aref q)^4 + \ldots  
\\ \nonumber
& - & \frac{3}{3!} \frac{\Gamma(4)}{t^4} \  \dref  \  (\aref q)^6 
  +   \frac{1}{3!} \frac{216}{35} \frac{\Gamma(7/2)}{t^{7/2}} \  c  \   (\aref q)^6 + \ldots 
\\ \nonumber
& + & \ldots 
\end{eqnarray}
where we have used Euler's Gamma function $\Gamma(\alpha)$ \cite{abramowitz}.
The first leading term at each order in $q^2$ --- being proportional to the coefficient $\dref$ 
--- ensures the renormalization of the effective bond length $\bref \to b$. The next term scaling 
with the coefficient $c$ corresponds to the leading finite strand size correction.
Performing the inverse Laplace transformation 
$\Gamma(\alpha)/t^{\alpha} \rightarrow s^{\alpha-1}$
and adding the Gaussian reference distribution $G_0(q,s)$
this yields the $A_{2p}$-coefficients for the expansion of $G(q,s)$
around $q=0$:
\begin{eqnarray}
\nonumber
A_0 & = & 1 \\ \nonumber
A_2 & = & - \aref^2 s \left(1 + \dref - \frac{c}{\sqrt{s}}  \right) 
\\ \nonumber
A_4 & = & \frac{1}{2} \aref^4 s^2 \left(1 + 2 \dref - \frac{16}{5} \frac{c}{\sqrt{s}} \right) 
\\ \nonumber 
A_6 & = & -\frac{1}{6} \aref^6 s^3 \left(1 + 3 \dref - \frac{216}{35} \frac{c}{\sqrt{s}} \right) 
\\ 
A_8 & = & \ldots 
\label{eq_app_A2pfirst}
\end{eqnarray}
More generally, one finds
\begin{equation}
A_{2p} = \frac{(-1)^p}{p!} (s \aref^2)^p \left( 1 +
p \dref - \frac{3 (2^p p! p)^2}{2 (2p+1)!} \frac{c}{\sqrt{s}} 
\right) 
\label{eq_app_A2pgeneral}
\end{equation}
From this result and using Eq.~(\ref{eq_app_r2pA2p}) one immediately verifies that
%
\begin{equation}
\la \rvec^{2p} \ra = 
\frac{(2p+1)!}{6^pp!} (b^2 s)^p 
\left( 1 - \frac{3 (2^p p! p)^2}{2 (2p+1)!} \frac{\ce}{\sqrt{s}} 
\ \left( \frac{\bref}{b} \right)^{2p-3}  \right)
\label{eq_app_Rppert}
\end{equation}
where we have defined 
\begin{equation}
b^2 \equiv \bref^2 \left(1 + \frac{12}{\pi} \frac{\veff \xi}{\bref^4} \right)
= \bref^2 \left(1 + \frac{\sqrt{12}}{\pi} \Giref \right)
\label{eq_app_bepert}
\end{equation}
for the effective bond length with $\Giref \equiv \sqrt{\veff \rho} / \bref^3 \rho$.
%
We remind that $\bref$ refers here to the bond length of the Gaussian reference chain
while $b$ is the effective bond length for asymptotically long chains. 
Since our key interest is not as for Edwards to predict $b$ \cite{DoiEdwardsBook,Edwards82}
but to describe the deviations for finite $s$ from the Gaussian limit for $s \to N$ which 
sets the relevant reference length scale. Setting thus $\bref=b$ this leads to 
Eq.~(\ref{eq_theo_Rps_perturb}) with $b$ being a fit parameter.
Using Eq.~(\ref{eq_app_alphapAp}) one justifies similarly 
Eq.~(\ref{eq_theo_alphaps_perturb}) for the Gaussianity parameter $\alpha_p$. 

The moments Eq.~(\ref{eq_app_Rppert}) completely determine the segmental distribution $G(r,s)$ 
which is indicated in Eq.~(\ref{eq_theo_Grspert}). While at least in principle this may be done 
directly by inverse Fourier-Laplace transformation of the correction $\delta \Ghat(q,t)$ 
to the generating function it is helpful to simplify further Eq.~(\ref{eq_app_deltaGqt}).
We observe first that $\delta \Ghat(q,t)$ does diverge for strictly incompressible systems 
($\veff \rightarrow \infty$) and one must keep $\veff$ finite in the effective 
potential whenever necessary to ensure convergence (actually everywhere but in the 
diagram corresponding to the interaction between two external monomers). Since
we are not interested in the wave vectors larger than $1/\xi$ we expand 
$\delta \Ghat(q,t)$ for $\xi\rightarrow 0$ which leads to the much simpler expression
\begin{eqnarray}
\delta \Ghat(q,t) & \approx &
-\frac{\veff \xi q^2}{3 \pi \aref^2 Q^2} + 
\frac{\veff \xi^2}{4\pi \aref^6}\frac{\aref\sqrt{t}(3 (\aref q)^2+t)}{Q^2} 
\nonumber \\
&- & \frac{\veff \xi^2}{4\pi \aref^6}\frac{\mbox{Arctan}[\frac{\aref q}{\sqrt{t}}]}{q} + 
\Ocal(\veff \xi^3)
\label{eq_app_Gdl}
\end{eqnarray}
with $Q= (\aref q)^2 + t$.
The first term diverges as $\sqrt \veff$ for diverging $\veff$. It renormalizes the 
effective bond length in the zero order term which is indicated in the first
line of Eq.~(\ref{eq_theo_Grspert}).
The next two terms scale both as $\veff^0$. Subsequent terms must all vanish for diverging $\veff$ 
and can be discarded. It is then easy to perform an inverse Fourier-Laplace transformation
of the two relevant $\veff^0$-terms. This yields
\begin{equation}
\delta G(x,s) =
G_0(x,s) \frac{c}{\sqrt s}
\frac{3\sqrt{\pi}}{4} \left(-\frac{2}{x}+\frac{3x}{2}
- \frac{x^3}{8}\right)
\label{eq_app_Greal}
\end{equation}
with $x=r/\aref \sqrt{s} = \sqrt{6} n$. This is consistent with the expression given in
Eq.~(\ref{eq_theo_Grspert}). 
%

\subsection{Intramolecular form factor $F(q)$}
\label{app_stat_Fq}
The deviation $\delta F(q) = F(q) - F_0(q)$ of the intramolecular form factor
from the Gaussian reference $F_0(q)$ can be readily obtained from the deviation
$\delta \Ghat(q,t)$ of the subchain size distribution in Fourier-Laplace space.
Since $G(q,s)=\la \exp(-{\rm i} \qvec\cdot \rvec) \ra$ we have for 
asymptotically long chains 
\begin{equation}
\delta F(q) = 2 \int ds \ \delta G(q,s) = 2 \ \delta \Ghat(q,t=0)
\label{eq_app_deltaFq1}
\end{equation}
where we focus on the intermediate scale-free wavevector regime which does not depend 
on the chain length distribution $\pchain$.
Using Eq.~(\ref{eq_app_Gdl}) derived in Appendix~\ref{app_stat_Grs} one obtains
\begin{equation}
\delta F(q) \approx - 2 \frac{\veff \xi^2}{4 \pi a^6} \frac{\pi/2}{q}
= - \frac{9}{4} \frac{1}{b^3 \rho} \frac{1}{b q},
\label{eq_app_deltaFq2}
\end{equation}
where only the third term of Eq.~(\ref{eq_app_Gdl}) contributes.\footnote{The first term in 
Eq.~(\ref{eq_app_Gdl}) must be discarded as before since it only renormalizes the effective 
bond length and would also contribute to the reference form factor $F_0(q)$.} 
Note that $b=\sqrt{6}a$ stands now for {\em both} the measured effective bond length and 
the bond length of the Gaussian reference chain.
It follows from Eq.~(\ref{eq_app_deltaFq2}) that within first-order perturbation theory
\begin{equation}
F(q) = F_0(q) + \delta F(q) 
\approx F_0(q) \left( 1 - \frac{3}{8} \frac{b q}{b^3\rho} \right)
\label{eq_app_deltaFq3}
\end{equation}
which is equivalent to the prediction Eq.~(\ref{eq_intro_Fq}) made in the Introduction.
Note that this perturbation result is consistent with the renormalization group calculations 
of semidilute solutions by Sch\"afer \cite{SchaferBook,MM00}. This is of course expected
since semidilute solutions may be considered as incompressible melts of (semidilute) blobs
\cite{DegennesBook}.

\subsection{Angular correlations $\Pone(s)$}
\label{app_stat_Ps}
\begin{figure}[t]
\centerline{
\resizebox{0.9\columnwidth}{!}{\includegraphics{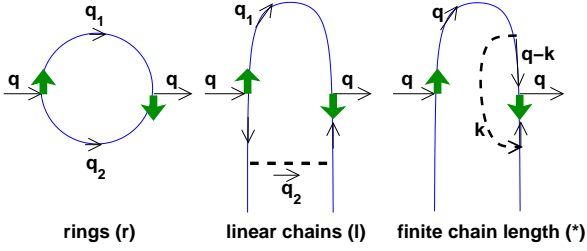}}
}
\caption{Sketch of interaction diagrams in reciprocal space used for the
computation of the bond-bond correlation function $\Pone(r)$ summing over all intrachain
bond pairs at constant $r=|\rvec|$. The injected wavevector $\qvec$
is conjugated to the distance $\rvec$.
The bold vertical arrows represent the bond vectors in reciprocal space, Eq.~(\ref{eq_theo_bondFourier}),
dashed lines the effective monomer interactions $\vpot(q)$, and the thin lines the Fourier-Laplace
transformed Gaussian propagators $\Ghat(q,t)$ with $t$ being conjugated to the curvilinear distance $s$.
Cyclic rings are described by the diagram $(r)$, the behavior of asymptotically long linear chains by the diagrams $(l)$.
The last diagram $(\star)$ describes the finite-size corrections for $\Pone(r)$ relevant for large distances
$r \gg \rstar$.
\label{fig_app_Pr}
}
\end{figure}

We present here the direct perturbation calculation of the bond-bond correlation function 
$\Pone(s)$ of Flory-distributed linear chains in reciprocal space following the discussion in 
Sec.~\ref{theo_perturb_bondbond}. As sketched in panel (b) of Fig.~\ref{fig_theo_graphD3}, one only needs
to compute the interaction diagram between the mono\-mers of the two dangling tails if the
two bonds $\lonevec$ and $\ltwovec$ are placed outside the subchain of length $s$
connecting the head of the first bond $\lonevec$ with the tail of the second bond $\ltwovec$.
To simplify the notations we set immediately $\bref = b$ and $\aref = a$ and do not distinguish 
between the bond $l$ of the computer model connecting the monomers and the effective Gaussian 
bond length $b$, i.e. we set $\cinf = (b/l)^2 = 1$.
Restating Eq.~(\ref{eq_theo_Cone2Pone}) the relevant interaction diagram $\Iouter(s)$ reads\footnote{The diagram
$\Iouter(q) = (-1) G_0(q,s) \Acal(\qvec,\qvec) \vpot(q) w(q)$ in reciprocal space is very similar to the diagram $(l)$ 
indicated in Fig.~\ref{fig_app_Pr} for the calculation of the bond-bond correlation function $\Pone(r)$ as a function 
of distance $r$. Since for $\Pone(s)$ we want
to sample at constant $s$ between the bond pairs irrespective of the distance between the bonds no momentum needs 
to be injected in the diagram and the momentum flowing along the diagram is constant everywhere.}
\begin{equation}
\Pone(s) = \Iouter(s) = (-1) \int \frac{{\rm d}\qvec}{(2\pi)^d} G_0(q,s) \Acal(\qvec,\qvec) \vpot(q) w(q). 
\label{eq_app_Ps1}
\end{equation}
The negative sign in front of the integral is due to the negative sign of the general perturbation calculation
formula Eq.~(\ref{eq_theo_Bcalav}). The first factor $G_0(q,s) = \exp(-(a q)^2 s)$ in the integral
over the wavevector $q$ stands for the Fourier transformed Gaussian propagator $G_0(r,s)$
between the two bonds indicated in Fig.~\ref{fig_theo_graphD3}(b). Using Eq.~(\ref{eq_theo_AcalFourier}) the
scalar product of the bond vectors in real space is represented by the scalar product of the
wavevectors flowing through the bonds. Note that the  momentum flows in the same direction
as the two bonds and the operator thus reads $\Acal(\qvec,\qvec) = - (b q /d)^2$.
We assume that chains are Flory-distributed and that the effective interaction $\vpot(q)$ between
two monomers in the two dangling ends is described by Eq.~(\ref{eq_theo_vpot_xi_Flory}).
The combinatorics between the interacting monomers --- corresponding to the second line in Eq.~(\ref{eq_theo_Cone_r}) ---
leads using Eq.~(\ref{eq_theo_Gqt0_Flory}) to an additional weight factor 
\begin{equation}
w(q) = \frac{1}{((a q)^2 +\mu)^2}
\label{eq_app_Ps_wq}
\end{equation}
for the two Flory-distributed dangling tails. The task is thus to compute
\begin{equation}
\Pone(s) = 
\int \frac{{\rm d}\qvec}{(2\pi)^d} e^{-(a q)^2 s} \left(\frac{b q}{d}\right)^2 
\frac{1}{\gT \rho} \frac{1}{((a q)^2 +\mu) ((a q)^2 + 2/\gT)} 
\label{eq_app_Ps2}
\end{equation}
where we have used that $\xi^2= a^2 \gT/2 = a^2/2 \veff \rho$, Eq.~(\ref{eq_theo_xidef}).
Rewriting the last factor in Eq.~(\ref{eq_app_Ps2}) as
\begin{equation}
\frac{\gT}{2} \times \frac{1}{1-\mu \ \gT/2} 
\times \left(\frac{1}{(a q)^2 +\mu} - \frac{1}{(a q)^2 + 2/\gT)} \right)
\label{eq_app_Ps3}
\end{equation}
and defining the integral
\begin{equation}
A(x) = \int \frac{\exp(-v^2 )}{v^2 + x} v^2 \frac{{\rm d}\vvec}{(2\pi)^d}
\label{eq_app_Ps4}
\end{equation}
one may rewrite the interaction integral as
\begin{equation}
\Pone(s) = \frac{1}{d \rho a^d} \frac{1}{s^{d/2}} \frac{1}{1-\mu \gT/2} \left[A(x=\mu s) - A(x=2 s/\gT) \right]
\label{eq_app_Ps5}
\end{equation}
We remind that the integral $A(x)$ takes the asymptotics 
\begin{equation}
\lim_{x\to 0} A(x) = \frac{1}{(4\pi)^{d/2}} 
\ \mbox{ and } \
\lim_{x \to \infty} A(x) \sim \lim_{x \to \infty} 1/x = 0.
\label{eq_app_Ps_limits}
\end{equation}
Considering thus the limit of infinite Flory-distributed polymers ($x= \mu s \to 0$) and
incompressible solutions ($x=2s/\gT \to \infty$), Eq.~(\ref{eq_app_Ps5}) reduces to
\begin{equation}
\Pone(s) = \frac{1}{d \rho a^d} \frac{1}{s^{d/2}} \left[ \frac{1}{(4\pi)^{d/2}} - 0\right]
= \frac{1}{2\omega} \left(\frac{\omega}{\pi} \right)^{\omega} \frac{1}{b^d\rho} \frac{1}{s^{\omega}}
\label{eq_app_Ps_infty}
\end{equation}
with $\omega \equiv d/2$ in agreement with Eq.~(\ref{eq_intro_PsD}).
Evaluating the integral $A(x)$ over the wavevector in $d=3$ dimensions one obtains for general $x$
\begin{equation}
A(x) = \frac{1}{8\pi^{3/2}} \left( 1- 2 x + 2 \sqrt{\pi} x^{3/2} e^{x} \mbox{erfc}(\sqrt{x})  \right)
\label{eq_app_Ps6}
\end{equation}
with $\mbox{erfc}(x)$ being the complementary error function \cite{abramowitz}.
Setting $\cP \equiv \sqrt{24/\pi^3}/8(b^3 \rho)$ in agreement with the power-law amplitude
indicated in Eq.~(\ref{eq_intro_keystat}) and using $u = s/\gT$ for the reduced subchain length
it follows from Eq.~(\ref{eq_app_Ps5}) that
\begin{eqnarray}
\Pone(s) & = &  \frac{\cP}{\gT^{3/2}} \frac{1}{1-\mu \ \gT/2}
\left[ \frac{4}{\sqrt{u}} (1-\mu \gT/2) \right. \nonumber \\
& - & 
\left. 
 4 \sqrt{2 \pi} e^{2u} \mbox{erfc}(\sqrt{2u})
+ 2 (\mu \gT)^{3/2} \sqrt{\pi} e^{\mu s} \mbox{erfc}(\sqrt{\mu s})
\right]. 
\label{eq_app_Ps7}
\end{eqnarray}
Since implicit to the effective interaction potential Eq.~(\ref{eq_theo_vpot_xi_Flory}) we have $\mu \gT/2 \ll 1$
and since we focus on short subchains with $\mu s = s/\Nav \ll 1$, this further simplifies to
\begin{equation}
\Pone(s) = \frac{\cP}{\gT^{3/2}} \left[
\frac{4}{\sqrt{u}} - 4 \sqrt{2 \pi} e^{2u} \mbox{erfc}(\sqrt{2u}) + 2 \sqrt{\pi} (\mu \gT)^{3/2} \right].
\label{eq_app_Ps8}
\end{equation}
For large chains,  $\mu \to 0$, the last term in the bracket decays rapidly as $1/\Nav^{3/2}$ and 
can be omitted for reasonable (mean) chain lengths. This leads to Eq.~(\ref{eq_theo_Ponesoft}) which we have
checked numerically in Fig.~\ref{fig_conf_Ps_eps} for monodisperse chains.
The first term in the bracket dominates only for small $u$ when the structure within a (very) large thermal blob is probed.
The bracket reduces to $4/\sqrt{u} - (4/\sqrt{u} - 1/u^{3/2} + \ldots) \approx 1/u^{3/2}$ in the opposite large-$u$ limit 
as may be seen by expanding the error function \cite{abramowitz}.
Eq.~(\ref{eq_app_Ps8}) thus reduces to Eq.~(\ref{eq_app_Ps_infty}) for $d=3$. 
This thus confirms the key relation Eq.~(\ref{eq_intro_keystat}) for the power-law decay of the bond-bond
correlation function for all $\gT$ on scales larger than the thermal blob ($s/\gT \gg 1$).

\subsection{Distance dependence of angular correlations}
\label{app_stat_Pr}

\subsubsection{Interaction diagrams}
\label{app_stat_Pr_diagrams}
We focus now on the low-wave\-vector regime where the melt can be considered as incompressible
(at least in terms of thermal blobs) and assume a Flory-distributed bath of a given inverse mean 
chain length $\mu$. The corresponding results for infinite chains may be obtained by setting $\mu=0$.
The bond-bond correlation function $\Pone(r) = \la \lonevec \cdot \ltwovec \ra/l^2$ is obtained
by averaging over all intrachain pairs of bonds $\lonevec$ and $\ltwovec$ at a given distance $r = |\rvec|$ 
irrespective of their curvilinear distance $s$. Setting thus $t=0$ for the Laplace variable conjugated to $s$, 
the summed up Fourier-Laplace propagator becomes according to Eq.~(\ref{eq_theo_Gqt0_Flory})
\begin{equation}
\Gtilde(k) \equiv \left. \Ghat_0(k,t) \right|_{t=0}  = \frac{1}{(a k)^2 + \mu}.
\label{eq_app_Pr_Gtilde}
\end{equation}
Using Eq.~(\ref{eq_theo_vpot_lowq}) and Eq.~(\ref{eq_theo_vpot_Flory}) the effective interaction potential
thus reads \cite{WJO10}
\begin{equation}
\vpot(k) \Gtilde(k) = \frac{\Gtilde(k)}{F_0(k) \rho} = \frac{1}{2\rho}
\label{eq_app_Pr_vpot}
\end{equation}
on large scales $k \ll 1/\sigma$ and $k \ll 1/\xi$.
The momentum $\qvec$ inserted in the interaction diagrams shown in Fig.~\ref{fig_app_Pr}
is conjugated to the distance $\rvec$ between both bonds. Momentum is a conserved quantity flowing
from one correlated point to the other. The wavevector associated with the first bond at the origin
is called $\qonevec$, the wavevector through the second bond $\qtwovec$.
As may be better seen from the sketch included in Fig.~\ref{fig_conf_Pr_EP}, the first bond is now within the $s$-subchain 
while the second bond is chosen to be outside.
Putting both bonds within the subchain leads to trivial correlations even for Gaussian chains,
putting both bonds outside the subchain leads to a vanishing correlation 
for $\xi \to 0$ as already remarked in Footnote \ref{foot_theo_Cone} for the correlation
function $C_1(r)$ \cite{ANS10a}.
The first two diagrams in Fig.~\ref{fig_app_Pr} for closed rings $(r)$ and linear chains $(l)$ 
are given by the convolution integrals
\begin{eqnarray}
\Icyc(\qvec) & = & 
   \int_{\qonevec+\qtwovec=\qvec} \Gtilde(\qonevec) \Acal(\qonevec,-\qtwovec) \Gtilde(\qtwovec) 
   \label{eq_app_Pr_Icyc} \\
\Ione(\qvec) & = &  (-1) 
   \int_{\qonevec+\qtwovec=\qvec} \Gtilde(\qonevec) \Acal(\qonevec,-\qtwovec) \Gtilde(\qtwovec) 
   \vpot(\qtwovec) \Gtilde(\qtwovec) \label{eq_app_Pr_Ione} 
\end{eqnarray}
We remind that for closed cycles the bond-bond correlation function does not vanish even
for perfect Gaussian statistics and it is the corresponding zero order average
$\la \Acal \ra \approx \la \Acal \ra_0$ which is computed by the first integral.
The minus sign in front of the second integral for linear chains stems from the minus sign implied by 
the first order perturbation, Eq.~(\ref{eq_theo_Bcalav}).
Using Eq.~(\ref{eq_theo_AcalFourier}) and assuming Eq.~(\ref{eq_app_Pr_vpot})
the integrals simplify considerably 
\begin{eqnarray}
\Ione(\qvec) & = & - \frac{1}{2 \rho} \ \Icyc(\qvec) 
   \label{eq_app_Pr_Icyc_a} \\
             & = & - \frac{(b/d)^2}{2\rho} \int_{\qonevec+\qtwovec=\qvec}  
                    \qonevec \Gtilde(\qonevec) \cdot \qtwovec \Gtilde(\qtwovec).
   \label{eq_app_Pr_Ione_a} 
\end{eqnarray}
Using the well-known theorem for the (inverse) Fourier transformation of convolutions and that
$\Fcal[\partial_r f(\rvec)] = {\rm i} \qvec f(\qvec)$, the inverse Fourier transforms are thus
\begin{eqnarray}
\Ione(r) & = & - \frac{1}{2 \rho} \ \Icyc(r)
   \label{eq_app_Pr_Icyc_b} \\
   & = & \frac{(b/d)^2}{2\rho} (\partial_r \Gtilde(r))^2
   \label{eq_app_Pr_Ione_b} 
\end{eqnarray}
with $\Gtilde(r)$ being the probability distribution to find another monomer of the chain around a reference monomer
as given by Eq.~(\ref{eq_theo_Grt0_gauss}).
Interestingly, up to a constant prefactor the integrals $\Icyc$ and $\Ione$ are thus equal on large scales ($r \gg \xi$).

\subsubsection{Normalization and bond-bond correlations}
\label{app_stat_Pr_bondbond}
Focusing from now on the 3D case the bond-bond correlation function of closed rings $\Ponecyc(r)$ 
is obtained from $\Icyc(r)$ after normalization with $\Gtilde(r)$. Using Eq.~(\ref{eq_theo_Grt0_gauss}) this yields
\begin{equation}
\Ponecyc(r) = \frac{\Icyc(r)}{\Gtilde^2(r)} 
= - \left(\frac{b}{3r}\right)^2 \ ( 1 + 2 x)^2
\label{eq_app_Pr_Pcyc_norm}
\end{equation}
with $x = \sqrt{\mu} r/ 2 a$ comparing the distance $r$ to the typical ($z$-averaged) size $\Rgyrz \approx a \sqrt{\Nav}$ 
of Flory-distributed chains.
The reason for the normalization factor $\Gtilde^2(r)$ is that
for $\Ponecyc(r)$ both bonds are {\em known} to be bonds of the same polymer ring
while the interaction integral Eq.~(\ref{eq_app_Pr_Icyc}) corresponds only to a probability
$\Gtilde(r)$ for both bonds being in the same chain {\em times}
a probability $\Gtilde(r)$ that this chain is closed.
That $\Ponecyc(r)$ is negative is of course due to the closure constraint
which corresponds to an entropic spring force bending the second bond back to the origin. 
For large chain lengths the factor $(1 + 2x)^2$ can be neglected and the
bond-bond correlation function $\Ponecyc(r)$ becomes a scale-free power law.

For linear chains it follows immediately from Eq.~(\ref{eq_app_Pr_Ione_b}) that 
\begin{equation}
\Pone(r) = \frac{\Ione(r)}{\Gtilde(r)} = \Poneasym(r) h(x) 
\ \mbox{ with } \Poneasym(r) \equiv \frac{1}{12\pi} \frac{1}{\rho r^3}
\label{eq_app_Pr_Plin_norm}
\end{equation}
being the limit for asymptotically long chains and using the scaling function
$h(x) = (1+2x)^2 \exp(-2x)$ for the finite-$\mu$ corrections.
The normalization factor $\Gtilde(r)$ is due to the fact
that for $\Pone(r)$ both bonds are known to belong to the same chain
while the interaction integral Eq.~(\ref{eq_app_Pr_Ione}) corresponds only to a probability
$\Gtilde(r)$ for both bonds to be on the same chain.
As compared to the closed cycles the correlation has the opposite sign
since the attractive spring of the ring closure
indicated by $\Gtilde(\qonevec)$ in Eq.~(\ref{eq_app_Pr_Icyc_a})
has been replaced by the effective {\em repulsion}
indicated by $-\Gtilde(\qonevec) \vpot(\qonevec) \Gtilde(\qonevec) = - \Gtilde(\qonevec)/2\rho$ in 
Eq.~(\ref{eq_app_Pr_Ione_a}).
This repulsion bends the second bond away from the origin increasing
thus the bond-bond correlation function.

\subsubsection{Sum rule and geometrical interpretation}
\label{app_stat_Pr_sumrule}
Interestingly, the perturbation result, Eq.~(\ref{eq_app_Pr_Icyc_b}), for Flory-distributed chains
may be rewritten as
\begin{equation}
\Pone(r)  + \frac{\Gtilde(r)}{2\rho} \Ponecyc(r) =  0
\label{eq_app_Pr_sumrule}
\end{equation}
where we have used the normalization factors mentioned above.
This ``sum rule" suggests a geometrical interpretation of the observed
relation between infinite linear chains and closed cycles which may
remain valid beyond the one-loop approximation used here.\footnote{A similar sum rule is obtained 
for the bond-bond correlation function $\Pone(s)$ as a function of arc-length $s$. One verifies
readily that $\Ione(s) = - \Icyc(s)/2\rho$ and thus $$\Pone(s) + \frac{\Gtilde(s)}{2\rho} \Ponecyc(s) = 0$$
with $\Ponecyc(s) \sim -1/s$ and $\Gtilde(s) \sim s^{1-d\nu}$ being the density of chains of length $N > s$
in a Flory-distributed bath which would return after $N$ steps to the origin. This is the most elegant way to
demonstrate that $\Pone(s) \sim 1/s^{d \nu}$.}
The idea is that in an hypothetical ideal melt containing both
linear chains and closed cycles all correlations disappear
(on distances much smaller than the typical chain sizes)
when summed up over the contributions of both architectures.
The weight $(\Gtilde(r)/2)/\rho$ corresponds to the fraction of bond pairs
in closed loops.\footnote{Note that $\Gtilde(r)$ is the density of the monomers of {\em both} 
strands the reference monomer is connected to. We know for linear chains as for cycles
that both bonds are connected by a first strand. The probability
for both bonds to be in a closed loop is given by the density
$\Gtilde(r)/2$ of the second strand.
The factor $1/2$ is thus needed to avoid counting the same ring twice.
} 
Since the orientational correlations in ideal cycles are necessarily
long-ranged due the ring closure (Eq.~\ref{eq_app_Pr_Pcyc_norm}), it follows,
{\em assuming} the sum rule, that the same applies to bond pairs of linear chains.
Since bonds in closed cycles are anti-correlated ($\Ponecyc(r) < 0$),
they must be aligned ($\Pone(r)>0$) for linear chains.
Interestingly, 
if one turns the argument around {\em assuming} the sum rule
(rather than deriving it as we did) this imposes Eq.~(\ref{eq_app_Pr_vpot}) and, 
hence, the effective intrachain potential $\vpot(k) = ((a k)^2 + \mu)/2\rho$
for a Flory-distributed linear chain bath.

\subsubsection{Finite chain size effects for Flory-distributed chains}
\label{app_stat_Pr_finiteN}
We emphasize that the diagram $(l)$ shown in Fig.~\ref{fig_app_Pr} is not sufficient to characterize 
$\Pone(r)$ for larger distances since the last diagram $(\star)$ corresponding to the convolution integral
\begin{equation}
\Itwo(\qvec) = (-1) 
    \int \frac{d\kvec}{(2\pi)^3} G(\qvec) \Acal(\qvec,-\kvec)) \Gtilde(\qvec-\kvec)
   \vpot(\kvec) \Gtilde(\kvec)
   \label{eq_app_Pr_Itwo}
\end{equation}
provides, as we shall see, the actual cutoff of the power law in this limit.
Using again Eq.~(\ref{eq_theo_AcalFourier}) and Eq.~(\ref{eq_app_Pr_vpot})
the integral factorizes
\begin{eqnarray}
\Itwo(\qvec)
  & = & \frac{-1}{2\rho} \int \frac{d\kvec}{(2\pi)^3} \Gtilde(\kvec)
       \times \underline{\Gtilde(\qvec) \frac{(b q)^2}{9}}
       \label{eq_app_Pr_Itwo_a}
\\
 & \equiv & - \ctwo \times \ (a \qvec)^2 \Gtilde(\qvec)
   \label{eq_app_Pr_Itwo_b}
\end{eqnarray}
where we have introduced in the last line the convenient dimensionless constant
\begin{equation}
\ctwo = \frac{(b/a)^2}{18\rho a^3} \int \frac{d\kvec}{(2\pi)^3} a^3 \Gtilde(\kvec)
\label{eq_app_Pr_ctwo_def}
\end{equation}
in which we dump local physics at large wavevector $\kvec$.
Before evaluating the angular correlations in real space it is important to clarify the physics described by the diagram.
The underlined second factor in Eq.~(\ref{eq_app_Pr_Itwo_a}) characterizes the alignment of the bond vectors of the 
monomers $n_1$ and $n_2-1$ at a fixed distance $\rvec$ of the monomers $n_1$ and $n_2=n_1+s$
as shown by the sketch included in Fig.~\ref{fig_conf_Pr_EP}. Obviously, even for Gaussian chains
these two bonds become more and more aligned if the distance $r=|\rvec|$
gets larger than $b s^{1/2}$, i.e. when the chain segment becomes stretched.
For perfectly Gaussian chains the bonds $\lonevec$ and $\ltwovec$ at $n_1$ and $n_2$ would
still remain uncorrelated, however, since the second bond is outside
the chain segment on which we have imposed the distance constraint.
As indicated by the dashed line in the diagram, it is then due to the effective
interaction between the monomers within the stretched segment ($n < n_2$)
and the monomers outside ($n > n_2$) that the bonds at $n_2-1$ and $n_2$
get aligned and then in turn the two bonds at $n_1$ and $n_2$.
We note that, strictly speaking, $\ctwo$ depends on the mean chain
length $\Nav$, since $\Gtilde(k)$ is a function of $\mu$. However, one checks readily
that this effect can be neglected for reasonable mean chain lengths.
We also note that the constant $\ctwo$ is {\em finite}, since the UV divergence
which formally arises for large $k$ (where $\ctwo \sim k$) may be regularized
by local and, hence, model dependent physics.\footnote{For soft melts with weak 
bare excluded volume $\veff=1/\gT\rho$, i.e. $\xi \gg b$, it can be shown that an upper cutoff wavevector
$\kcut \approx 1/\xi$ regularizes the integral over $k$.
The coefficient becomes $\ctwo = 1/2\pi\rho b^2\xi$.
Using $b=3.244$ and $\xi = 0.5$ for $\rho=0.5/8$
this gives $\ctwo \approx 0.48$. This is not that far off
the best fit value $\ctwo \approx 0.14$ considering that for
small $\xi \le b$ one expects $\ctwo$ to be rather determined by
the model-depending stiffness between adjacent bonds.
}
We determine $\ctwo$ numerically from our simulations of self-assembled linear EP in Sec.~\ref{conf_Pr}.

Assuming a finite and chain length independent coefficient $\ctwo$ in Eq.~(\ref{eq_app_Pr_Itwo_b})
and inserting the propagator Eq.~(\ref{eq_app_Pr_Gtilde}) we obtain by inverse Fourier transformation
\begin{equation}
\Itwo(r) = \ctwo \ (\mu \Gtilde(r) - \delta(r))
\label{eq_app_Pr_Itwo_c}
\end{equation}
for the interaction integral in real space.
Normalizing $\Itwo(r)$ as before with $G(r)$ and summing over both diagrams for linear chains this yields
\begin{equation}
\Pone(r)  =  \Poneasym(r) h(x) + \ctwo \mu
\label{eq_app_Pr_EPcorr}
\end{equation}
for $r \gg \xi > 0$.
Comparing both terms in Eq.~(\ref{eq_app_Pr_EPcorr}) one verifies that a crossover
occurs at $\rstar \approx b \Nav^{1/3}$ in agreement with Eq.~(\ref{eq_conf_rstar})
stated in the main text.
The bond-bond correlation function $\Pone(r)$ of an incompressible solution of
Flory distributed polymers becomes thus constant for $r \gg \rstar$.
This remarkable result is essentially due to the polydispersity.
This allows to find for all distances $r$ pairs of bonds $\lonevec$ and $\ltwovec$
stemming from segments which are slightly stretched by an energy of order $\mu \ll 1$ and
which are, hence, slightly shorter than a unstretched segment of length $s \approx (r/b)^2$.
Since there are more shorter chains and chain segments this just compensates the
decay of the weight due to the weak stretching.
Although the number of such slightly stretched segments decays strongly with distance,
their {\em relative} effect with respect to the typical unstretched segments,
$e^{\mu}-1 \approx \mu$, remains constant for all $r$.
It is for this reason that the chemical potential appears in the second term of Eq.~(\ref{eq_app_Pr_EPcorr}).
Please note that bond pairs from strongly stretched segments (corresponding to an energy much larger than $\mu$) 
are, however, still exponentially suppressed and can be neglected.
As explained in detail in Ref.~\cite{WJO10}, this is qualitatively different for monodisperse chains 
where strongly stretched chain segments contribute increasingly to the average for large distances.

\subsection{Non-extensivity of the chemical potential}
\label{app_stat_muEP}

We take again as a reference for the perturbation calculation a melt of Gaussian chains 
where the bond length $b$ is set by the effective bond length of asymptotically long chains.
Averages performed over this unperturbed reference system are 
labeled by an index $0$. The task is now to compute the ratio $Q(n)/Q_0(n)$
of the perturbed to the unperturbed partition function of the test chain of length $n$ plugged 
into the bath of $N$-chains,
\begin{eqnarray}
1 - \frac{Q(n)}{Q_{0}(n)}
& = & 1- \la e^{-\upot} \ra_0 \approx
\nonumber \label{eq_muEP_task1} \\
\la \upot \ra_0
& = & \sum_{s=0}^n (n-s)
\int d{\bm r} \ G_0(r,s) \vpot(r)
\label{eq_muEP_task2}
\end{eqnarray}
with the perturbation potential $\upot$ being the sum of the effective mono\-mer interactions
$\vpot(r)$ of all pairs of monomers of the test chain.
The factor $n-s$ in Eq.~(\ref{eq_muEP_task2}) counts the number of equivalent monomer pairs
separated by an arc-length $s$.
The deviation $\dmuchainn$ from Flory's hypothesis is then
given by the contribution to $\la \upot \ra_0$ which is non-linear in $n$.
The calculation of Eq.~(\ref{eq_muEP_task2}) in $d$ dimensions is most readily performed
in Fourier-Laplace space with $q$ being  the wavevector conjugated to the monomer distance $r$
and $t$ the Laplace variable conjugated to the chain length $n$.
The Laplace transformed averaged perturbation potential reads
\begin{equation}
\utau \equiv \int_{n=0}^{\infty} dn \la \upot \ra_0 e^{-n t}
= \int \frac{{\rm d}^dq}{(2\pi)^d} G_0(q,t) \vpot(q) w(t)
\label{eq_muEP_task3}
\end{equation}
where $G_0(q,t)$ represents the Fourier-Laplace transformed Gaussian pro\-pa\-gator $G_0(r,s)$
as given in Eq.~(\ref{eq_theo_Gqt_gauss}) and the weight factor $w(t) = 1/t^2$ accounts again for 
the combinatorics between the interaction monomers --- just as in Eq.~(\ref{eq_app_Ps1}). 

%
The effective interaction potential $\vpot(q)$ being given by Eq.~(\ref{eq_theo_vpot_def}) 
one realizes that the naive perturbation calculation using Eq.~(\ref{eq_muEP_task3}) is formally 
diverging at high wavevectors in three dimensions (becoming only regular below $d=2$). 
This is due to mono\-mer self-interactions which must be first subtracted.
Using Eq.~(\ref{eq_theo_vpot_q2}) instead of Eq.~(\ref{eq_theo_vpot_def}) even makes things worse
due to an additional divergency associated with the self-interactions of the blobs
whose size was set to zero ($\gT \to 0$). However, since we are not interested in
(possibly diverging) contributions linear in the length of the test chain or independent of it,
we can freely subtract linear terms (i.e., terms $\sim 1/t^2$ in Laplace space)
or constant terms (i.e., terms $\sim 1/t$) to regularize and to simplify $\utau$.
Such a transformation leads to
\begin{equation}
\utau =
\int \frac{d^dq}{(2\pi)^d} \frac{1}{t^2} \frac{2 F_0^{-1}-(a q)^2 - t}{(a q)^2+t}
\frac{v}{2(v\rho+F_0^{-1})} + \ldots
\label{eq_muEP_task4}
\end{equation}
where ``$\ldots$" stands for the linear and constant contributions we do not compute.
This converges now for incompressible melts in $d < 2$ dimensions.
%
%
Applying Eq.~(\ref{eq_muEP_task4}) to incompressible Flory-distributed melts,
i.e. assuming Eq.~(\ref{eq_theo_Fq_Flory}), this yields
\begin{eqnarray}
\utau & =& \frac{1}{2\rho} \frac{\mu-t}{t^2}
\int \frac{d^dq}{(2\pi)^d} G(q,t)
+ \ldots
\label{eq_muEP_FD1} \\
& = & \frac{1}{2\rho} (\mu/t^2 -1/t) G(r=0,t) + \ldots
\label{eq_muEP_FD1b}
\end{eqnarray}
where we have read Eq.~(\ref{eq_muEP_FD1}) as an inverse Fourier transform taken at $r=0$.
Remembering that a factor $1/t$ in $t$-space stands for an integral $\int_0^n \rm{d}s$ in $n$-space,
the inverse Laplace transform of $\utau$ can be expressed in terms of integrals of the return probability
$G(r=0,s) = (4\pi s a^2)^{-d/2}$. We obtain, hence, in $n$-space
\begin{equation}
\dmuchainn =
\frac{1}{(d-2)(4\pi)^{d/2}}\frac{1}{ \rho a^d}\left(n^{1-d/2} - \mu \frac{ n^{2-d/2}}{2-d/2}\right)
\label{eq_muEP_FD2}
\end{equation}
where $\dmuchainn$ stands for the non-extensive contribution to $\la \upot \ra_0$.
Note that the first term in the brackets scales as the correlation hole in $d$ dimension.
Its marginal dimension is $d=2$. The second term characterizes the effective two-body interaction
of the test chain with itself. As one expects \cite{DegennesBook}, its marginal dimension is $d=4$.
Although Eq.~(\ref{eq_muEP_FD2}) is formally obtained for $d<2$ it applies to higher dimensions
by analytic continuation.
In three dimensions Eq.~(\ref{eq_muEP_FD2}) becomes
\begin{equation}
\dmuchainn = \frac{1}{(4\pi)^{3/2}}\frac{1}{ \rho a^3} \left(n^{-1/2} - 2\mu n^{1/2}\right)
\label{eq_muEP_FD3}
\end{equation}
which demonstrates finally the non-extensive correction to the ideal polymer chain
chemical potential announced in Eq.~(\ref{eq_muEP_claim1}) and represented by the
solid lines in Fig.~\ref{fig_muEP_sketch}.
The chemical potential for monodisperse chains is obtained from Eq.~(\ref{eq_muEP_claim1})
within the Pad\'e approximation, Eq.~(\ref{eq_theo_Fq_Pade}), where $\mu$ is replaced by $2/N$. 
This result is indicated by the dash-dotted lines in Fig.~\ref{fig_muEP_sketch}. 
A more precise calculation
for monodisperse chains using the full Debye function is given in Ref.~\cite{WJC10}.


\bibliographystyle{spphys}       

\end{document}